\documentclass{aastex63}
\usepackage{here}
\usepackage{array}
\usepackage{bm}
\usepackage{ulem}

\received{}
\revised{}
\accepted{03-11-2020}

\shorttitle{Superflares on Solar-type Stars from All the {\it Kepler} Data}
\shortauthors{Okamoto et al.}

\begin{document}

\title{Statistical Properties of Superflares on Solar-type Stars: Results Using All of the {\it Kepler} Primary Mission Data}

\correspondingauthor{Soshi Okamoto}
\email{sokamoto@kusastro.kyoto-u.ac.jp}

\author{Soshi Okamoto}
\affil{Department of Astronomy, Kyoto University, Sakyo, Kyoto 606-8502, Japan}

\author[0000-0002-0412-0849]{Yuta Notsu}
\altaffiliation{JSPS Overseas Research Fellow}
\affil{Laboratory for Atmospheric and Space Physics, University of Colorado Boulder, 3665 Discovery Drive, Boulder, CO 80303, USA}
\affil{National Solar Observatory, 3665 Discovery Drive, Boulder, CO 80303, USA}

\author[0000-0003-0332-0811]{Hiroyuki Maehara}
\affil{Okayama Branch Office, Subaru Telescope, National Astronomical Observatory of Japan, NINS, Kamogata, Asakuchi, Okayama 719-0232, Japan}
\affil{Okayama Observatory, Kyoto University, Kamogata, Asakuchi, Okayama 719-0232, Japan}

\author[0000-0002-1297-9485]{Kosuke Namekata}
\altaffiliation{JSPS Research Fellow DC1}
\affil{Department of Astronomy, Kyoto University, Sakyo, Kyoto 606-8502, Japan}

\author{Satoshi Honda}
\affil{Nishi-Harima Astronomical Observatory, Center for Astronomy, University of Hyogo, Sayo, Hyogo 679-5313, Japan}

\author{Kai Ikuta}
\affil{Department of Astronomy, Kyoto University, Sakyo, Kyoto 606-8502, Japan}

\author{Daisaku Nogami}
\affil{Department of Astronomy, Kyoto University, Sakyo, Kyoto 606-8502, Japan}

\author{Kazunari Shibata}
\affil{Kwasan Observatory,Yamashina, Kyoto 607-8471, Japan}

\begin{abstract}
We report the latest statistical analyses of superflares on solar-type (G-type main-sequence; effective temperature is 5100 -- 6000 K) stars using all of the $Kepler$ primary mission data, and $Gaia$-DR2 (Data Release 2) catalog.
We updated the flare detection method from our previous studies by using high-pass filter to remove rotational variations caused by starspots.
We also examined the sample biases on the frequency of superflares, taking into account gyrochronology and flare detection completeness.
The sample size of solar-type stars and Sun-like stars (effective temperature is 5600 -- 6000 K and rotation period is over 20 days in solar-type stars) are $\sim$4 and $\sim$12 times, respectively, compared with Notsu et al. (2019, ApJ, 876, 58).
As a result, we found 2341 superflares on 265 solar-type stars, and 26 superflares on 15 Sun-like stars: the former increased from 527 to 2341 and the latter from 3 to 26 events compared with our previous study.
This enabled us to have a more well-established view on the statistical properties of superflares.
The observed upper limit of the flare energy decreases as the rotation period increases in solar-type stars.
The frequency of superflares decreases as the stellar rotation period increases.
The maximum energy we found on Sun-like stars is $4 \times 10^{34}$ erg.
Our analysis of Sun-like stars suggest that the Sun can cause superflares with energies of $\sim 7 \times 10^{33}$ erg ($\sim$X700-class flares) and $\sim 1 \times 10^{34}$ erg ($\sim$X1000-class flares) once every 
$\sim$3,000
years and
$\sim$6,000 
years, respectively.
\end{abstract}

\keywords{stars: flare  -- stars: activity -- stars: starspots -- stars: rotation -- stars: solar-type, solar: flare
}

\section{Introduction}\label{sec:intro}
 Flares are explosions in the stellar atmosphere with intense release of magnetic energy stored around starspots (e.g., \citealt{Shibata+2011}).
 The typical total bolometric energy of solar flares ranges from \(10^{29}\) to \(10^{32}\) erg (e.g., \citealt{Aschwanden+2000}; \citealt{Crosby+1993}; \citealt{Shibata+Yokoyama+2002}) and the duration from a few minutes to a few hours (e.g., \citealt{Veronig+2002}; \citealt{Namekata+2017}).
 Many other stars besides the Sun also show magnetic activities such as flares.
 In particular, young rapidly-rotating stars, close binary stars, and dMe stars show much higher levels of magnetic activities than the Sun (e.g., \citealt{Gershberg+2005}; \citealt{Reid+2005}; \citealt{Benz+2010}; \citealt{Kowalski+2010}; \citealt{Osten+2016}; \citealt{Linsky2019}; \citealt{Namekata+2020_PASJ}).
 They often causes ``superflares'', which have total bolometric energies $10^1$ -- $10^6$ times more energetic than the largest solar flares ($\sim10^{32}$ erg; \citealt{Emslie+2012}).
 On the other hand, since the Sun is old and rotates slowly (rotation period \(P_{\rm{rot}} \sim 25 \) days), it had been thought that the Sun and slowly-rotating Sun-like stars (\(P_{\rm{rot}} > 20\) days) would not show superflares.
 However, many solar-type (G-type main-sequence) stars including slowly-rotating Sun-like stars that showed superflares have been recently reported by using {\it Kepler} space telescope
 (\citealt{Maehara+2012}\&\citeyear{Maehara+2015}; \citealt{Shibayama+2013}; \citealt{Candelaresi+2014}; \citealt{Wu+2015}; \citealt{Balona+2015}; \citealt{Davenport+2016}; \citealt{Vandoorsselaere+2017}; \citealt{Notsu+2019}) and by {\it TESS} (\citealt{Tu+2020}; \citealt{Doyle+2020})
 \footnote{
  We note here for reference that superflares on solar-type stars have been recently reported not only with {\it Kepler}\&{\it TESS} (optical photometry in space) but also with X-ray space telescope observations (e.g., \citealt{Pye+2015}),
  UV space telescope observations (e.g., \citealt{Brasseur+2019}), and ground-based photometric observations (e.g., \citealt{Jackman+2018}),
  though currently, the number of the observed events are much smaller than the {\it Kepler}\&{\it TESS} results.
  }.

 Since solar flares sometimes lead to magnetic storms on the Earth, serious damage may be caused in our modern lives if a superflare occurs on the Sun (e.g., \citealt{Baker+2004}; \citealt{Takahashi+2016}; \citealt{Riley+2018}; \citealt{Battersby+2019}).
 Electrical currents induced by geomagnetic storms can have significant damage on the ground electrical transmission equipment, leading to a widespread blackout.
 For example, on 1989 March 13, in Quebec, Canada, about 6 million people suffered an electrical power blackout caused by a terrible geomagnetic storm \citep{Allen+1989}.
 One of the largest solar flares, named the Carrington flare \citep{Carrington+1859}, caused one of the most extreme space weather events in the last 200 years (\citealt{Tsurutani+2003}; \citealt{Hayakawa+2019}), leading to a failure of telegraph systems in Europe and North America \citep{Loomis+1861}.
 The possibility of much more extreme solar storms in a history of solar activity over $\sim$1000 year have been also recently discussed from cosmogenic isotope measurements and low-latitude auroral drawings (e.g., \citealt{Miyake+2012}, \citeyear{Miyake+2013}, \& \citeyear{Miyake+2019}; \citealt{Usoskin+2017}; \citealt{Hayakawa+2017b} \& \citeyear{Hayakawa+2017a}).
 Thus, it is important to study the superflares on solar-type stars, especially Sun-like stars (main-sequence stars with surface effective temperature $T_{\rm{eff}}$ of 5600 -- 6000 K and rotation periods $P_{\rm{rot}}$ of more than 20 days), and investigate whether our Sun can really generate superflares or not (\citealt{Shibata+2013}; \citealt{Aulanier+2013}; \citealt{Schmieder+2018}; \citealt{Battersby+2019}).
 Superflares can also affect climate and habitability of various planets (e.g., \citealt{Airapetian+2016}\&\citeyear{Airapetian+2020}; \citealt{Kay+2019}; \citealt{Yamashiki+2019}), and it is important to know how superflare activities of the Sun and stars change over the stellar age.

 A lot of statistical properties of superflares on solar-type stars have been recently revealed through the analyses of the high-precision photometric data of $Kepler$ and the follow-up spectroscopic observations using Subaru \& APO 3.5m telescopes.
 The details of these latest results, which are summarized in the following, are described in our study \citet{Notsu+2019}.

  \begin{enumerate}
    \renewcommand{\labelenumi}{(\roman{enumi})}
     \item
    High-dispersion spectroscopic observations of solar-type superflare stars show that more than half of their observed superflare stars
    are single solar-type stars (\citealt{Notsu+2015a}\&\citeyear{Notsu+2019}).
    This suggests that stars having spectroscopic properties similar to the Sun can generate superflares (see also \citealt{Nogami+2014}).

    \item
    The occurrence frequency of superflares as a function of the flare energy on solar-type stars shows a power-law distribution $dN/dE \propto E^{-1.5 \sim -1.9}$, and this is consistent with that of solar flares (\citealt{Maehara+2012}\&\citeyear{Maehara+2015}; \citealt{Shibayama+2013}).
    This power-law distribution is also confirmed by \citet{Tu+2020} using  $TESS$ data.

    \item
    The superflare stars show quasi-periodic brightness variations with the typical amplitude of $0.1 \% \sim 10 \%$.
    These variations are assumed to be explained by the rotation of the stars with large starspots (\citealt{Notsu+2013}; \citealt{Maehara+2017}; \citealt{Reinhold+2020}).
    These assumptions are confirmed by spectroscopic observations (\citealt{Notsu+2015b}\&\citeyear{Notsu+2019}; \citealt{Karoff+2016}).
    The occurrence frequency of starspots size derived from these variations are consistent with those of sunspots, indicating the common generation process of spots (\citealt{Maehara+2017}; \citealt{Notsu+2019}).
    Also, the temporal evolution of these large starspots are found to be explained by the same process as sunspots (\citealt{Namekata+2019} \& \citeyear{Namekata+2020_ApJ}).

    \item
    Superflare energies are consistent with the stored magnetic energy estimated from the starspots coverage, indicating the superflares release the magnetic energy around the large starspots (\citealt{Notsu+2013}\&\citeyear{Notsu+2019}).
    Also, the timescale of the superflares are explained by the magnetically-driven energy-release mechanism (i.e., magnetic reconnection; \citealt{Maehara+2015}; \citealt{Namekata+2017})

    \item
    Superflares with up to $5 \times 10^{34}$ erg can occur on slowly-rotating Sun-like stars ($T_{\rm eff}=$5600 -- 6000 K, $P_{\rm rot}\sim$25 days, and Age$\sim$4.6 Gyr) once every few thousands of years. In contrast, young rapidly-rotating stars with $P_{\rm rot}\sim$ a few days have superflares up to $10^{36}$ erg, and the flare frequency of such rapidly rotating stars is $\gtrsim$100 times higher compared with the above old, slowly rotating Sun-like stars  (\citealt{Notsu+2019}).

  \end{enumerate}

  These results have clarified a lot of properties of superflares on Sun-like stars, and we have shown some insights into the important question, “Can our Sun have superflares?”.
  However, there remains a large problem that the number of slowly-rotating Sun-like superflare stars ($T_{\rm eff}$=5600 -- 6000 K and $P_{\rm rot}\sim$25 days) in \citet{Notsu+2019} are very small because of the following reasons.
  \citet{Notsu+2019} only used the data originally found as solar-type superflare stars from {\it Kepler} 30-minute (long) time-cadence data of the first $\sim$500 days in \citet{Shibayama+2013}.
  {\it Kepler} prime mission continued for 4 years, so remains the data of $\sim$1000 days of data, which were not used in the previous studies.
  Moreover, \citet{Notsu+2019} only investigated superflare stars that are identified as solar-type (G-type main-sequence) stars both in the original $Kepler$ input catalog \citep{Brown+2011} and in the updated {\it Gaia}-DR2 (Data Release 2) catalog \citep{Berger+2018}, since they applied $Gaia$-DR2 stellar radius updates to superflare stars originally reported in \citet{Shibayama+2013} in order to remove contamination of sub-giant stars.
  Thus, a certain number of stars newly identified as solar-type stars in the updated $Gaia$-DR2 catalog \citep{Berger+2018} are not included for statistical discussions of superflare stars in \citet{Notsu+2019}.
  In summary, the statistical survey of superflares with $Kepler$ 4-year primary data was not completed, and we needed to conduct further statistical analysis of solar-type stars using the full dataset of $Kepler$ primary mission, in order to increase the number of Sun-like superflares stars and reveal their statistical properties.

  In this study, we searched for superflares using all of the 30-minute (long) time cadence data of $Kepler$ primary mission covering $\sim$1500 days, including the targets newly identified as solar-type stars in the $Gaia$-DR2 catalog \citep{Berger+2018}.
  Through this, we aim to create the complete catalog of superflares on solar-type stars using $Kepler$ primary mission data and aim to show more well-established results on the properties of superflare stars.
  In Section \ref{sec:method}, we describe the method of analysis, and in Section \ref{subsec:example} we show the detected superflares on solar-type stars (in Appendix \ref{app:subgiant} we also briefly show and discuss superflares on subgiants).
  The flare detection method is based on the ones in \citet{Shibayama+2013} and \citet{Maehara+2015}, but not completely the same as them.
  We use high-pass Butterworth filter to remove the stellar rotation signals so that we can detect smaller flares in rapidly-rotating stars. Processes of removing rotation signals for flare detection are done in many studies (e.g., \citealt{Davenport+2014}).
  Using these detected superflares, we discuss the latest view on the statistical properties of superflares on solar-type stars in Sections \ref{subsec:energy_rotation} -- \ref{subsec:exoplanets}.
  Different from the previous studies, we also include the effect of gyrochronology and sensitivity of flare detection, when discussing flare frequency distributions in Sections \ref{subsec:frequency} \&  \ref{subsec:superflare_sunlike_sun}.

\section{Method}\label{sec:method}
 \subsection{Data and Analysis}\label{subsec:analysis}
  We searched for superflares using the $Kepler$ 30 minutes (long) time cadence data \citep{Koch+2010} that were taken from 2009 May to 2013 May (quarters 0 -- 17).
  We retrieved the data of the $Kepler$ Date Release 25 \citep{Thompson+2016} from the Multimission Archive at the Space Telescope (MAST).
  We selected solar-type (G-type main-sequence stars) stars, based on the evolutionary state classification and effective temperature ($T_{\rm eff}$) values listed in \citet{Berger+2018}.
  \citet{Berger+2018} is the catalog of all the 177911 {\it Kepler} stars combining new stellar radius estimates ($R_{\rm Gaia}$) from Gaia Data Release 2 ($Gaia$-DR2) parallaxes with the stellar parameters from the DR25 Kepler Stellar Properties Catalog (DR25-KSPC; \citealt{Mathur+2017}), which incorporates the revised method of $T_{\rm eff}$ estimation \citep{Pinsonneault+2012}.
  They report the evolutionary state classifications (Main Sequence (MS)/ Subgiants / Red giants/ Cool main-sequence binaries), on the basis of these $R_{\rm Gaia}$ and $T_{\rm eff}$ values (cf. Table 5 of \citealt{Berger+2018}).
  Using the values in \citet{Berger+2018}, we identified the main sequence (MS) stars with $T_{\rm eff}$= 5100 -- 6000 K as solar-type stars.
  In total, 49305 solar-type stars are selected (see (2) in Table \ref{table:num_flares}).
  Among these stars, 11601 stars with brightness variations amplitude ($Amp$) and rotation period ($P_{\rm{rot}}$) values reported in \citet{McQuillan+2014} (see (3) in Table \ref{table:num_flares}) are finally used for the flare search process.

  There are various methods detecting flares from light curves (e.g., \citealt{Shibayama+2013}; \citealt{Davenport+2014}; \citealt{Gunther+2020}).
  In this study, we follow the flare detection method that is based on our previous studies (\citealt{Maehara+2012}\&\citeyear{Maehara+2015}; \citealt{Shibayama+2013}), but include some updates by referring to the methods in several studies (e.g., \citealt{Gao+2016}; \citealt{Yang+2017}).
  The advantage of the following method we use is that we do not need to assume any models of flare lightcurve shapes to detect flares.
  Because of this, it is expected we can detect most of the flares even if they have complicated shapes (e.g., a flare with multiple peaks) in the light curve.

  Stellar rotation, which causes periodic variation to lightcurves, can prevent us from detecting smaller flares adequately (cf. \citealt{Yang+2017}; \citealt{Davenport+2020}; \citealt{Feinstein+2020}).
  In particular, this effect becomes larger for detecting flares on rapidly-rotating stars ($P_{\rm rot} \sim 1$ day), since the timescales of flares approximately equals the rotation periods.
  In order to reduce this rotation effect, we use high-pass Butterworth filter.
  The pass-band edge frequency ($W_{\rm p}$) is $2 \times 1/P_{\rm rot}$, and the stop-band edge frequency ($W_{\rm s}$) is $3 \times 1/P_{\rm rot}$.
  If $P_{\rm rot}$ is smaller than 0.2 days, we set $W_{\rm p} = 2 \times 1/0.2$ day$^{-1}$, and $W_{\rm s} = 3 \times 1/0.2$ day$^{-1}$, because the high-pass filter does not work.
  After the lightcurves are corrected with the high-pass filter method, we calculated the distributions of brightness variations between all pairs of 2 consecutive data points, as done in our previous studies (e.g., \citealt{Shibayama+2013}).
  We use this distribution for the flare detection.
  The threshold of the flare detection was determined to be
  three times the value at the value of the top 1\% of this distribution ($1\% \times 3$).
  This threshold was chosen as a result of test runs so as not to falsely detect other brightness variations.
  We also use this to define the end time of the flare.

\begin{figure*}[ht!]
    \gridline{
    \fig{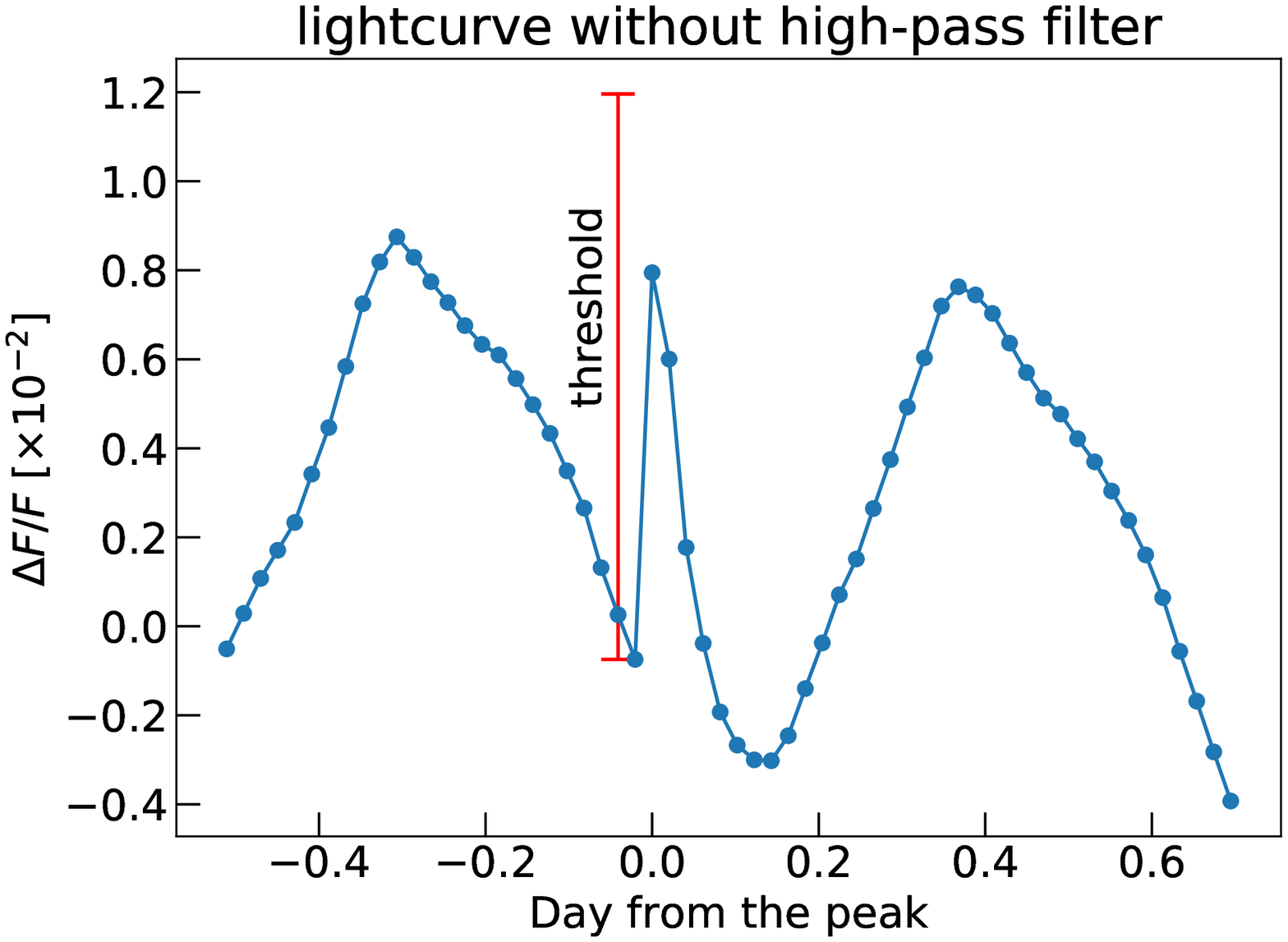}{0.49\textwidth}{\vspace{0mm}(a)}
    \fig{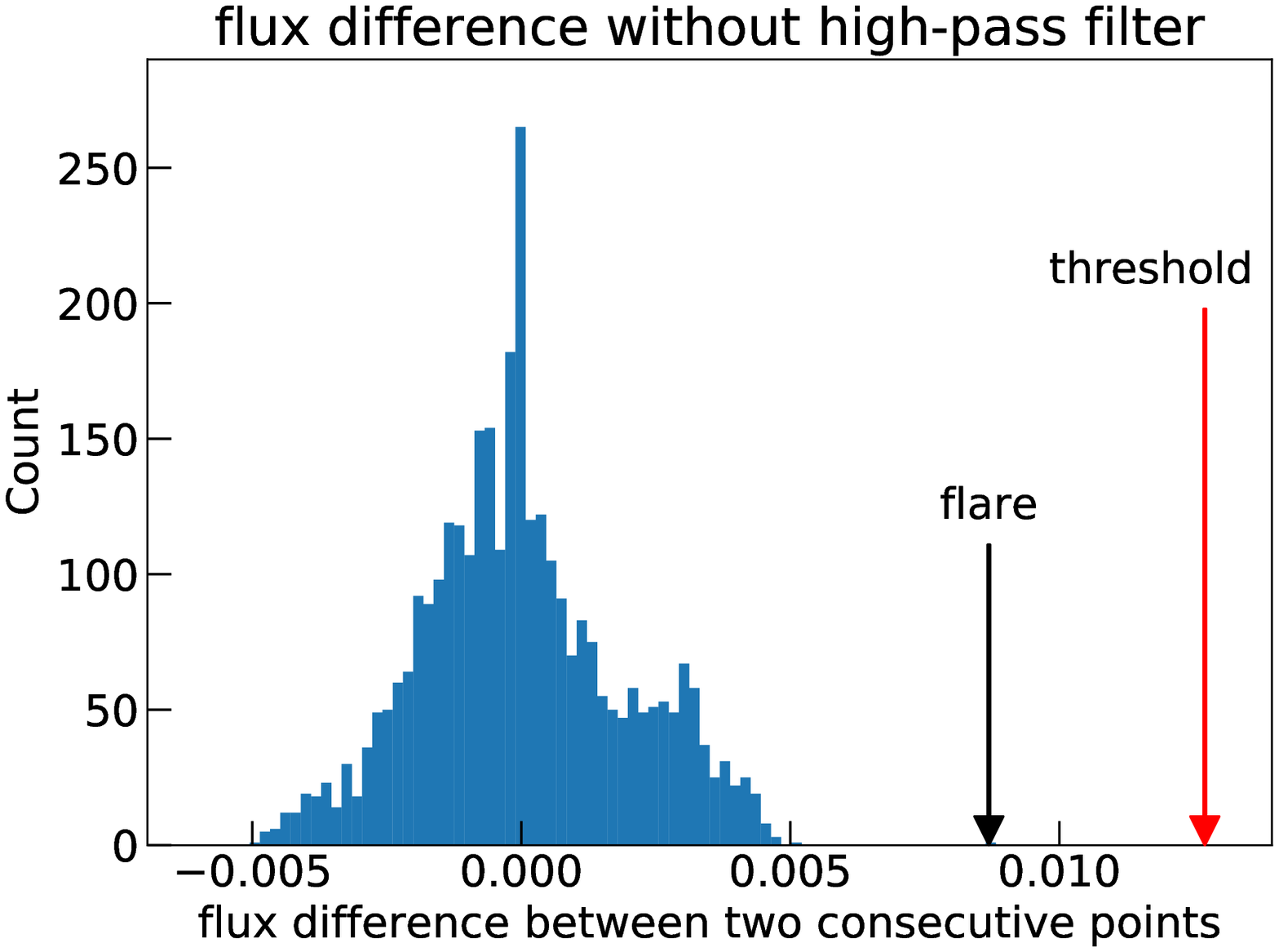}{0.49\textwidth}{\vspace{0mm}(b)}
    }
    \gridline{
    \fig{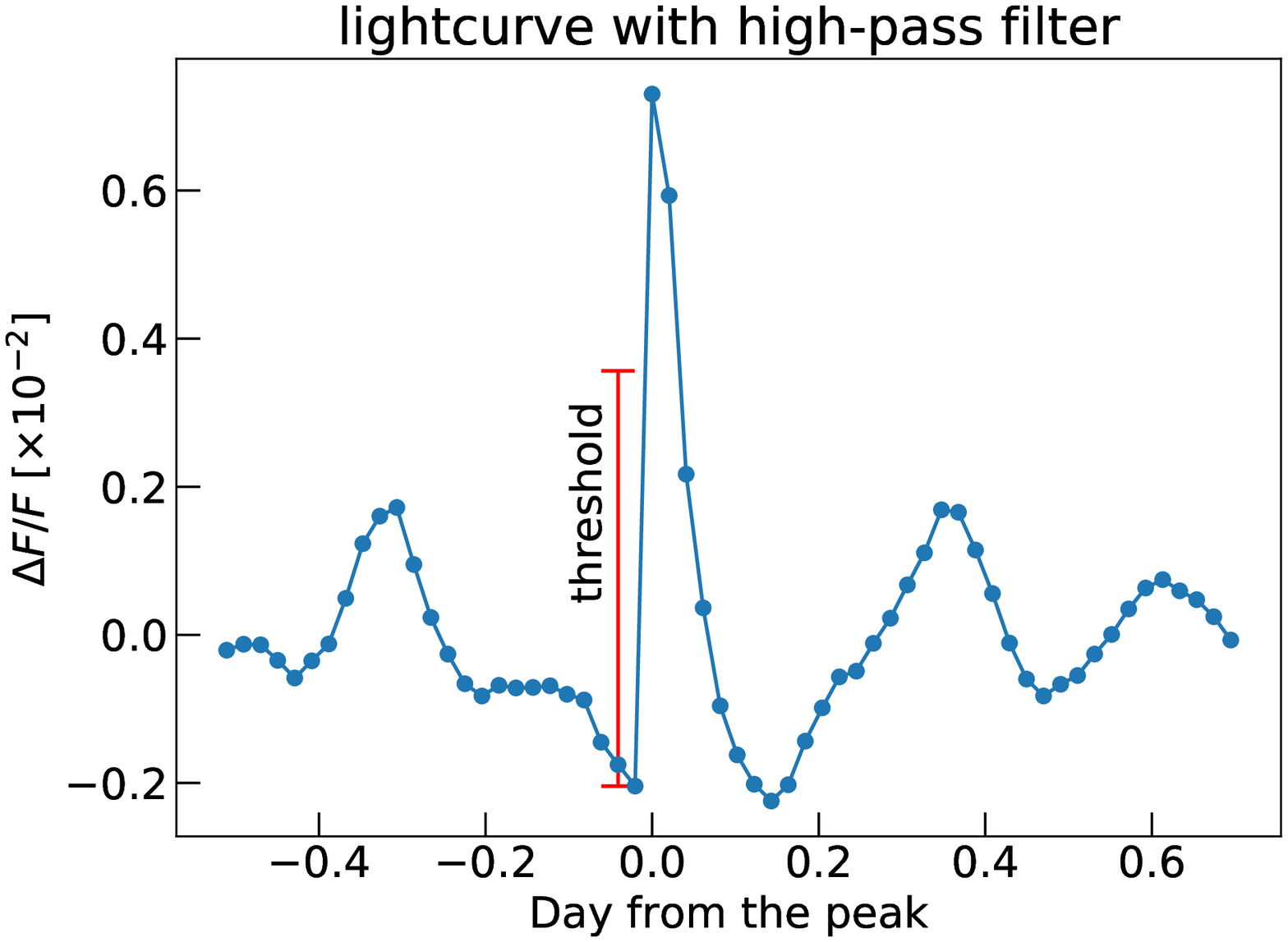}{0.49\textwidth}{\vspace{0mm}(c)}
    \fig{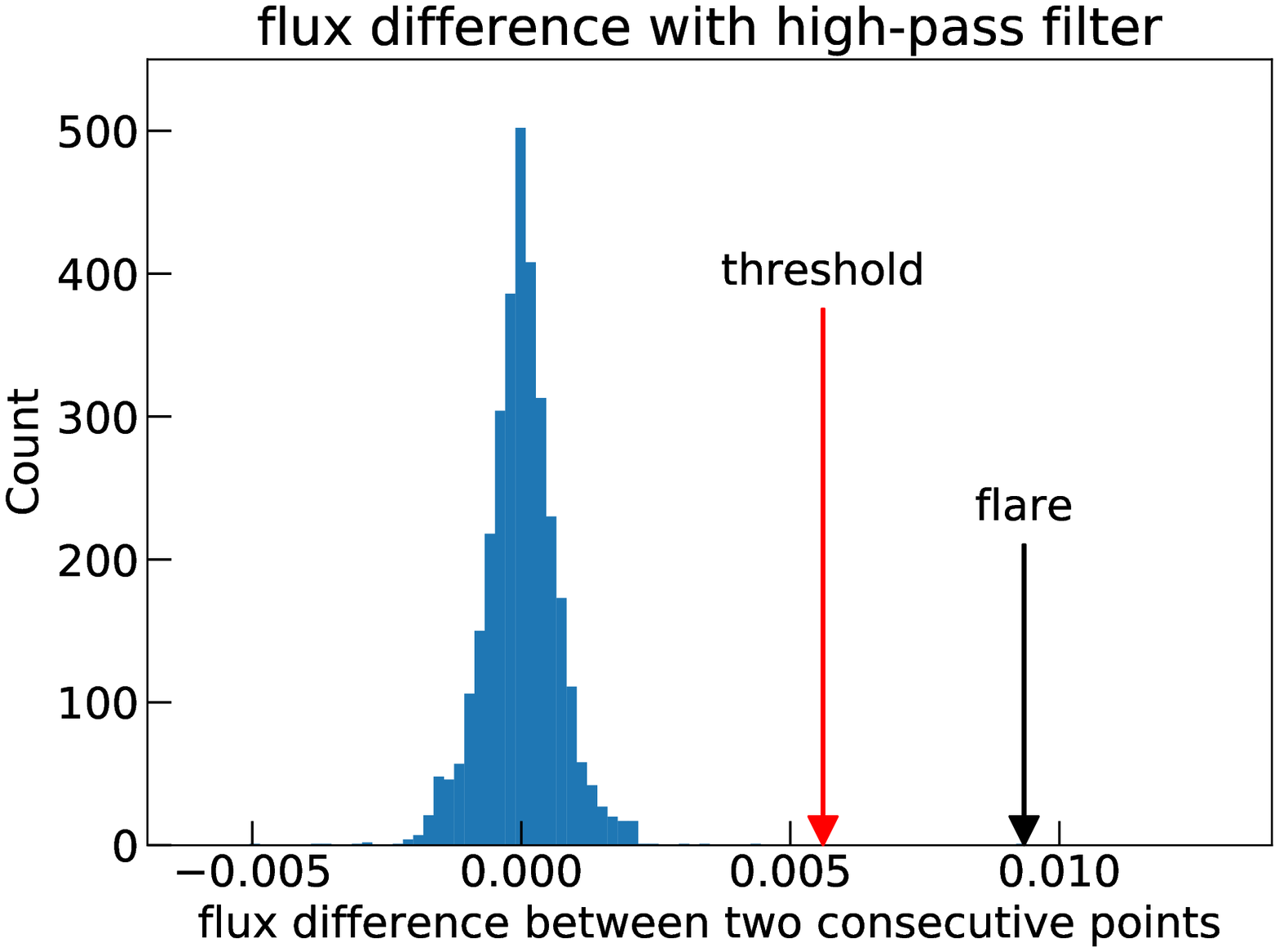}{0.49\textwidth}{\vspace{0mm}(d)}
    }
    \caption{
    Explanatory figures of our flare-detection method using the lightcurve data of KIC 003836772.
    (a) The original lightcurve without high-pass filter.
    (b) The distribution of brightness between all pairs of data points in (a).
    (c) The filtered lightcurve with high-pass filter.
    (d) The same as (b) but the data for (c).
    The threshold of the flare detection is
    three times the value at the value of the top 1\% of all pairs distribution ($1\% \times 3$).
    Red arrows (written as ``threshold") in right panels ((b)\&(d)) show the above threshold
    ($1\% \times 3$)
    in the distribution, and they are also shown as red bars in the left panels ((a)\&(c)).
    Black arrows (written as ``flare")  in right panels ((b)\&(d)) indicate the maximum flux difference between two consecutive points during a flare.
     \label{fig:method_filter}}
  \end{figure*}

  Figure \ref{fig:method_filter} shows an example of our flare detection method using the lightcurve data of KIC 003836772.
  The rotation period of this star is $\sim$0.6 days.
  Figure \ref{fig:method_filter} (a) is an original light curve without the high-pass filter and (b) is the distribution of brightness variations between consecutive data points.
  Figure \ref{fig:method_filter} (c) and (d) are the same as (a) and (b) but with the high-pass filter.
  Red arrows (written as ``threshold") in the right panels ((b)\&(d)) show the threshold (
  $1\% \times 3$
  ) in the distribution, and they are also shown as red bars in the left panels ((a)\&(c)).
  Black arrows (written as ``flare") in the right panels ((b)\&(d)) indicate the maximum flux difference between two consecutive points during a flare shown in the left panels ((a)\&(c)).
  Although the flare amplitudes in (a)\&(c) are almost equal, the ``threshold" value in (c) is smaller than that in (a), thanks to the high-pass filter.
  Without the high-pass filter, this flare is not detected as a flare since the ``threshold" value (the red arrow in (b)) is smaller than maximum flux difference during a flare (the black arrow in (a)).
  As shown in this example, this high-pass filter method can help us not to overlook or falsely detect flares.
  However, since the rotational modulations are not completely removed, the detection threshold of flares can be still affected by the rotational modulations.
  In particular, rapidly-rotating stars tend to have larger effects, and we discuss this point again in Section \ref{subsec:sensitivity}.

  The flare start time is defined as the time 30-min before the flux first exceeds the detection threshold, considering the time cadence of the {\it Kepler} data used here is 30-min.
  We removed long-term brightness variations around the flare by fitting with the spline curves of three points in order to determine the flare end time.
  Three points are determined by the data points just before the beginning of the flare and 5 hr and 8 hr after the peak of the flare.
  The flare end time was determined by the time the flux residuals become smaller than the standard deviation of the distribution ($\sigma_{\rm diff}$).
  Two consecutive data points must exceed the threshold so that the event is defined as a flare.
  We also excluded flare candidates with a shorter decline phase than the increase phase.
  This is because most lightcurves of general stellar flares in optical show the impulsive brightness increase phase and subsequent more gradual decrease phase (e.g.,``classical flares" shown in Figure 4 of \citealt{Davenport+2014}).
  It has been reported that short-duration near-ultraviolet (NUV) flares show different shape of lightcurves (e.g., flares whose brightness increase time is comparable to or shorter than the decay time), but the total duration is only less than few minutes \citep{Brasseur+2019}.
  We can assume it is almost impossible to detect these short-duration flares with $Kepler$ 30-min time cadence data used in this study, though we need more studies on the detailed physical properties on these short-duration flares.
  Because of these things, we exclude flare candidates with the shorter decline phase than the increase phase, considering these events are caused by instrumental effects or unknown astrophysical events different from flares on solar-type stars.

  The pixel scale of the $Kepler$ CCDs is around 4 arcsec and the typical photometric aperture for a 12 mag star is about 30 pixel \citet{vanCleve+2016}.
  This brightness variations of neighboring stars can affect the flux of the target star, and we removed this effect as done in \citet{Shibayama+2013} and \citet{Maehara+2015}.
  If a flare star is near the target and large amplitude flares occur, $Kepler$ would detect fake flares on the target star.
  To eliminate such events, we chose stars without neighbor stars within 12 arcsec for the analysis, using the $Kepler$ Input Catalog.
  We also excluded pairs of flares that occurred at the same time and the distance between the flares is less than 24 arcsec.
  By this selection, $\sim$64 \% of the detected superflares were excluded.
 We note here for caution that it is still possible that there remain some faint stars ($>19-20$ mag: fainter than typical limiting magnitude of Kepler Input Catalog) even after this selection process.
  After eliminating the candidates of flares on neighboring stars, we checked the light curves of the flare candidates by eye.
  Finally, we checked the flux-weighted centroid of flares in order to investigate whether the flux-weighted centroid in the quiescent phase and the flare phase are the same.
  If the difference of the flux-weighted centroid at the time of the flare is 3 times larger than the standard deviation of the difference of 10 moving averages, we judge that the flare does not occur on the target star.
  We compared the periodic flux-weighted centroid moving with periodic lightcurve amplitudes of the targets in order to eliminate binaries.
  If the correlation coefficient of consecutive data points of brightness variations and those of the flux-weighted centroid is more than 0.5, we eliminated the targets from our flare detection.
\newpage
 \subsection{Flare Energy Estimation}\label{subsec:flare-energy-estimation}
  We use the methods described in \citet{Shibayama+2013} to estimate the total energy of each flare.
  We assumed that the spectrum of white-light flares can be described by a blackbody radiation with an effective temperature ($T_{\rm flare}$) of 10,000 K (\citealt{Mochnacki+1980}; \citealt{Hawley+1992}).
  Assuming that the star is a blackbody radiator, the relationship between the bolometric flare luminosity ($L_{\rm flare}$), the flare effective temperature ($T_{\rm flare}$), and the area of flare ($A_{\rm flare}$) is written as the following equation with the Stefan-Boltzmann constant ($\sigma_{\rm SB}$).

  \begin{equation}
    L_{\rm flare} = \sigma_{\rm SB} T^4_{\rm flare} A_{\rm flare}
    \label{eq:L_flare}
  \end{equation}

  We estimate $A_{\rm flare}$ with the observed luminosity of the star ($L_{\rm star}^{\prime}$) and flare ($L_{\rm flare}^{\prime}$), using Equations (2) -- (5) of \citet{Shibayama+2013}.
  The total bolometric energy of superflare ($E_{\rm flare}$) is written as an integral of $L_{\rm flare}(t)$ during the flare duration,

  \begin{equation}
    E_{\rm flare} = \int_{\rm flare} L_{\rm flare}(t) dt
  \end{equation}

  This estimate of $E_{\rm flare}$ is affected by several types of uncertainties.
  Errors of stellar effective temperature ($T_{\rm eff}$) and stellar radius ($R$) affect $E_{\rm flare}$ values, and these are typically $\sim$3 \% and $\sim$7 \%, respectively \citep{Berger+2018}.
  The determination errors of flare start/end points and quiescent levels have also effects on flare amplitude values and consequently on $E_{\rm flare}$ values (typically $\sim 30$ \%).
  If the black-body temperature of a flare emission changes to 6000 -- 7000 K as in solar white-light flares, the flare energy can change by a factor of 0.5 \citep{Namekata+2017}.
  Moreover, since both stellar quiescent radiation and flare emission may not be complete blackbody radiation, $E_{\rm flare}$ values may have an error of a few tens of percent.
  In total,  $E_{\rm flare}$ values can have error values of a few tens of percent.
  We must note these potential error values, when we discuss the exact $E_{\rm flare}$ values reported in the following sections.
  However, the overall statistical discussions on the relations between superflares and stellar properties (e.g., rotation period, starspot coverage) do not change since the errors are smaller than one order of magnitude.

\clearpage
\section{Results and Discussion}\label{sec:result+discussion}
 \subsection{Detected Superflares}\label{subsec:example}
  We detected 2344 superflares on 266 solar-type stars from $Kepler$ 30-min (long) time cadence data of $\sim$4 years (quarters 0-17).
  Among them, we detected 29 superflares on 16 Sun-like stars ($T_{\rm eff}$ = 5600 -- 6000 K and $P_{\rm rot}>$ 20 days).
  The number of superflares ($N_{\rm flare}$), superflare stars ($N_{\rm flarestar}$), and all stars analyzed ($N_{\rm star}$) are listed in Table \ref{table:num_flares}.
  The sample size (the number of stars $\times$ observation period) of solar-type stars in this study is about 4 times larger than that of \citet{Notsu+2019}, and the sample size on Sun-like stars is about 12 times larger.
  The number of superflares on Sun-like stars detected in this study is $>$12 times larger than \citet{Notsu+2019}, since we improved the methods to detect superflares (Section \ref{subsec:analysis}).
  We note again that our previous study \citet{Notsu+2019} did not include stars newly identified as solar-type stars in the updated $Gaia$-DR2 catalog \citep{Berger+2018}.
  \citet{Notsu+2019} applied $Gaia$-DR2 stellar radius values only to the stars
  that had been identified as solar-type stars in \citet{Shibayama+2013},
  which used the original $Kepler$ input catalog \citet{Brown+2011} for selecting solar-type stars.
  The effective temperature in the original $Kepler$ input catalog \citep{Brown+2011} is about 250 K lower than that in $Gaia$-DR2 catalog  \citep{Berger+2018} for most of solar-type stars.
  This is the main reason why the sample increase of Sun-like stars ($T_{\rm eff}=5600$ -- $6000$ K and $P_{\rm rot} > 20$ day) is 12 times, while that of the solar-type stars ($T_{\rm eff}=5100$ -- $6000$ K) is only 4 times.

  Stellar and flare parameters of all the Sun-like superflare stars detected in this study are listed in Table \ref{table:params_sunlike}.
  The data of all 2344 superflares on 266 solar-type superflare stars detected in this study are available in the online-only table.
  Some example lightcurves of superflares on the Sun-like stars are shown in Figure \ref{fig:eg_sunlikeflare}, and all superflares on Sun-like stars are shown in Appendix \ref{app:allflare_sunlike}.
  In addition to superflares on solar-type (G-type main-sequence) stars, we also detected superflares on subgiants, and these are briefly shown and discussed in Appendix \ref{app:subgiant}.

 \begin{deluxetable*}{lcccc}
   \tablecaption{The number of superflares ($N_{\rm flare}$), superflare stars ($N_{\rm flarestar}$), and all stars we analyzed ($N_{\rm star}$).}
   \tablewidth{0pt}
   \tablehead{
     \colhead{} & \colhead{$N_{\rm flare}$} & \colhead{$N_{\rm flarestar}$} & \colhead{$N_{\rm star}$}
   }
   \startdata
   (1) All $Kepler$ stars having $T_{\rm eff}$ and $R_{\rm Gaia}$ values in \citet{Berger+2018} & & & 177911 \\
   (2) All solar-type stars (main-sequence (MS) stars with $T_{\rm eff}$ = 5100 -- 6000 K) & & & 49305 \\
   (3) Solar-type stars that have $P_{\rm rot}$ and $Amp$ values reported in \citet{McQuillan+2014} & 2341 (2344) & 265 (266) & 11601 \\
   (4) Stars with $T_{\rm eff} = 5100$ -- $5600$ K among (3) & 1412 & 148 & 6527 \\
   (5) Stars with $T_{\rm eff} = 5600$ -- $6000$ K among (3) & 929 (932) & 117 (118) & 5074 \\
   (6) Sun-like stars ($T_{\rm eff} = 5600$ -- $6000$ K and $P_{\rm rot} > 20$ days) among (5) & 26 (29) & 15 (16) & 1641 \\
   \enddata
   \tablecomments{
     $T_{\rm eff}$ values and evolutionary state classifications (Main Sequence (MS)/Subgiants/Red giants/Cool MS binaries) in \citet{Berger+2018} (cf. Figure 5 of \citealt{Berger+2018}) are used for the classification in this table. $P_{\rm rot}$ values in \citet{McQuillan+2014} are used for the classification of Sun-like stars in (6).
     The number in the parentheses include one superflare star KIC 007772296, which is suspected as a binary candidate in Table \ref{table:params_sunlike} and is not used in the main statistical discussions of this paper (after Section \ref{subsec:energy_rotation})
   }
   \label{table:num_flares}
 \end{deluxetable*}

  It is generally difficult to accurately detect rotation period values of stars as slow as the Sun ($P_{\rm rot}>$ 20 days), so we double-checked the rotation periods of all the Sun-like superflare stars (detailed in Appendix \ref{app:allflare_sunlike}).
  As a result, some stars may have rotation period values smaller than the values reported in \citet{McQuillan+2014}.
  These five stars are flagged as ``1" in Table \ref{table:params_sunlike}, while they are used in the statistical discussions in the following part of this paper, because we cannot
  finally judge only from the periodogram analyses in Appendix \ref{app:allflare_sunlike}.
  KIC 007772296, flagged as ``2' in Table \ref{table:params_sunlike} is suspected as a binary candidate.
  The lightcurve and periodogram analyses suggest that this star also shows brightness variations with the period of $\sim$1 day and this short period might be related to the orbital motion of a binary system (for details, see Appendix \ref{app:allflare_sunlike}).
  Because of this, KIC 007772296 is not used in the statistical discussions in the following part of this paper.
  We also note again that it is still possible that there remain some faint stars ($>19-20$ mag) in Kepler photometric aperture (12 arcsec).

  As a result of this paper, we detected 2341 superflares on 265 solar-type stars, and among them, we detected 26 superflares on 15 Sun-like stars ($T_{\rm eff}$ = 5600 -- 6000 K and $P_{\rm rot}>$20 days).
  These superflare stars are used in the statistical discussions in the following part of this paper.

 \begin{longrotatetable}
 \begin{deluxetable*}{llccccccccccc}
   \tablecaption{Parameters of Sun-like superflare stars}
   \tabletypesize{\footnotesize}
   \tablehead{
    \colhead{KIC} & \colhead{$T_{\rm{eff}}$\tablenotemark{\dag}} &
    \colhead{$R$\tablenotemark{\dag}} & \colhead{\(P_{\rm{rot}}\)\tablenotemark{\ddag}} &
  \colhead{$Amp$\tablenotemark{\ddag}} & \colhead{$m_{\rm Kepler}$\tablenotemark{\#}} & \colhead{$N_{\rm flare}$\tablenotemark{\P}} & \colhead{$E_{\rm max}$\tablenotemark{\P}} & \colhead{$A_{\rm spot}$\tablenotemark{*}} & \colhead{Flag\tablenotemark{\%}} \\
    \colhead{} & \colhead{(K)} & \colhead{($R_{\sun}$)} & \colhead{(day)} & \colhead{(ppm)} & \colhead{(mag)} & \colhead{} & \colhead{(erg)} & \colhead{($1/2 \times A_{\sun}$)} & \colhead{}
   }
   \startdata
   005648294       & $5653 \pm 198$ & $1.140^{+0.091}_{-0.082}$ & $22.336 \pm 0.085$ & 2543.81 & 14.757 & 1 & $2.4 \times 10^{34} $ & $4.3 \times 10^{-3}$ & 1 \\
   005695372       & $5791 \pm 203$ & $0.886^{+0.066}_{-0.060}$ & $21.195 \pm 0.130$ & 8881.39 & 14.054 & 3 & $7.7 \times 10^{33} $ & $8.9 \times 10^{-3}$ & - \\
   006347656$^{+}$ & $5623 \pm 197$ & $0.823^{+0.062}_{-0.057}$ & $28.441 \pm 0.598$ & 4110.04 & 14.860 & 4 & $2.2 \times 10^{34} $ & $3.6 \times 10^{-3}$ & 1 \\
   006932164       & $5731 \pm 201$ & $1.038^{+0.104}_{-0.093}$ & $22.631 \pm 0.077$ & 8538.25 & 15.883 & 1 & $4.0 \times 10^{34} $ & $1.3 \times 10^{-2}$ & - \\
   007435701       & $5812 \pm 203$ & $0.809^{+0.063}_{-0.057}$ & $20.648 \pm 0.053$ & 8039.18 & 15.090 & 1 & $9.1 \times 10^{33} $ & $6.7 \times 10^{-3}$ & - \\
   007772296       & $5616 \pm 197$ & $0.897^{+0.070}_{-0.063}$ & $23.257 \pm 0.083$ & 4136.53 & 14.880 & 3 & $1.1 \times 10^{34} $ & $4.3 \times 10^{-3}$ & 2 \\
   007886115       & $5729 \pm 201$ & $0.873^{+0.066}_{-0.060}$ & $20.351 \pm 0.110$ & 5280.17 & 14.960 & 1 & $9.0 \times 10^{33} $ & $5.2 \times 10^{-3}$ & - \\
   008090349       & $5760 \pm 202$ & $1.169^{+0.092}_{-0.083}$ & $24.898 \pm 3.822$ & 2482.89 & 14.675 & 1 & $1.8 \times 10^{34} $ & $4.4 \times 10^{-3}$ & - \\
   008416788       & $5889 \pm 206$ & $0.908^{+0.068}_{-0.061}$ & $22.519 \pm 0.042$ & 3300.19 & 13.582 & 5 & $1.7 \times 10^{34} $ & $3.4 \times 10^{-3}$ & - \\
   009520338       & $5947 \pm 208$ & $0.805^{+0.064}_{-0.058}$ & $23.857 \pm 0.093$ & 4138.47 & 15.270 & 1 & $9.9 \times 10^{33} $ & $3.4 \times 10^{-3}$ & 1 \\
   010011070$^{+}$ & $5669 \pm 198$ & $0.770^{+0.059}_{-0.053}$ & $23.970 \pm 0.623$ & 4423.92 & 14.949 & 2 & $1.6 \times 10^{34} $ & $3.4 \times 10^{-3}$ & - \\
   010275962       & $5782 \pm 202$ & $0.789^{+0.064}_{-0.058}$ & $26.117 \pm 1.426$ & 2323.69 & 15.296 & 2 & $1.9 \times 10^{34} $ & $1.9 \times 10^{-3}$ & 1 \\
   011075480       & $5953 \pm 208$ & $0.795^{+0.063}_{-0.057}$ & $21.651 \pm 0.278$ & 2001.46 & 15.108 & 1 & $9.3 \times 10^{33} $ & $1.6 \times 10^{-3}$ & - \\
   011141091       & $5756 \pm 201$ & $0.879^{+0.068}_{-0.062}$ & $23.446 \pm 0.373$ & 2345.32 & 14.982 & 1 & $1.5 \times 10^{34} $ & $2.3 \times 10^{-3}$ & - \\
   011413690       & $5886 \pm 206$ & $0.818^{+0.069}_{-0.062}$ & $24.238 \pm 0.246$ & 3232.87 & 15.725 & 1 & $3.0 \times 10^{34} $ & $2.7 \times 10^{-3}$ & 1 \\
   012053270       & $5971 \pm 209$ & $0.744^{+0.058}_{-0.053}$ & $21.796 \pm 1.587$ & 4338.80 & 15.017 & 1 & $3.1 \times 10^{34} $ & $3.0 \times 10^{-3}$ & - \\
   \enddata
   \tablecomments{
    The data of all the 2344 superflares on all the 266 solar-type superflare stars detected in this study are available in the online-only table.
    }
   \tablenotetext{\dag}{
    Effective temperature and stellar radius values are taken from $Gaia$-DR2 in \citet{Berger+2018}.
    stellar radius are shown in the unit of the solar radius ($R_{\sun}$)
   }
   \tablenotetext{+}{
    These 2 stars are also reported in \citet{Notsu+2019}.
    See further details in Appendix \ref{app:sunlike-spec-previous}.
   }
   \tablenotetext{\ddag}{
    Rotation periods and average rotational variability are reported in \citet{McQuillan+2014}.
   }
   \tablenotetext{\#}{
    $Kepler$-band magnitude \citep{Thompson+2016}.
   }
   \tablenotetext{\P}{
   $N_{\rm flare}$ and $E_{\rm max}$ are the number of superflares and maximum energy of superflares of each Sun-like star, respectively.
   }
   \tablenotetext{*}{
   Area of the starspots on each Sun-like star in the unit of the area of the solar hemisphere ($1/2\times A_{\sun}\sim 3\times 10^{22}$cm$^{2}$).
   }
   \tablenotemark{\%}{
    Flag 1: It is possible the real rotation period value of the star is different from those reported in \citet{McQuillan+2014} and used in this table, on the basis of our extra analyses (see  Appendix \ref{app:allflare_sunlike} and Figures \ref{fig:eg_sunlikeflare_flag1}\&\ref{fig:lomb_flag1} therein). 
    Then the rotation period value of these stars should be treated with caution, and 
    it is possible some of these stars are not really Sun-like stars,
    though more investigations are needed (see Appendix \ref{app:allflare_sunlike} for details).
    \\
    Flag 2: This star is suspected to have binary (see Figures \ref{fig:eg_sunlikeflare_flag2}\&\ref{fig:lomb_flag2} in Appendix \ref{app:allflare_sunlike}).
    This star is not used in the statistical discussions after Section \ref{subsec:energy_spot}.
   }
   \label{table:params_sunlike}
 \end{deluxetable*}
 \end{longrotatetable}

 \begin{figure*}[ht!]
   \plottwo{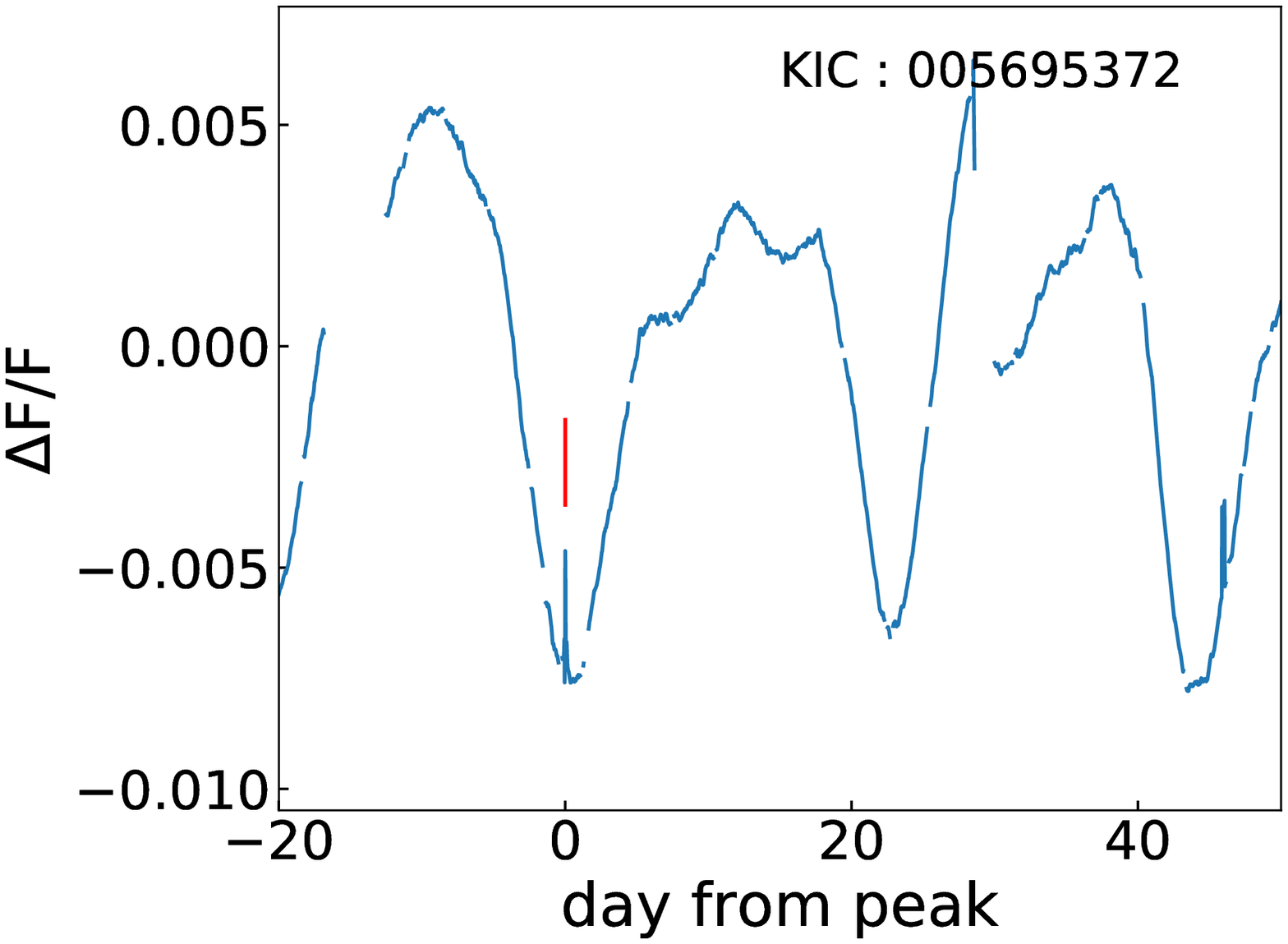}{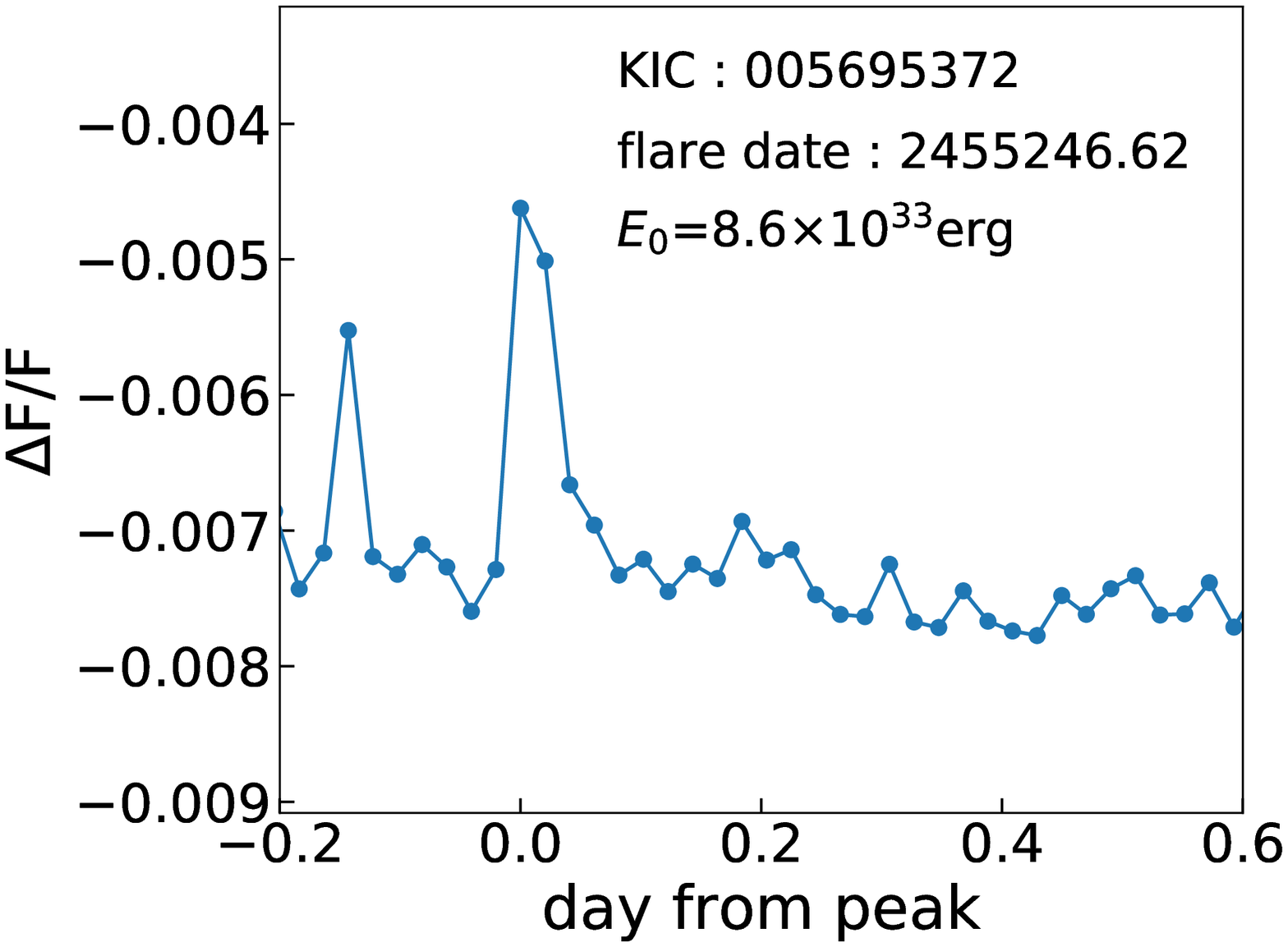}
   \plottwo{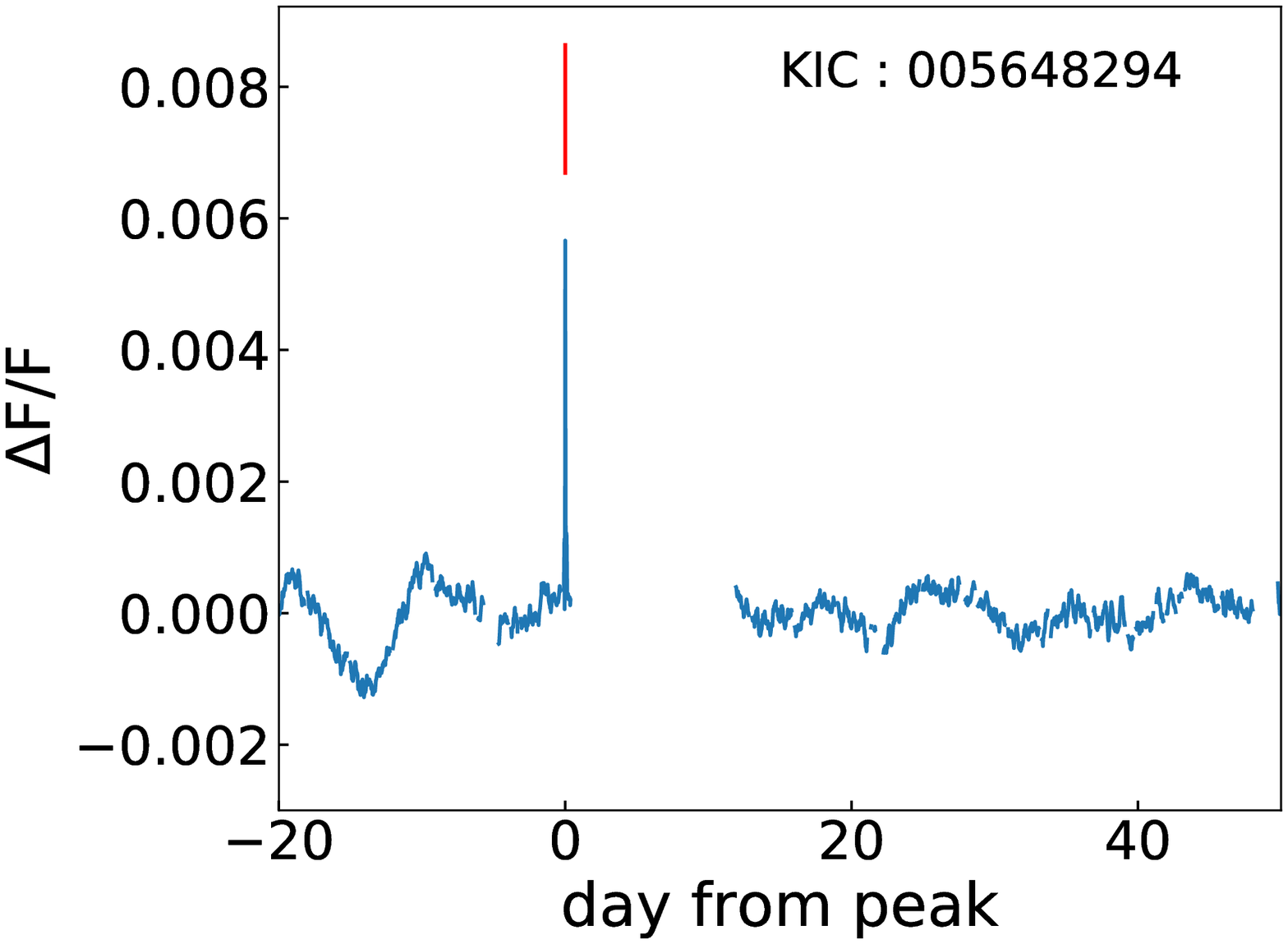}{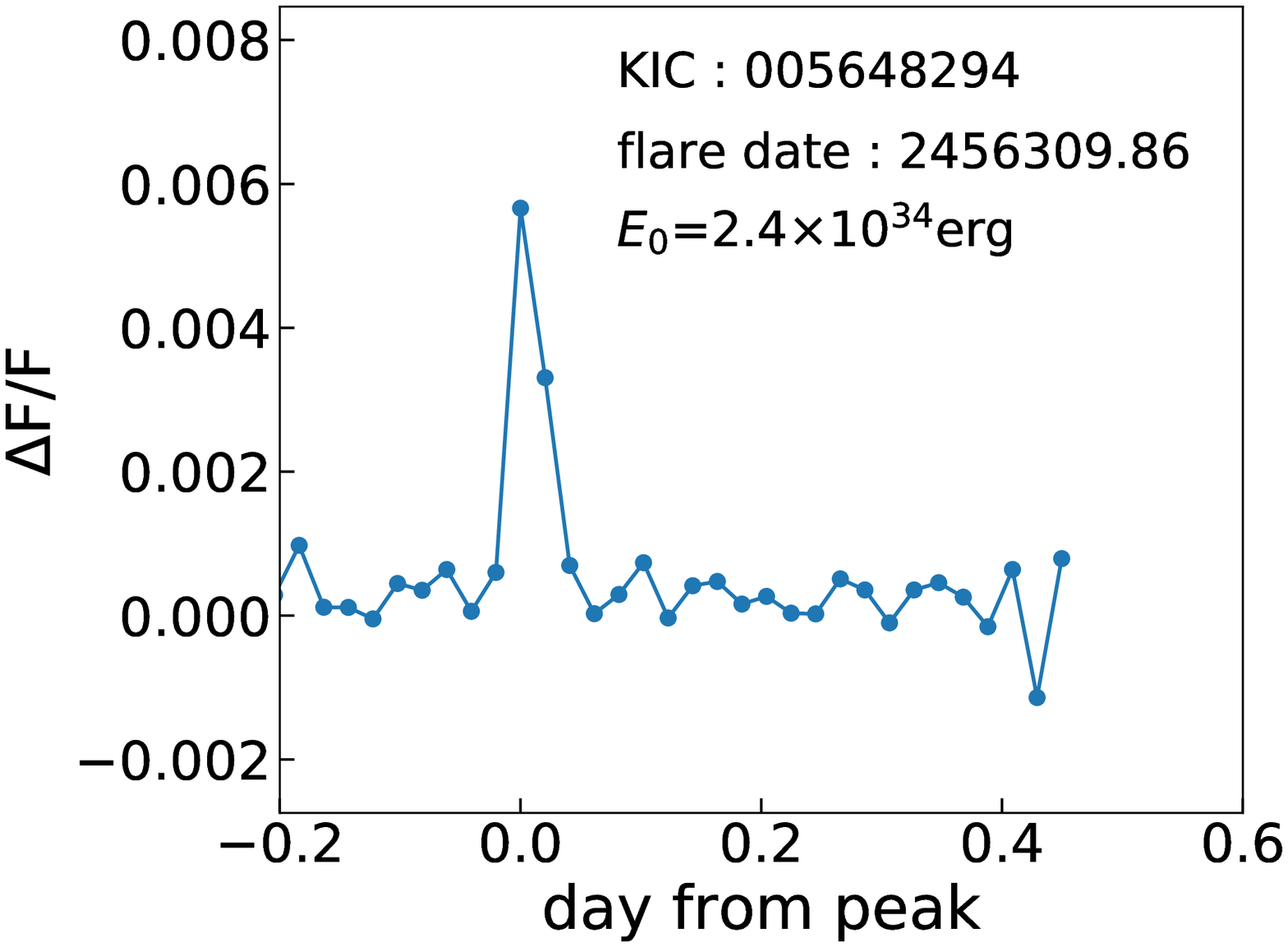}
   \label{fig:eg_sunlikeflare}
   \caption{
   Light curves of the example superflares on Sun-like stars.
   Horizontal and vertical axes correspond to days from the flare peak and stellar brightness normalized by the average brightness during the observation quarter of $Kepler$.
   Panels on the left side show the 70 days time variation of stellar brightness to see the periodic brightness variations caused by stellar rotation.
   Panels on the right side show the detailed brightness variations of a flare.
   The star ID ($Kepler$ ID), Barycentric Julian Date of flare peak, and released total bolometric energy are shown in the right panels.
   }
 \end{figure*}

 \subsection{Dependence of Superflare Energy on Rotation Period}\label{subsec:energy_rotation}
   Previous survey observations of stellar activity levels (e.g., X-ray quiescent luminosity, Ca II H\&K index) have shown that the stellar magnetic level decreases as rotation period increases (\citealt{Noyes+1984};  \citealt{Gudel+2007}; \citealt{Wright+2011}).
   Since stellar age has a strong correlation with rotation, young rapidly-rotating stars show the higher activity levels, and slowly-rotating stars, like the Sun, show lower activity levels.
   Following this, we expected superflare activities to depend on the rotation period.
   \citet{Notsu+2019} suggested that the maximum superflare energy continuously decreases as the rotation period increases and there is roughly a one-order-of-magnitude difference between the maximum flare energy on rapidly-rotating stars ($P_{\rm rot}\sim$ a few days) and on slowly-rotating stars ($P_{\rm rot}>$ 20 days).
   However, a large problem remained that there were some uncertainties on the upper limit of flare energy on Sun-like ($T_{\rm{eff}} = 5600$ -- $6000$ K and $P_{\rm rot}>$ 20 days) stars, because the number of Sun-like superflare stars in \citet{Notsu+2019} was very small.

\begin{figure*}[ht!]
   \gridline{
    \fig{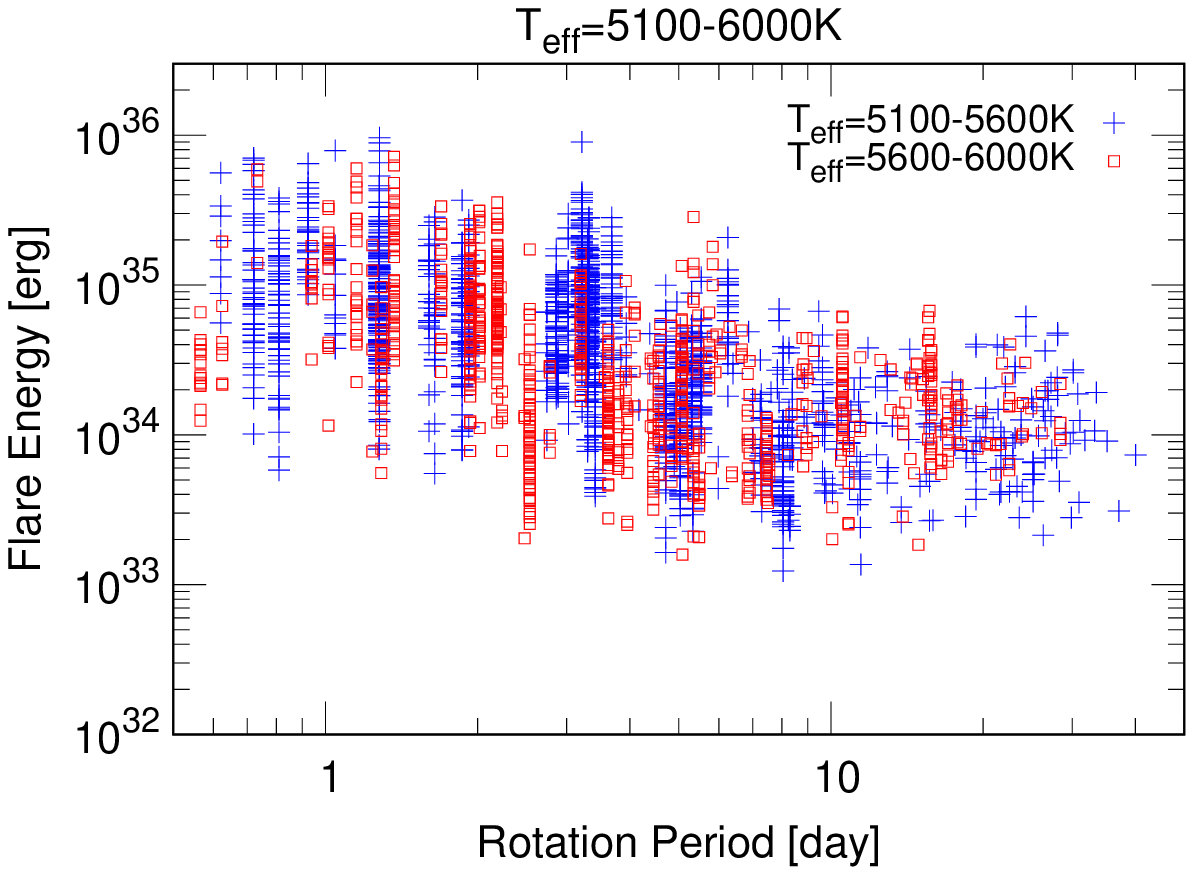}{0.49\textwidth}{\vspace{0mm}(a)}
    \fig{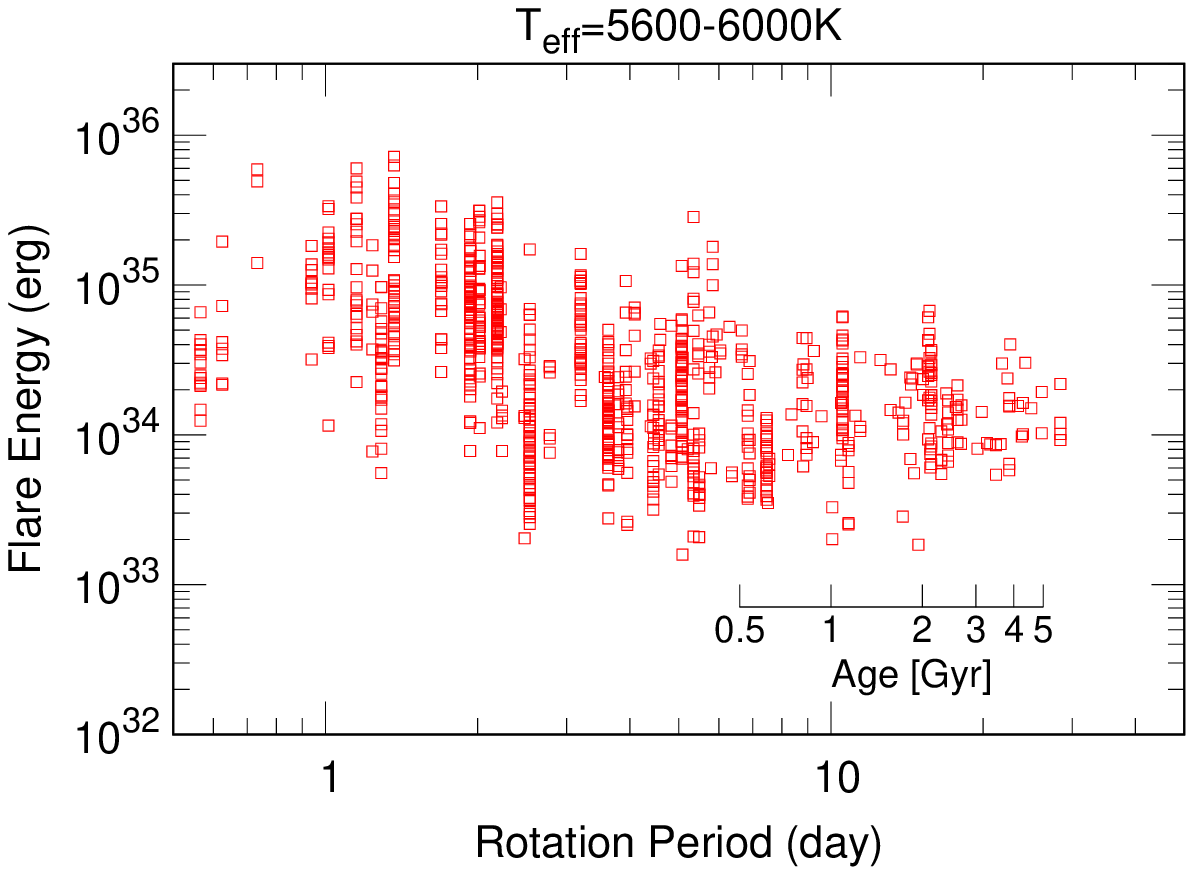}{0.49\textwidth}{\vspace{0mm}(b)}
   }
   \caption{
      Scatter plot of the superflare energy ($E_{\rm flare}$) vs. the rotation period ($P_{\rm rot}$).
      $P_{\rm rot}$ was estimated from the brightness variations period \citep{McQuillan+2014}, and superflares are detected from all the $Kepler$ 4-year 30-min cadence data.
      Blue crosses indicate superflares detected on solar-type stars with $T_{\rm eff} = 5100$ -- $ 5600$ K, while red squares those with $T_{\rm eff} = 5600-6000$ K.
      Only for $T_{\rm eff} = 5600-6000$ K in (b), we added the scale of stellar age ($t$) based on the gyrochronology relation of solar-type star ($P_{\rm rot} \propto t^{0.6}$; \citealt{Ayres+1997}), as also used in the previous study \citep{Notsu+2019}.
      We note that there is an apparent negative correlation between the rotation period and the lower limit of the flare energy, but this trend only reflects the detection limit, and this detection limit is related to the rotation period, as discussed in Section \ref{subsec:sensitivity}.
    }
    \label{fig:erg_prot_solartype}
  \end{figure*}

  We then investigated this relation using a much larger number of superflares newly detected in this study (e.g., $\sim$12 times larger sample of Sun-like stars, as described in Section \ref{subsec:example}).
  Figure \ref{fig:erg_prot_solartype} shows the relationship between the flare energy ($E_{\rm{flare}}$) and the rotation period ($P_{\rm rot}$).
  The data of the stars with a temperature range ($T_{\rm eff} = 5600$ -- $6000$ K) close to the solar temperature ($T_{\rm eff} \sim 5800$ K) are plotted with red open squares, while those of late G-type main-sequence stars ($T_{\rm eff} = 5100$ -- $5600$ K) are blue crosses.
  Similar to the results of \citet{Notsu+2019}, the upper limit of $E_{\rm{flare}}$ in each period bin has a continuous decreasing trend with the rotation period.
  There is at least one-order-of-magnitude difference between the maximum flare energy on rapidly-rotating stars ($P_{\rm rot}\sim$ a few days) and that on slowly-rotating stars ($P_{\rm rot}>$ 20 days).
  This decreasing trend, which is confirmed in \citet{Notsu+2019} and this study, was not reported in our initial study, \citet{Notsu+2013}.
  This is because some fraction ($\sim$40\%) of superflare stars in our initial studies (\citealt{Maehara+2012}; \citealt{Shibayama+2013}; \citealt{Notsu+2013}) are now found to be subgiants \citep{Notsu+2019}, and this contamination of subgiants affected the statistics (see Appendix \ref{app:subgiant} for details).

  In Figure \ref{fig:erg_prot_solartype}, we added the stellar age ($t$) using the gyrochronological relation ($P_{\rm rot} \propto t^{0.6}$: \citealt{Ayres+1997})
  \footnote{
   We note here that the relation $P_{\rm rot} \propto t^{0.6}$ was incorrectly written in the text of \citet{Notsu+2019} (as $t \propto P_{\rm rot}^{0.6}$), though this did not affect on our discussions and conclusions in \citet{Notsu+2019} since the figures were plotted correctly.
  }
  for $T_{\rm eff} = 5600 $ -- $ 6000$ K, in order to compare the age of superflare stars with that of the Sun ($t\sim$4.6 Gyr).
  With this scale, as a result of this Figure \ref{fig:erg_prot_solartype}, we confirm again that old slowly-rotating Sun-like stars ($T_{\rm{eff}}=$5600 -- 6000 K, $P_{\rm rot}>$ 20 days, and Age$\sim$4.6 Gyr) can have superflares with $\lesssim 4 \times 10^{34}$ erg, while young rapidly-rotating stars have superflares with up to $\sim 10^{36}$ erg.
  The scale of $t$ is only plotted in the limited age range $t=$0.5 -- 5 Gyr for the following two reasons:
  (1) As for young solar-type stars with $t\lesssim$0.5 -- 0.6 Gyr, a large scatter in the age–rotation relation has been reported from young cluster observations (e.g., \citealt{Soderblom+1993}; \citealt{Ayres+1997}; \citealt{Tu+2015}).
  (2) As for old solar-type stars beyond solar age ($t\sim$4.6 Gyr), a breakdown of the gyrochronology relations has been recently suggested (\citealt{vanSaders+2016}; \citealt{Metcalfe+2019}). 

  We note that in Figure \ref{fig:erg_prot_solartype}, there is an apparent negative correlation between the rotation period and the lower limit of the flare energy.
  However, this trend only reflects the detection limit affected by the rotation period, as discussed in Section \ref{subsec:sensitivity}.
  We also note that in Figure \ref{fig:erg_prot_solartype}, the sample size of stars in each $P_{\rm rot}$ bin is different.
  The number of samples of slowly-rotating stars are larger than rapidly-rotating stars.
  It is then possible that rare energetic flares approaching the upper limit value for each $P_{\rm rot}$ bin can be a bit more easily detected on slowly-rotating stars than rapidly-rotating stars, assuming the power-law flare frequency distributions shown in Section \ref{subsec:frequency}.
  This might be related to the tendency that the upper limit of flare energy is roughly constant in the range of $P_{\rm rot}>$ 10 days.
  For the same reason, the upper limit $E_{\rm{flare}}$ value of rapidly-rotating stars ($P_{\rm rot}<$ 20 days) can be a bit larger ($>10^{36}$erg).
  We will discuss this point in more detail together with the flare frequency distributions in Section \ref{subsec:frequency}.

 \subsection{Starspot Size and Superflare Energy}\label{subsec:energy_spot}
\begin{figure*}[ht!]
   \gridline{
    \fig{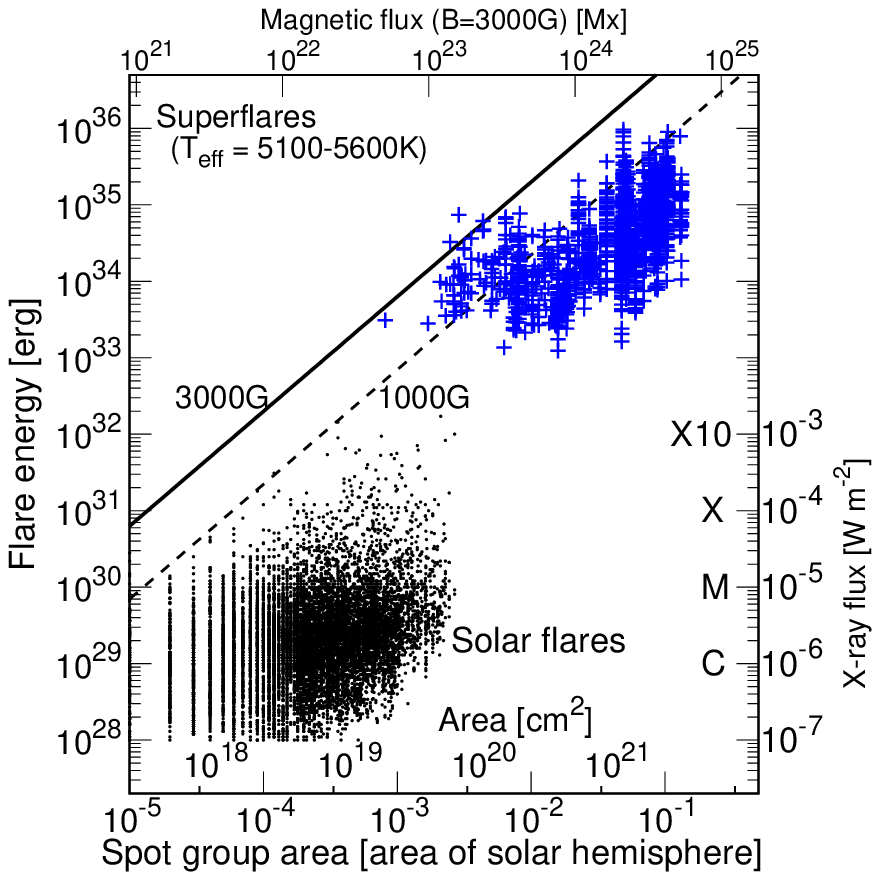}{0.49\textwidth}{\vspace{0mm}(a)}
    \fig{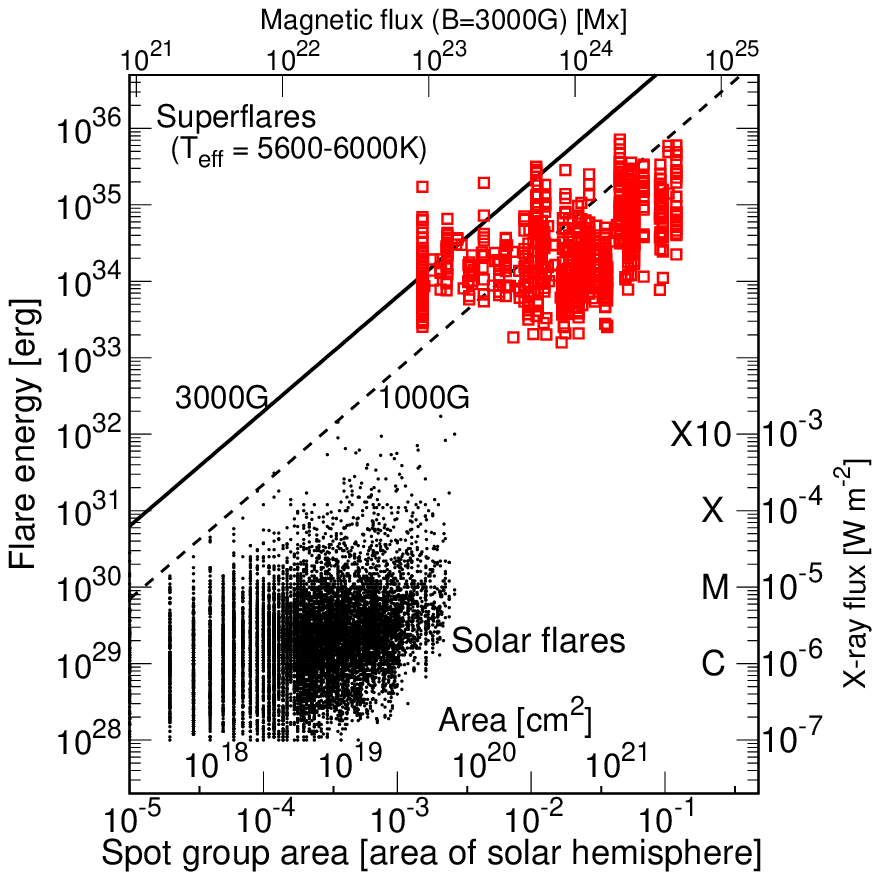}{0.49\textwidth}{\vspace{0mm}(b)}
   }
   \caption{
    Scatter plot of the superflare energy (\(E_{\rm{flare}}\)) and spot group area (\(A_{\rm{spot}}\)) of solar flares and superflares.
    The lower and upper horizontal axes show the area of the spot group in the unit of the solar hemisphere ($1/2\times A_{\sun}\sim 3\times 10^{22}$cm$^{2}$) and the magnetic flux for $B=3000$G.
    The left vertical axis shows the bolometric energy released by each flare.
    The data of solar flares are exactly the same as those in our previous studies (e.g., Figure 6 of \citealt{Notsu+2019}).
    We assumed that bolometric energies of B, C, M, X, and X10 class solar flares are $10^{28}, 10^{29}, 10^{30}, 10^{31}$, and $10^{32}$ erg from observational estimates of solar flare energies.
    The black solid and dashed lines correspond to the relationship between $E_{\rm flare}$ and $A_{spot}$ for $B=3000$ G and 1000 G, respectively, Equation  (\ref{eq:erg_aspt}). \\
    (a) Superflares on solar-type stars with $T_{\rm eff} = 5100$ -- $ 5600$ K are shown with blue crosses. \\
    (b) Superflares on solar-type stars with $T_{\rm eff} = 5600-6000$ K are shown with red open squares.
    }
    \label{fig:erg_aspt}
  \end{figure*}

  Most superflare stars show large-amplitude brightness variations, and they suggest that the surfaces of superflare stars are covered by large starspots.
  Figure \ref{fig:erg_aspt} shows a scatter plot of flare energy (\(E_{\rm{flare}}\)) as a function of the spot group area (\(A_{\rm{spot}}\)) of solar flares and superflare stars.
  $A_{\rm spot}$ values are estimated by using the same way in our previous studies (\citealt{Shibata+2013}; \citealt{Notsu+2019}).
  In these studies, we used the following equation (cf. Equation (3) of \citealt{Notsu+2019}):

  \begin{equation} \label{eq:aspt}
    A_{\rm spot} = \frac{\Delta F}{F}A_{\rm star} \left[ 1 - \left( \frac{T_{\rm spot}}{T_{\rm star}} \right)^4 \right]^{-1} \ ,
  \end{equation}

  where $A_{\rm star}$, $T_{\rm spot}$, and $T_{\rm star}$ are the apparent area of the star ($\pi R_{\rm star}^2$), the temperature of the starspot, and that of the photosphere, respectively.
  In this study, $T_{\rm eff}$ values from \citet{Berger+2018} are used for $T_{\rm star}$ values.
  $T_{\rm spot}$ values are estimated from $T_{\rm star}$ values by using the empirical relation (Equation (4) of \citealt{Notsu+2019}) deduced from \citet{Berdyugina+2005}.
  The total energy released by the flare ($E_{\rm flare}$) must be smaller than (or equal to) the magnetic energy stored around starspots ($E_{\rm mag}$).
  Our previous papers  (e.g., \citealt{Shibata+2013}; \citealt{Notsu+2019}) suggested that the upper limit of flare energy ($E_{\rm flare}$) can be determined by the simple scaling law (cf. Equation (5) of \citealt{Notsu+2019}):

  \begin{eqnarray} \label{eq:erg_aspt}
    E_{\mathrm{flare}} \approx fE_{\mathrm{mag}} \approx \frac{B^2L^3}{8\pi} \approx 7 \times 10^{32} (\mathrm{erg}) \left( \frac{f}{0.1} \right)
    \left(\frac{B}{10^3\mathrm{G}}\right)^2 \left( \frac{A_{\rm spot}/(2 \pi R_{\sun}^2)}{0.001} \right)^{3/2} \ ,
  \end{eqnarray}

  where $f$ is the fraction of magnetic energy that can be released as flare energy, $B$ and $L$ are the magnetic field strength and size of the spot, and $R_{\sun}$ is the solar radius.
  We note that analyses of large solar flares in \citet{Emslie+2012} suggest bolometric flare energies are $\sim$0.1 of the available magnetic energies (see Figure 4 of \citet{Emslie+2012}).

 \begin{figure}[ht!]
  \gridline{
    \fig{}{0.25\textwidth}{\vspace{0mm}}
    \fig{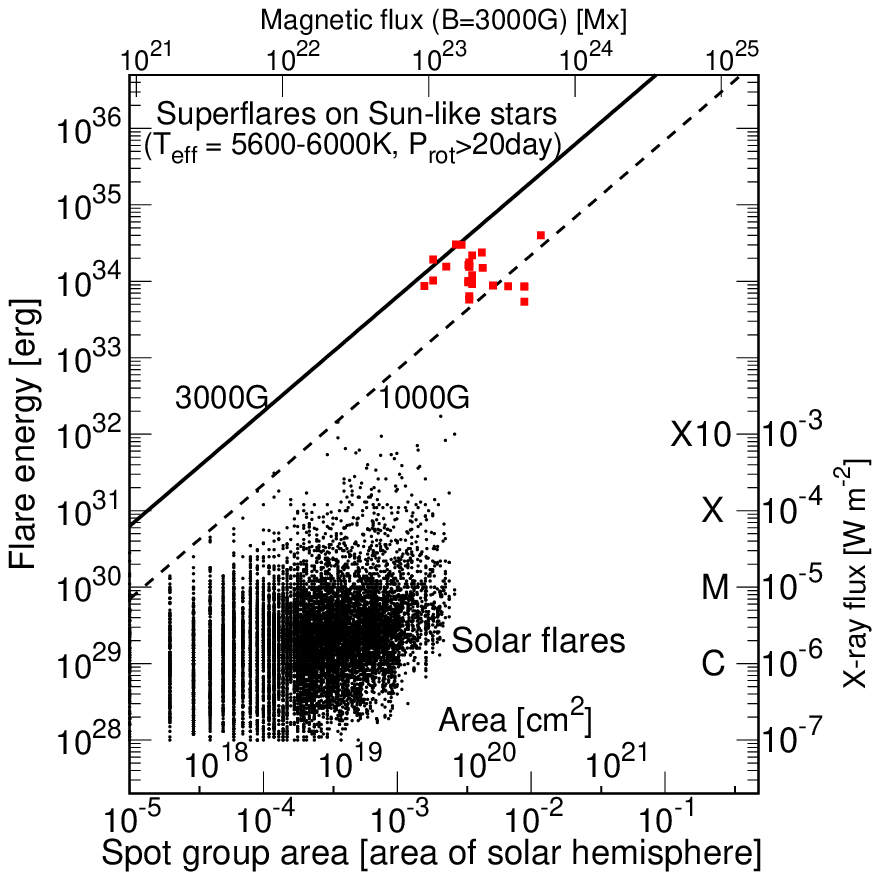}{0.49\textwidth}{\vspace{0mm}}
    \fig{figure/white.eps}{0.25\textwidth}{\vspace{0mm}}
   }
     \caption{
     Same as Figure \ref{fig:erg_aspt}, but for Sun-like stars.
    }
   \label{fig:erg_aspt_sunlike}
 \end{figure}

  In Figure \ref{fig:erg_aspt}, almost all the data points of superflares are below the line of Equation (\ref{eq:erg_aspt}) for $B=3000$G, and the upper limit of flare energies tend to increase as the spot area increase.
  This means that the superflare is the magnetic energy release stored around the starspots, and the process is the same as that of solar flares.
  In Figure \ref{fig:erg_aspt_sunlike}, we also compare the data of Sun-like stars with that of the Sun for reference.

  We note that the starspot group area estimated here can be smaller than the actual values if the stars have a low inclination angle or have starspots around the pole region (see also \citealt{Notsu+2015b} \& \citeyear{Notsu+2019} for details).
  Besides the real size of the spot group when a superflare occurs can be larger than the size estimated from the brightness amplitude values, since we only use time-averaged brightness amplitude ($Amp$) values reported in \citet{McQuillan+2014}.
  Some evolution/decay of starspot coverage, which can occur in the time scale of $\sim$100 days (e.g., \citealt{Namekata+2019} \& \citeyear{Namekata+2020_ApJ}), is time-averaged.
  These can be the reason why some data points of superflares locate above the line of Equation (\ref{eq:erg_aspt}) (for $B=3000$G) in Figure \ref{fig:erg_aspt}.
  As a result, we can conclude that the upper limit of the flare energy is consistent with the magnetic energy stored around the starspots, as also confirmed in \citet{Notsu+2019}.

  \newpage
  \subsection{Starspot Size versus Rotation Period of Solar-type Stars, and Implications for Superflare Energy}\label{subsec:spot_rotation}
  In Section \ref{subsec:energy_rotation}, we confirmed that the upper limit of the superflare energy has a decreasing trend with rotation period.
  In Section \ref{subsec:energy_spot}, we also confirmed the upper limit of flare energy is consistent with the starspot coverage, considering magnetic energy is stored around starspots.
  Then, how is starspot coverage related to rotation period, and what implications exist for superflare energy?

\begin{figure*}[ht!]
    \gridline{
    \fig{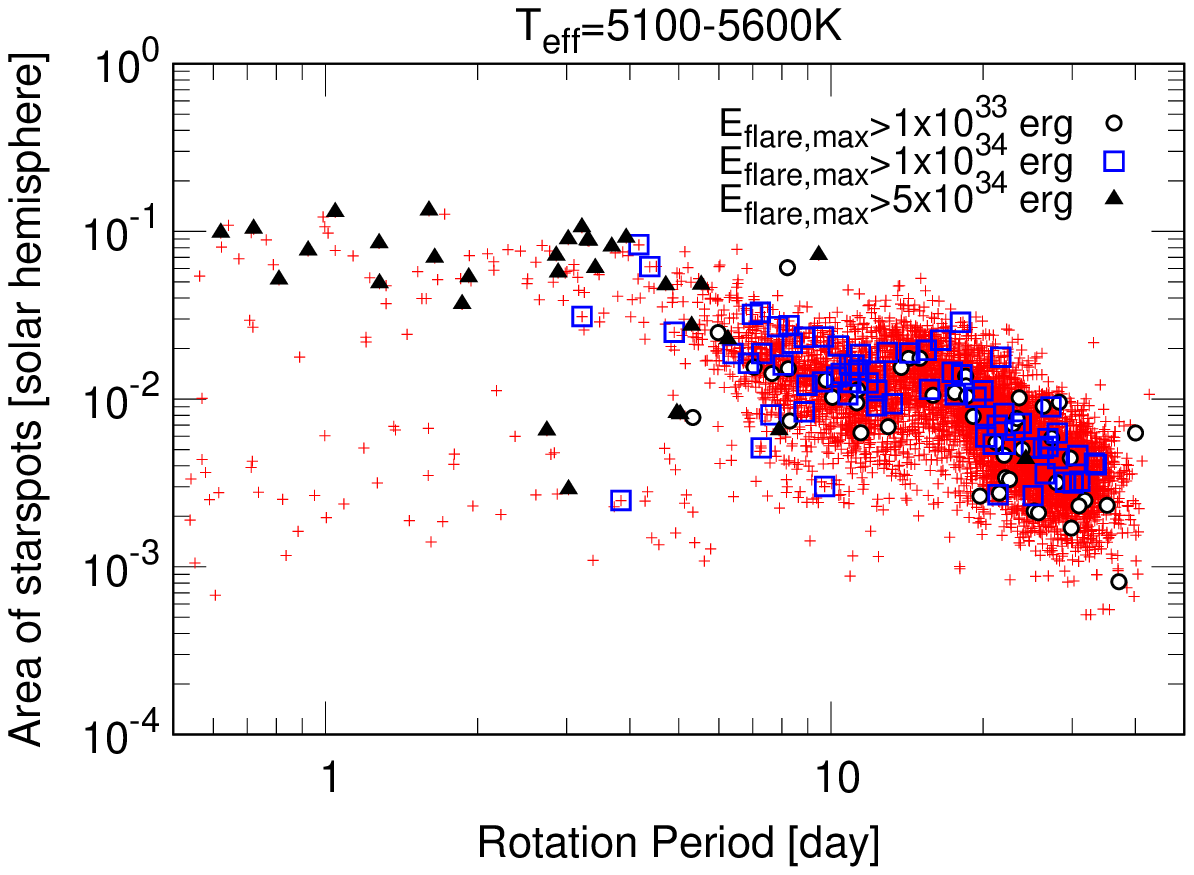}{0.49\textwidth}{\vspace{0mm}(a)}
    \fig{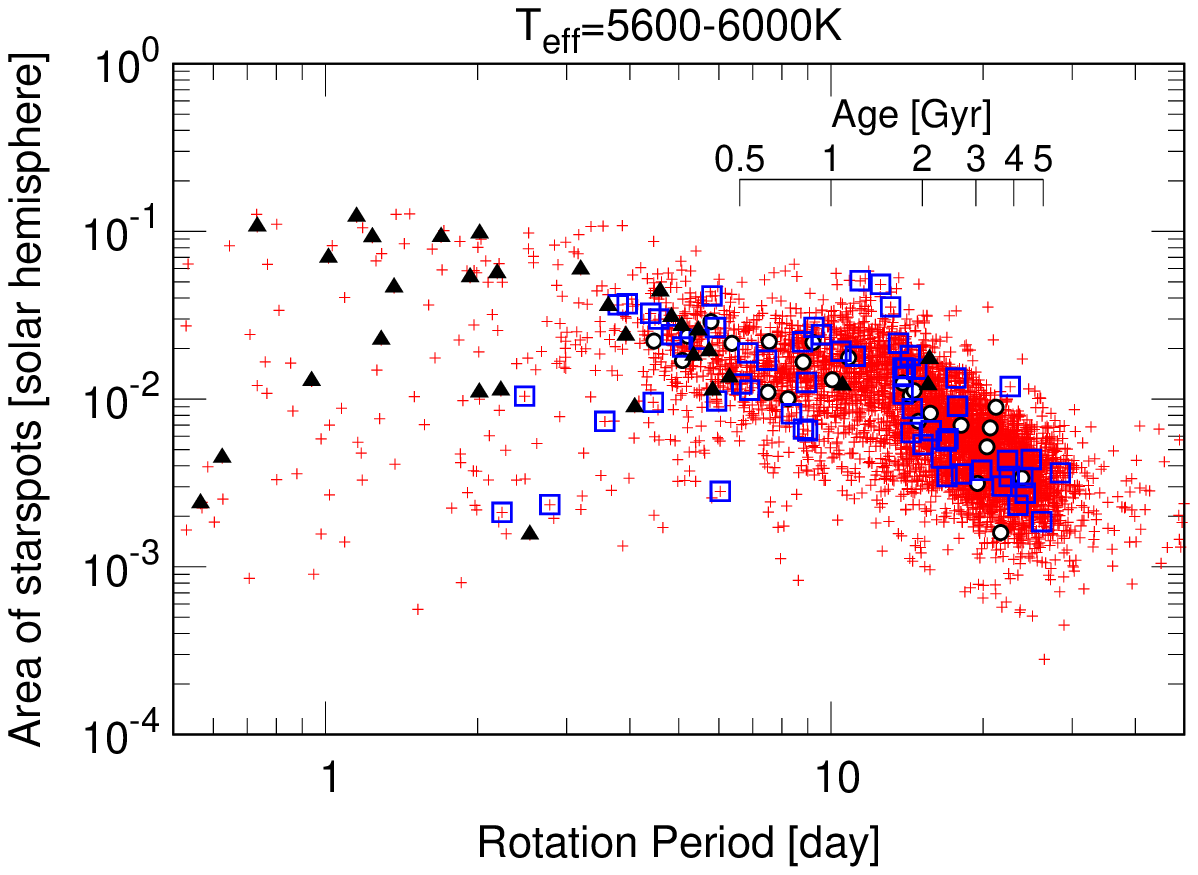}{0.49\textwidth}{\vspace{0mm}(b)}
    }
   \caption{
   Scatter plot of the spot group area of solar-type stars ($A_{\rm{spot}}$) as a function of the rotation period ($P_{\rm{rot}}$).
   The vertical axis represents $A_{\rm{spot}}$ in units of the area of solar hemisphere ($A_{\sun}\sim 3\times 10^{22}$ cm$^{2}$).
   Black open circles, blue open squares, and black filled triangles indicate solar-type stars that have superflares with the energy values of their most energetic flares $E_{\rm flare,max}>1\times10^{33}$ erg, $E_{\rm flare,max}>1\times10^{34}$ erg, and $E_{\rm flare,max}>5\times10^{34}$ erg, respectively.
   Small red cross indicates all solar-type stars.
   The plotted data are separated on the basis of the temperature values:
   (a) $T_{\rm{eff}} = 5100$ -- $ 5600$ K and (b) $T_{\rm{eff}} = 5600-6000$ K.
   Only in (b), we added the scale of stellar age ($t$) based on the gyrochronology relation for solar-type stars ($P_{\rm rot} \propto t^{0.6}$; \citealt{Ayres+1997}; see Section \ref{subsec:energy_rotation} for details).
   }
   \label{fig:aspot_prot_solartype}
 \end{figure*}

  Our previous paper, \citet{Maehara+2017}, investigated the statistical properties of starspots on solar-type stars by using the starspot size ($A_{\rm{spot}}$) and rotation period ($P_{\rm{rot}}$) estimated from the brightness variations of $Kepler$ data, and \citet{Notsu+2019} updated the results by using $Gaia$-DR2 stellar radius.
  The resultant values of $A_{\rm{spot}}$ and $P_{\rm{rot}}$ are shown in Figure \ref{fig:aspot_prot_solartype}, with the increased data of superflare stars newly identified in this paper.
  As also shown in \citet{Notsu+2019}, the stars showing more energetic superflares tend to have shorter rotation periods (younger ages) and larger starspot areas.

  Figure \ref{fig:aspot_prot_solartype} shows the largest area of starspots on solar-type stars in a given $P_{\rm{rot}}$ bin has a roughly constant or very gentle decreasing trend in the period range of $P_{\rm{rot}}\lesssim$ 12 days (age: $t\lesssim$1.4 Gyr) for the solar-type stars with $T_{\rm{eff}} = 5600-6000$ K and $P_{\rm{rot}}\lesssim$ 14 days for those with $T_{\rm{eff}} = 5100$ -- $ 5600$ K.
  However, in the period range of $P_{\rm{rot}}\gtrsim$ 12 days (age: $t\gtrsim$ 1.4 Gyr) for the stars with $T_{\rm{eff}} = 5600-6000$ K) and $P_{\rm{rot}} \gtrsim$ 14 days for those with $T_{\rm{eff}} = 5100$ -- $ 5600$ K, the largest starspot area of the stars steeply decreases as the rotation period increases.
  This decreasing trend of maximum area of starspots can be related to the decreasing trend of maximum flare energy confirmed in Section \ref{subsec:energy_rotation} (see Figure \ref{fig:erg_prot_solartype}).
  This is because the maximum area of starspots determines well the upper limit of flare energy, as indicated in Section \ref{subsec:energy_spot}.
  In the case of slowly-rotating Sun-like stars with $T_{\rm{eff}} = 5600-6000$ K and $P_{\rm{rot}} \sim 25$ days, Figure \ref{fig:aspot_prot_solartype}(b) shows that the maximum size of starspots is a few percents of the solar hemisphere.
  These values correspond to $10^{34}-10^{35}$ erg by Equation (\ref{eq:erg_aspt}), and the upper limit of superflare energy of these Sun-like stars in Figure \ref{fig:erg_prot_solartype} is roughly in the same range.

  It is difficult to conclude whether $A_{\rm{spot}}$ values are constant or decrease in the short period range because of several factors.
  As indicated in Section \ref{subsec:energy_spot}, the $A_{\rm{spot}}$ values estimated from the brightness variations can be smaller than the actual values.
  For example, the $A_{\rm{spot}}$ values estimated from the brightness variations can be smaller than the actual values if the stars have a low inclination angle or have starspots around the pole-region (see also \citealt{Notsu+2015b}\&\citeyear{Notsu+2019}).
  In addition, more active stars, such as rapidly-rotating stars, can have multiple starspots on the surface, and this also causes the $A_{\rm{spot}}$ estimated from the brightness variations is smaller than the total spot coverage of the star \citep{Namekata+2020_ApJ}. 

  However, there is a difference in a bit strict sense between the decreasing trends of the maximum superflare energy in Figure \ref{fig:erg_prot_solartype} and the maximum area of starspots in Figure \ref{fig:aspot_prot_solartype}.
  The maximum superflare energy continuously decreases as the rotation period increases (the star becomes older) in Figure \ref{fig:erg_prot_solartype}, but the maximum area of starspots does not show such continuous decreasing trend in  Figure \ref{fig:aspot_prot_solartype}.
  On the other hand, the maximum area of starspots is roughly constant or very gently decreases in the short-period range ($P_{\rm{rot}}\lesssim$ 14 days), but it steeply decreases as the period increases in the longer range ($P_{\rm{rot}}\gtrsim$ 14 days).
  This difference was also suggested in our previous paper \citet{Notsu+2019}, but this was not as clear since the number of superflare events was small.
  If this difference between the decreasing trends in Figures \ref{fig:erg_prot_solartype} and \ref{fig:aspot_prot_solartype} is true, there can be a possibility that the flare energy is determined not only by the starspot area but also by other important factors, though the starspot area is still a necessary condition to determine the flare energy (see Section \ref{subsec:energy_spot}).
  Considering the correlation between the flare activity and the magnetic structure of sunspot groups (see \citealt{Sammis+2000}; \citealt{Toriumi+2019}), one of the possible factors might be the effect of the magnetic structure of starspots.
  More complex spots can generate more frequent and more energetic flares according to solar observations.
  If the magnetic structure (complexity) of spots also has a correlation with the rotation period, the upper limit of flare energy can depend on rotation period even if the starspot size is roughly constant.
  We need to conduct more detailed studies on starspot properties to clarify such possibilities (e.g., observations using exoplanet transits in \citealt{Namekata+2020_ApJ}).

  In addition, as suggested in \citet{Notsu+2019}, the constant and decreasing trends of maximum starspot coverage can be compared with the relation between X-ray flux and rotation period (e.g., \citealt{Wright+2011}).
  The X-ray fluxes of solar-type stars are also known to show the constant regime (or so-called ``saturation" regime) in the period range of $P_{\rm{rot}}\lesssim 2-3$ days, but they decrease constantly as the $P_{\rm{rot}}$ values increase in the range of $P_{\rm{rot}}\gtrsim 2-3$ days.
  The changing point of this X-ray trend ($P_{\rm{rot}}\lesssim 2-3$ days) is different from that of maximum spot size values ($P_{\rm{rot}}\lesssim 12-14$ days).
  These similarities and differences can be interesting and helpful when considering the relation between stellar activity (including starspots, flares, X-ray steady emissions) and rotation period in more detail, though a detailed study on this point is beyond the scope of this paper.


\subsection{Differences between the number of stars having rotation period values and the number of stars in Kepler field} \label{subsec:gyrochronology}
 In Appendix B of \citet{Notsu+2019}, we investigated the potential differences between the number of stars with $P_{\rm rot}$ values in \citet{McQuillan+2014} and the number of ``real" sample of $Kepler$ field stars, by using the gyrochronological relation.
 As also shown in Table \ref{table:num_stars_gyro}, approximately $\sim$69\% (=$(20893-6527)/20893$) and $\sim$82 \% (=$(28412-5064)/28412$)  of the solar-type stars with $T_{\rm{eff}} = 5100$ -- $5600$ K and $T_{\rm{eff}} = 5600$ -- $6000$ K, respectively, have no $P_{\rm rot}$ values in our sample, and they are not plotted in Figure \ref{fig:aspot_prot_solartype}.
 This is because the brightness variations amplitudes of these ``inactive" solar-type stars is smaller than the detection limit.
 We need to pay attention to biases caused by these ``inactive'' solar-type stars when discussing the relation between the rotation period and superflare properties, such as the occurrence frequency of superflares.
 The number of stars having $P_{\rm{rot}}$ values in each $P_{\rm{rot}}$ bin of Table \ref{table:num_stars_gyro} does not show the actual $P_{\rm{rot}}$ distribution of the $Kepler$ field.
 This is because the ``inactive” stars with no $P_{\rm{rot}}$ values are expected to be dominated by old slowly-rotating stars, and this can be a big problem especially for discussing the flare frequencies of slowly-rotating stars.

 In order to roughly evaluate the potential differences caused by the above points, our previous paper \citet{Notsu+2019} estimated the number fraction of solar-type stars in specific $P_{\rm{rot}}$ bins, by using the empirical gyrochronology relation (\citealt{Ayres+1997}; \citealt{Mamajek+2008}).
 The number of stars with $P_{\rm rot}$ values larger than $P_0$ ($N_{\rm star}(P_{\rm rot} \geq P_0)$) can be estimated from the duration of the main-sequence phase ($\tau_{\rm MS}$), the gyrochronological age of the star ($t_{\rm gyro}(P_0)$), and the total number of stars ($N_{\rm all}$):

  \begin{equation}\label{eq:gyrochronology}
   N_{\rm star}(P_{\rm rot} \geq P_0) = \left( 1 - \frac{t_{\rm gyro}(P_0)}{\tau_{MS}} \right) N_{\rm all} \ .
 \end{equation}

 We assumed that the star formation rate around the $Kepler$ field has been roughly constant over $\tau_{\rm MS}$.

 Using Equations (12) -- (14) of \citet{Mamajek+2008}, we roughly estimated the ages ($t_{\rm gyro}(P_0)$) of solar-type stars with $T_{\rm{eff}}\sim$5800 and 5350 K ($B-V\sim 0.65$ and 0.80 from Equation (2) of \citet{Valenti+2005}) and $P_{\rm rot}\sim$5, 10, and 20 days.
 Using these $t_{\rm gyro}(P_0)$ values and Equation (\ref{eq:gyrochronology}), we also estimated the fraction of stars as listed in Table \ref{table:num_stars_gyro}.
 We assume that $N_{\rm star}$ ($P_{\rm rot}\geq20$ days) is the same as $N_{\rm star}$ ($P_{\rm rot}=20$ -- $40$ days), when we compare it with $N_{P}$($P_{\rm rot}=20$ -- $40$ days) in Table \ref{table:num_stars_gyro}.
 This is because according to the gyrochronology relation, the rotation period of solar-type stars cannot be larger than 40 days in the duration of the main-sequence phase, and the relation has a breakdown in older, slowly-rotating solar-type stars (e.g., $t>5.0$ Gyr) (\citealt{vanSaders+2016}; \citealt{Metcalfe+2019}).

  \begin{table}[h!]
  \centering
  \caption{
   Differences between the number of $Kepler$ solar-type stars with $T_{\rm eff} = 5100$ -- $ 5600$ K and $T_{\rm eff} = 5600-6000$ K that have $P_{\rm rot}$ values reported in \citet{McQuillan+2014}, and the number of field stars estimated from the gyrochronological relation (See Equation (\ref{eq:gyrochronology})).
  }
  \label{table:num_stars_gyro}
  \begin{tabular}{lcc}
   \hline\hline
  $T_{\rm eff}=5100$ -- $ 5600$ K & $N_{P}$\tablenotemark{\dag} (having $P_{\rm rot}$ value) & $N_{\rm star}$\tablenotemark{\ddag} (from gyrochronology) \\
   \hline
       all                       & 6527        & 20893\\
       $P_{\rm rot} < 5$ day     & 175 (3\%)   & 221 (1\%)\\ 
       $P_{\rm rot} = 5 $ -- $10$ day & 734 (11\%)  & 620 (3\%)\\ 
       $P_{\rm rot} = 10$ -- $20$ day & 2235 (34\%) & 2359 (11\%)\\ 
       $P_{\rm rot} = 20$ -- $40$ day & 3374 (52\%) & 17693 (85\%)\\
     \hline\hline
  $T_{\rm eff}=5600-6000$ K & $N_{P}$\tablenotemark{\dag} (having $P_{\rm rot}$ value) & $N_{\rm star}$\tablenotemark{\ddag} (from gyrochronology) \\
   \hline
            all                  & 5064        & 28412\\
       $P_{\rm rot} < 5$ day     & 314 (6\%)   & 507 (2\%)\\ 
       $P_{\rm rot} = 5 $ -- $10$ day  & 785 (16\%) & 1422 (5\%)\\ 
       $P_{\rm rot} = 10$ -- $20$ day & 2325 (46\%) & 5407 (19\%)\\ 
       $P_{\rm rot} = 20$ -- $40$ day & 1640 (32\%) & 21075 (74\%)\\
   \hline
  \end{tabular}
     \tablenotetext{\dag}{
     Number of stars that have $P_{\rm rot}$ value and $Amp$ values reported in \citet{McQuillan+2014}.
     }
     \tablenotetext{\ddag}{
     Number of stars estimated from the gyrochronology relation.
     The number in the first column corresponds to the number of all $Kepler$ solar-type stars with $T_{\rm eff} = 5100$ -- $5600$ K and $T_{\rm eff} = 5600$ -- $6000$ K (cf. (2) in Table \ref{table:num_flares}).
     }
  \end{table}

 As seen from Table \ref{table:num_stars_gyro}, there are differences between the number fractions of slowly/rapidly rotating stars in the $Kepler$ sample from \citet{McQuillan+2014} ($N_{P}$) and those estimated from the gyrochronology relation ($N_{\rm{star}}$).
 In the case of Sun-like stars with $T_{\rm eff} = 5600$ -- $6000$ K, the fraction of stars with $P_{\rm rot}=20$ -- $40$ days among all the sample stars has roughly a factor of two difference:
 $N_{P}$($P_{\rm rot}=20$ -- $40$ days) $/$ $N_{P,\rm{all}}\sim 32$\% and $N_{\rm{star, all}}$($P_{\rm rot}=20$ -- $40$ days) $/$ $N_{\rm{star, all}}\sim70$\%.
 This means that the flare frequency value of Sun-like stars ($T_{\rm{eff}}=5600$ -- 6000 K and $P_{\rm rot}>$ 20 days) can become a factor of two smaller than the real value if we only consider the Sun-like stars in \citet{McQuillan+2014}.
 We have already showed these differences in Appendix B of \citet{Notsu+2019} only as a caution.
 In this study, we consider the differences when discussing flare frequencies in the following sections.
 Through this, we aim to discuss more accurate values of flare frequencies, especially for slowly-rotating Sun-like stars.

 Before applying these ``gyrochronology corrections'' to flare frequency discussions in the following section, we note several possible errors that the gyrochronology can have.
 One example is that the age–rotation relation (gyrochronology relation) of young ($t \lesssim 0.5$ -- $0.6$ Gyr) and old ($t$ $\gtrsim 5.0$ Gyr) solar-type stars can have a large scatter (e.g., \citet{Soderblom+1993}; \citealt{Ayres+1997}; \citealt{Tu+2015}) and breakdown (\citealt{vanSaders+2016}; \citealt{Metcalfe+2019}), respectively, as mentioned in Section \ref{subsec:energy_rotation}.
 We also assumed that the star formation rate around the $Kepler$ field has been roughly constant over $\tau_{\rm{MS}}$ in the above estimation, but this assumption is not necessarily correct.
 We should keep these potential errors in mind.

\subsection{Flare detection limit}\label{subsec:sensitivity}
 As shown in our previous study \citet{Shibayama+2013}, the threshold to detect superflares used in our method (See Section \ref{subsec:analysis}) tended to be larger in rapidly-rotating stars than in slowly-rotating stars.
 We used the high-pass filter described in Section \ref{subsec:analysis} to eliminate this tendency.
 In this section, we evaluate how this tendency is changed by the filter and how much this tendency remains in the filtered data to discuss the relations between rotation period and flare frequency in the following sections.
 As described in Section \ref{subsec:analysis}, a flare is detected if the assumed flare amplitude ($FAmp_{\rm assumed}$) is larger than the flare detection threshold ($Threshold$).
 The $Threshold$ value is calculated as three times the value at the value of the top 1\% of the brightness distribution of each star.
  Here we check how large amplitude flares can be detected ($FAmp_{\rm assumed} \geq Threshold$) for each star.
  We count the number of stars that satisfies ``$FAmp_{\rm assumed} \geq Threshold$'' for several period ranges ($P_{\rm rot}=$ 0 -- 5, 5 -- 10, 10 -- 20, 20 -- 40 days) as a function of $FAmp_{\rm assumed}$: $N_{P}^{FAmp_{\rm assumed} \geq Threshold}$.
  Then we can define the detection completeness ($DC_{\rm filter}$) as:

 \begin{equation}
      DC_{\rm filter} = \frac{N_{P}^{FAmp_{\rm assumed} \geq Threshold}}{N_{P}} ,
 \end{equation}

 where $N_{P}$ is the total number of solar-type stars having $P_{\rm rot}$ values in the considered $P_{\rm rot}$ and temperature range shown in Table \ref{table:num_stars_gyro}.
 $DC_{\rm filter}$ means the fraction of stars from which flares can be surely detected since the flare amplitude is larger than the detection threshold ($FAmp_{\rm assumed} \geq Threshold$).

\begin{figure*}[ht!]
    \gridline{
    \fig{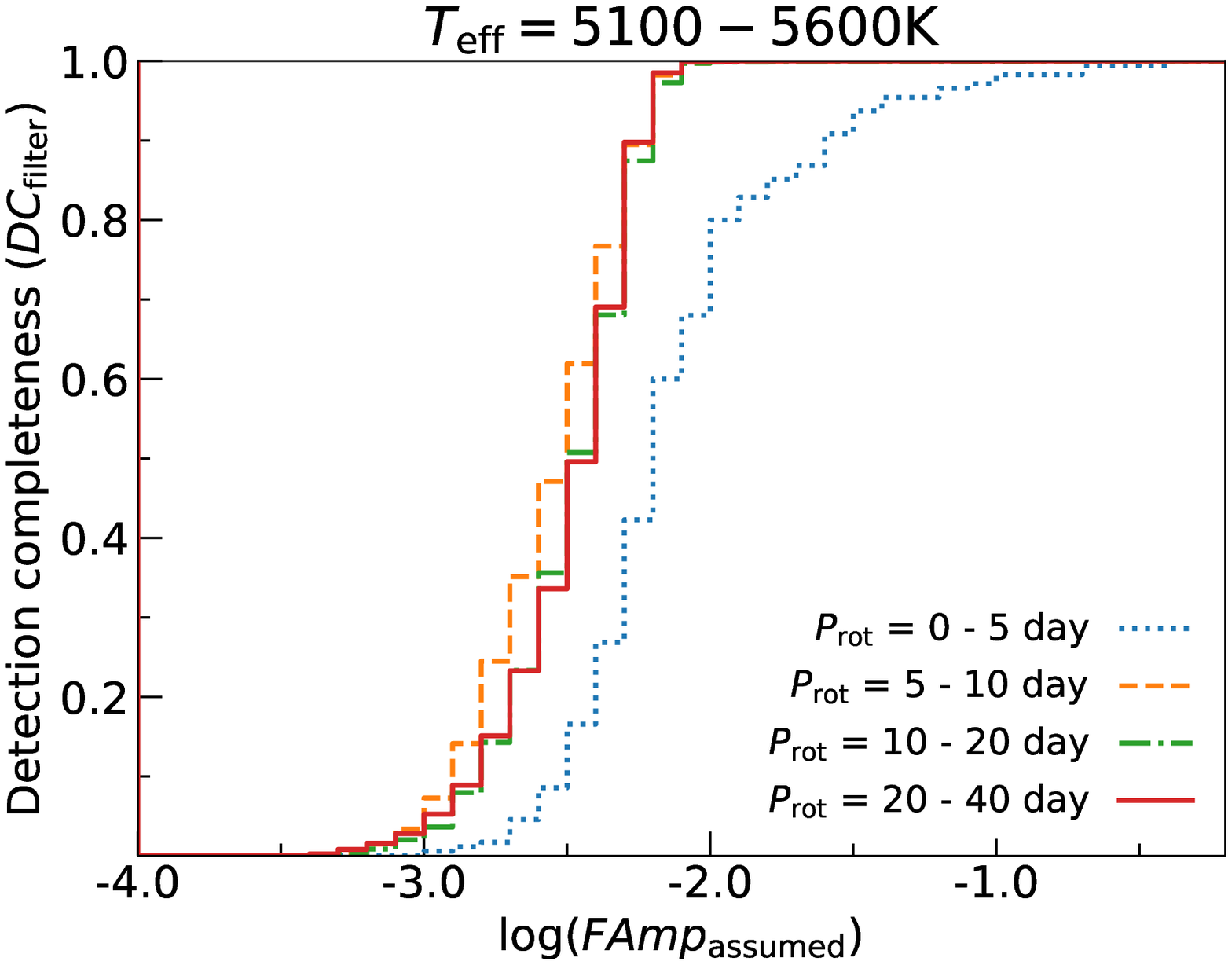}{0.49\textwidth}{\vspace{0mm}(a)}
    \fig{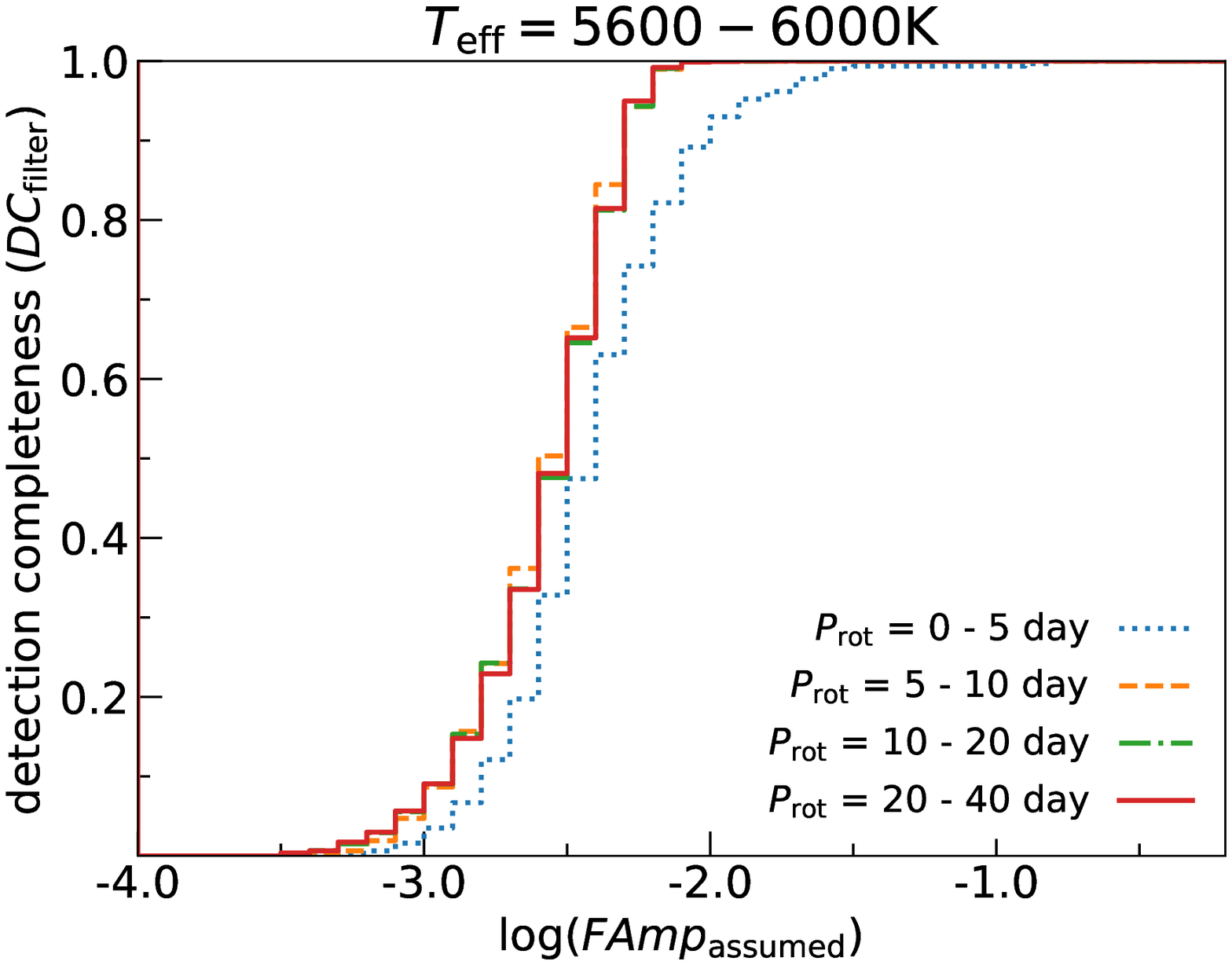}{0.49\textwidth}{\vspace{0mm}(b)}
    }
   \caption{
    The relationship between $DC_{\rm filter}$ and the assumed flare amplitude ($FAmp_{\rm assumed}$) for several $P_{\rm{rot}}$ ranges, estimated from the filtered lightcurve data (cf. Figure \ref{fig:method_filter} (c)\&(d)) of the solar-type stars with the brightness variations period and amplitude ($P_{\rm rot}$ and $Amp$) values detected in \citet{McQuillan+2014} (cf. $N_{P}$ values in Table \ref{table:num_stars_gyro}).
    The horizontal axis is the logarithm of assumed amplitude of flares ($FAmp_{\rm assumed}$).
    The data are divided in (a) and (b) on the basis of the stellar temperature: (a) $T_{\rm eff} = 5100$ -- $ 5600$ K and (b) $T_{\rm eff} = 5600-6000$ K.
    Different lines correspond to $P_{\rm{rot}}$ ranges:
    blue dotted lines are the values of $P_{\rm{rot}}=0$ -- $5$ days,
    orange dashed lines are those of $P_{\rm{rot}}=5$ -- $10$ days,
    green dash dotted lines are those of $P_{\rm{rot}}=5$ -- $10$ days,
    and red filled lines are those of $P_{\rm{rot}}=5$ -- $10$ days.
   }
   \label{fig:detlim_logamp_prot_solartype}
 \end{figure*}

 Figure \ref{fig:detlim_logamp_prot_solartype} shows the relationship between $DC_{\rm filter}$ and the flare amplitude for several period ranges ($P_{\rm rot}=$ 0 -- 5, 5--10, 10 -- 20, 20 -- 40 days), for the filtered lightcurve data (cf. Figure \ref{fig:method_filter} (c)\&(d)).
 There is a tendency that the detection completeness $DC_{\rm filter}$ becomes larger as the flare amplitude becomes larger.
 However, the tendencies are different between rapidly-rotating ($P_{\rm rot}<5$ days) and slowly-rotating stars ($P_{\rm rot} \geq 5$ days).
 As for the stars with $P_{\rm rot}\geq5$ days, the detection completeness $DC_{\rm filter}$ becomes almost $\sim$1.0 at the flare amplitude of $\sim 10^{-2}$.
 This means that we can detect almost all the superflares with amplitude value of $10^{-2}$ occurred on stars with $P_{\rm rot}\geq5$ days. 
 We note this flare amplitude value $10^{-2}$ corresponds to the flare energy with $10^{34.5}$ erg in the case of typical duration superflares in our sample.
 In contrast, for the rapidly-rotating stars with $P_{\rm rot}<5$ days, the detection completeness $DC_{\rm filter}$ is still $\sim$0.7 at the flare amplitude of $\sim 10^{-2}$.
 This means we miss 30\% of superflares with amplitude value of $10^{-2}$ occurred on stars with $P_{\rm rot}<5$ days, while we can detect 70\% superflares with amplitude value of $10^{-2}$ occurred on stars with $P_{\rm rot}<5$ days.
 
\begin{figure*}[ht!]
    \gridline{
    \fig{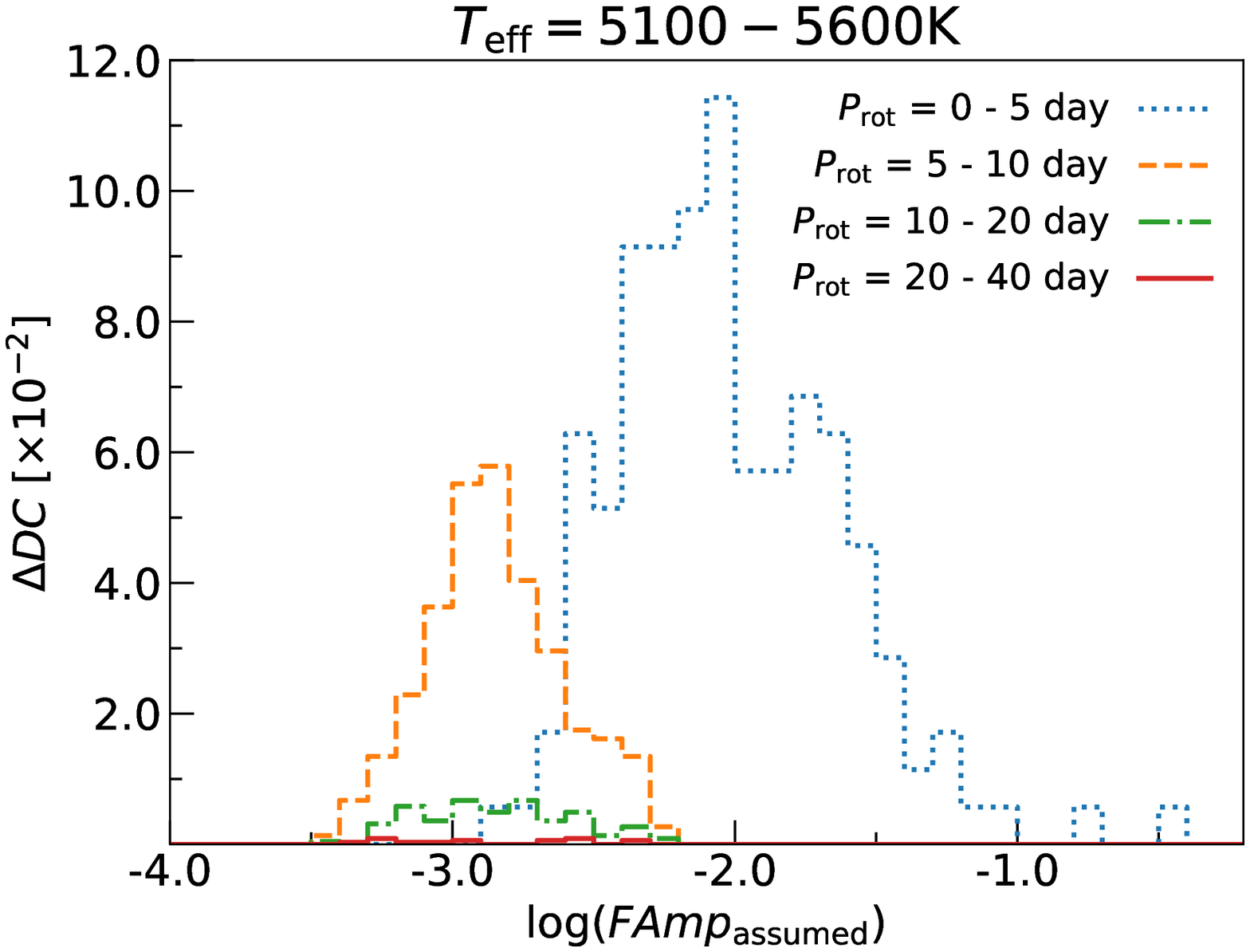}{0.49\textwidth}{\vspace{0mm}(a)}
    \fig{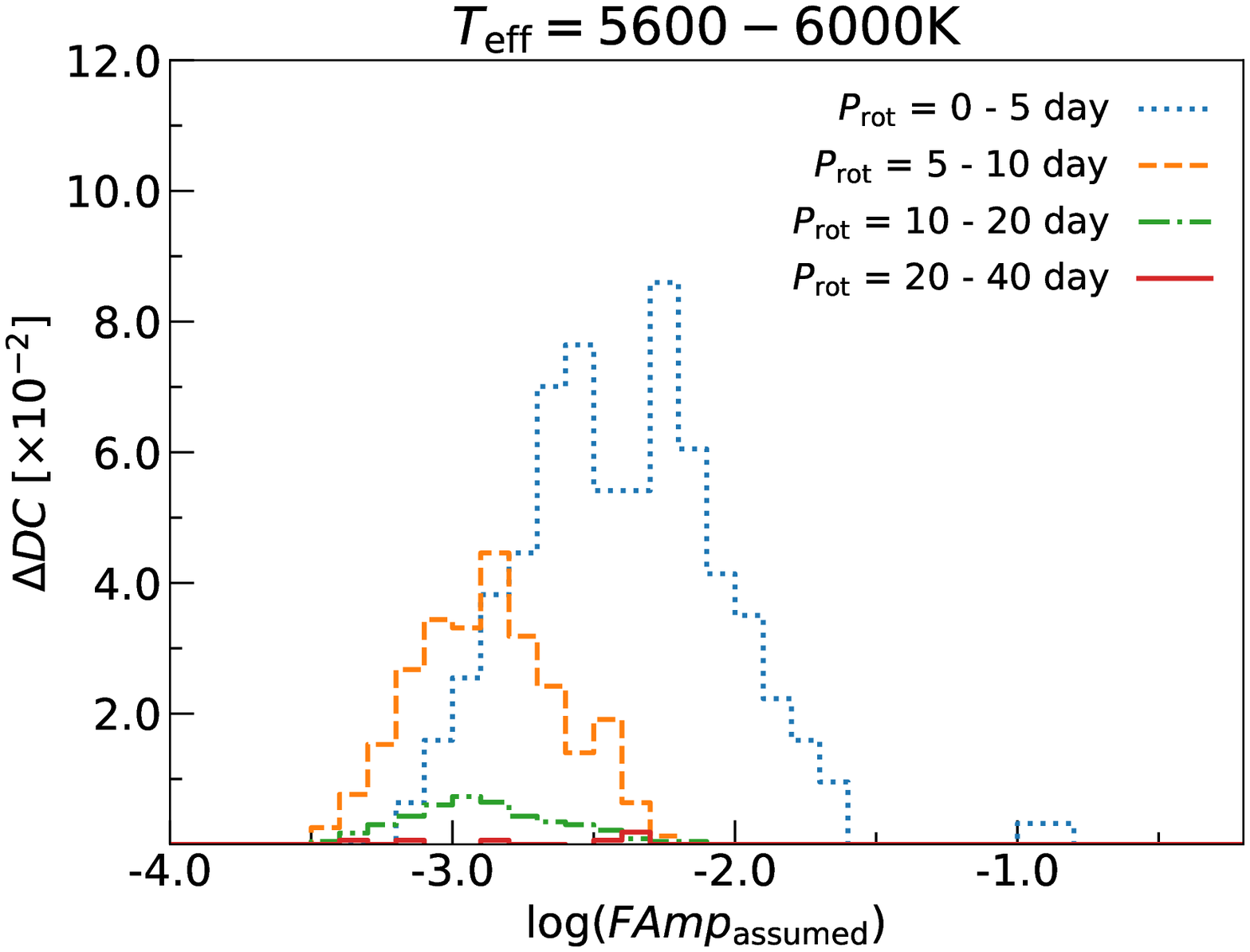}{0.49\textwidth}{\vspace{0mm}(b)}
    }
   \caption{
    Histograms showing how much the high-pass filter has improved the detection completeness ($DC$).
    The vertical axis ($\Delta DC$) is the difference between the detection completeness from the high-pass filtered data ($DC_{\rm filter}$) and without high-pass filter ($DC_{\rm original}$).
    The data are divided in (a) and (b) on the basis of the stellar temperature: (a) $T_{\rm eff} = 5100$ -- $5600$ K and (b) $T_{\rm eff} = 5600$ -- $6000$ K.
   }
   \label{fig:detlim_logamp_prot_filter_solartype}
 \end{figure*}

 This difference means that it is difficult to detect smaller superflares on rapidly-rotating stars and corresponds to the apparent negative correlation between the rotation period and the lower limit of the flare energy seen in Figure \ref{fig:erg_prot_solartype}.
 This difference also can be explained by that rapidly-rotating stars tend to have a larger amplitude of the brightness variations (e.g., Figure \ref{fig:aspot_prot_solartype}) and the $threshold$ values for the flare detection can be larger.
 In addition to Figure \ref{fig:detlim_logamp_prot_solartype} using the filtered lightcurve data, we also calculated $DC_{\rm filter}$ values for the original lightcurve data without the high-pass filter process (cf. Figure \ref{fig:method_filter} (a)\&(b)), $DC_{\rm original}$.
 $\Delta DC$ is then defined as the difference between the detection completeness from the high-pass filtered data ($DC_{\rm filter}$) and without high-pass filter ($DC_{\rm original}$).
 As shown in Figure \ref{fig:detlim_logamp_prot_filter_solartype}, the high-pass filter has caused the detection completeness to  be improved with a few percent especially for the rapidly-rotating stars ($P<5$ days), though large effects of the brightness variations remain as already seen in Figure \ref{fig:detlim_logamp_prot_solartype}.

 We take into account these dependences of the detection completeness (or ``missing rate of flares") on the rotation period and the flare amplitude when we discuss the relationship between the rotation period and superflare frequency in the following Section \ref{subsec:frequency}.

\newpage

\subsection{Frequency distribution of superflares and the dependence on the rotation period} \label{subsec:frequency}
\begin{figure*}[ht!]
    \hspace{0mm}
    \gridline{
    \fig{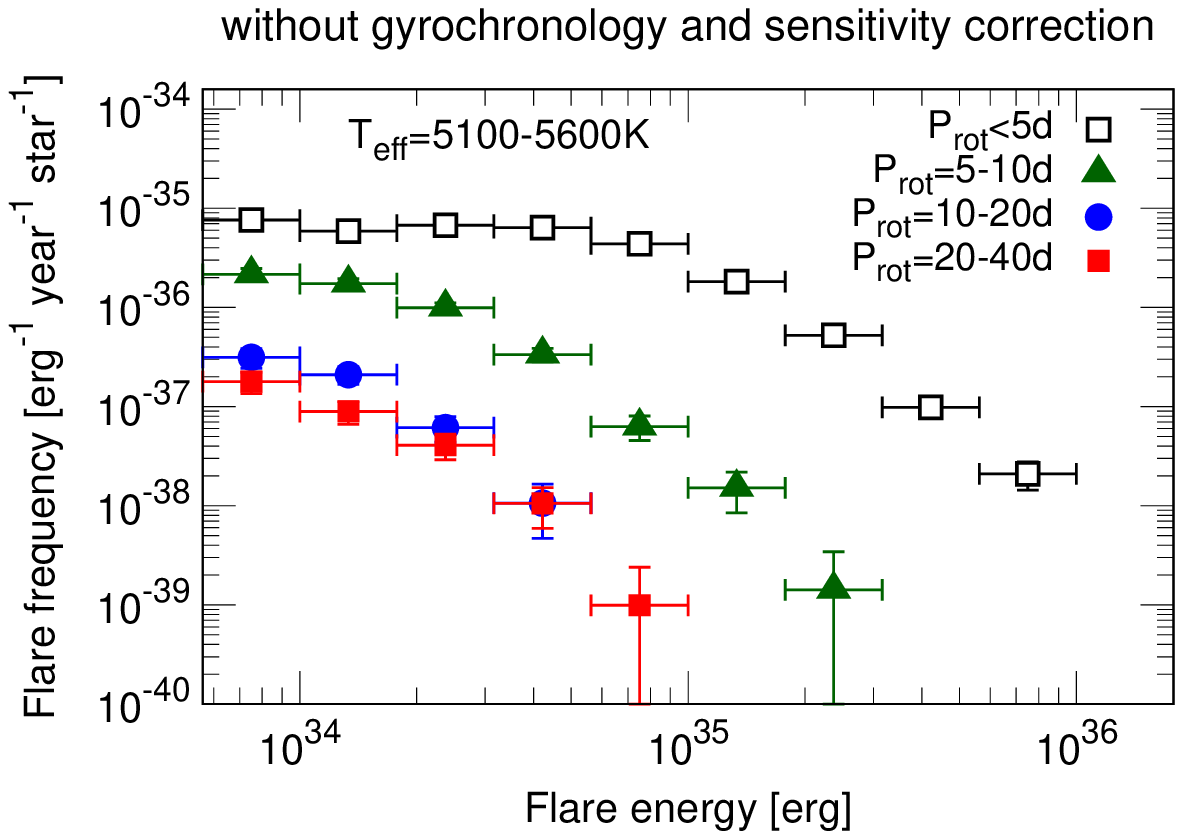}{0.49\textwidth}{\vspace{0mm}(a)}
    \fig{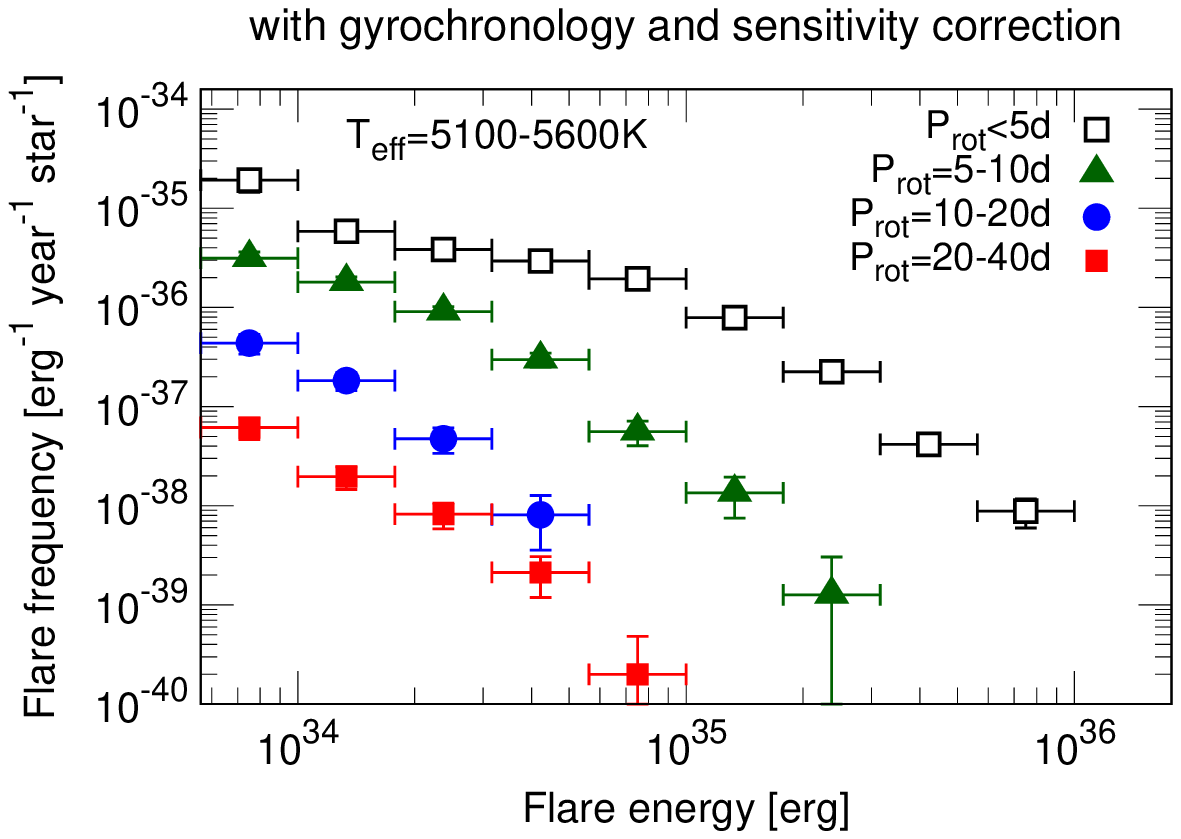}{0.49\textwidth}{\vspace{0mm}(b)}
    }
   \caption{
   Occurrence frequency distribution of superflares on solar-type stars with $T_{\rm{eff}} = 5100$ -- $5600$ K, using the superflare data that are found from $Kepler$ 30-min cadence data of 4-years (Q0-17).
   The horizontal axis is the flare energy, and the error bar in the horizontal axis indicates each energy bin.
   The vertical axis indicates the number of superflares per star, per year, and per unit energy in each energy bin.
   The error bars in the vertical axis are explained in the main text.
   The symbols are classified with rotation period ($P_{\rm rot}$) values:
    open squares for $P_{\rm rot}<5$ days,
    green triangles for $P_{\rm rot}=5$ -- $10$ days,
    blue circles for $P_{\rm rot}=10$ -- $20$ days,
    and the red filled squares for $P_{\rm rot}=20$ -- $40$ days.
    The potential errors from the gyrochronology and the flare detection completeness are not taken into consideration in (a), while in (b) the frequency values are calculated by taking into consideration the effect of gyrochronology and the flare detection completeness.
   }
   \label{fig:Ffreq_Eflr_5100_5600}
 \end{figure*}

\begin{figure*}[ht!]
    \hspace{0mm}
    \gridline{
    \fig{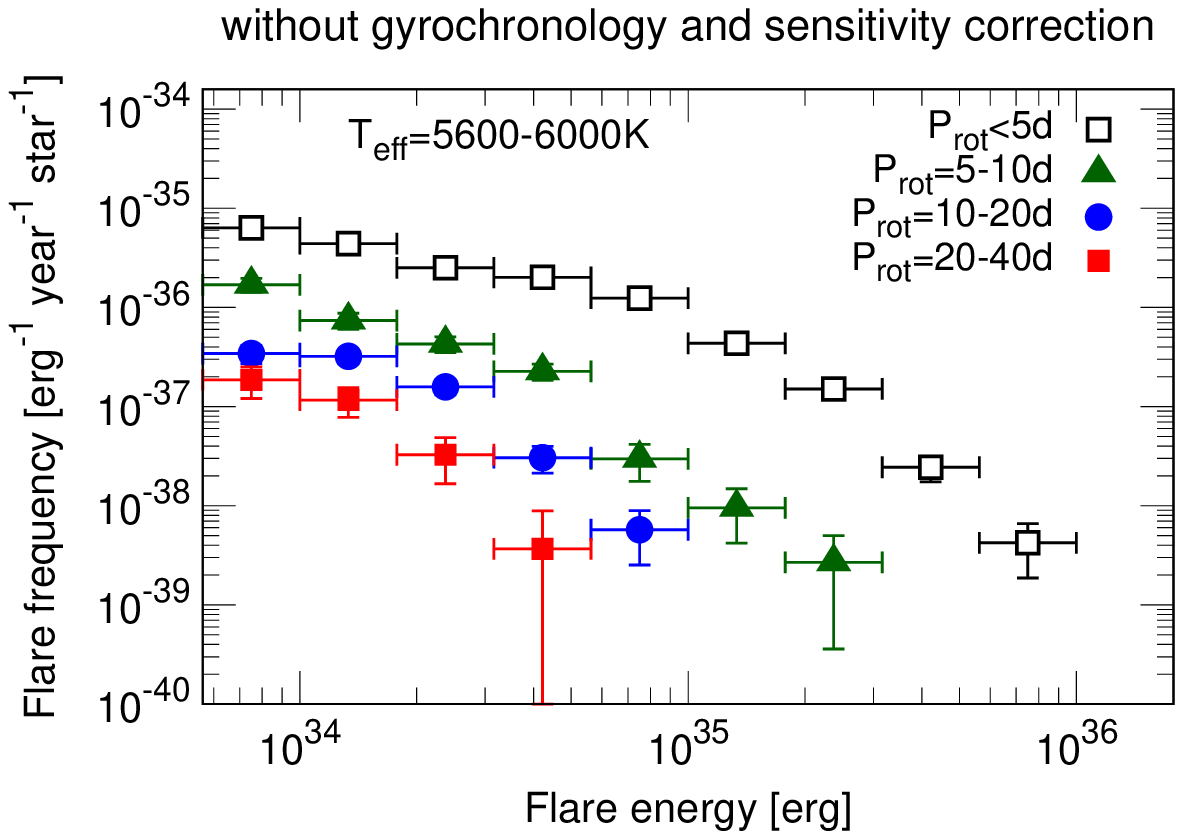}{0.49\textwidth}{\vspace{0mm}(a)}
    \fig{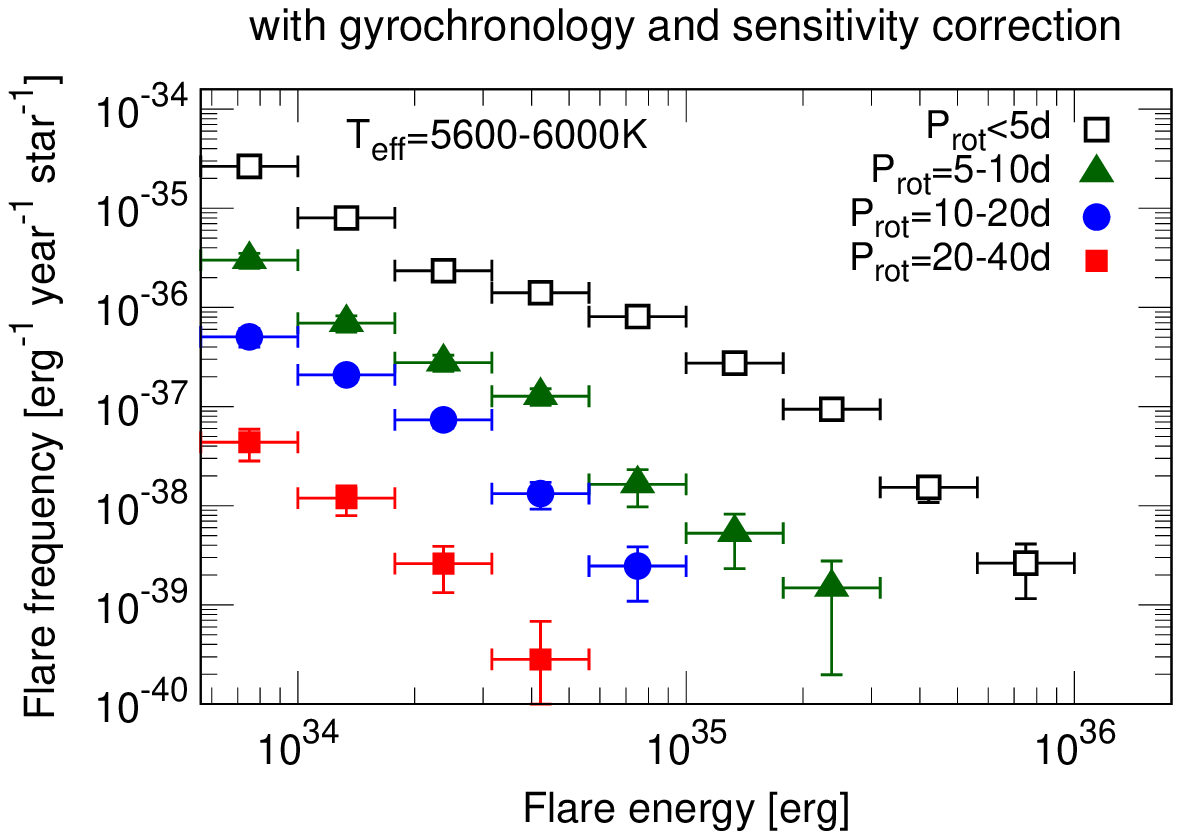}{0.49\textwidth}{\vspace{0mm}(b)}
    }
   \caption{
   Same as Figure \ref{fig:Ffreq_Eflr_5100_5600}, but for $T_{\rm{eff}} = 5600$ -- $6000$ K.
   }
   \label{fig:Ffreq_Eflr_5600_6000}
 \end{figure*}

  We then discuss the flare frequency distributions, and its relation with rotation period, by using the updated sample of superflares from $Kepler$ 4-year data (Q0-17) in this study.
  Figures \ref{fig:Ffreq_Eflr_5100_5600}\&\ref{fig:Ffreq_Eflr_5600_6000} represent the occurrence frequency distributions of superflares on solar-type stars with $T_{\rm{eff}}=5600$ -- $6000$ K and $T_{\rm{eff}}=5600$ -- $6000$ K, respectively, for the $P_{\rm rot}$ ranges of $<5$, 5 -- 10, 10 -- 20, and 20 -- 40 days.
  These occurrence rate values of superflares are estimated for each $P_{\rm rot}$ bin from the number of observed superflares, the number of observed stars, and the total length of the observational period.
  The potential effects from the gyrochronology correction (Section \ref{subsec:gyrochronology}) and the flare detection completeness correction (Section \ref{subsec:sensitivity}) are not taken into consideration in the left panels of Figures \ref{fig:Ffreq_Eflr_5100_5600}(a) and \ref{fig:Ffreq_Eflr_5600_6000}(a).
  These are the same definition of flare frequencies as in our previous studies (e.g., \citealt{Shibayama+2013}; \citealt{Notsu+2019}).
  In the right panels of Figures \ref{fig:Ffreq_Eflr_5100_5600}(b) and \ref{fig:Ffreq_Eflr_5600_6000}(b), the frequency values are calculated by incorporating the effect of the gyrochronology correction (Section \ref{subsec:gyrochronology}) and the flare detection completeness correction (Section \ref{subsec:sensitivity}).
  In order to calculate the frequency of superflares incorporating the effect of gyrochronology correction, we used $N_{\rm{star}}$ instead of $N_{P}$ in Table \ref{table:num_stars_gyro}, as a value of the number of observed stars in each $P_{\rm rot}$ bin.
  Through this, we aim to discuss more accurate values of flare frequencies, especially for slowly-rotating Sun-like stars.
  In order to calculate the frequency of superflares incorporating the effect of the flare detection completeness correction, we correct the number of superflares in each energy and $P_{\rm rot}$ bin by using the relationship among $DC_{\rm filter}$ (the detection completeness), flare amplitude, and $P_{\rm rot}$, shown in Figure \ref{fig:detlim_logamp_prot_solartype}.
  From the value $DC_{\rm filter}$ in Figure \ref{fig:detlim_logamp_prot_solartype}, we can estimate how many flares are missed for each $P_{\rm rot}$ and $E_{\rm flare}$ (flare energy) range.
  When we estimate the flare frequency, we correct the number of superflares in each $P_{\rm rot}$ and $E_{\rm flare}$ by replacing the number of flares with ``(the number of flares)$/DC_{\rm filter}$".
  In this process, we assume the average relationship among flare amplitude, flare duration, and $E_{\rm flare}$ for solar-type stars (cf. energy vs. duration discussed in \citealt{Maehara+2015}).
  It is assumed that first data point with the flare the same value as the $threshold$, the second point is half of the $threshold$, and the flare does not continue longer than an hour.
  This does not include various flares such as long, small flare or superflares with sharp declines.

  We also estimate the potential errors of superflare frequency.
  The number of superflares found in the study is followed by the Poisson statistics and the estimated $\sigma$ is $\sqrt{N_{\rm flare}+1}$, where $N_{\rm flare}$ is the number of flares in each $P_{\rm rot}$ bin.
  We include these errors in Figure \ref{fig:Ffreq_Eflr_5100_5600}\&\ref{fig:Ffreq_Eflr_5600_6000}(a).
  Additionally, we include the following errors in Figure \ref{fig:Ffreq_Eflr_5100_5600}\&\ref{fig:Ffreq_Eflr_5600_6000}(b).
  The error from the sensitivity correction is from the number of stars that satisfies 
  ``$FAmp_{\rm assumed} \geq Threshold$" ($N_{P}^{FAmp_{\rm assumed} \geq Threshold}$).
  We estimate $\sigma$ is $\sqrt{N_{P}^{FAmp_{\rm assumed} \geq Threshold}}$.
  The gyrochronology relation error is from the error values of Equations (12)-(14) of \citet{Mamajek+2008}.

 Comparing the results without and with corrections in Figures \ref{fig:Ffreq_Eflr_5100_5600} and \ref{fig:Ffreq_Eflr_5600_6000} shows are two clear changes.
 First, because of the sensitivity correction, the smaller amplitude events, which was missed originally as shown in Figure \ref{fig:detlim_logamp_prot_solartype}, are now included in the calculation of the flare frequency, and the frequency value in the smaller energy superflares ($\lesssim 10^{34}$ erg) becomes larger.
 The difference is larger especially for rapidly-rotating stars ($P_{\rm rot}<5$ days) since the detection completeness of rapidly-rotating stars is smaller than that of slowly-rotating stars, as shown in Figure \ref{fig:detlim_logamp_prot_solartype}.
 For example, the frequency value of flares with $10^{33.75}$ -- $10^{34.0}$ erg on stars with $T_{\rm{eff}} = 5600$ -- $6000$ K and $P_{\rm rot}<5$ day in Figure \ref{fig:Ffreq_Eflr_5600_6000}(a) (without correction) is $<10^{-35}$ erg$^{-1}$ year$^{-1}$ star$^{-1}$, but that in Figure \ref{fig:Ffreq_Eflr_5600_6000}(b) (with correction) is $>10^{-35}$ erg$^{-1}$ year$^{-1}$ star$^{-1}$.
 Second, because of the gyrochronology correction (using $N_{\rm{star}}$ instead of $N_{P}$ in Table \ref{table:num_stars_gyro}, as a value of the number of observed stars in each $P_{\rm rot}$ bin), the exact values of flare frequencies changes, while the shape of overall distributions (e.g., the power-law distributions in the following) are not affected from this correction.
 For example, the frequency of slowly-rotating Sun-like stars ($T_{\rm{eff}} = 5600$ -- $6000$ K and $P_{\rm rot}=20$ -- $40$ days) becomes about half because of the the gyrochronology correction:  $N_{P}$($P_{\rm rot}=20$ -- $40$ days) $/$ $N_{P,\rm{all}}\sim 32$\% and $N_{\rm{star, all}}$($P_{\rm rot}=20$ -- $40$ days) $/$ $N_{\rm{star, all}}\sim70$\% in Table \ref{table:num_stars_gyro}.

  As our previous studies (e.g., \citealt{Shibayama+2013}; \citealt{Notsu+2019}) presented, these distributions can be described by a power-law ($dN/dE\propto E^{\alpha}$) and rapidly rotating stars tend to have larger frequency values.
  The power-law indexes ($\alpha$) of the distributions in Figures \ref{fig:Ffreq_Eflr_5100_5600}(b)\&\ref{fig:Ffreq_Eflr_5600_6000}(b) are listed in Table \ref{table:power_law_index}.
  We estimated the power-law relation ($dN/dE\propto E^{\alpha}$) using all points from the $P_{\rm rot}$ bins shown in the figures, and the error is shown as the vertical error bars.

   \begin{deluxetable*}{lcc}
   \tablecaption{
    Power-law index ($dE/dN \propto E^\alpha$) of flare frequency distributions in each rotation bin.
   }
   \tablewidth{0pt}
   \tablehead{
     \colhead{} & \colhead{$T_{\rm eff}=5100$ -- $ 5600$ K} & \colhead{$T_{\rm eff}=5600-6000$ K}
   }
   \startdata
     $P_{\rm rot} < 5$           day &  -1.5 $\pm$ 0.1 (-2.9 $\pm$ 0.1) &  -1.8 $\pm$ 0.1 (-3.1 $\pm$ 0.1)\\
     $P_{\rm rot} = 5 $ -- $10$  day &  -1.9 $\pm$ 0.2 (-2.9 $\pm$ 0.4) & -2.1 $\pm$ 0.1 (-2.1 $\pm$ 0.1)\\
     $P_{\rm rot} = 10 $ -- $20$ day & -2.2 $\pm$ 0.2 (-2.6 $\pm$ 0.2) & -2.2 $\pm$ 0.1 (-3.0 $\pm$ 0.1)\\
     $P_{\rm rot} = 20 $ -- $40$ day & -2.1 $\pm$ 0.2 (-2.9 $\pm$ 0.5) & -2.7 $\pm$ 0.1 (-3.0 $\pm$ 0.3)\\
   \enddata
   \tablecomments{
    We calculate the power-law index from Figure \ref{fig:Ffreq_Eflr_5100_5600} (b) and Figure \ref{fig:Ffreq_Eflr_5600_6000} (b).
    The power-law index without brackets is calculated from all points in each $P_{\rm rot}$ bin, while the numbers within brackets show the power-law index using three larger flare energy points in each $P_{\rm rot}$ bin.
   }
   \label{table:power_law_index}
 \end{deluxetable*}

  The indexes are roughly around $\alpha \sim -2$, which can be consistent with previous studies of superflares on solar-type stars (e.g., \citealt{Shibata+2013}; \citealt{Shibayama+2013}; \citealt{Maehara+2015}; \citealt{Notsu+2019}; \citealt{Tu+2020}).
  We should note that the sensitivity correction in this study (cf. Figure \ref{fig:detlim_logamp_prot_solartype}) is a very simple one only discussing the flare amplitude, and can be an incomplete correction, as we described above.
  Because of this, the power-law distributions in the smaller energy region ($\lesssim 10^{34}$ erg) can have some systematic errors that are not included in the error bars in Figures \ref{fig:Ffreq_Eflr_5100_5600}(b)\&\ref{fig:Ffreq_Eflr_5600_6000}(b).
  We should keep in mind when we see the power-law indexes listed in Table \ref{table:power_law_index}.

  The upper-limit values of flare energy roughly depend on rotation period, as already seen in Figure \ref{fig:erg_prot_solartype}.
  As also seen in the index values listed in Table \ref{table:power_law_index}, we can see that the power-law distribution becomes a bit steeper as the flare energy becomes larger.
  For example, the power-law index of superflares on Sun-like stars ($T_{\rm{eff}} = 5600$ -- $6000$ K and $P_{\rm rot}=20$ -- $40$ days) in Figure \ref{fig:Ffreq_Eflr_5600_6000}(b) is 
  $\alpha \sim$ -2.7 
  if we calculate use the data of $10^{33.75} < E_{\rm flare} < 10^{34.75}$ erg.
 However, if we use the data of $10^{34.0} < E_{\rm flare} < 10^{34.75}$, the power-law index $\alpha$ is
 $\sim$ -3.0.
 Similar tendencies can be also seen for the stars with $P_{\rm rot}<20$ days in Table \ref{table:power_law_index}, while we should note that the number of larger energy flares is small and the exact values of $\alpha$ should be treated with large caution.
 These tendencies might be related to the maximum flare energy cutoff of solar-type stars since there is a correlation between the maximum spot coverage and the rotation period (Figure \ref{fig:aspot_prot_solartype}) and the spot size can explain the flare energy (Figure \ref{fig:erg_aspt}).
 In the case of old (age $\sim$ 4.6 Gyr) slowly-rotating Sun-like stars with $T_{\rm{eff}} = 5600-6000$ K and $P_{\rm{rot}}\sim 25$ days, as also described in Section \ref{subsec:spot_rotation}, Figures \ref{fig:aspot_prot_solartype}(b) show that the maximum size of starspots is a few \% of the solar hemisphere.
 This corresponds to $10^{34}-10^{35}$ erg on the basis of Equation (\ref{eq:erg_aspt}), and the upper limit of superflare energy of these Sun-like stars in Figure \ref{fig:erg_prot_solartype} and \ref{fig:Ffreq_Eflr_5600_6000}(b) is roughly in the same range.

\begin{figure*}[ht]
    \gridline{
    \fig{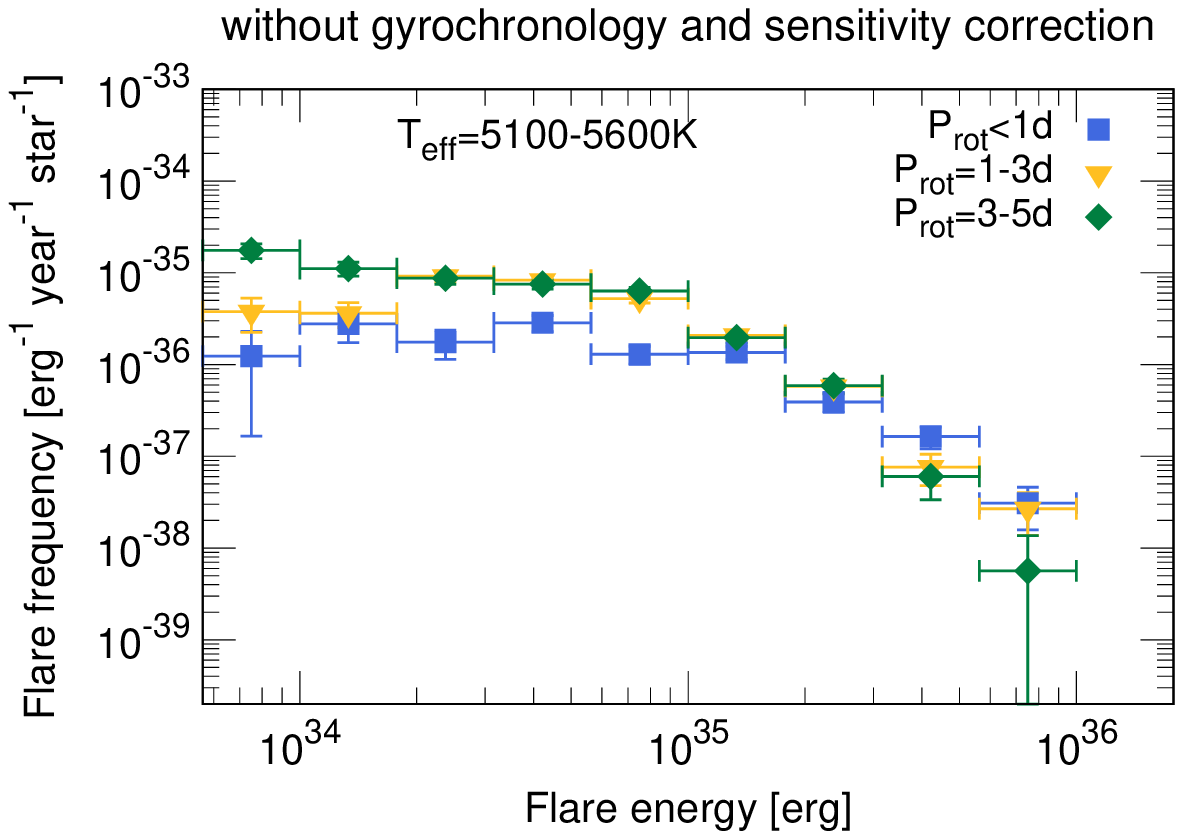}{0.49\textwidth}{\vspace{0mm}(a)}
    \fig{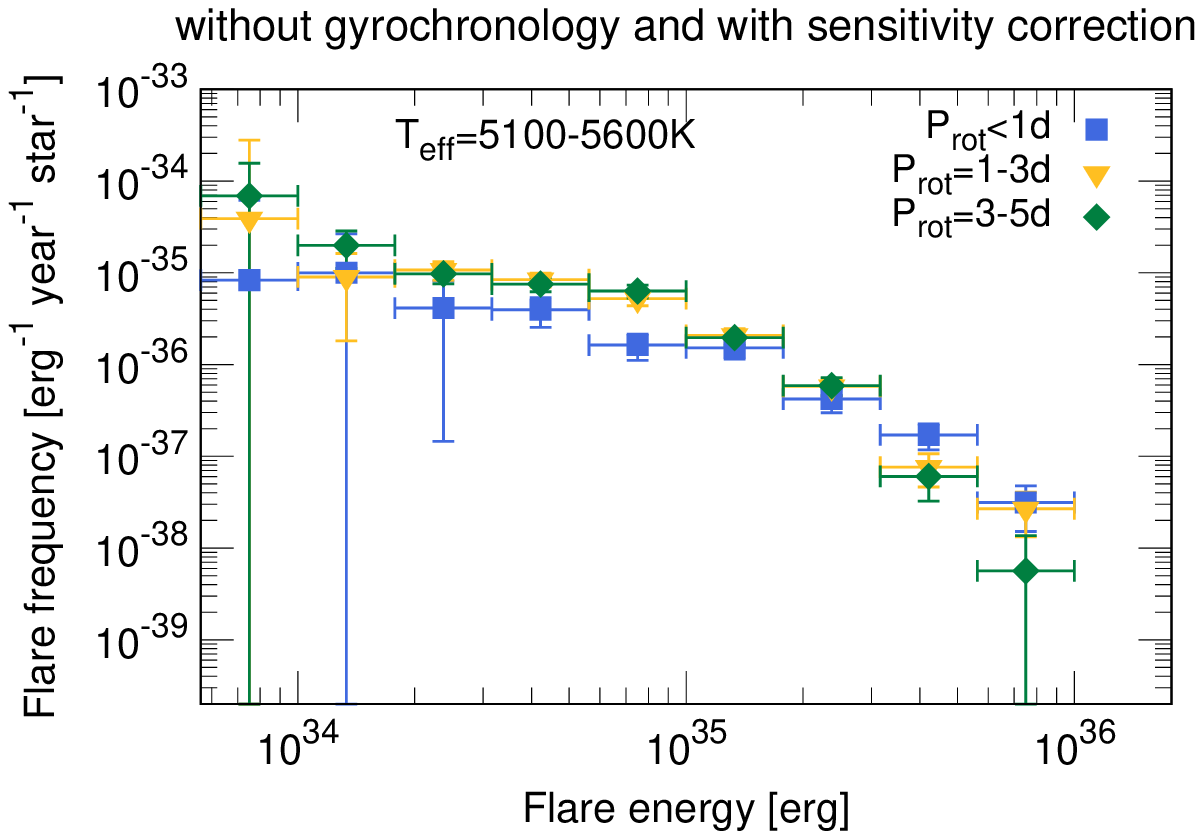}{0.49\textwidth}{\vspace{0mm}(b)}
    }
   \caption{
    Occurrence frequency distribution of superflares on solar-type stars with $T_{\rm{eff}} = 5100$ -- $5600$ K and the short rotation periods ($P_{\rm rot}<5$ days).
    The symbols are classified with rotation period ($P_{\rm rot}$) values:
    blue squares for $P_{\rm rot}<1$ days,
    orange triangles for $P_{\rm rot}=1$ -- $3$ days,
    green diamonds for $P_{\rm rot}=3$ -- $5$ days.
    The horizontal axes, the vertical axes, definitions of error bars are the same as Figure \ref{fig:Ffreq_Eflr_5100_5600}, while we note the range of the vertical axes is a bit different from Figure \ref{fig:Ffreq_Eflr_5100_5600}.
    The potential errors from the gyrochronology and the flare detection completeness are not taken into consideration in (a), while in (b) the frequency values are calculated by taking into consideration the effect of the flare detection completeness.
    Different from Figure \ref{fig:Ffreq_Eflr_5100_5600}(b), the gyrochronology effect is not considered in (b) of this figure, since the age–rotation relation (gyrochronology relation) of young solar-type stars can have a large scatter (e.g., \citet{Soderblom+1993}; \citealt{Ayres+1997}; \citealt{Tu+2015}), as mentioned in Section \ref{subsec:energy_rotation}.
   }
   \label{fig:Ffreq_Eflr_short_5100_5600}
 \end{figure*}

\begin{figure*}[ht!]
    \gridline{
    \fig{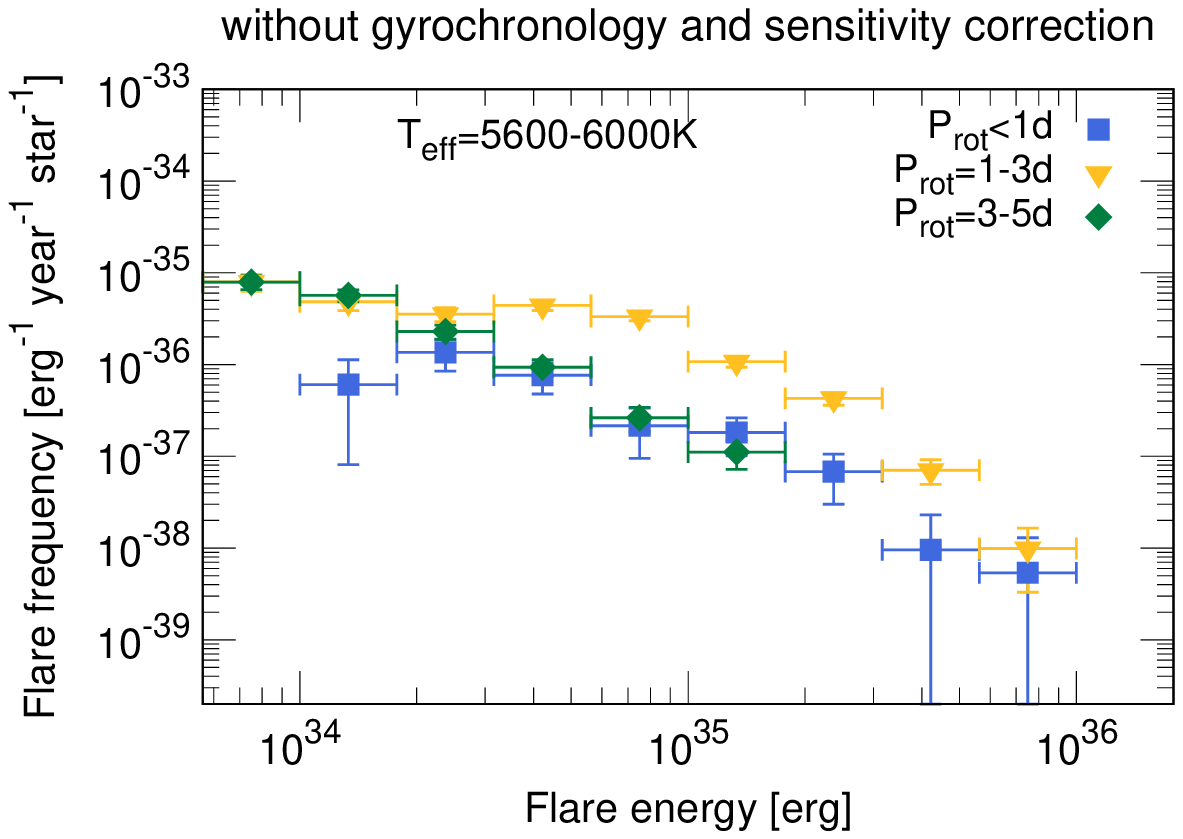}{0.49\textwidth}{\vspace{0mm}(a)}
    \fig{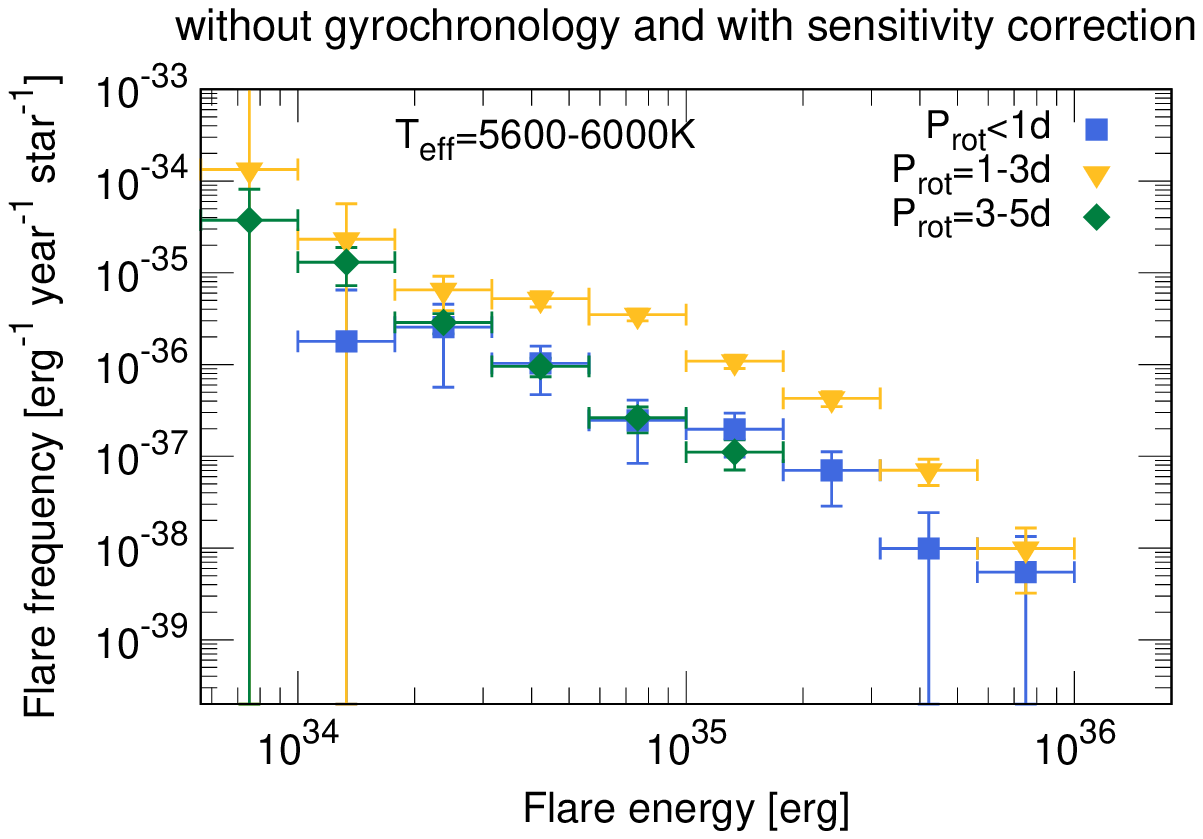}{0.49\textwidth}{\vspace{0mm}(b)}
    }
   \caption{
    Same as Figure \ref{fig:Ffreq_Eflr_short_5100_5600}, but for the stars with $T_{\rm{eff}} = 5600$ -- $6000$ K.
   }
   \label{fig:Ffreq_Eflr_short_5600_6000}
 \end{figure*}

 Next for reference, we see the flare frequency distributions of rapidly-rotating stars with $P_{\rm{rot}}<5$ days in a bit more detail.
 Like Figures \ref{fig:Ffreq_Eflr_5100_5600}\&\ref{fig:Ffreq_Eflr_5600_6000}, Figures \ref{fig:Ffreq_Eflr_short_5100_5600}(a)\& \ref{fig:Ffreq_Eflr_short_5600_6000}(a) are the flare frequency distributions of $P_{\rm{rot}}<1$, $1$ -- $3$, $3$ -- $5$ days without the gyrochronology and sensitivity corrections, and Figures \ref{fig:Ffreq_Eflr_short_5100_5600}(b)\& \ref{fig:Ffreq_Eflr_short_5600_6000}(b) are those with only the sensitivity correction.
 Since the gyrochronology relation has a larger scatter in younger age ($\lesssim 0.5$ -- $0.6$ Gyr) as described in Section \ref{subsec:energy_rotation}, the gyrochronology correction is not included in Figures \ref{fig:Ffreq_Eflr_short_5100_5600}(b)\& \ref{fig:Ffreq_Eflr_short_5600_6000}(b).
 Although the original sample size of rapidly-rotating stars is small (see $N_{P}$ values in Table \ref{table:num_stars_gyro}), we can still see the power-law distributions for very rapidly-rotating stars in the range of $E_{\rm flare} \gtrsim 10^{34.5}$ erg (e.g., $P_{\rm{rot}}<1$ day).

\clearpage

\begin{figure*}[ht!]
    \hspace{0mm}
    \gridline{
    \fig{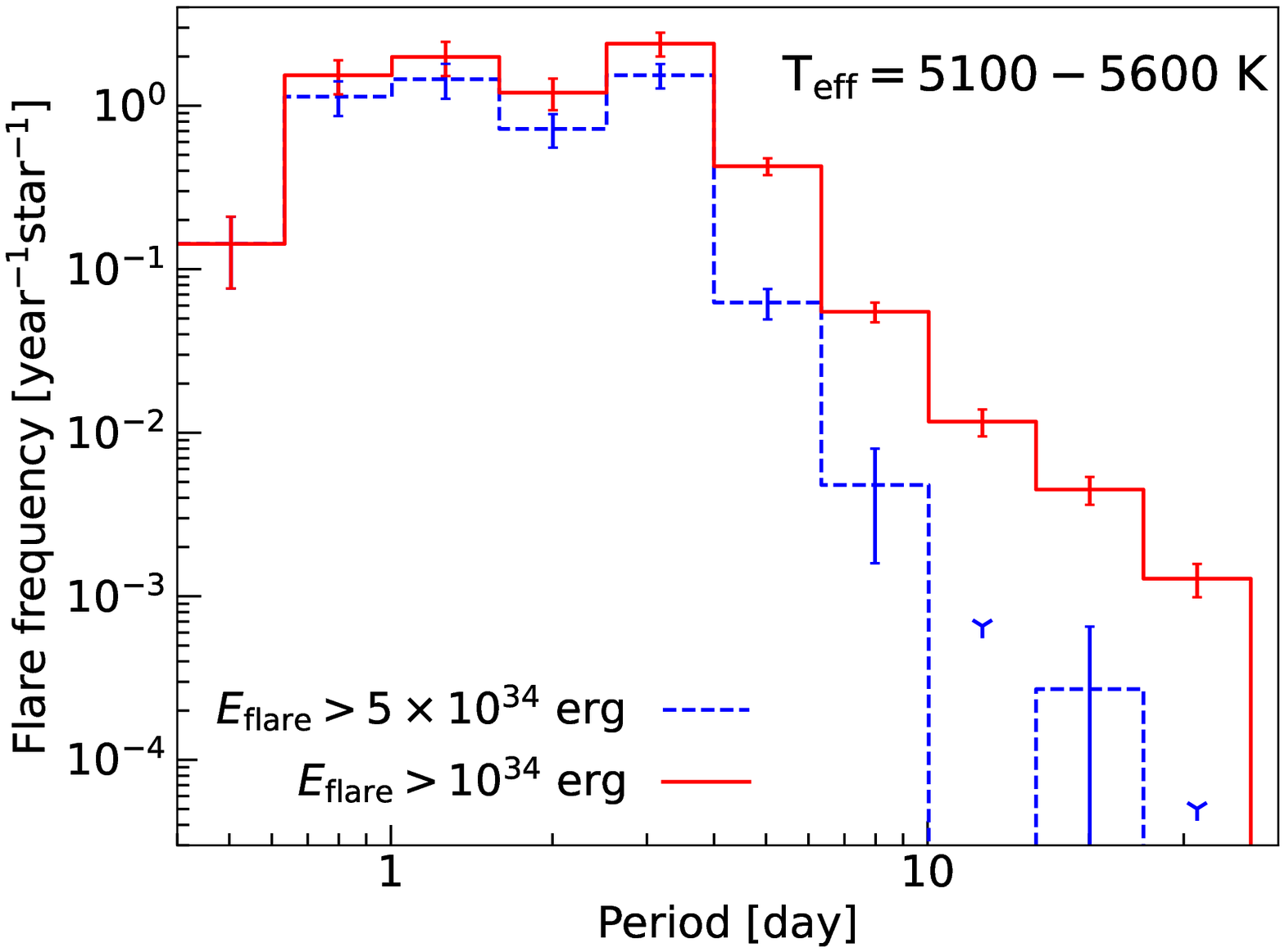}{0.49\textwidth}{\vspace{0mm}(a)}
    \fig{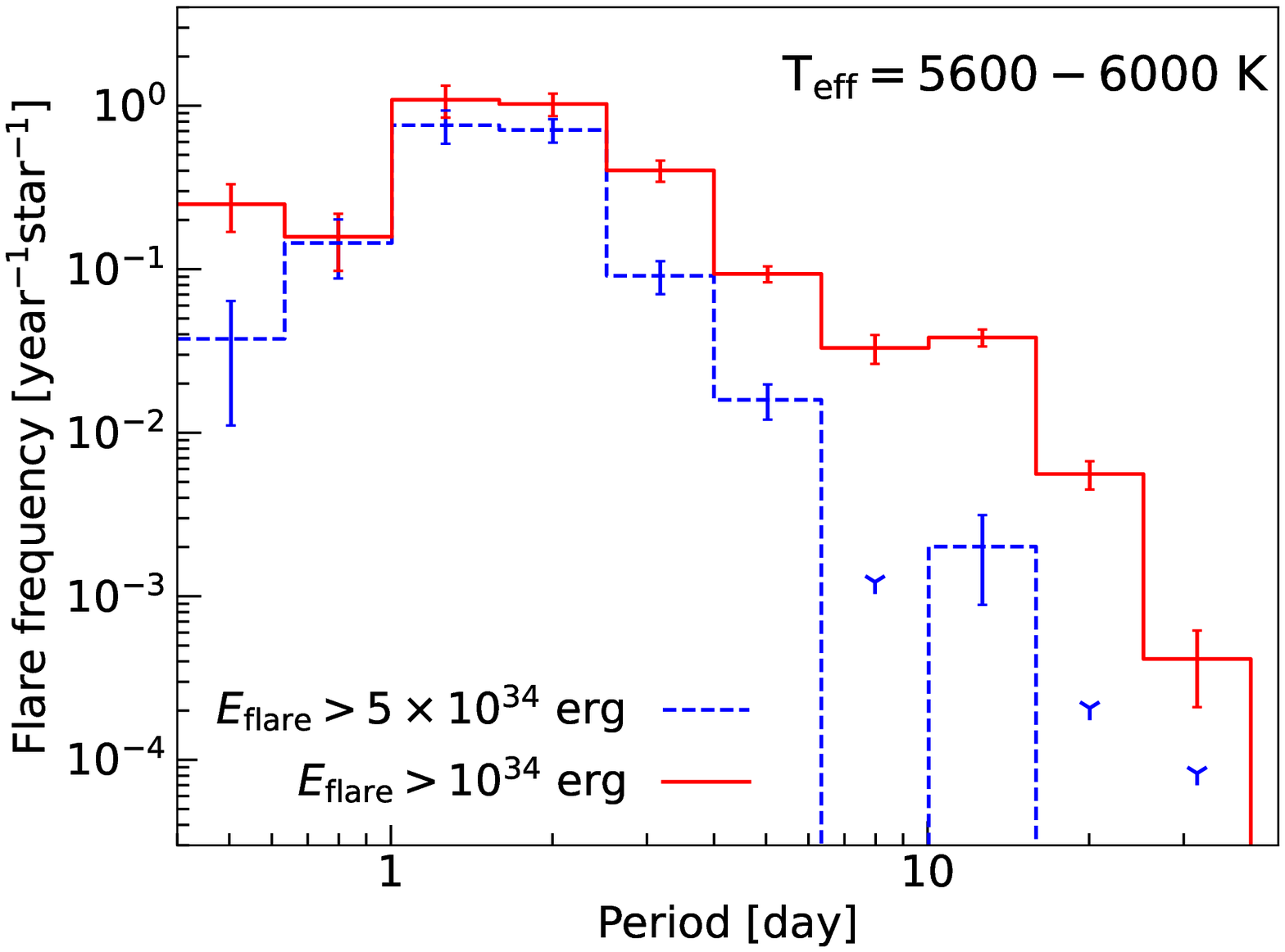}{0.49\textwidth}{\vspace{0mm}(b)}
    }
   \caption{
    Occurrence frequency distributions of superflares as a function of the rotation period ($P_{\rm rot}$), using the superflare data that are found from $Kepler$ 30-min cadence data of 4-years (Q0-17).
    The figures are separated into two temperature ranges: (a) $T_{\rm{eff}} = 5100$ -- $5600$ K and (b) $T_{\rm{eff}} = 5600$ -- $6000$ K.
    The vertical axes indicate the number of superflares with energy $>5\times 10^{34}$ erg (blue dashed lines) and $>10^{34}$ erg (red filled lines) per star and year.
    Error bars in the vertical axis represent 1-$\sigma$ uncertainty of the frequency, which is estimated in the same way as Figures \ref{fig:Ffreq_Eflr_5100_5600}\&\ref{fig:Ffreq_Eflr_5600_6000}.
    In the long period range ($P_{\rm rot} \geq 5$ day), the frequency values are calculated by taking into consideration the effect of gyrochronology and the flare detection completeness, as done in Figures \ref{fig:Ffreq_Eflr_5100_5600}(b)\&\ref{fig:Ffreq_Eflr_5600_6000}(b).
    In the short period range  ($P_{\rm rot} < 5$ days) only the flare detection completeness is considered, as done in Figures \ref{fig:Ffreq_Eflr_short_5100_5600}(b)\&\ref{fig:Ffreq_Eflr_short_5600_6000}(b).
    As for the blue dashed lines, in the case of no events in a period bin, the upper-limit values are shown with the blue Y-shape points assuming that less than one event occurs in each bin.
   }
   \label{fig:freq_prot_solartype_hist}
 \end{figure*}

 We then see the relation between flare frequency and rotation period, by using the superflare data that include the above gyrochronology and sensitivity corrections.
 Figure \ref{fig:freq_prot_solartype_hist} shows that the average flare frequency in a given period bin tends to decrease as the period increases in the range of $P_{\rm rot}$ longer than a few days, as we have already seen the same tendency in Figures \ref{fig:Ffreq_Eflr_5100_5600}(b)\&\ref{fig:Ffreq_Eflr_5600_6000}(b).
 This result is basically the same as that presented in our previous study \citep{Notsu+2019}, but the sample size is much larger especially for slowly-rotating stars (cf. Section \ref{subsec:example}).
 The frequency of superflares on young, rapidly rotating stars ($P_{\rm rot} = $1 -- 3 days) is $\sim$100 times higher compared with old, slowly rotating stars ($P_{\rm rot} > 20$ days) and this indicates that as a star evolves (and its rotational period increases), the frequency of superflares decreases.
 In contrast, the flare frequency is roughly constant in the range of $P_{\rm rot} \lesssim 3$ day.
 We note that in this range of $P_{\rm rot} \lesssim 3$ day, flare frequency distributions still show similar power-law distributions (Figures \ref{fig:Ffreq_Eflr_short_5100_5600}(b)\&\ref{fig:Ffreq_Eflr_short_5600_6000}(b)).
 We can now interpret that this correlation between the rotation period (roughly corresponding to age) and flare frequency is consistent with the correlation between the rotation period (age) and the quiescent stellar activity level such as the average X-ray luminosity (e.g., \citealt{Noyes+1984}; \citealt{Gudel+2007}; \citealt{Wright+2011}).
 The correlation between the rotation period and the X-ray luminosity of solar-type stars also shows a ``saturation" regime in the range of $P_{\rm rot} \lesssim 3$ days, and ``decreasing trend" in the range of $P_{\rm rot} \gtrsim 3$ days.

\subsection{Superflare Frequency on Sun-like Stars and Implications for the Sun} \label{subsec:superflare_sunlike_sun}
\begin{figure*}[ht!]
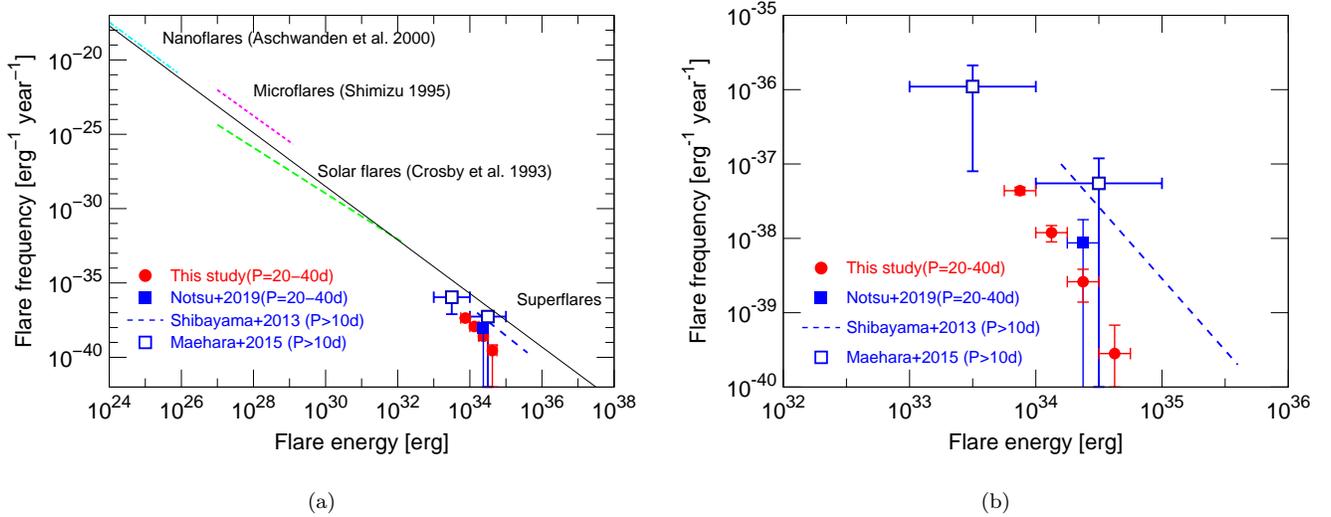

    \gridline{
    \fig{figure/freq_erg_sunlike/sunstar_flare_freq_shibayama_notsu_okamoto_filter.eps}{0.49\textwidth}{\vspace{0mm}(a)}
    \fig{figure/freq_erg_sunlike/star_flare_freq_shibayama_notsu_okamoto_zoom_filter.eps}{0.49\textwidth}{\vspace{0mm}(b)}
    }
   \caption{
   (a)
   Comparison between the frequency distribution of superflares on Sun-like stars and solar flares.
   The red filled circles, blue filled square, blue dashed line, and blue open squares indicate the occurrence frequency distributions of superflares on Sun-like stars (slowly rotating solar-type stars with $T_{\rm{eff}}$=5600 -- 6000 K).
   The red filled circles correspond to the updated frequency values of superflares on Sun-like stars with $P_{\rm{rot}}$=20 -- 40 days, which are calculated with gyrochronology and sensitivity correction in this study and presented in Figure \ref{fig:Ffreq_Eflr_5600_6000} (b).
   Horizontal and vertical error bars are the same as those in Figure \ref{fig:Ffreq_Eflr_5600_6000} (b).
   For reference, the blue filled squares are the values of Sun-like stars with $P_{\rm{rot}}$=20 -- 40 days, which we presented in Figure 17 of \citet{Notsu+2019}.
   Also for reference, the blue dashed line and blue open squares are the values of superflares on the stars with $P_{\rm{rot}} >$ 10 days, which we presented in Figure 4 of \citet{Maehara+2015} on the basis of original superflare data using Kepler 30-minute cadence data \citep{Shibayama+2013} and 1-minute cadence data \citep{Maehara+2015}, respectively.
   Three dashed lines on the upper left side of this figure indicate the power-law frequency distribution of solar flares observed in hard X-ray \citep{Crosby+1993}, soft X-ray \citep{Shimizu+1995}, and EUV\citep{Aschwanden+2000}.
   Occurrence frequency distributions of superflares on Sun-like stars and solar flares are roughly on the same power-law line with an index of -1.8 (black solid line) for the wide energy range between $10^{24}$ and $10^{35}$ erg.
   (b) Same as (a) but only the data of superflares are plotted.
   }
   \label{fig:flarefreq_sunstar}
 \end{figure*}

 Our previous study (\citealt{Shibayama+2013}; \citealt{Maehara+2015}: \citealt{Notsu+2019}) pointed out that the frequency distribution of superflares on Sun-like stars is roughly on the same power-law line.
 However, the definition of Sun-like stars in \citet{Shibayama+2013} and \citet{Maehara+2015} ($T_{\rm{eff}}=5600-6000$ K and $P_{\rm{rot}}>10$ days) is different from that of this study ($T_{\rm{eff}}=5600$ -- $6000$ K and $P_{\rm{rot}}=20$ -- $40$ days).
 This indicates many younger stars are included (e.g., stars with $P_{\rm{rot}}\sim 10$ days have an age of $t\sim 1$ Gyr).
 \citet{Notsu+2019} used the same definition of Sun-like stars as this study in order to discuss only the data of stars rotating as slowly as the Sun ($P_{\rm{rot}}\sim 25$ days and $t\sim 4.6$ Gyr).
 However, as also described in the above of this paper, the sample size of \citet{Notsu+2019} was so small that we were not able to discuss the detailed properties accurately.
 Moreover, different from this study, our previous studies (\citealt{Shibayama+2013}; \citealt{Maehara+2015}: \citealt{Notsu+2019}) did not include the effect of gyrochronology and flare detection completeness (sensitivity) correction when discussing the flare frequency distribution.
 Then, in Figure \ref{fig:flarefreq_sunstar}, we newly plot the frequency value of superflares on Sun-like stars with $P_{\rm rot}=$20 -- 40 days ($t>3.2$ Gyr) taken from Figure \ref{fig:Ffreq_Eflr_5600_6000} (b), in addition to the data of solar flares and superflares shown in Figure 17 of \citet{Notsu+2019}.

 The newly added data points of superflares on Sun-like stars are roughly on the same power-law line.
 The exact value of the superflare frequency with $P_{\rm{rot}} = 20$ -- $40$ days in this study is a bit smaller than that of \citet{Notsu+2019} as clearly seen in Figure \ref{fig:flarefreq_sunstar}(b), mainly because the gyrochronology correction newly included in this study.
 As also mentioned in Section \ref{subsec:frequency}, it might be possible that the frequency distribution of superflares becomes steeper around the upper limit of the superflare energy ($E_{\rm{flare}}\sim 5\times10^{34}$ erg), but we cannot judge whether this possibility is true in the case of superflares on Sun-like stars, because of the limited number of superflares detected on Sun-like stars in this study (cf. Table \ref{table:params_sunlike}).
 We also note that some of Sun-like superflare stars plotted in Figure \ref{fig:flarefreq_sunstar} have ``Flag 1" in Table \ref{table:params_sunlike}.
 This means it is difficult to finally judge whether these stars really have rotation periods as slow as the Sun ($P_{\rm rot} = 20$ -- $40$ days) and we need some cautions (see Appendix \ref{app:allflare_sunlike}) to discuss the frequency of superflares on Sun-like stars in Figure \ref{fig:flarefreq_sunstar}.
 More investigations including spectroscopic measurements of rotation periods for all of these slowly-rotating Sun-like superflare stars (e.g., \citealt{Notsu+2015b}; \citealt{Notsu+2019}) are needed in the future.

 As a result of Figure \ref{fig:flarefreq_sunstar}, we can roughly remark that superflares with energy $E_{\rm flare} \sim 10^{33.75}$, $\sim 10^{34.0}$, $\sim 10^{34.25}$, and $\sim 10^{34.5}$ erg would be approximately once every 
 $\sim3,000$, $\sim6,000$, $\sim16,000$, and $\sim85,000$ yr 
 on old slowly-rotating Sun-like stars with $P_{\rm rot}\sim25$ days and $t\sim4.6$ Gyr.
 These results of the superflare frequency distribution of Sun-like superflare stars (e.g., Figure \ref{fig:flarefreq_sunstar}) have shown the possibility that superflares whose energy is 
 $\sim 7 \times 10^{33}$ erg ($\sim$X700-class flares; cf. Figure \ref{fig:erg_aspt}) can occur on our Sun once every 
 $\sim$3000 years, 
 those $\sim 1 \times 10^{34}$ erg ($\sim$X1000-class flares; cf. Figure \ref{fig:erg_aspt}) once every 
 $\sim$6000 years, 
 and those larger than $1 \times 10^{34}$ erg can occur.
 This possibility of superflares on our Sun is now supported more strongly in this study because of much larger number of  superflares on slowly-rotating Sun-like stars compared with our previous studies.

 In contrast, the possibility of superflares with energy $>10^{34}$ erg has been questioned by several studies.
 \citet{Aulanier+2013} suggested it is unlikely that the current solar convective dynamo can produce giant sunspot groups necessary for such large superflares (cf. Figure \ref{fig:erg_aspt}), since such giant sunspot groups have not been recorded in the previous solar observations over a few hundred years (cf. \citealt{Schrijver+2012}).
 \citet{Schmieder+2018} also supported the idea that with our Sun as it is today, it seems impossible to get larger sunspots and superflares with energy $>10^{34}$ erg.
 However, in Section \ref{subsec:spot_rotation}, we have already seen that slowly-rotating Sun-like stars ($T_{\rm eff} = 5600$ -- $6000$ K, $P_{\rm rot} \sim 25$ days, and $t \sim $4.6 Gyr) have starspots with the size of $\sim1$\% of the solar hemisphere (Figure \ref{fig:aspot_prot_solartype}).
 This spot size value is enough for $\gtrsim10^{34}$ erg superflares on the basis of Equation (\ref{eq:erg_aspt}), and the upper limit of superflare energy of these Sun-like stars in this study (e.g., Figures \ref{fig:erg_prot_solartype} \& \ref{fig:flarefreq_sunstar}) is roughly in the same range.
 The possible existence of such large starspots on Sun-like stars and the Sun are also suggested in the $Kepler$ data analyses (\citealt{Maehara+2017}; \citealt{Notsu+2019}).
 \citet{Shibata+2013} also suggested that even the current Sun can generate the large magnetic flux necessary for $10^{34}$ erg superflares in a typical dynamo model.

 Moreover, \citet{Maehara+2017} and \citet{Notsu+2019} also investigated the size–frequency distribution of these large starspot groups on Sun-like stars and that of sunspots, and revealed that both sunspots and larger starspots could be related to the same physical processes.
 In Figure 18 of \citet{Notsu+2019}, they found that the occurrence frequency decreases as the area of sunspots or starspots increases, and the distribution of sunspot groups for large sunspots is roughly on this power-law line, although the appearance frequency of sunspots with spot areas of $\sim$ 1\% of the solar hemisphere is about 10 times lower than that of starspots on Sun-like stars.
 The difference between the Sun and Sun-like stars might be caused by the lack of a ``superactive" phase on our Sun during the past 140 yr \citep{Schrijver+2012}.
 The upper limit of starspot size values of Sun-like stars in Figure 18 of \citet{Notsu+2019} is approximately a few percent of the solar hemisphere and the appearance frequency of these spots is approximately once every 2000–3000 yr.
 As a result, these results of starspot frequency in \citet{Maehara+2017} and \citet{Notsu+2019} can be consistent with those of superflare frequency in the above of this section (e.g., Figure \ref{fig:flarefreq_sunstar}).

 Finally, it is interesting to note that several potential candidates of extreme solar flare events, which can be bigger than the largest solar flare in the past 200 yr ($E_{\rm flare} \sim 10^{32}$ erg), have been reported from the data over the recent 1000 -- 2000 yr (e.g., \citealt{Usoskin+2017} and \citealt{Miyake+2019} for reviews).
 For example, significant radioisotope $^{14}$C and $^{10}$Be enhancements have been reported in tree rings and ice cores for the years AD 775, AD 994, and BC 660.
 They suggest extremely strong and rapid cosmic-ray increase events possibly caused by extreme solar flares (e.g., \citealt{Miyake+2012}; \citealt{Miyake+2013}; \citealt{Mekhaldi+2015}; \citealt{OHare+2019}).
 These potential high-energy particle events can be caused by the superflares whose energies are at least larger than $10^{33}$ erg \citep{Cliver+2020}, although there is a large scatter between the flare energy and flux of energetic particles on the surface of the Earth (e.g., \citealt{Takahashi+2016}) and more investigations are necessary.
 Besides, various potential low-latitude auroral drawings and large sunspots drawings have also been reported from the historical documents around the world, and they suggest the possibility that extreme solar flare events have occurred several times in the recent $\sim1000$ yr (e.g., \citealt{Hayakawa+2017b} \& \citeyear{Hayakawa+2017a}).
 In the future studies, the superflare frequency information discussed in this study (e.g., Figure \ref{fig:flarefreq_sunstar}) should be compared with these radioisotope and historical data in detail, so that we can discuss the possibility of extreme solar flare events from various points of view.

\subsection{Superflare stars having exoplanets} \label{subsec:exoplanets}
 \citet{Rubenstein+2000} argued that stars having a hot Jupiter-like companion are good candidates for superflare stars, since the hot Jupiter may play a role of a companion star in binary stars, such as in RS CVn stars.
 RS CVn stars are magnetically very active and produce many superflares.
 Since the Sun do not have hot Jupiters, \citet{Schaefer+2000} concluded our Sun has never generated superflares.
 In order to check the argument, we checked whether the superflare solar-type stars found in this study have any exoplanets, using the data retrieved from the NASA Exoplanet Archive
 \footnote{\url{https://exoplanetarchive.ipac.caltech.edu/}}
 on May 28th, 2020.
 As a result, the solar-type superflare stars found in this study have no confirmed exoplanets, while 3 candidate exoplanets
  \footnote{KIC 008946267, KIC 008040308 and KIC 008822421.}
 and 3 false-positive exoplanets
  \footnote{KIC 005991070, KIC 008285970 and KIC 012207117.}
 are found.
 This suggests that hot Jupiter is not a necessary condition for superflares on solar-type stars.
 In addition, \citet{Shibata+2013} also found from theoretical considerastions that hot Jupiters do not play any essential role in the generation of magnetic flux in the star itself, if we consider only the magnetic interaction between the star and the hot Jupiter, whereas the tidal interaction remains to be a possible cause of enhanced dynamo activity, though more detailed studies would be necessary.
 This seems to be consistent with the above observational finding that hot Jupiter is not a necessary condition for superflares on solar-type stars.

\section{Conclusion}
\label{sec:conclusion}
 In our previous studies (\citealt{Maehara+2012},\citeyear{Maehara+2015},\&\citeyear{Maehara+2017}; \citealt{Shibayama+2013}; \citealt{Notsu+2013},\citeyear{Notsu+2015b},\&\citeyear{Notsu+2019}), we investigated the statistical properties of solar-type and Sun-like superflare stars using photometric data of $Kepler$ space telescope.
 These studies, however, were not enough since they only had 3 Sun-like superflare stars, mainly because they only used the first $\sim$500 days of the $Kepler$ 4-year primary mission data.
 Then, in this study, we report the latest statistical analyses of superflares on solar-type (G-type main-sequence) stars using all the $Kepler$ 4-year primary mission data covering $\sim$1500 days, and {\it Gaia}-DR2 (Data Release 2) catalog.
 Since the catalog was updated to $Gaia$-DR2, the sample size of solar-type stars become $\sim$4 times larger and the sample size of Sun-like stars become $\sim$12 times larger than \citet{Notsu+2019}.
 We updated the flare detection method by using high-pass filter to remove rotational variations caused by starspots (Section \ref{subsec:analysis}).
 We also took into account the effect of sample biases on the frequency of superflares, by considering gyrochronology (Section \ref{subsec:gyrochronology}) and flare detection completeness (Section \ref{subsec:sensitivity}).
 These results enabled us to discuss more well-established view on statistical properties of superflares on Sun-like stars:

\begin{enumerate}
  \renewcommand{\labelenumi}{(\roman{enumi})}
  \item 
  We found 2341 superflares on 265 solar-type stars, and 26 superflares on 15 Sun-like stars ($T_{\rm eff}=5600$ -- $6000$ K and $P_{\rm rot}=20$ -- $40$ days) (Table \ref{table:num_flares}). 
  The number of superflares occurred on solar-type stars increased 527 to 2341, and that of Sun-like stars greatly increased from 3 to 26 events compared with \citet{Notsu+2019}.
  
  \item
  The observed upper limit of the flare energy decreases as the rotation period (stellar age) increases in solar-type stars (Figure \ref{fig:erg_prot_solartype}), while flare energy can be explained by the magnetic energy stored around starspots (Figure \ref{fig:erg_aspt}).
  These can be consistent with the result that the starspot coverage decrease as the rotation period increases (Figure \ref{fig:aspot_prot_solartype}).
  In the case of slowly rotating Sun-like stars ($T_{\rm eff}=5600-6000$ K, $P_{\rm rot}=20$ -- $40$ days, and age $t\sim4.6$ Gyr), superflares with energies up to $\sim 4  \times10^{34}$ erg can occur, while superflares up to $\sim10^{36}$ erg can occur on young, rapidly-rotating stars ($P_{\rm rot}\sim$ a few days, and $t\sim$  few hundred Myr).
  The maximum size of starspots on old slowly-rotating Sun-like stars and young rapidly-rotating stars are a few \% and $\sim10$\% of the solar hemisphere, respectively, and these correspond to the above upper limits of superflare energies.
  We notice that the starspot area can be smaller than actual (see Section \ref{subsec:energy_spot}) and the more energetic flare can occur since the sample size is limited.

  \item
  The frequency of superflares on young, rapidly rotating stars ($P_{\rm{rot}}=1$ -- $3$ days) is $\sim100$ times higher compared with old, slowly rotating stars ($P_{\rm{rot}}>20$ days) and this indicates that as a star evolves (and its rotational period increases), the frequency of superflares decreases (Figure \ref{fig:freq_prot_solartype_hist}).
  In contrast, the flare frequency is roughly constant in the range of $P_{\rm rot} \lesssim 3$ days.

  \item
  The flare frequency distributions of each $P_{\rm{rot}}$ show the power-law distributions ($dN/dE\propto E^{\alpha}$) (Section \ref{subsec:superflare_sunlike_sun}).
  The frequency distribution of superflares on Sun-like stars and that of solar flares are roughly on the same power-law line (Figure \ref{fig:flarefreq_sunstar}).

  \item
  From the analysis of Sun-like stars, superflares whose energy is $\sim 7\times10^{33}$ erg ($\sim$X700-class flares), $\sim 1 \times10^{34}$ erg ($\sim$X1000-class flares), can occur once every 
  $\sim$3,000 years, $\sim$6,000 years, 
  respectively (Figure \ref{fig:flarefreq_sunstar}) on  the Sun.
  The results of starspot frequency on Sun-like stars in \citet{Notsu+2019} also are consistent with the results of the superflare frequency in this study.
  These results strongly support the possibility of superflares on the Sun, and revealed the frequency of superflares on the Sun through the Sun-like stars.

  \item
  The solar-type superflare stars found in this study have no confirmed exoplanets, and this indicates that hot Jupiter is not a necessary condition for superflares on solar-type stars.
 \end{enumerate}

 From these results, we have shown great insights into the important question ``Can our Sun have superflares?".
 In future studies, we will clarify the properties of superflare stars on Sun-like stars and get a final conclusion of this question.
 We have found 15 slowly-rotating Sun-like superflare stars in this study, but the number of Sun-like superflare stars ($T_{\rm{eff}}$=5600 -- 6000 K, $P_{\rm{rot}}\sim$25 days, and $t\sim$4.6 Gyr) that have been investigated spectroscopically and confirmed to be ``single" Sun-like stars, are now 0, as described in Appendix \ref{app:sunlike-spec-previous}.
 Future spectroscopic observations of Sun-like superflare stars are necessary to investigate whether the Sun-like stars really have superflares.
 Although all of the 15 $Kepler$ Sun-like superflare stars listed in Table \ref{table:params_sunlike} are relatively faint (all are fainter than 13.5 mag in $Kepler$-band magnitude and most of them are fainter than 14.5 mag), it is very important to conduct spectroscopic observations of these 15 Sun-like $Kepler$ superflare stars in order to confirm the validity of statistical results of superflares on Sun-like stars discussed in this study using $Kepler$ data (e.g., Figure \ref{fig:flarefreq_sunstar}).
  In addition, nearby bright superflare stars that are found from $TESS$ (\citealt{Ricker+2015}; \citealt{Tu+2020}; \citealt{Doyle+2020})
  and that will be found from $PLATO$ \citep{Rauer+2014} can be also good targets for future spectroscopic studies.

 The statistical properties of superflares presented in this study can be combined with many topics to more fully understand the physics of superflares and effects in the related research fields.
 For example, mechanisms of white-light continuum emissions of superflares (e.g., \citealt{Katsova+2015}; \citealt{Namekata+2017}; \citealt{Heinzel+2018}; \citealt{Kowalski+2018}; \citealt{Nizamov+2019}
 ),
 chromospheric line profiles during superflares (e.g., \citealt{Houdebine+1993}; \citealt{Honda+2018}; \citealt{Namekata+2020_PASJ}), stellar mass ejections (e.g., CMEs) during superflares (e.g., \citealt{Takahashi+2016}; \citealt{Crosley+2018}; \citealt{Lynch+2019}; \citealt{Moschou+2019}; \citealt{Vida+2019}; \citealt{Leitzinger+2020}), impacts of superflares on planets, especially exoplanets (e.g., \citealt{Segura+2010}; \citealt{Airapetian+2016}\&\citeyear{Airapetian+2020}; \citealt{Atri+2017}; \citealt{Lingam+2017}; \citealt{Kay+2019}; \citealt{Linsky2019}; \citealt{Herbst+2019}; \citealt{Yamashiki+2019}),
 solar activities over $\sim$1000 years (e.g., \citealt{Miyake+2012},\citeyear{Miyake+2013},\&\citeyear{Miyake+2019}; \citealt{Usoskin+2017}; \citealt{Hayakawa+2017b}\&\citeyear{Hayakawa+2017a}),
 complexities of starspots and starspot distributions of superflare stars (e.g., \citealt{Maehara+2017}; \citealt{Toriumi+2019}; \citealt{Doyle+2018}; \citealt{Morris+2018}; \citealt{Roettenbacher+2018}; \citealt{Takasao+2020}),
 and large starspots lifetimes and formation/decay process (e.g., \citealt{Shibata+2013}; \citealt{Giles+2017}; \citealt{Namekata+2019}\&\citeyear{Namekata+2020_ApJ}; \citealt{Ikuta+2020}).
 For these points, we plan to conduct long-term spectroscopic monitoring observations of nearby superflare stars, by using 2 -- 4 m telescopes, especially including the Kyoto University Okayama 3.8 m Seimei telescope (\citealt{Kurita+2020}; \citealt{Namekata+2020_PASJ})

 \acknowledgments
 This paper includes data collected by the $Kepler$ mission.
 Funding for the $Kepler$ mission is provided by the NASA Science Mission Directorate.
 The $Kepler$ data presented in this paper were obtained from the Multimission Archive at STScl.
 This paper also has made use of data from the European Space Agency (ESA) mission $Gaia$ (https://www.cosmos.esa.int/gaia), processed by the Gaia Data Processing and Analysis Consortium (DPAC, https://www.cosmos.esa.int/web/gaia/dpac/consortium).
 Funding for the DPAC has been provided by national institutions, in particular the institutions participating in the Gaia Multilateral Agreement.
 This work was also supported by JSPS KAKENHI Grant Number of 17K05400, 20K04032, 20H05643 (H.M.), 18J20048 (K.N.).

 Y.N. is supported by JSPS Overseas Research Fellowship Program.
 Y.N. also acknowledges the International Space Science Institute and the supported International Team 464: The Role Of Solar And Stellar Energetic Particles On (Exo)Planetary Habitability (ETERNAL, \url{http://www.issibern.ch/teams/exoeternal/})."
 K.N. is supported by the JSPS Overseas Challenge Program for Young Researchers.
We are grateful to Mr. Isaiah Tristan (University of Colorado Boulder) for valuable suggestions to improve the manuscript.

\newpage
\appendix
\section{Lightcurves of all the superflares on Sun-like stars}\label{app:allflare_sunlike}
 Since one of the most important points of this paper is investigating whether slowly-rotating stars really have superflares, we show light curves of all the superflares that occurred on Sun-like stars (All the superflares occurred on solar-type stars listed in Table \ref{table:params_sunlike}).
 Lightcurves are separated into Figures \ref{fig:eg_sunlikeflare_noflag},  \ref{fig:eg_sunlikeflare_flag1}, and \ref{fig:eg_sunlikeflare_flag2}, as described in the following.
 We also show the LombScargle periodograms of all the Sun-like superflare stars in order to check whether the $P_{\rm{rot}}$ values that are taken from \citet{McQuillan+2014} and are listed in Table \ref{table:params_sunlike} are appropriate.
 Since \citet{McQuillan+2014} used a different method (ACF method; see \citealt{McQuillan+2014} for details) to estimate rotation period values, these LombScargle periodograms can be independent checking of $P_{\rm{rot}}$ values.
 These periodograms are estimated from the whole available lightcurve data of each superflare star.
 These periodograms are also separated into Figures \ref{fig:lomb_noflag}, \ref{fig:lomb_flag1}, \& \ref{fig:lomb_flag2}.

 Figures \ref{fig:eg_sunlikeflare_noflag} \& \ref{fig:lomb_noflag} are the lightcurves and the LombScargle periodograms of the 10 Sun-like superflare stars that have no flags in Table \ref{table:params_sunlike}.
 The lightcurves and the LombScargle periodograms of these 10 stars look consistent with that these stars show the slow rotation period values ($P_{\rm{rot}}>20$ days) taken from \citet{McQuillan+2014}.

 Figures \ref{fig:eg_sunlikeflare_flag1} \& \ref{fig:lomb_flag1} are the lightcurves and the LombScargle periodograms of the 5 Sun-like superflare stars that have Flag ``1" in Table \ref{table:params_sunlike}.
 As shown in Figures \ref{fig:lomb_flag1}, these 5 stars have power peaks in different $P_{\rm rot}$ values from \citet{McQuillan+2014}, so the real rotation period value of these stars might possibly be different from that used in Table \ref{table:params_sunlike}.
 However, the large peaks are at the half of the rotation period reported by \citet{McQuillan+2014}.
 It is quite possible that this does not contradict between the periodograms and \citet{McQuillan+2014}, because the analysis method used by \citet{McQuillan+2014} is resistant to harmonics.
 These five stars have Flag ``1" in Table \ref{table:params_sunlike} for caution, but they are still used in the statistical discussions in the main part of this paper since only from the simple lightcurve analyses, it is difficult to finally judge whether the slow rotation period values ($P_{\rm{rot}}>20$ days) taken from \citet{McQuillan+2014} are correct or not.
 More investigations including spectroscopic measurements of rotation periods (e.g., \citealt{Notsu+2015b}; \citealt{Notsu+2019}) are needed.

 Figures \ref{fig:eg_sunlikeflare_flag2} \& \ref{fig:lomb_flag2} are the lightcurves and the LombScargle periodgram of KIC007772296 that have Flag ``2" in Table \ref{table:params_sunlike}.
 In Figure \ref{fig:eg_sunlikeflare_flag2} left panels, the periodic brightness variations with the period of a few days is found, and also in Figure \ref{fig:lomb_flag2}, a sharp peak can be seen around a few days.
 From these points, we judged KIC007772296 as a binary candidate and eliminated KIC007772296 from statistical analyses in the study.

\begin{figure*}
   \plottwo{figure/lightcurve_sunlikeflarestar/flarelightcurve_005695372_2455246.62rd.eps}{figure/lightcurve_sunlikeflarestar/flarelightcurve_005695372_2455246.62la.eps}
   \plottwo{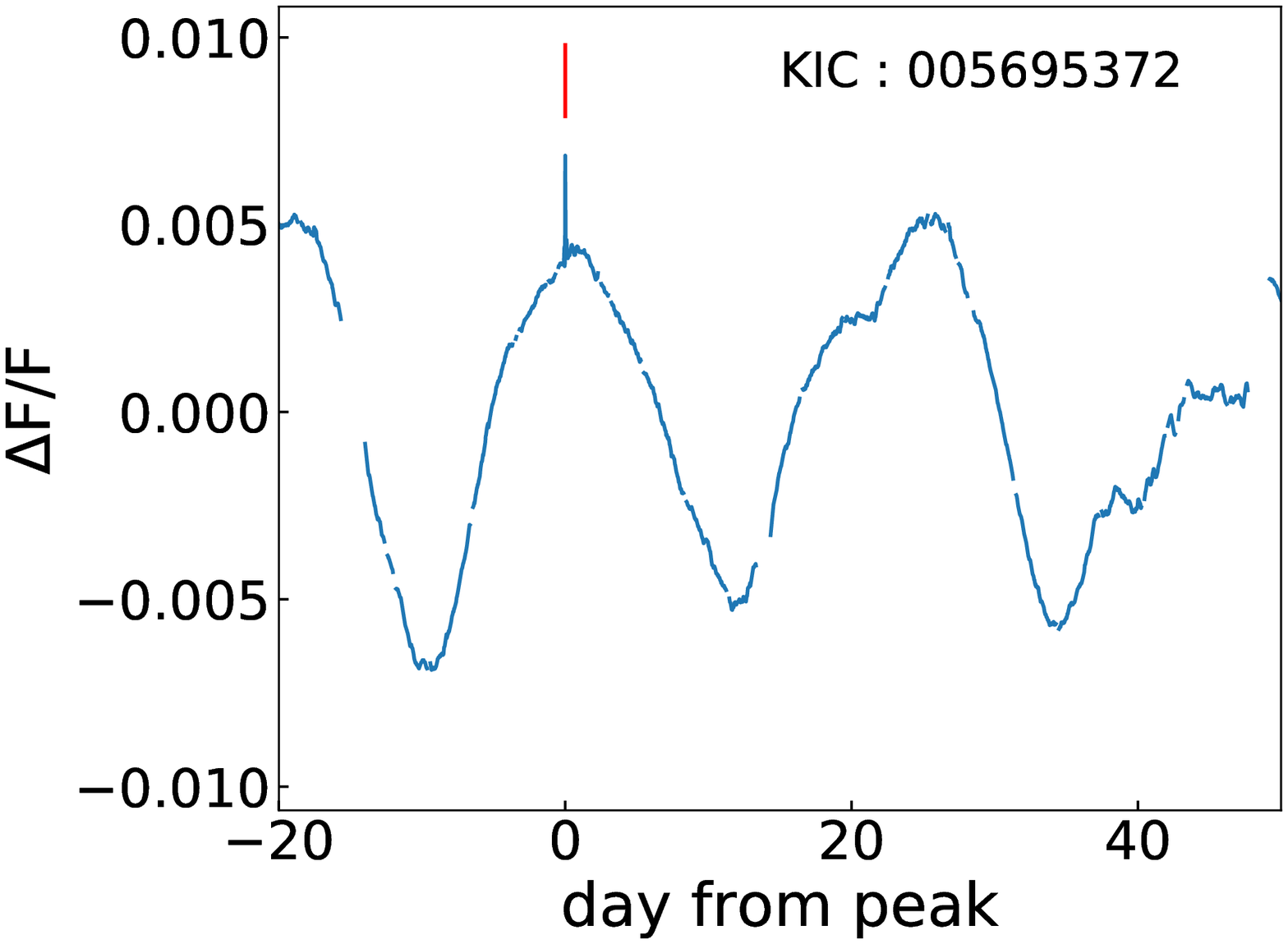}{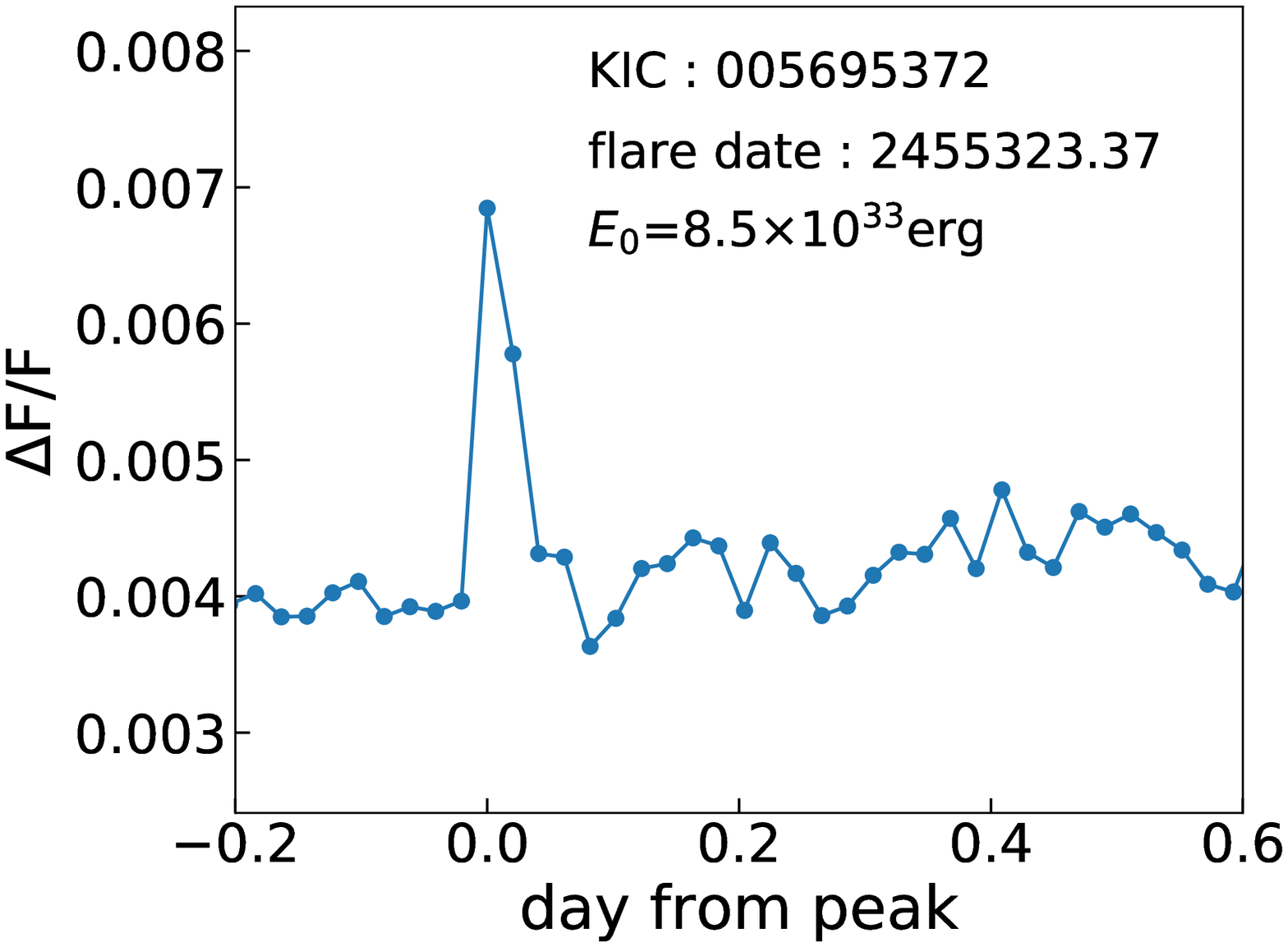}
   \plottwo{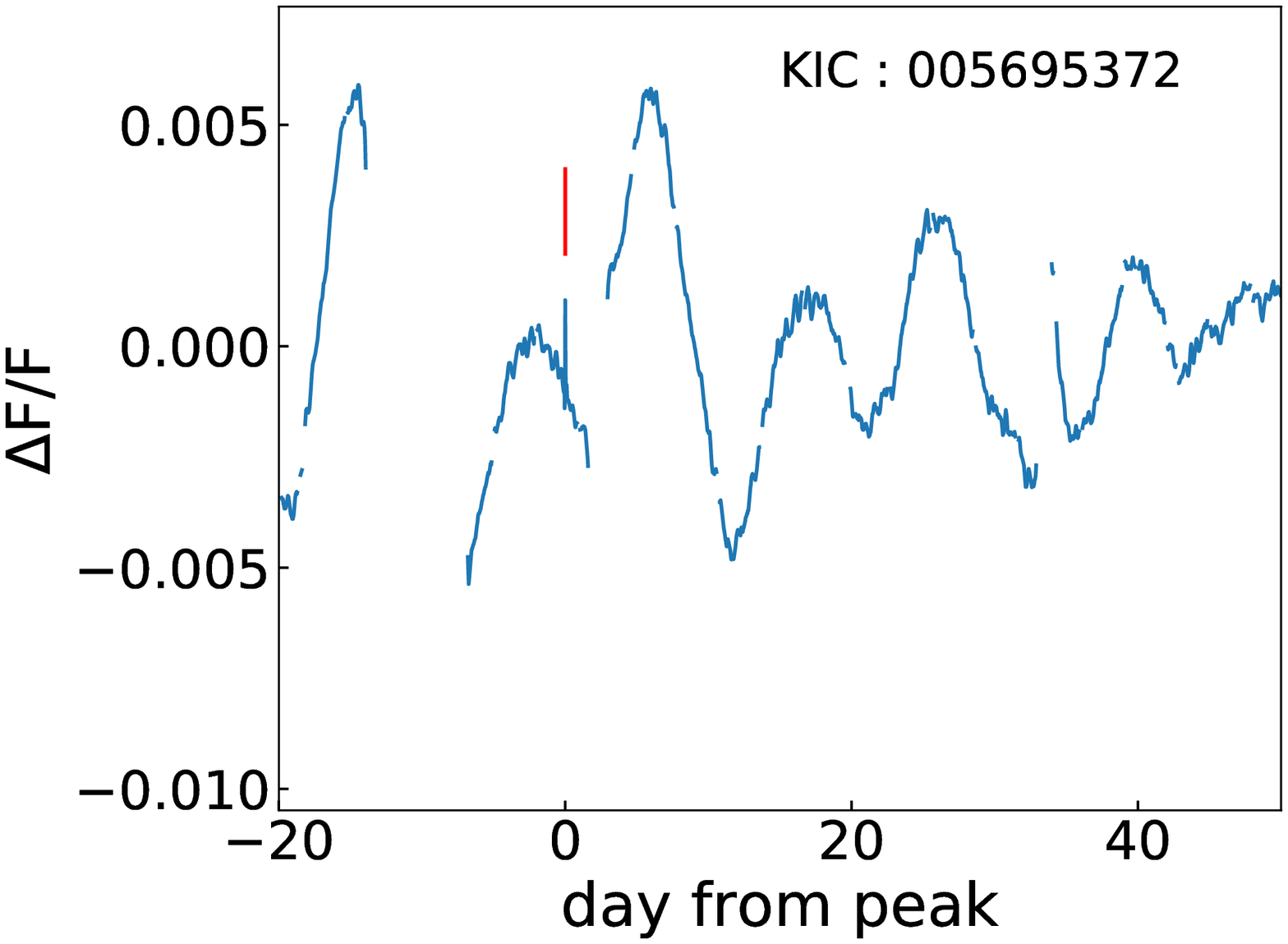}{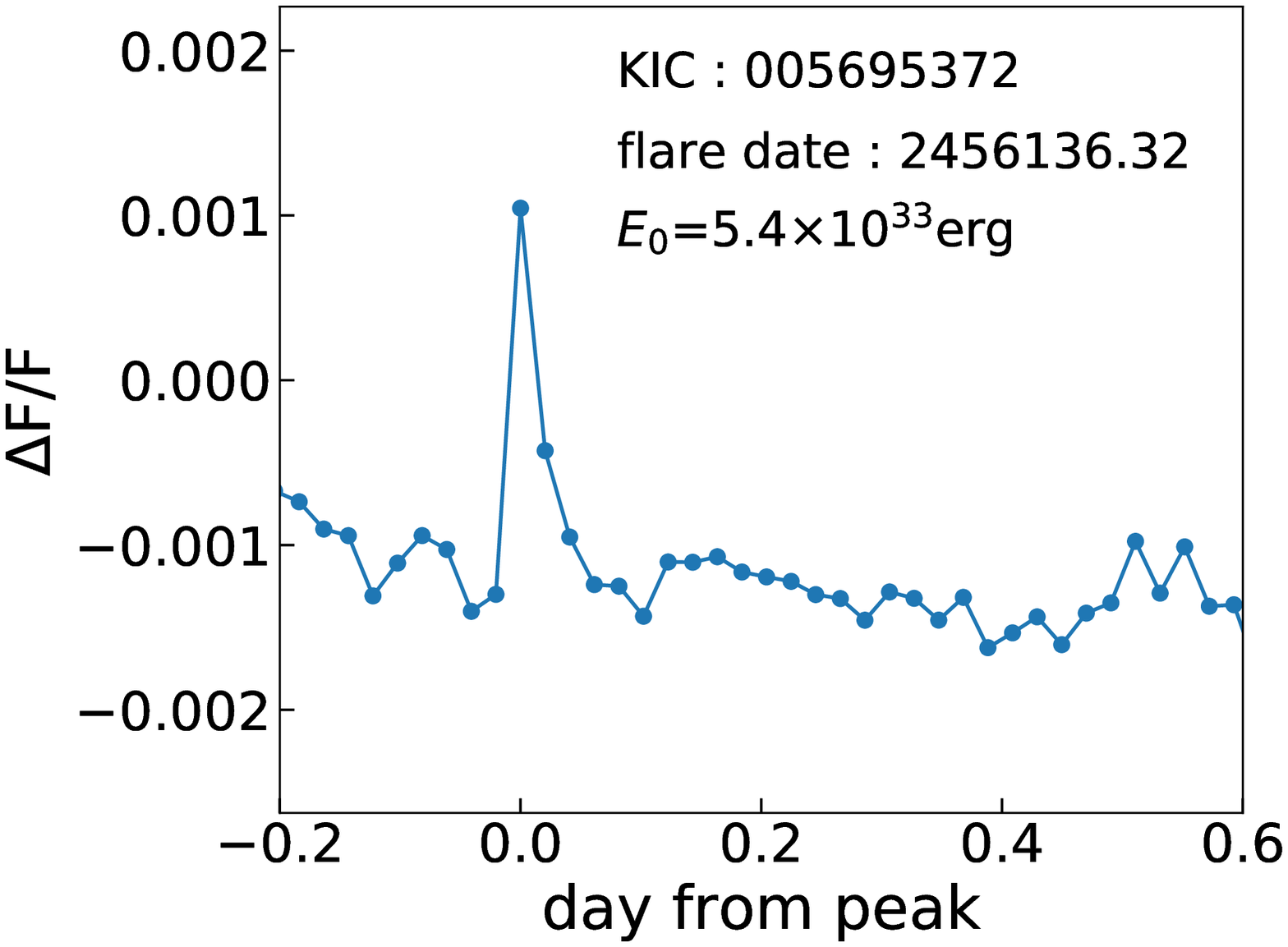}
   \caption{
   Light curves of all superflares on the 10 Sun-like superflare stars that have no flags in Table \ref{table:params_sunlike}.
   Horizontal and vertical axes correspond to days from the flare peak, and stellar brightness normalized by the average brightness during the observation quarter of $Kepler$.
   Panels on the left side show the 70 days time variation of stellar brightness to see the periodic brightness variations caused by stellar rotation.
   Right side panels show the detailed brightness variations of a flare.
   The star ID ($Kepler$ ID), the Barycentric Julian Date of flare peak, and released total bolometric energy are shown in the right panels.
   }
  \label{fig:eg_sunlikeflare_noflag}
 \end{figure*}
 
\addtocounter{figure}{-1}
 \begin{figure}
   \plottwo{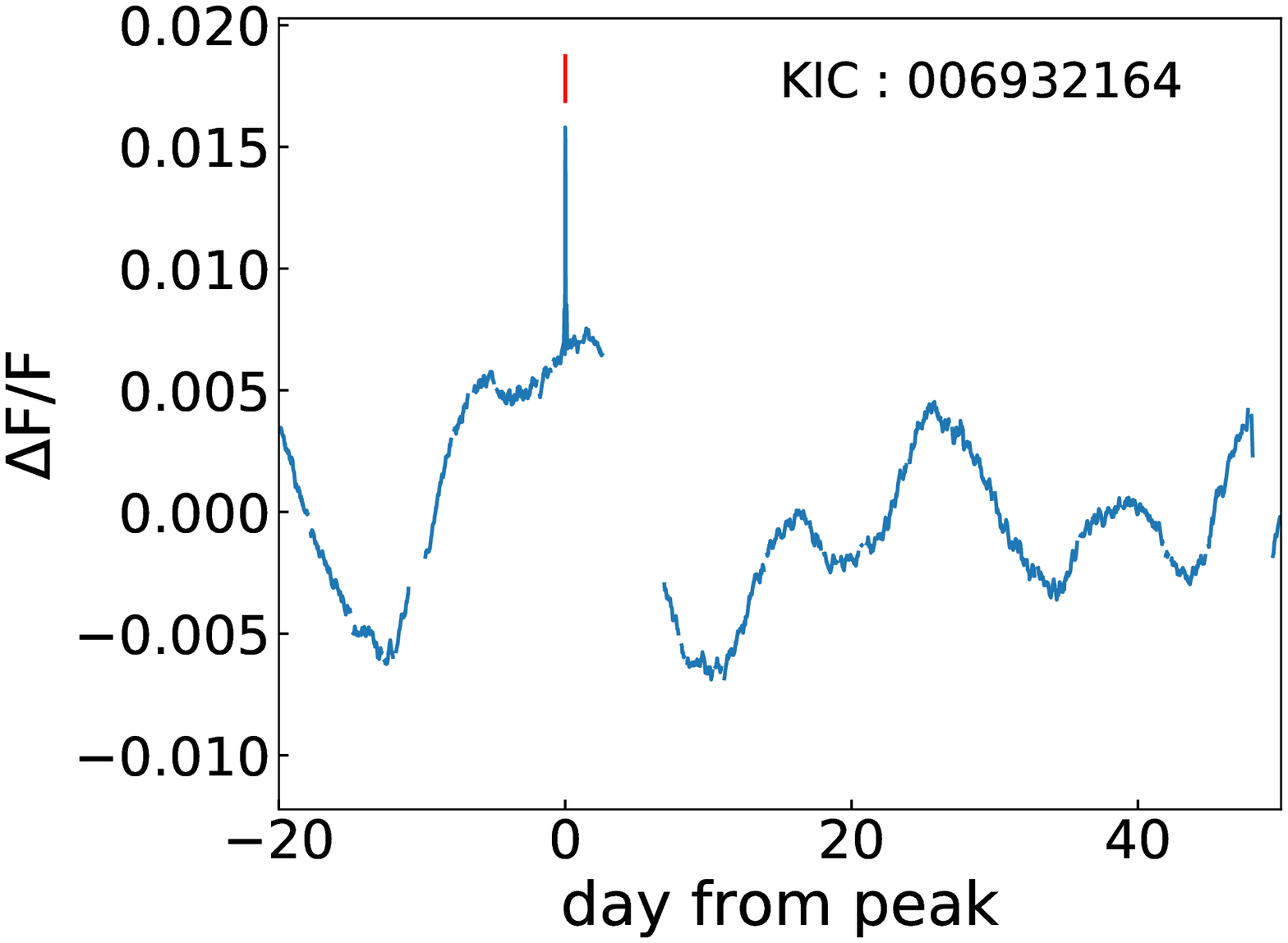}{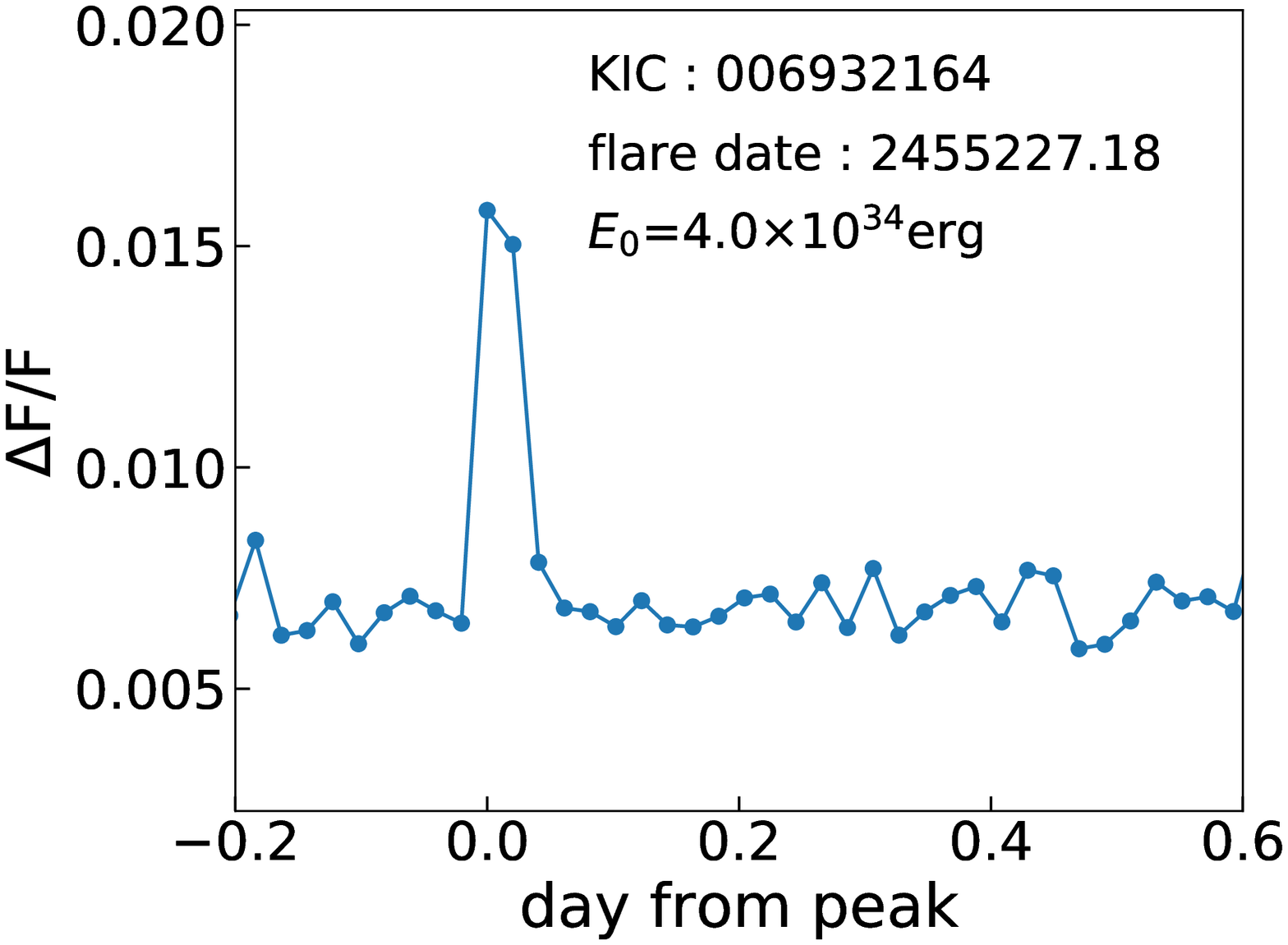}
   \plottwo{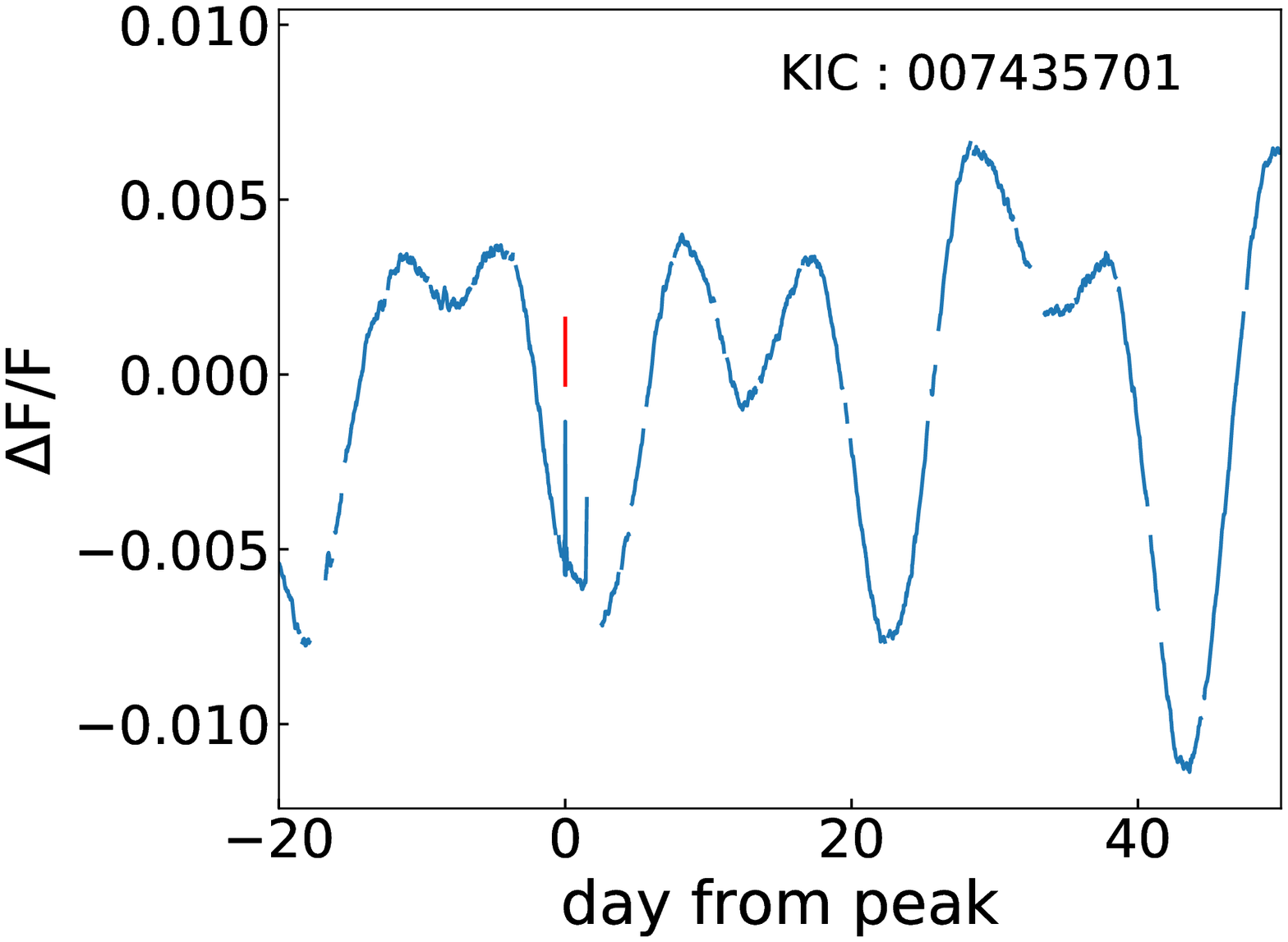}{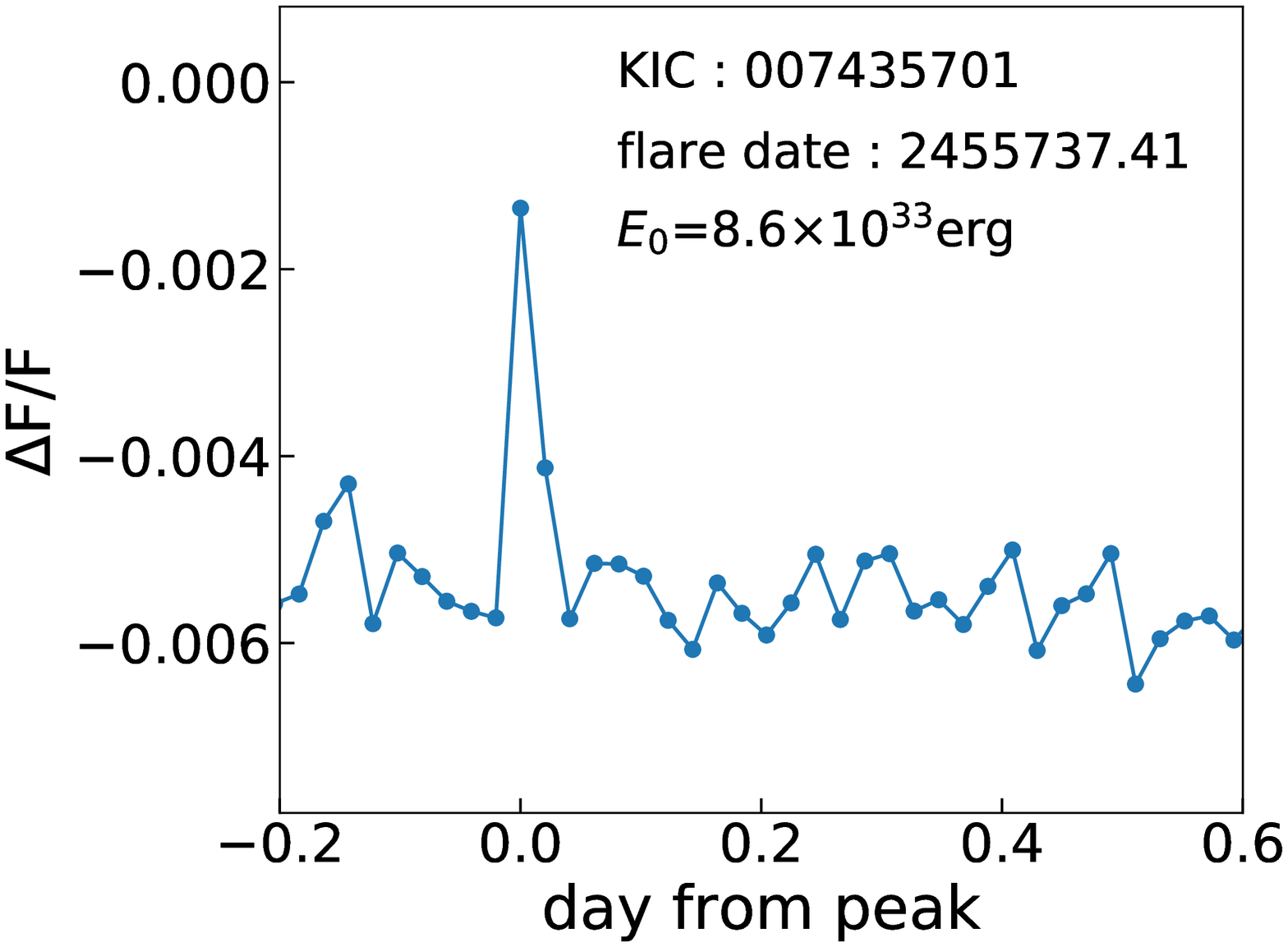}
   \plottwo{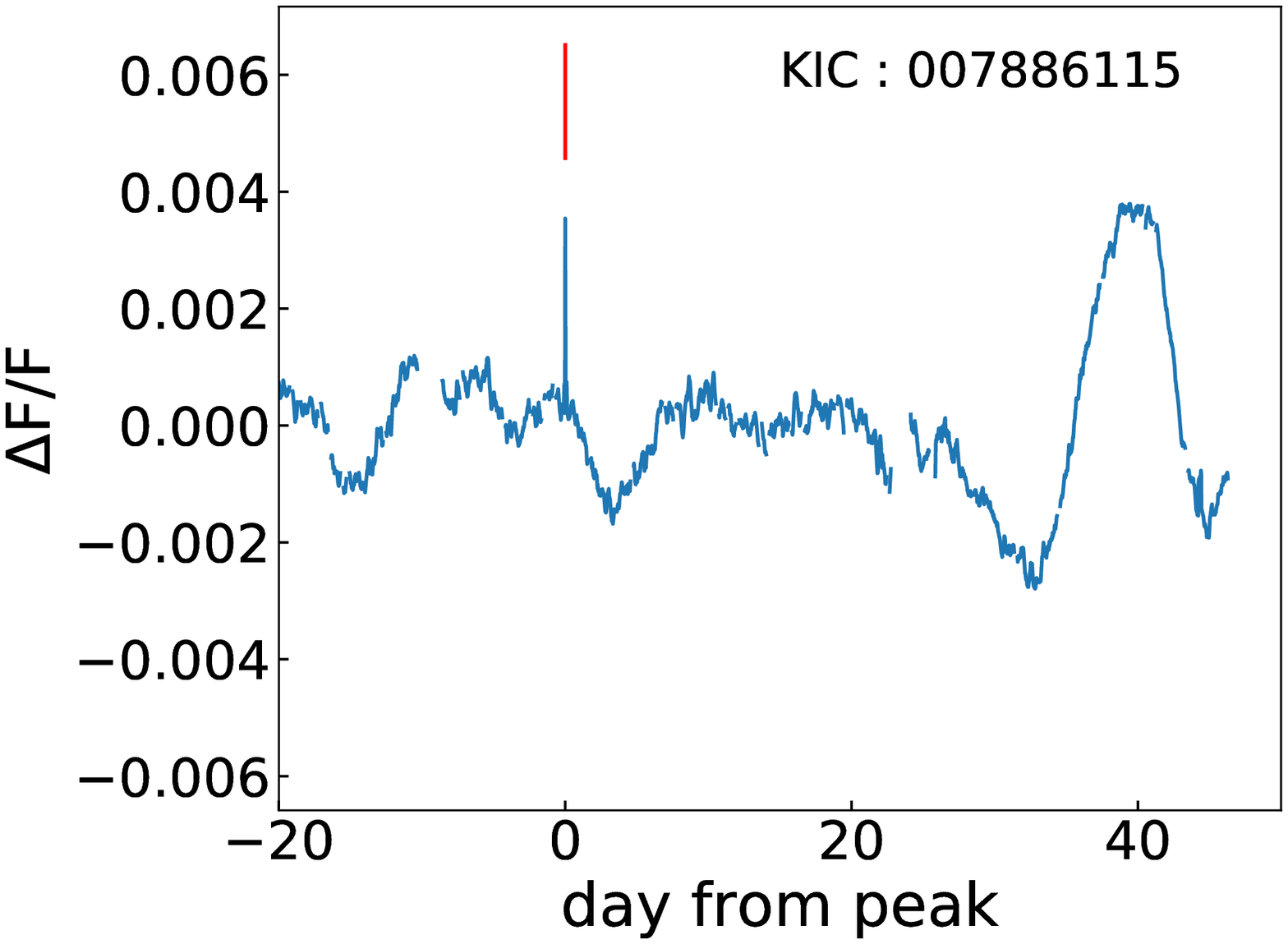}{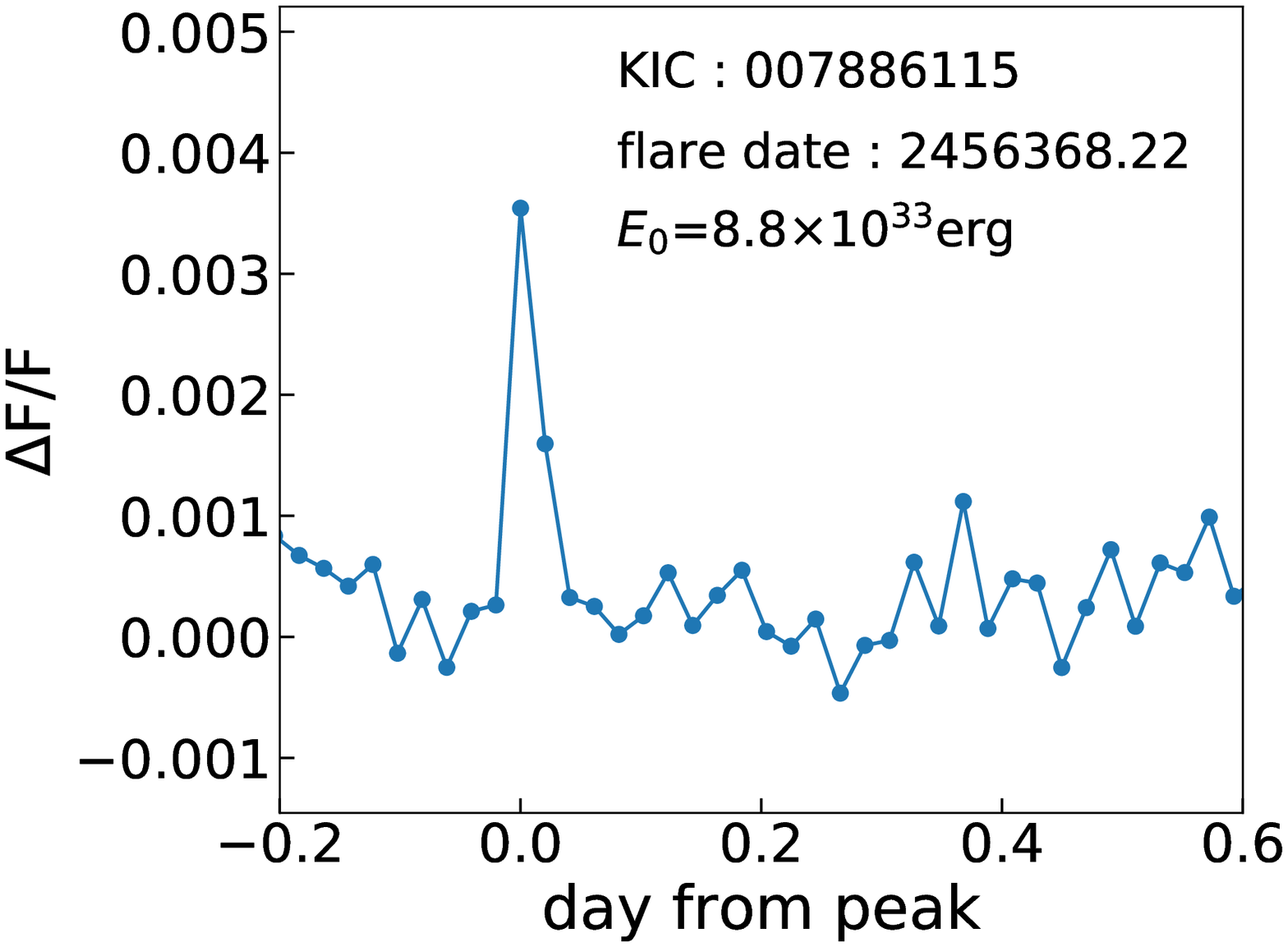}
    \caption{
   (Continued)
   }
 \end{figure}   
 \addtocounter{figure}{-1} 
 \begin{figure}
   \plottwo{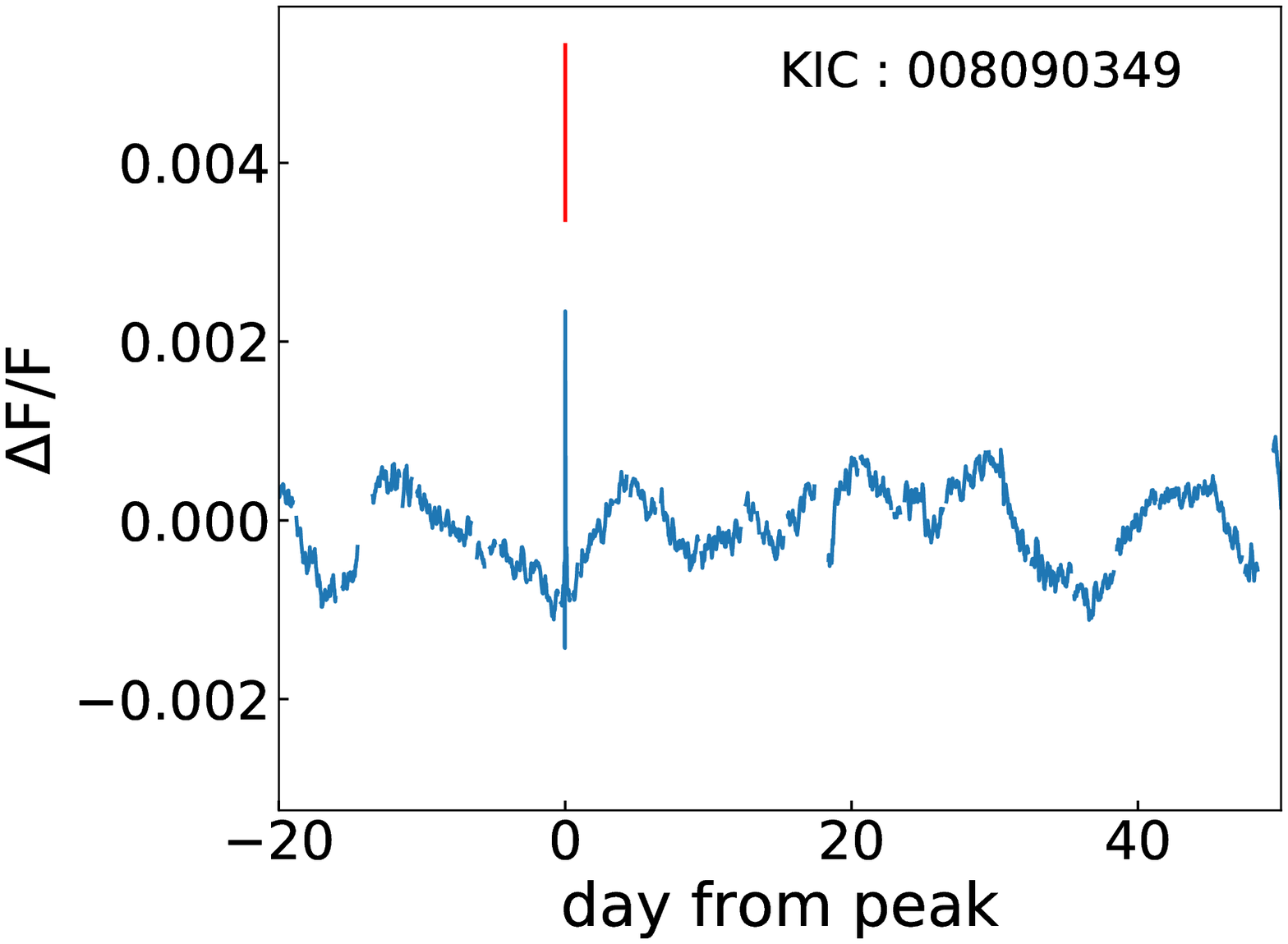}{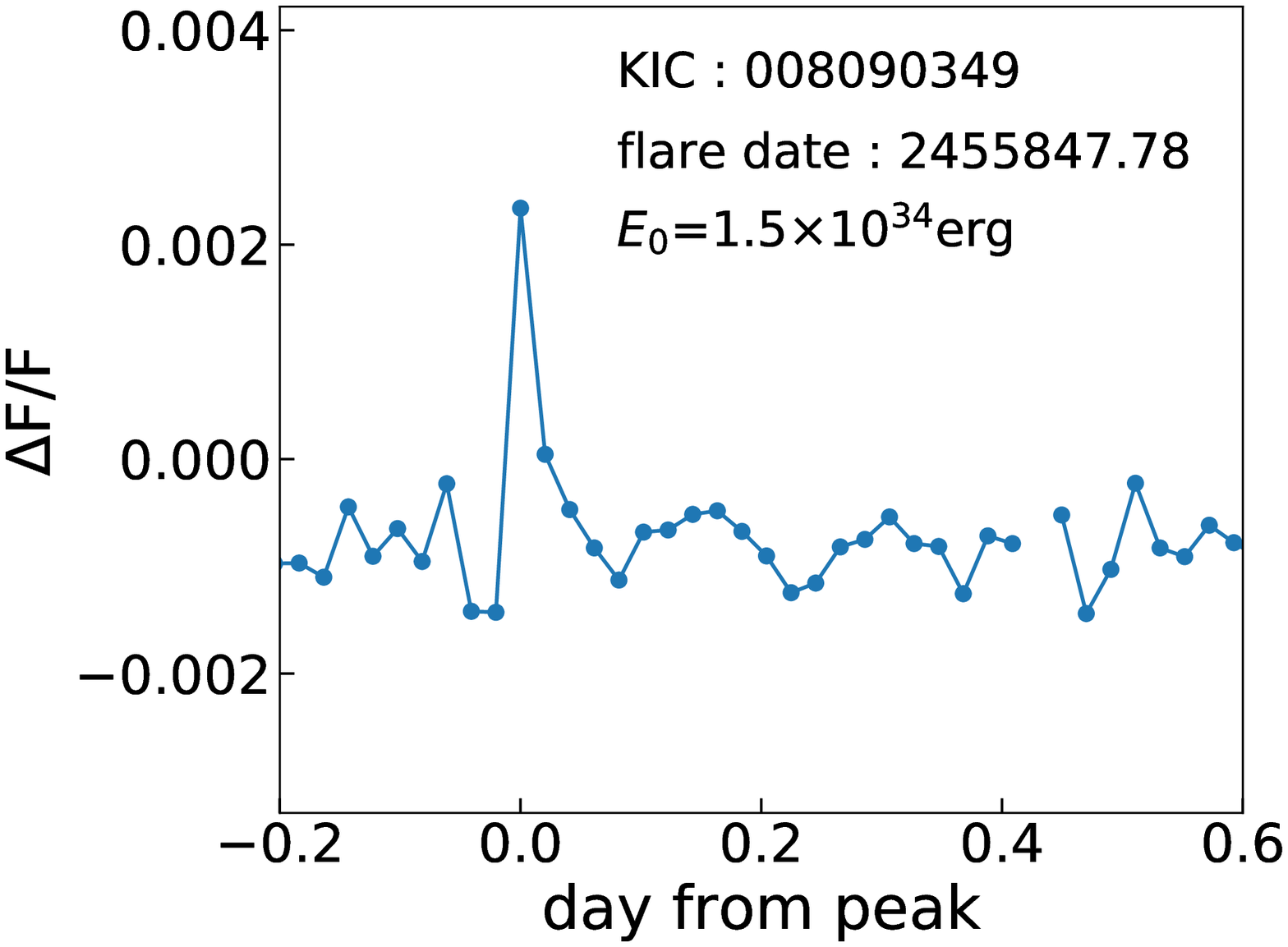}
   \plottwo{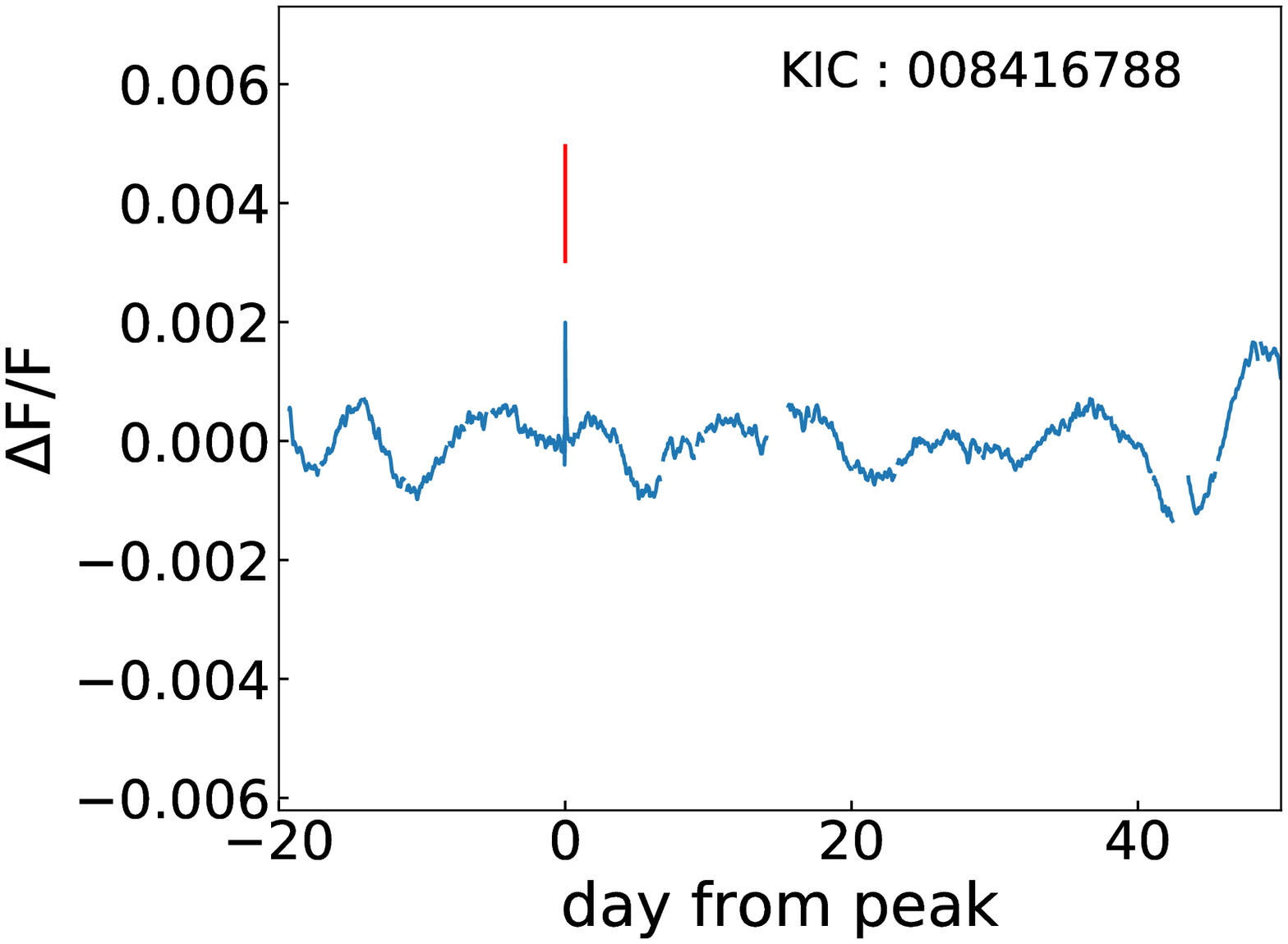}{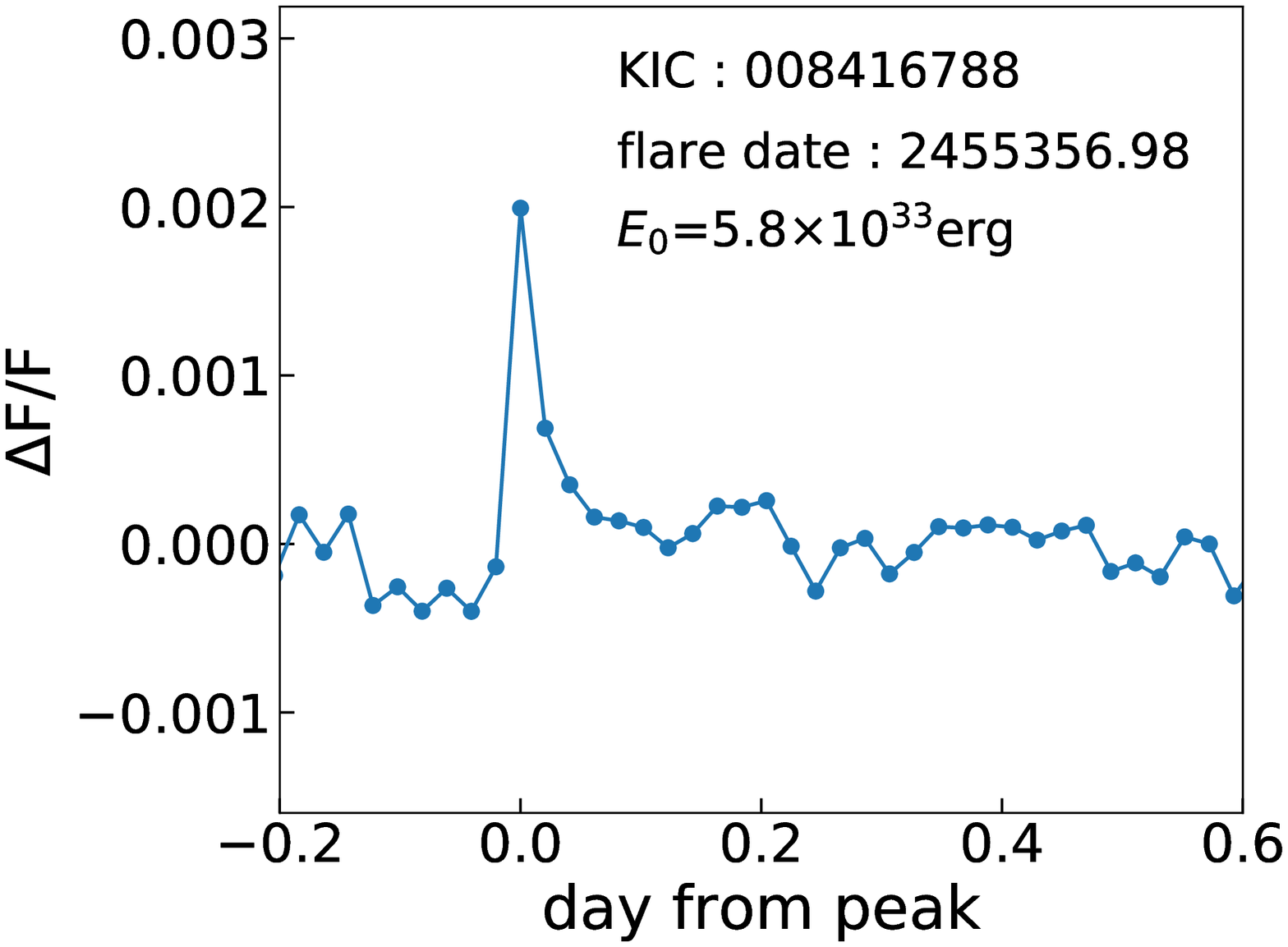}
   \plottwo{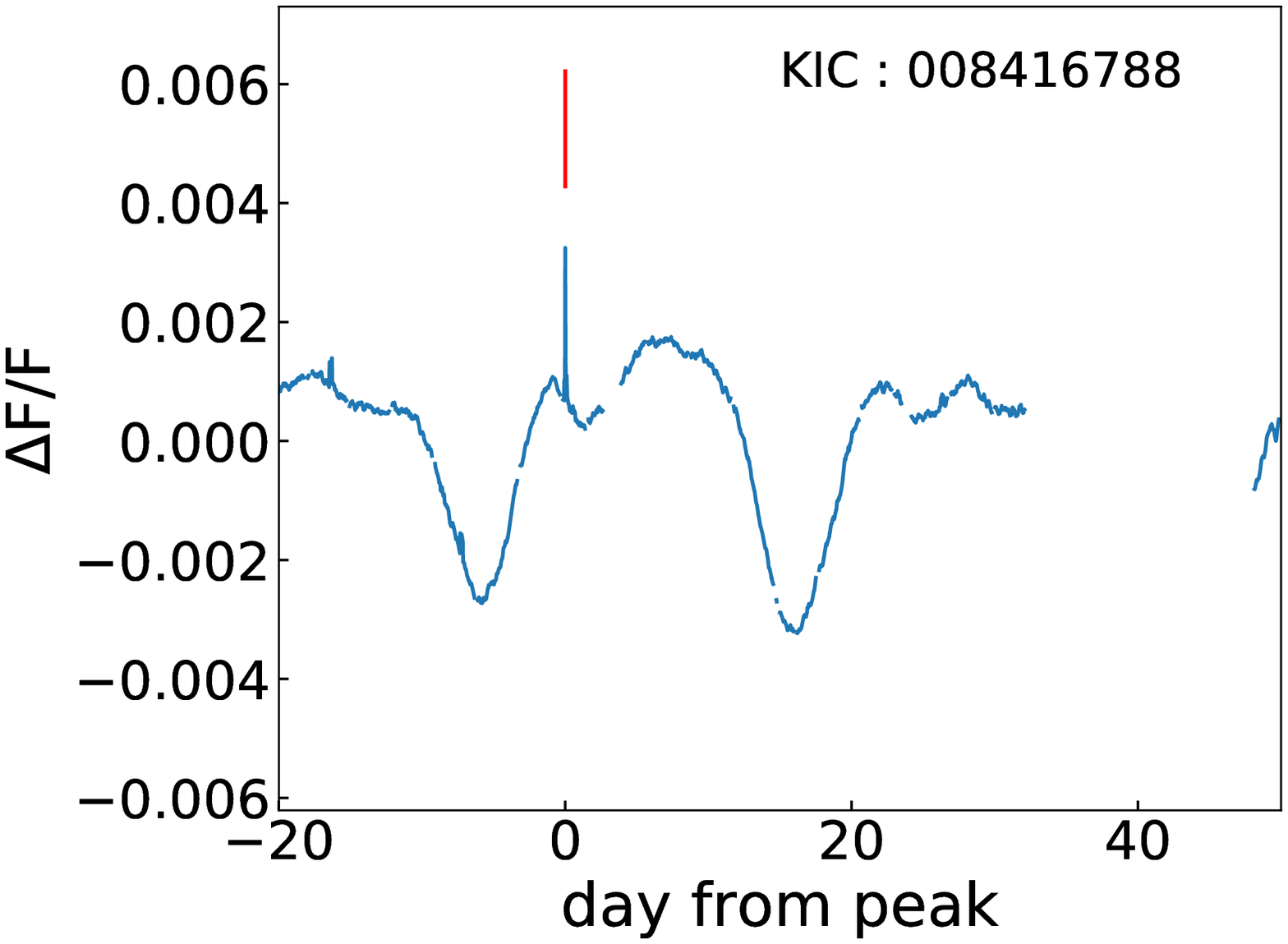}{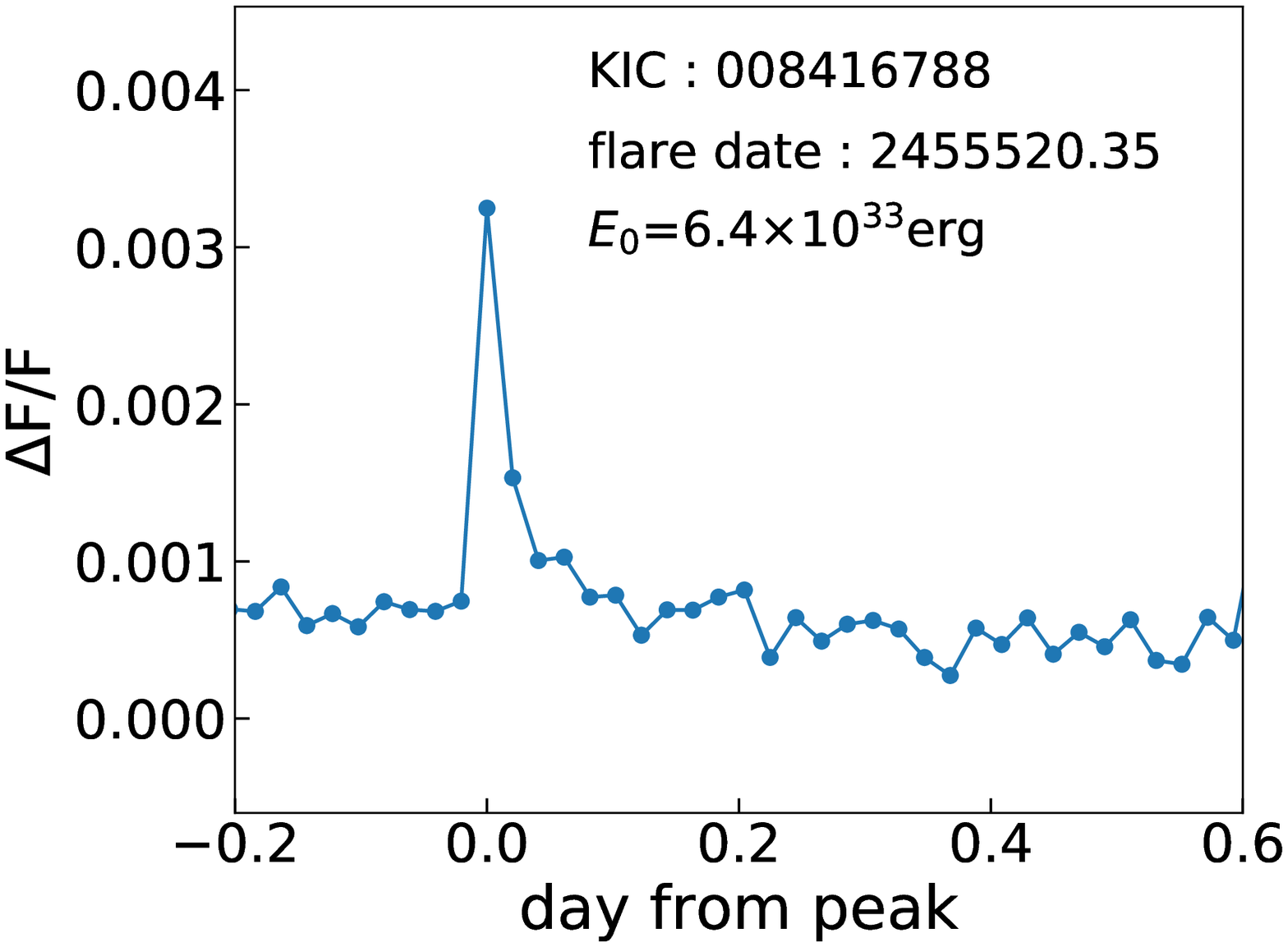}   
   \caption{
   (Continued)
   }
 \end{figure}
 \addtocounter{figure}{-1}
 \begin{figure}
   \plottwo{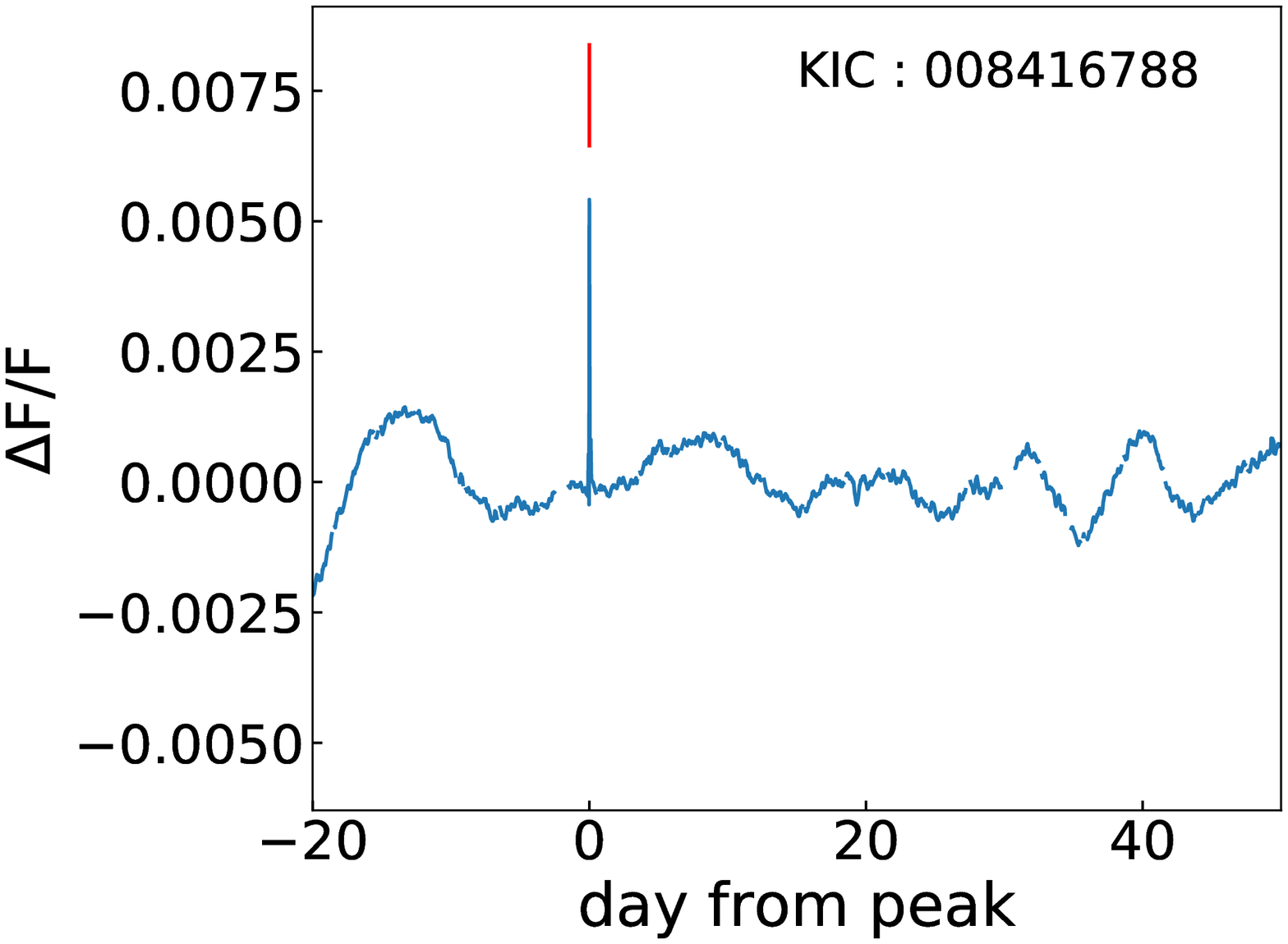}{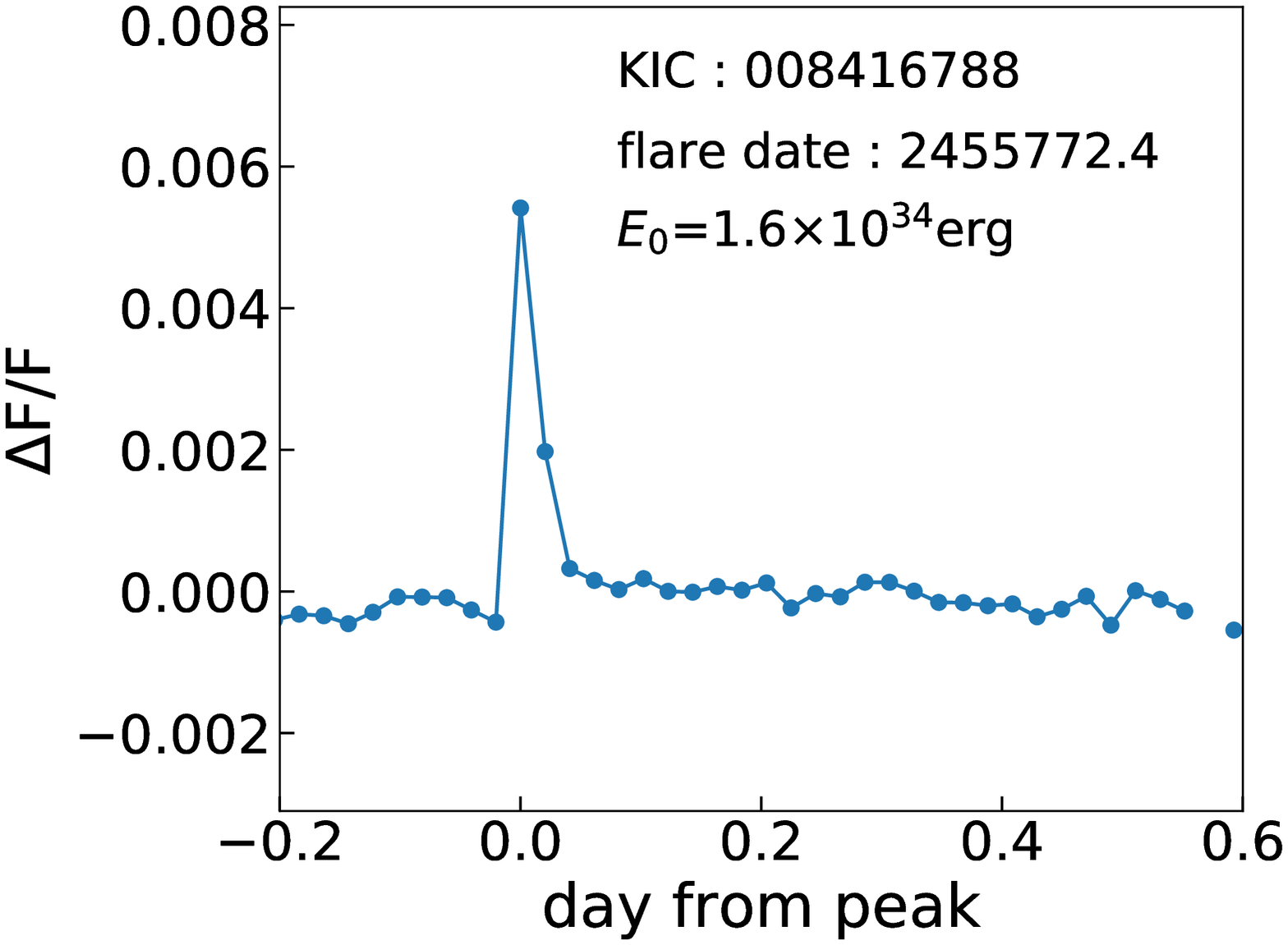}
   \plottwo{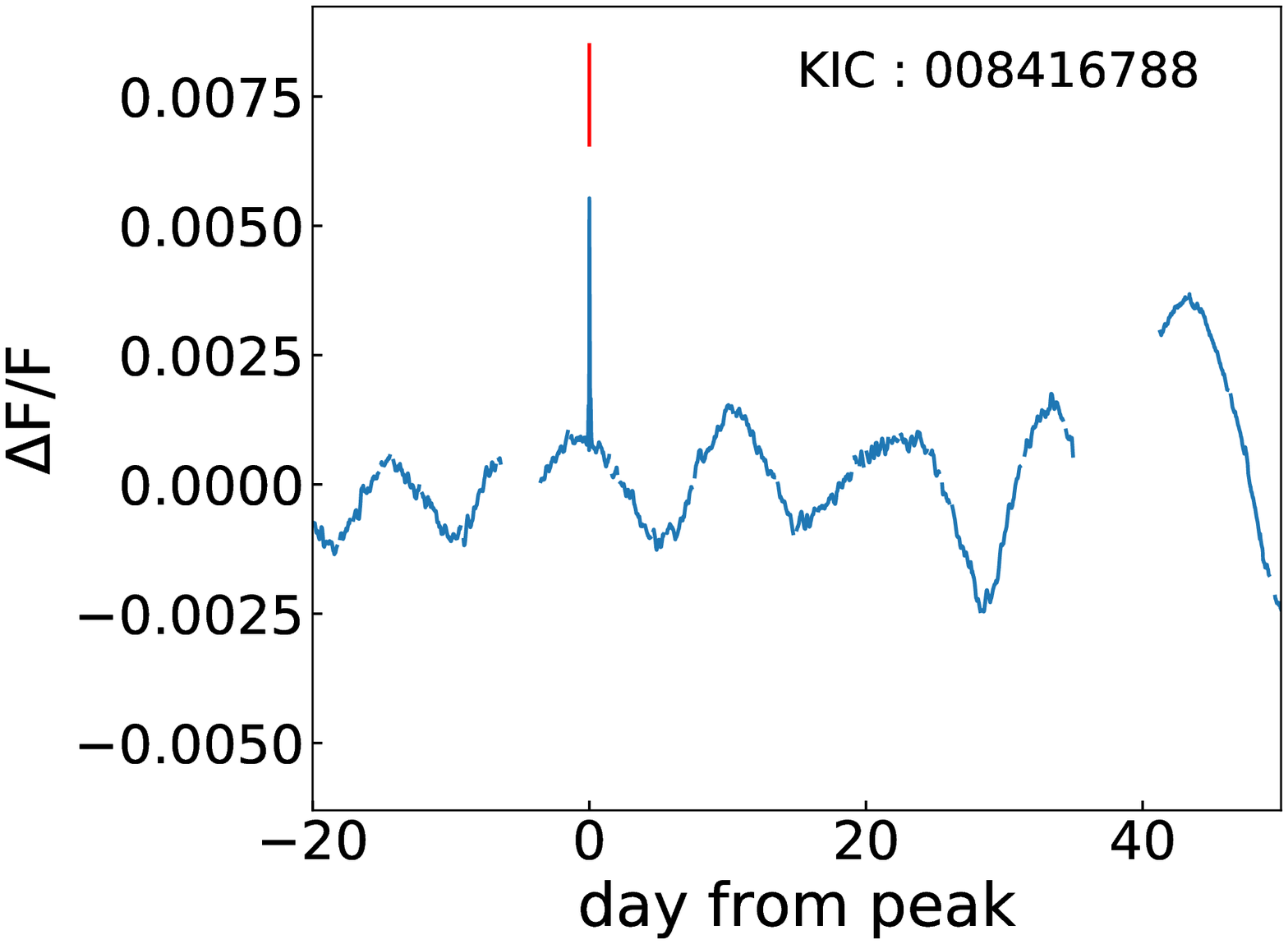}{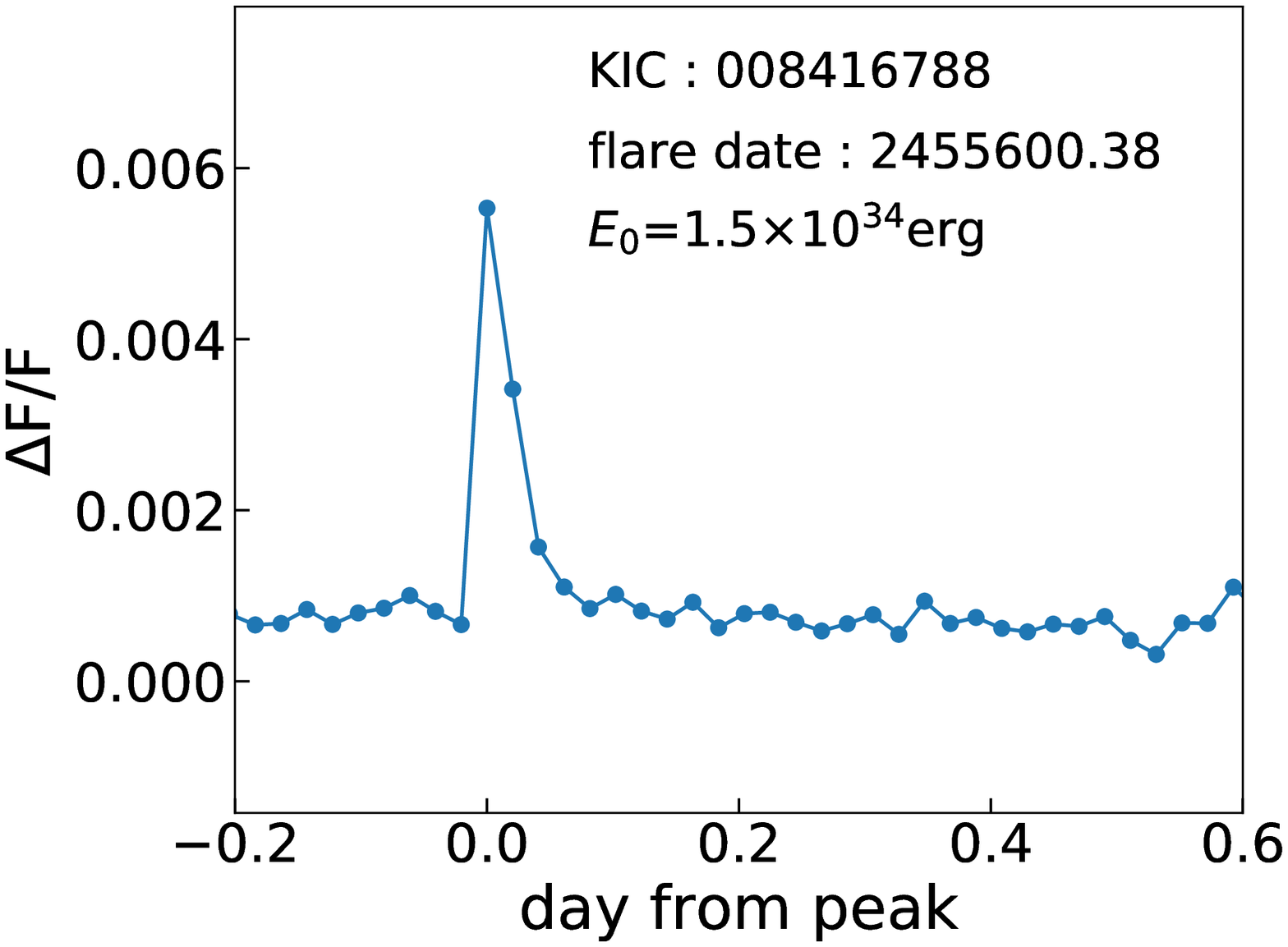}
   \plottwo{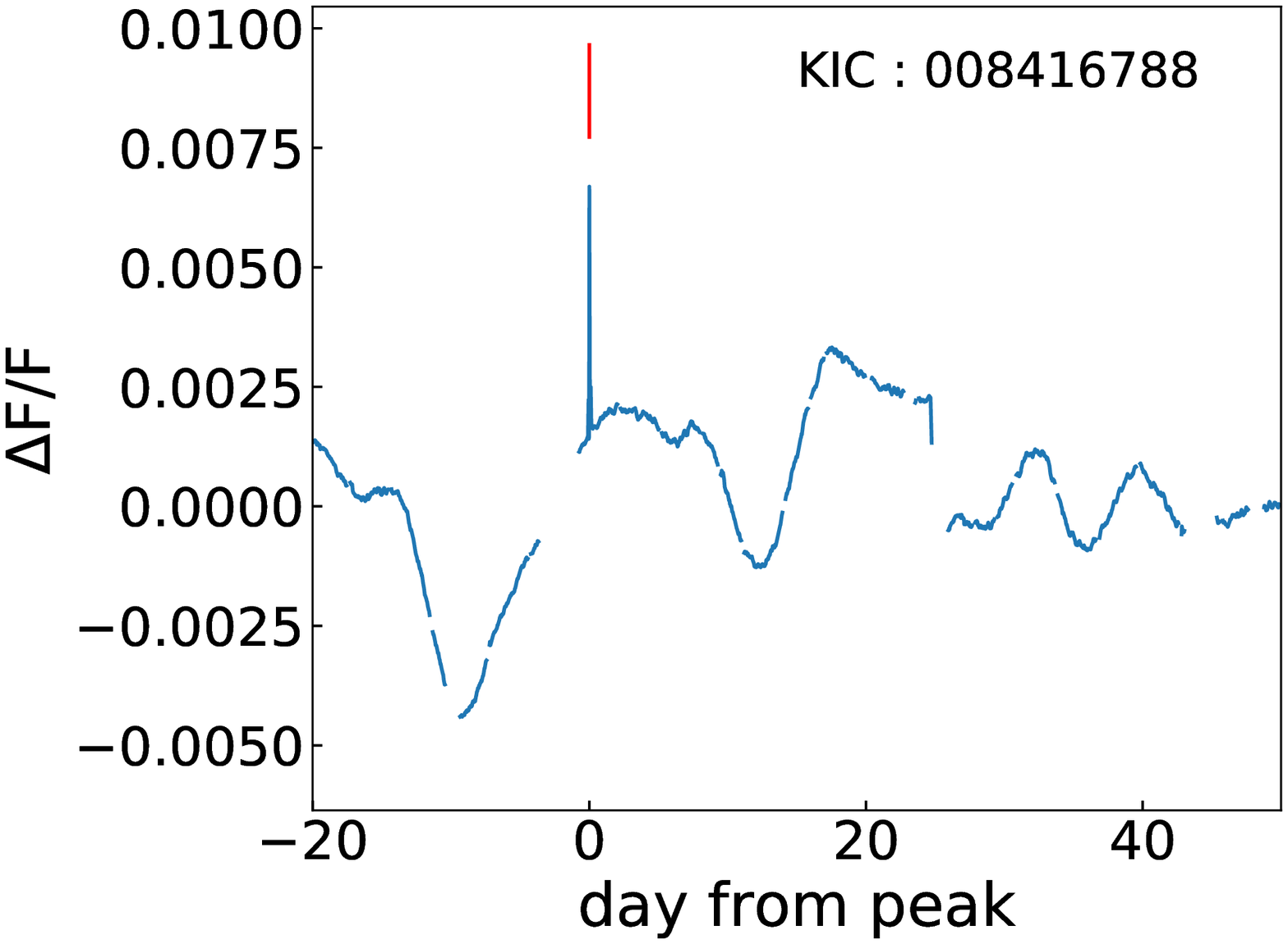}{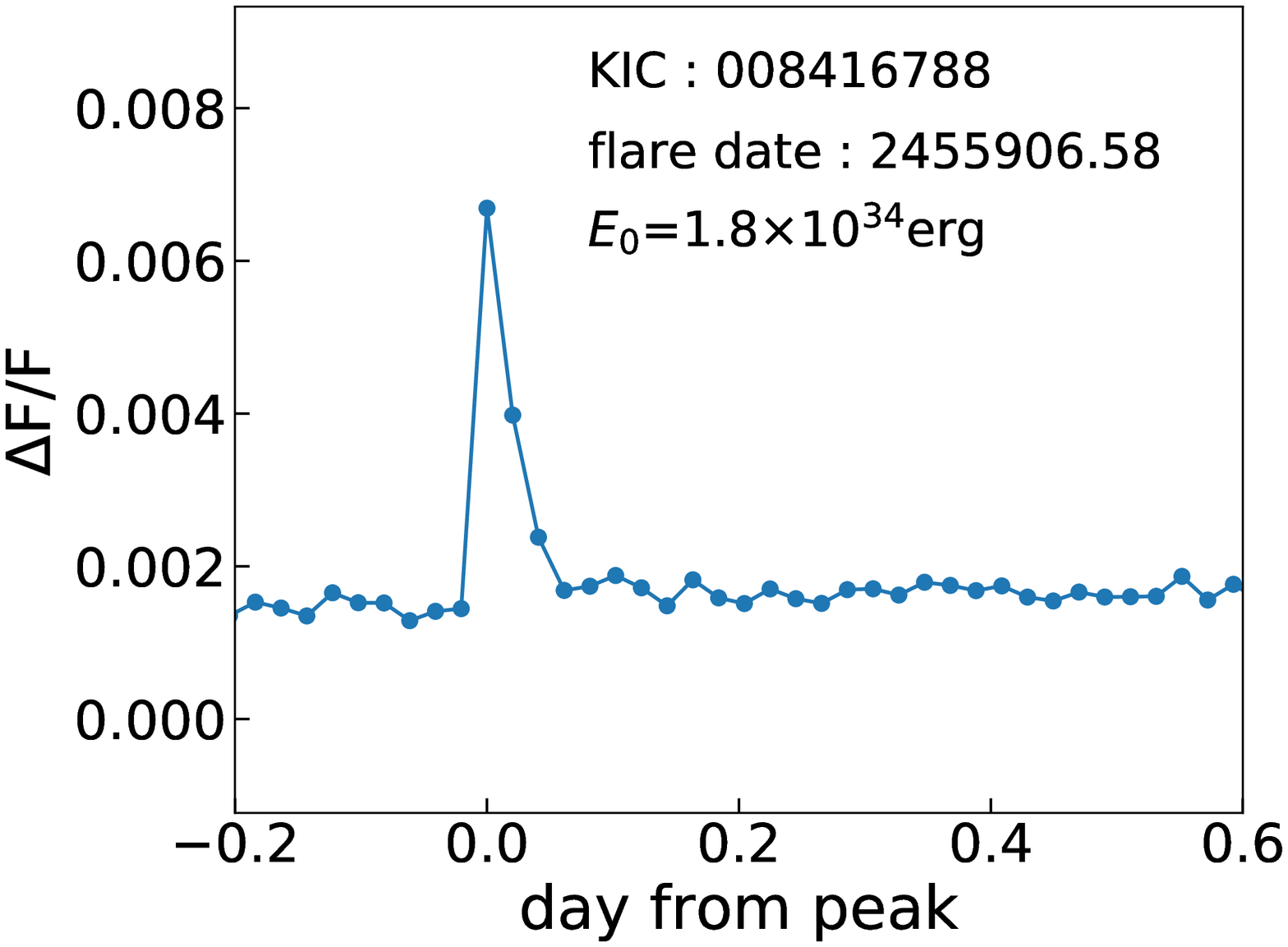}
   \caption{
   (Continued)
   }
 \end{figure}
  \addtocounter{figure}{-1}
   \begin{figure}
   \plottwo{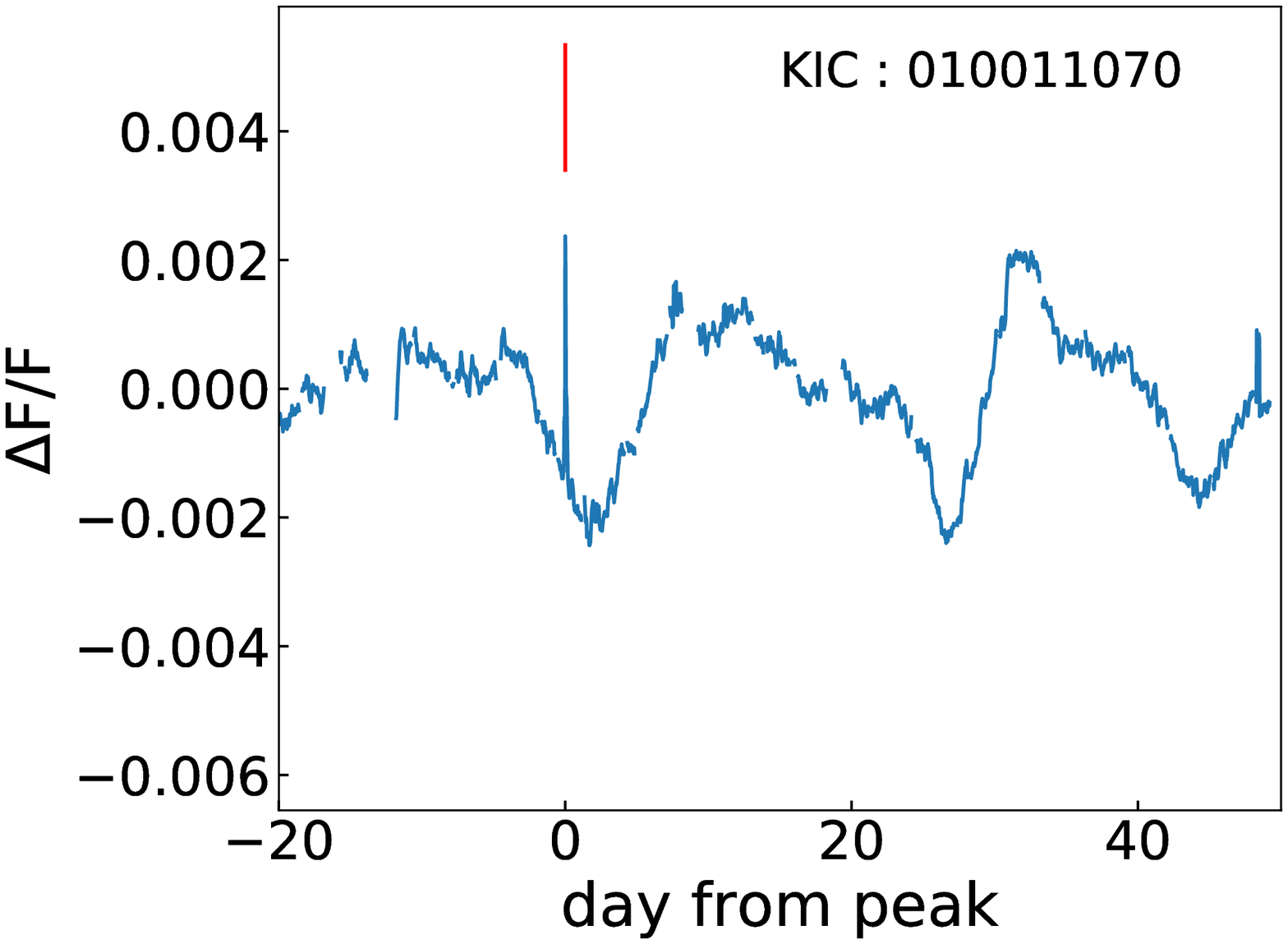}{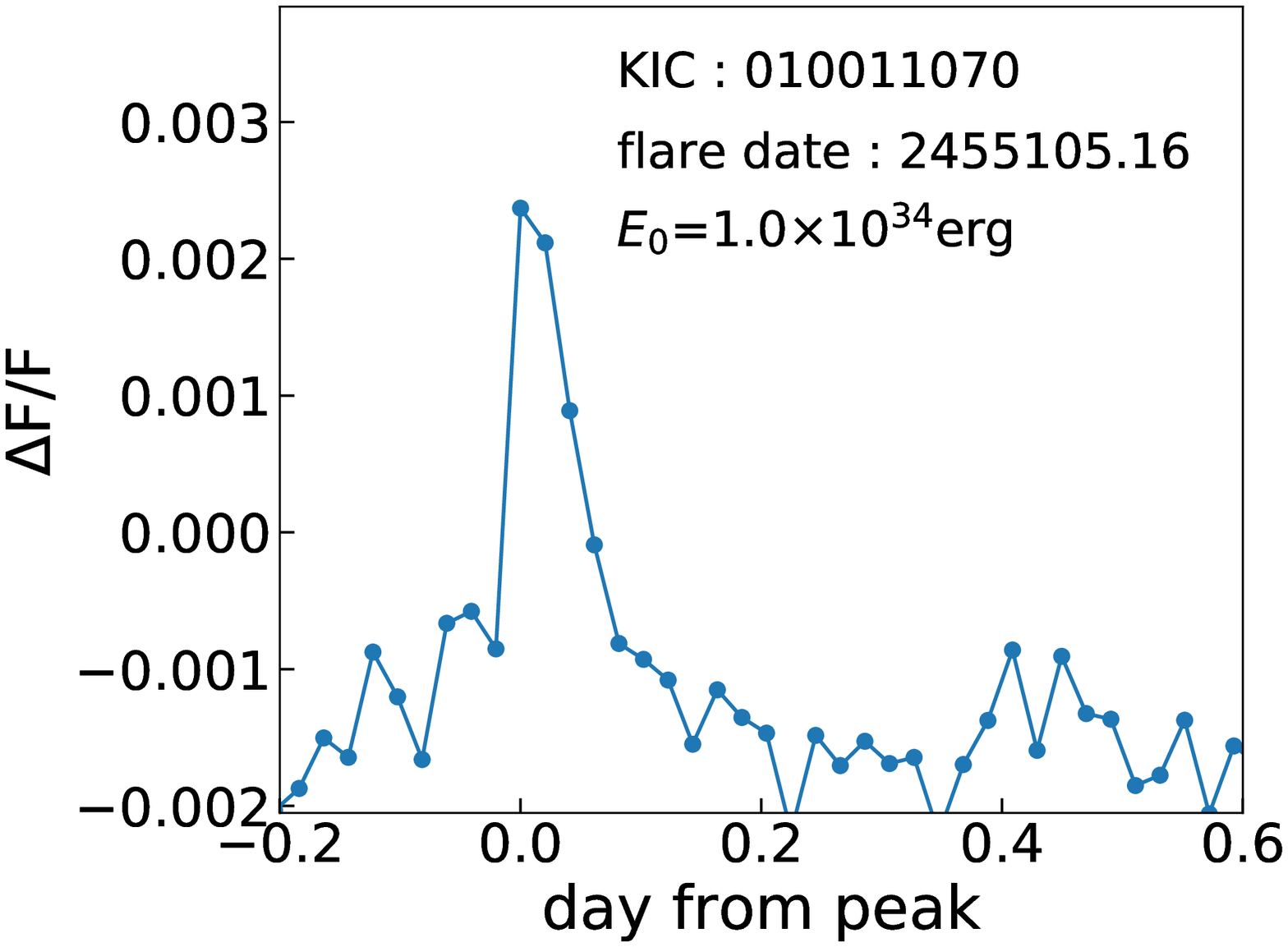}
   \plottwo{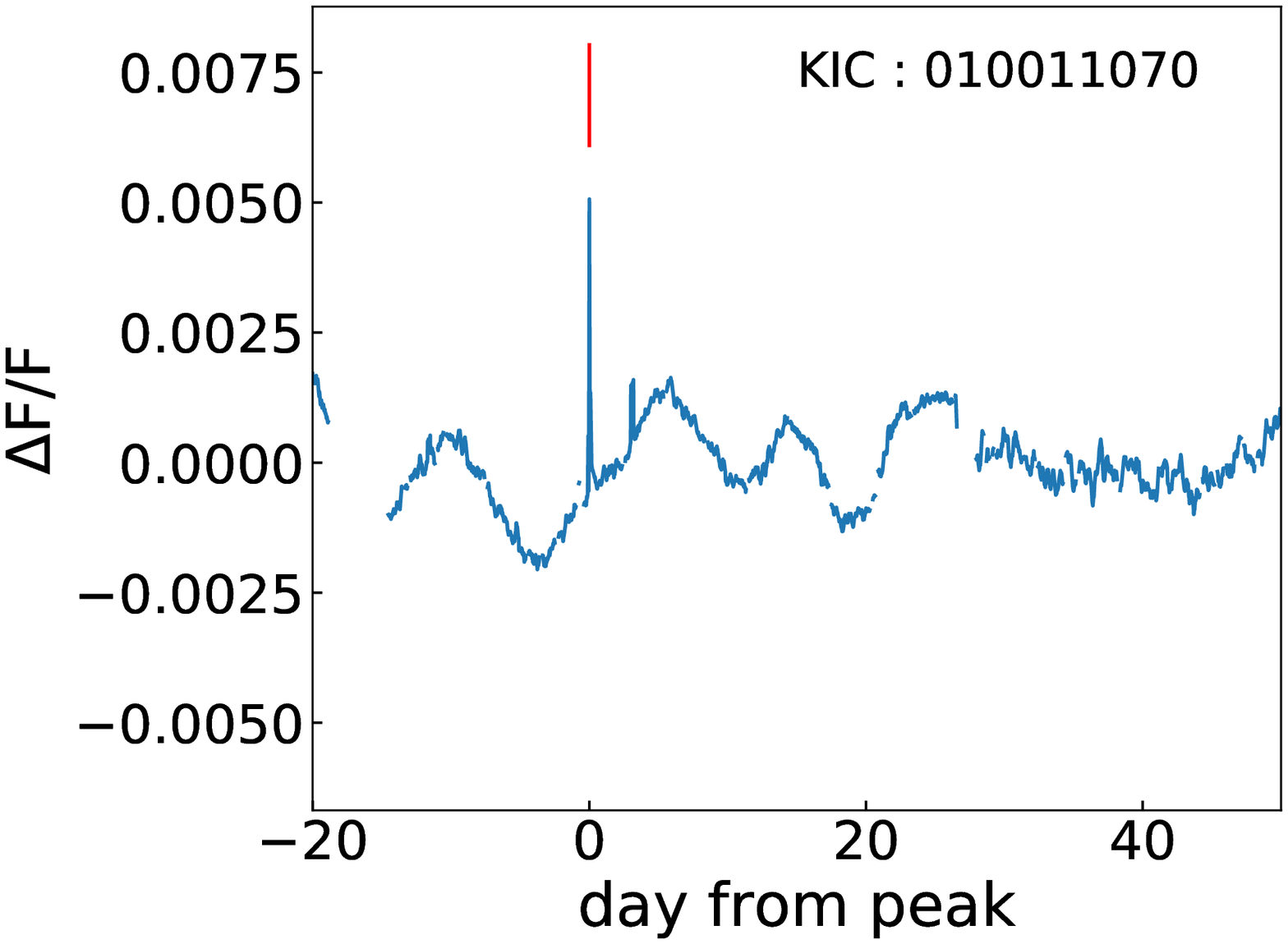}{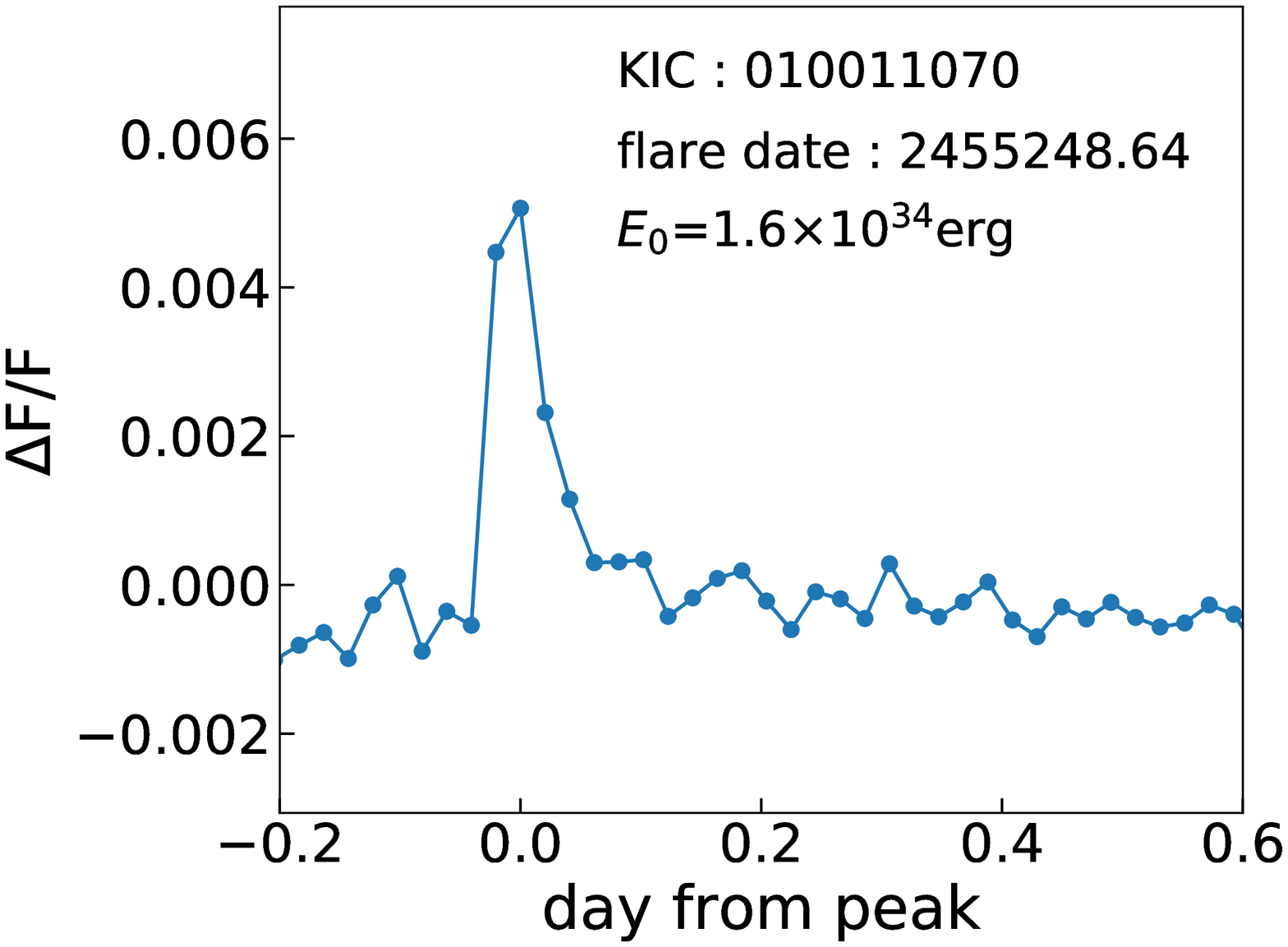}
   \plottwo{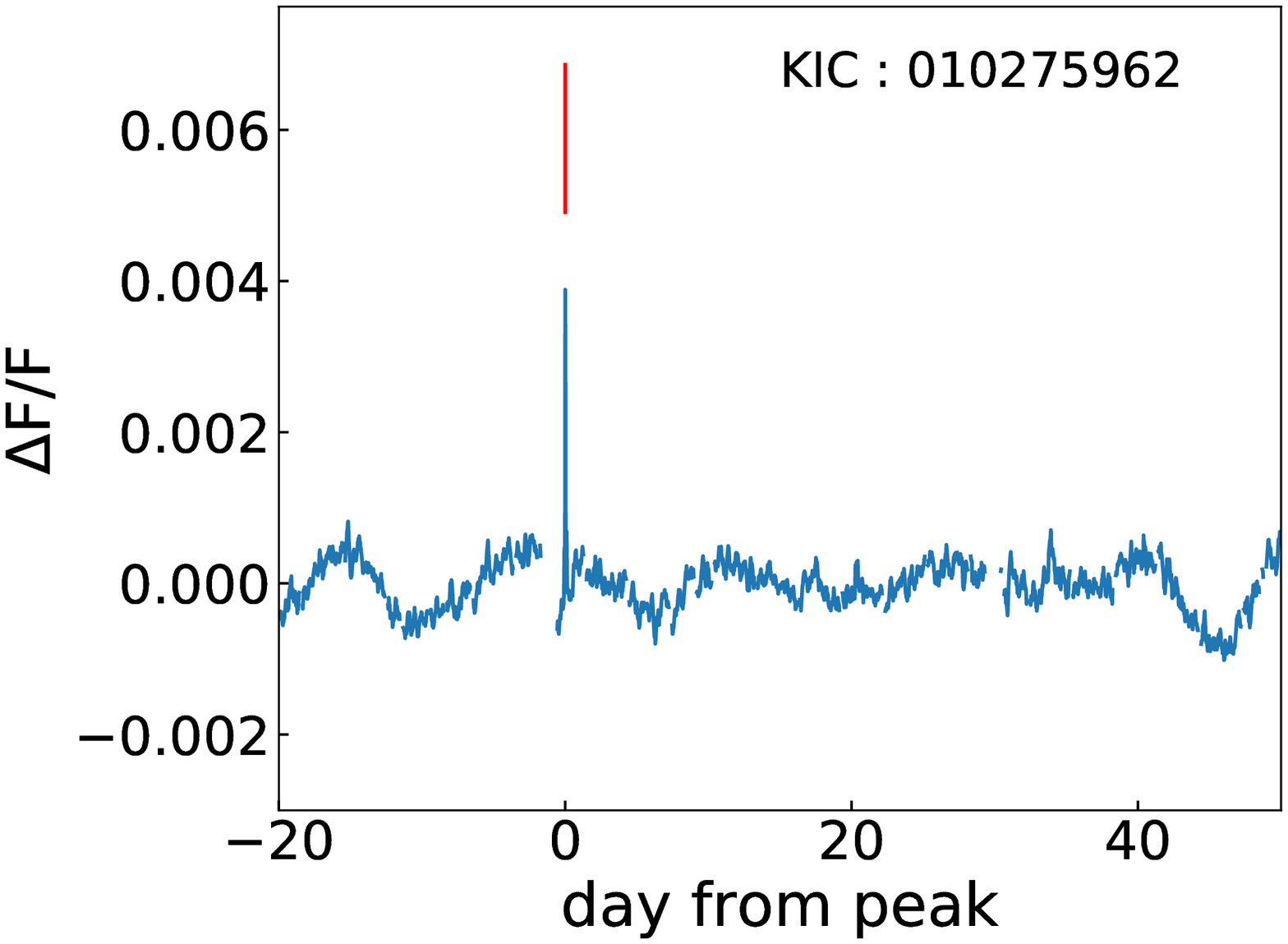}{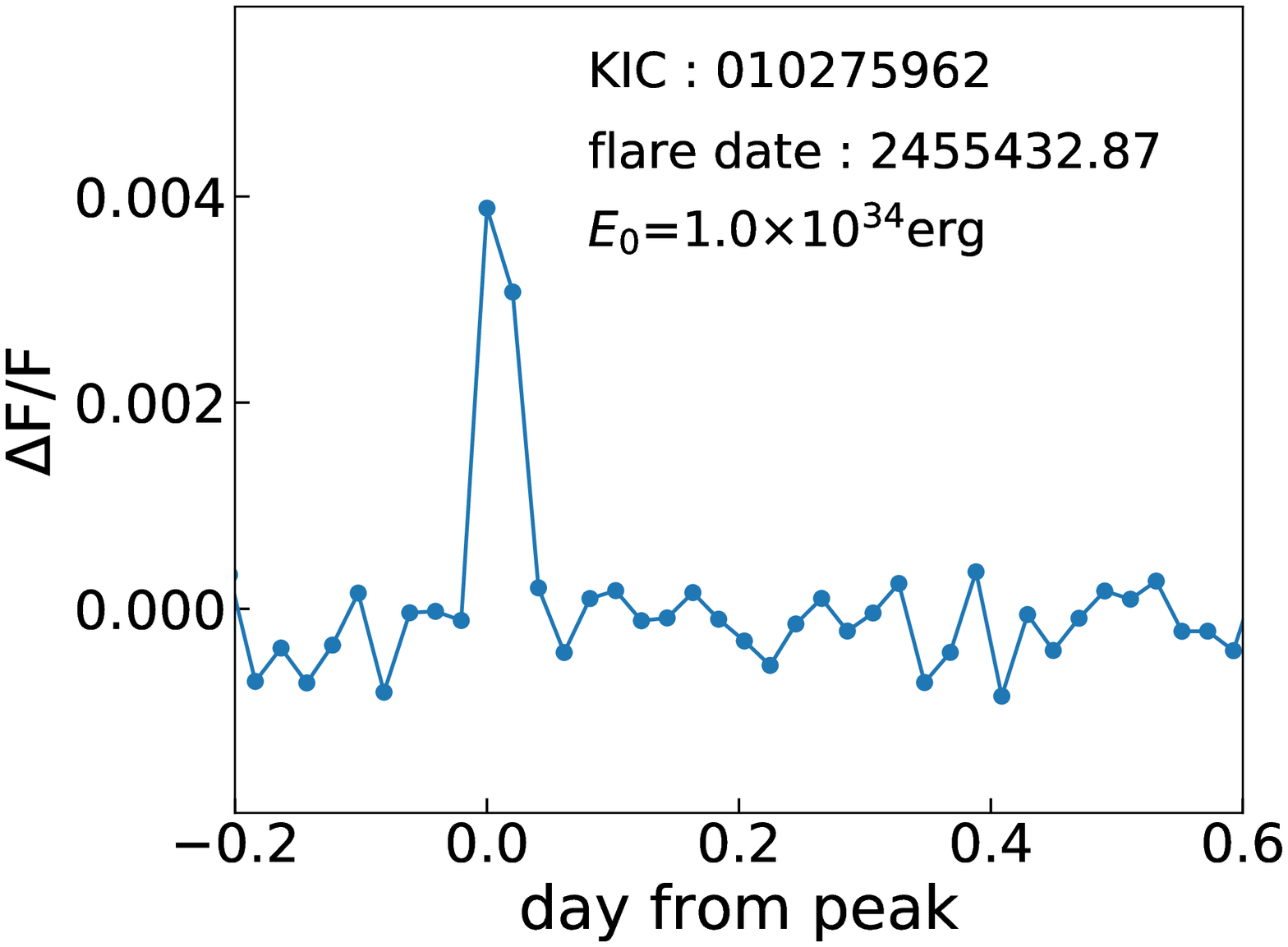}
   \caption{
    (Continued)
   }
 \end{figure}
 \addtocounter{figure}{-1}
 \begin{figure}
   \plottwo{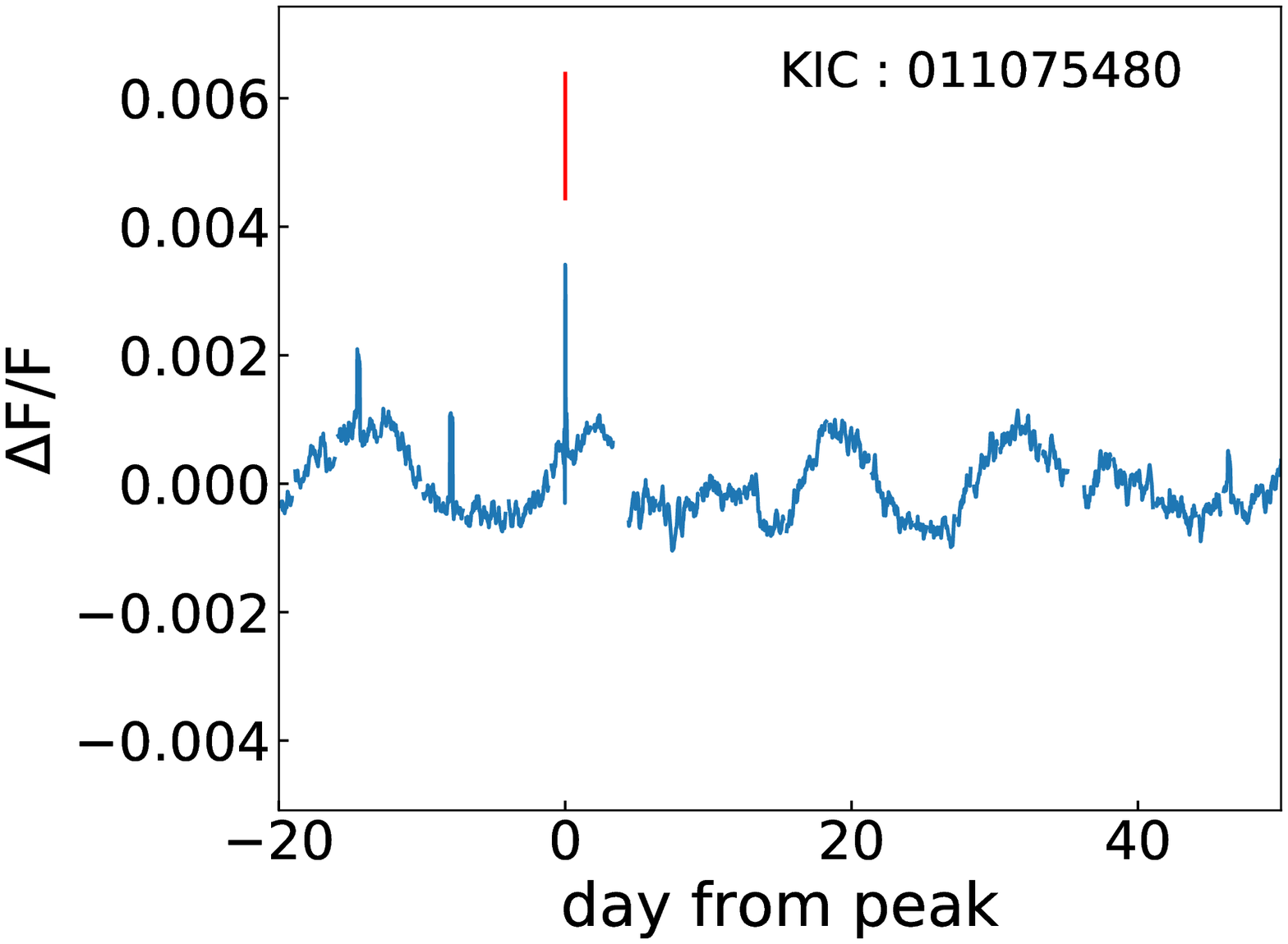}{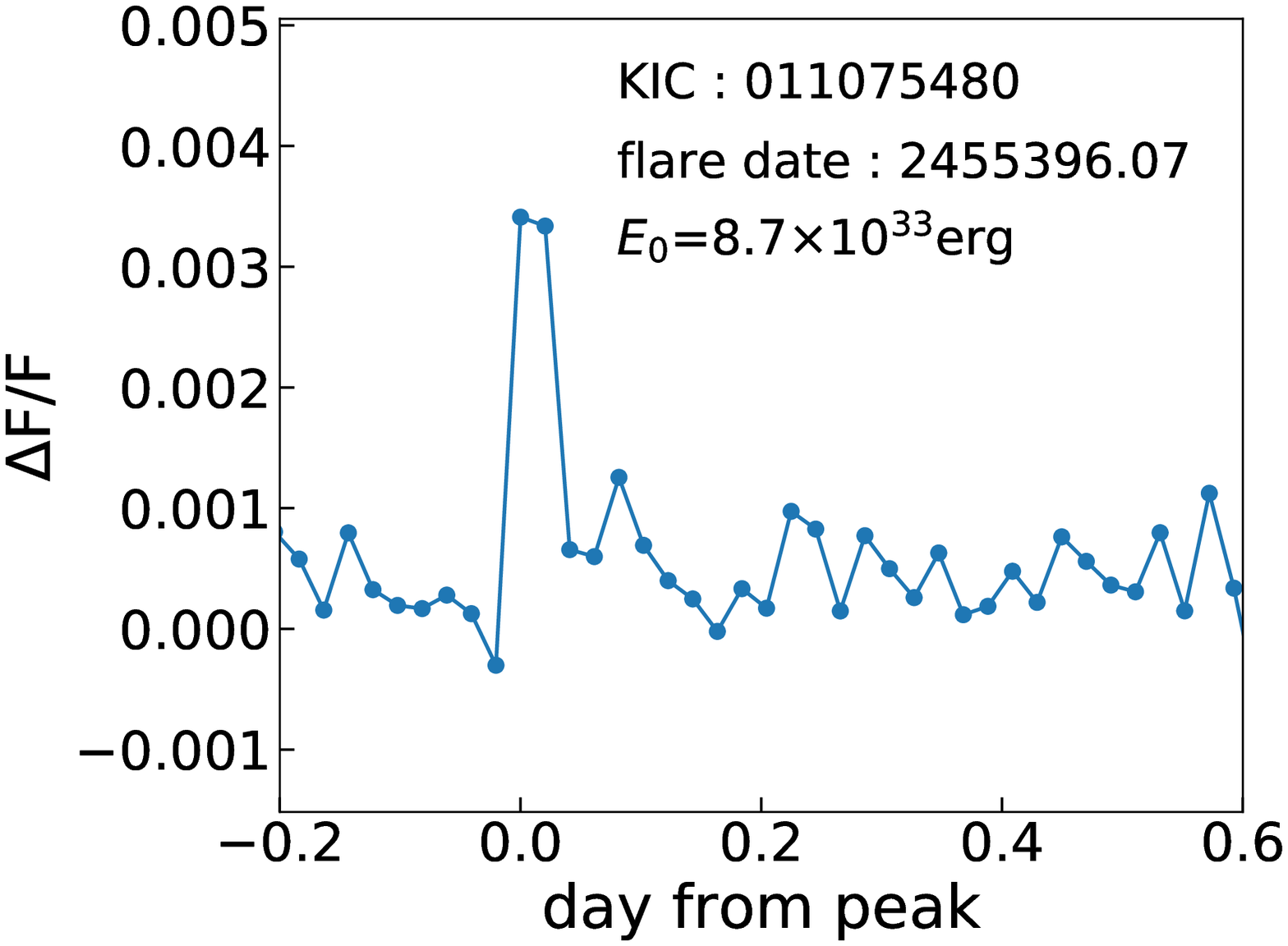}
   \plottwo{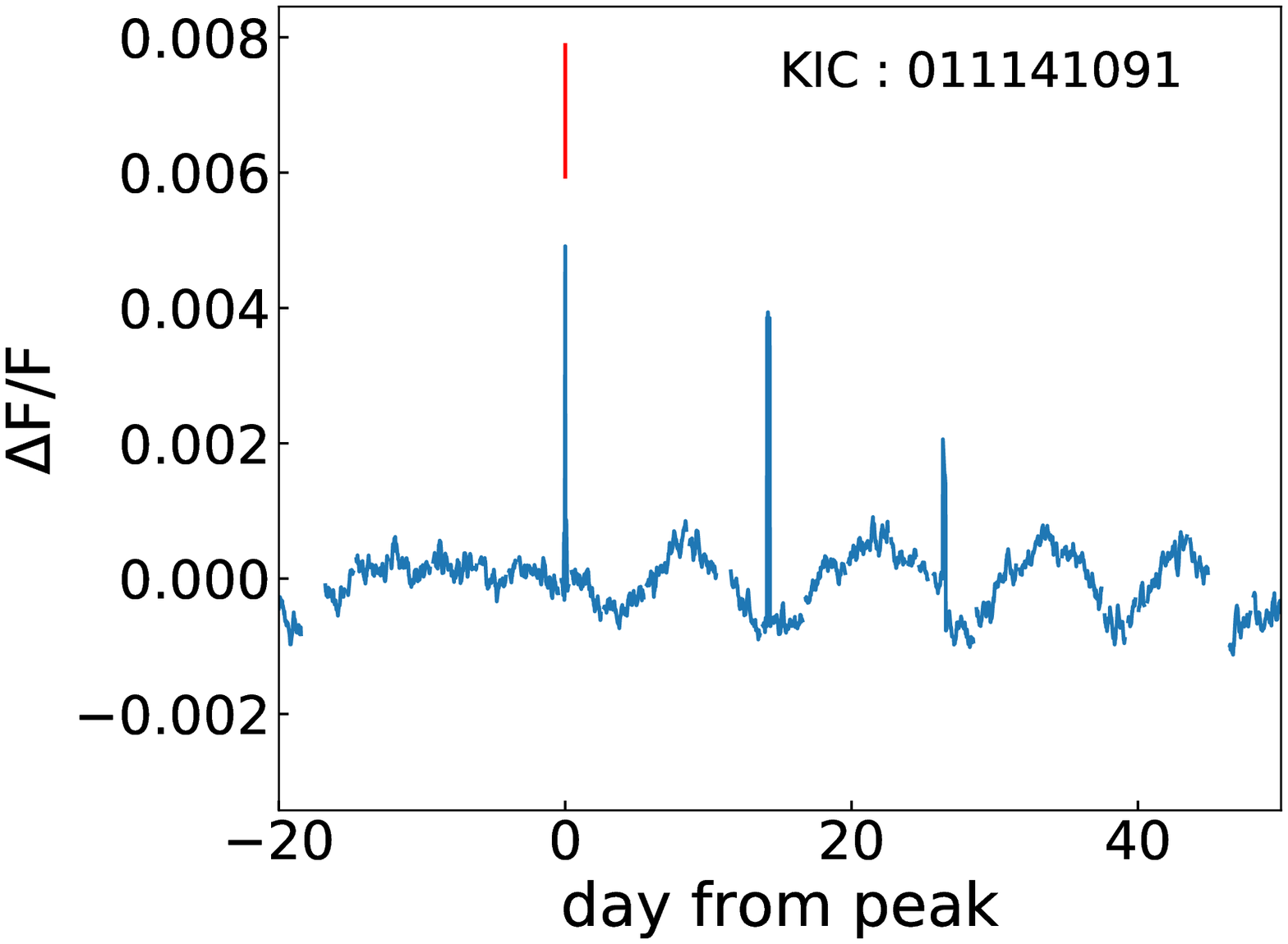}{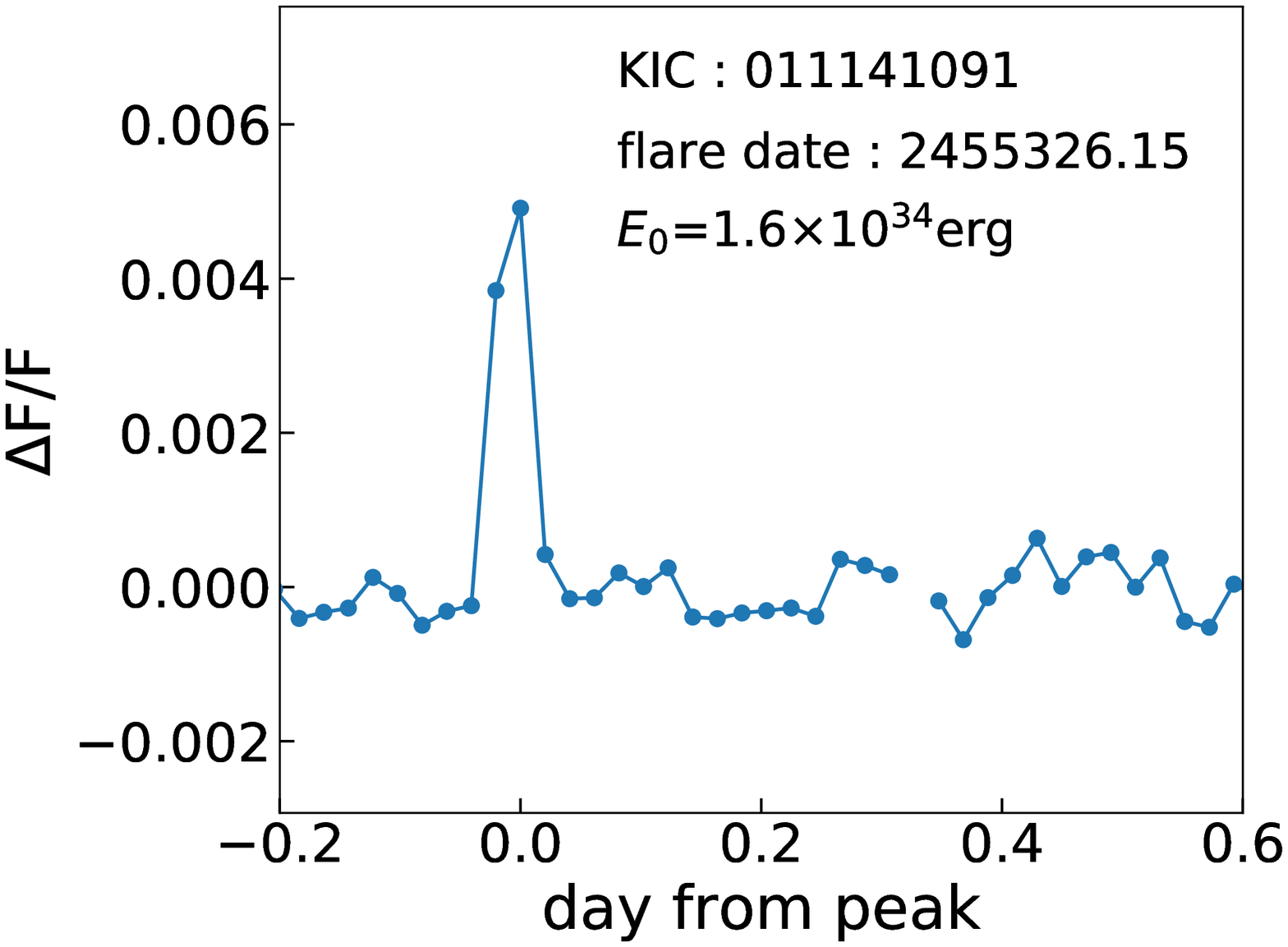}
   \plottwo{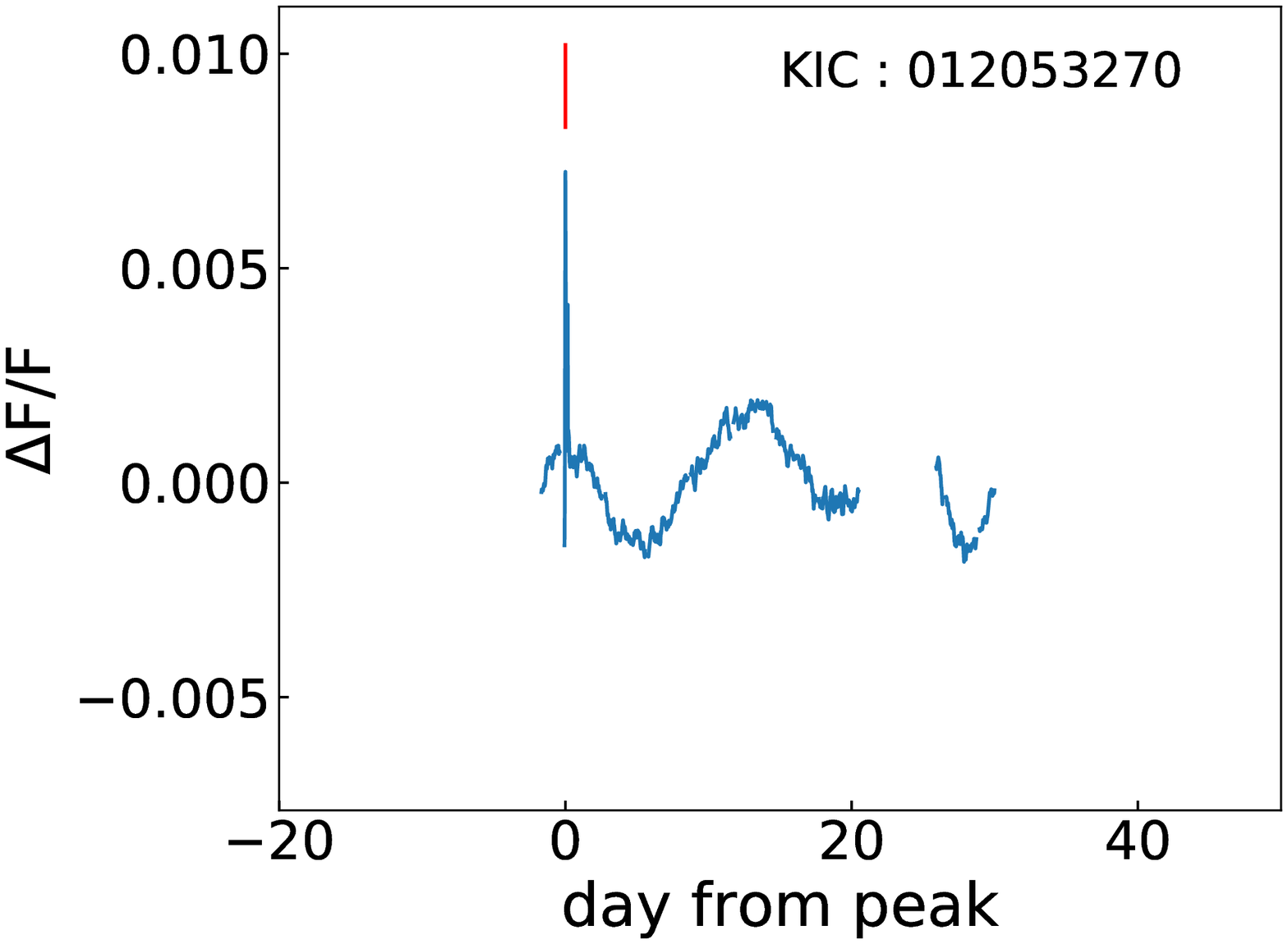}{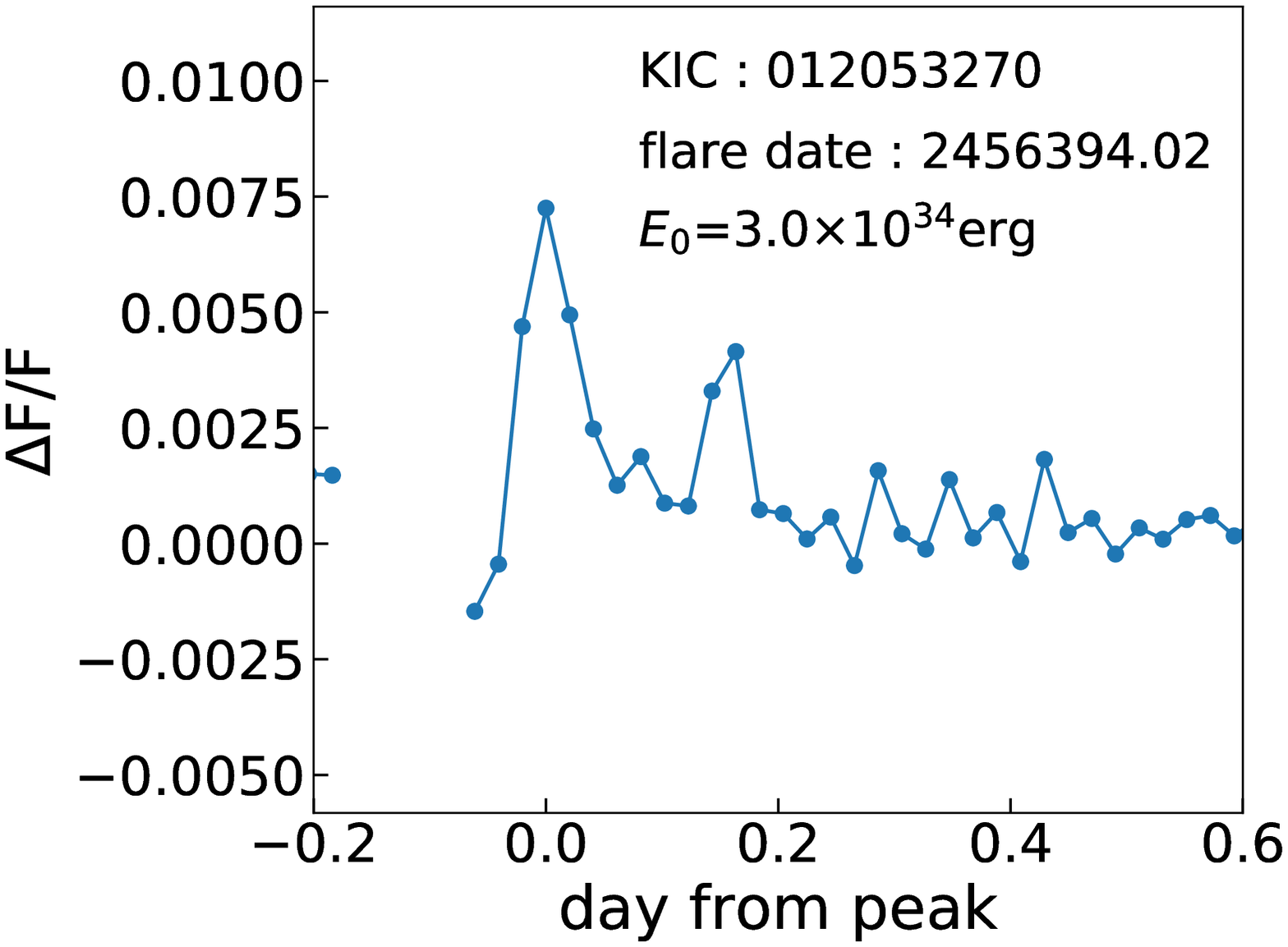}
   \caption{
   (Continued)
   }
\end{figure}

 \begin{figure}
   \plottwo{figure/lightcurve_sunlikeflarestar/flarelightcurve_005648294_2456309.86rd.eps}{figure/lightcurve_sunlikeflarestar/flarelightcurve_005648294_2456309.86la.eps}
   \plottwo{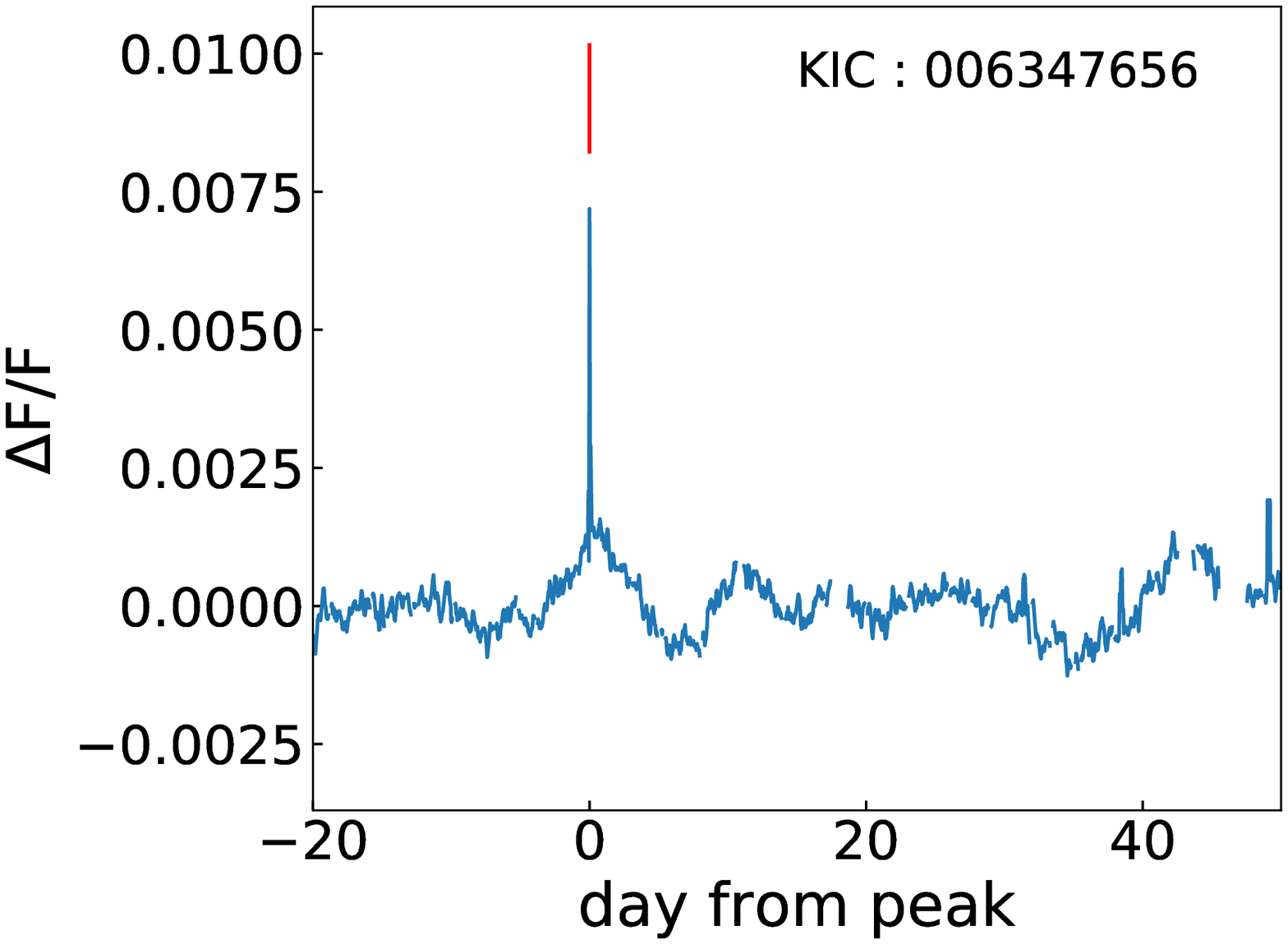}{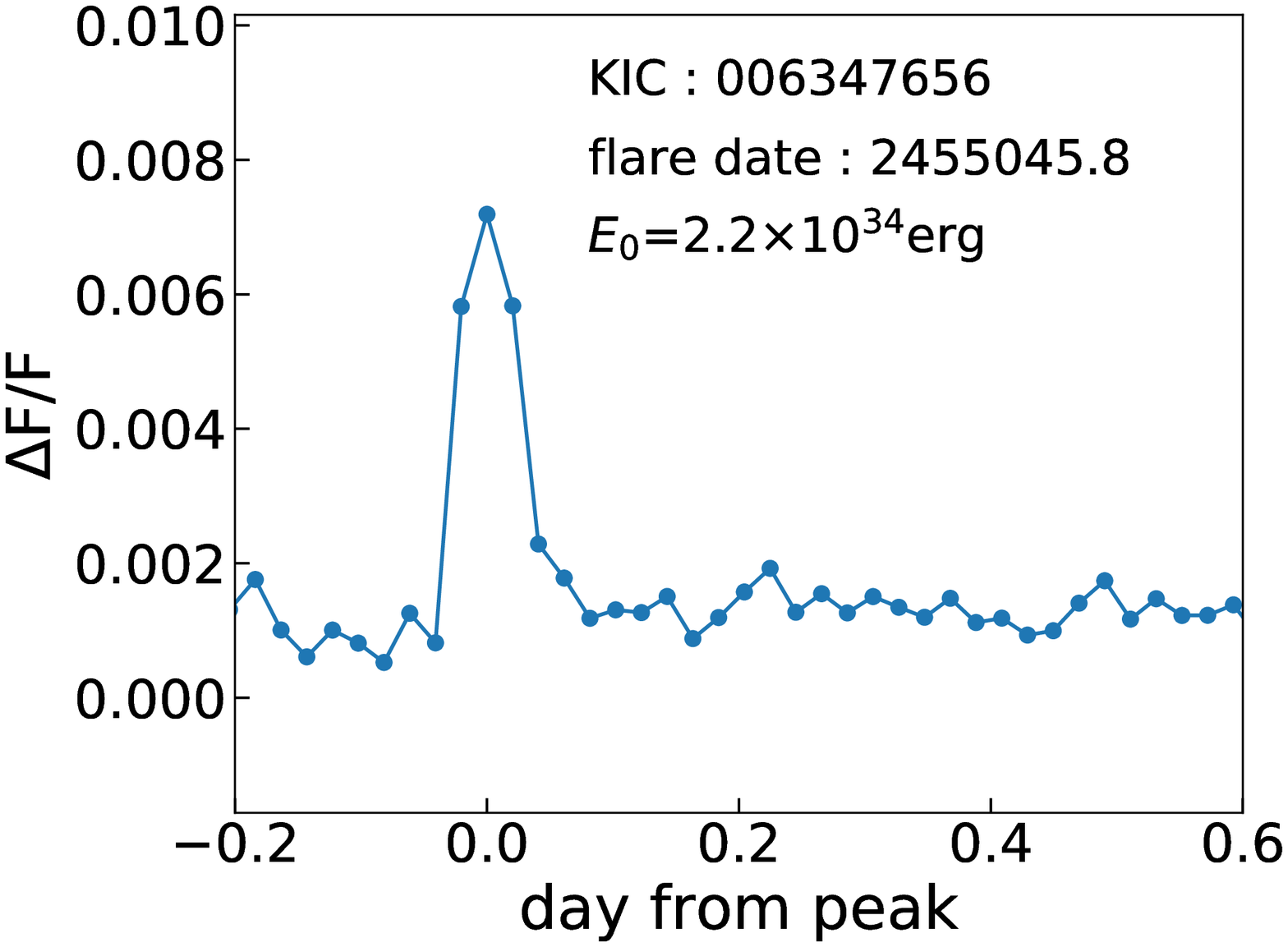}
   \plottwo{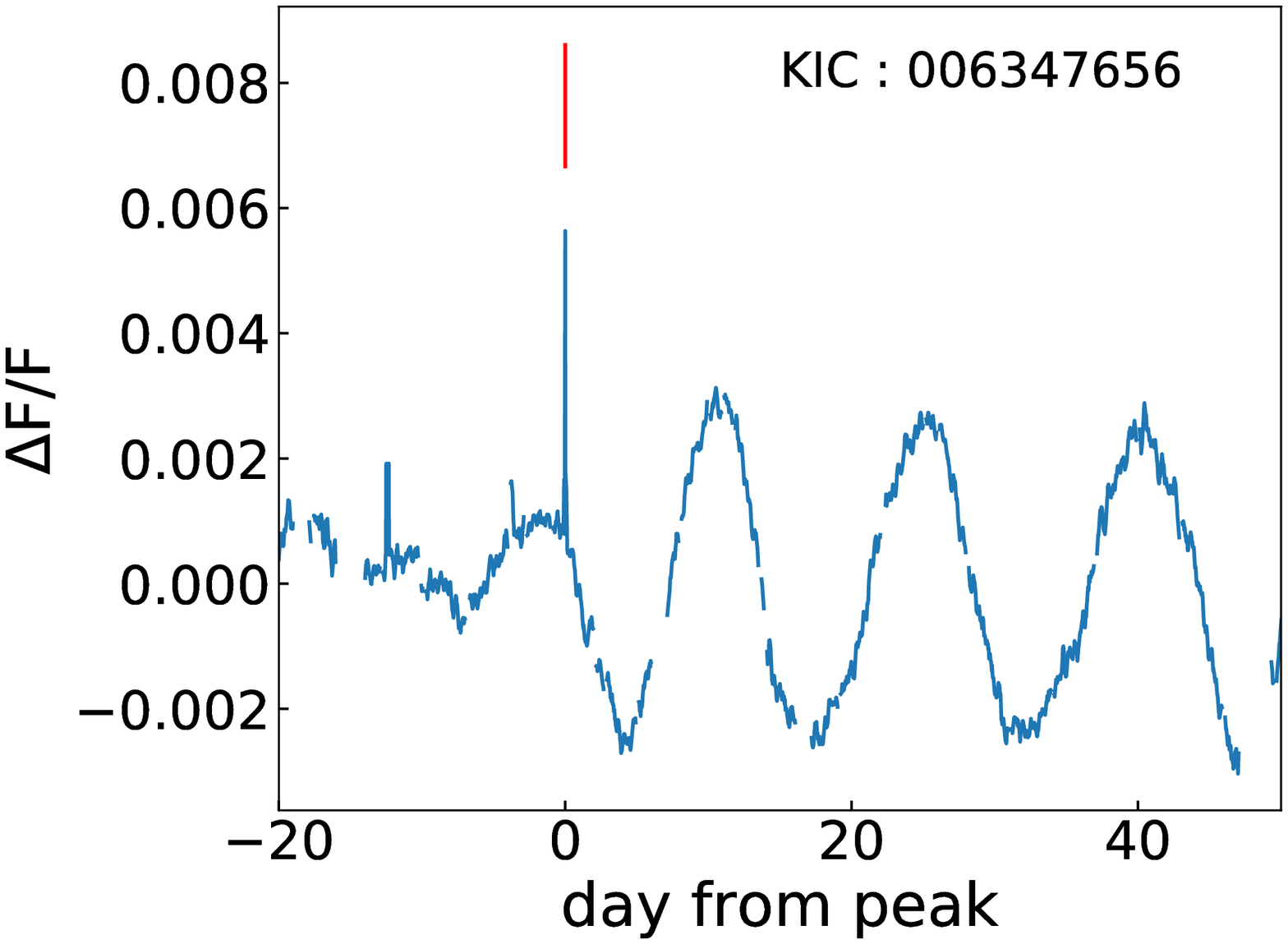}{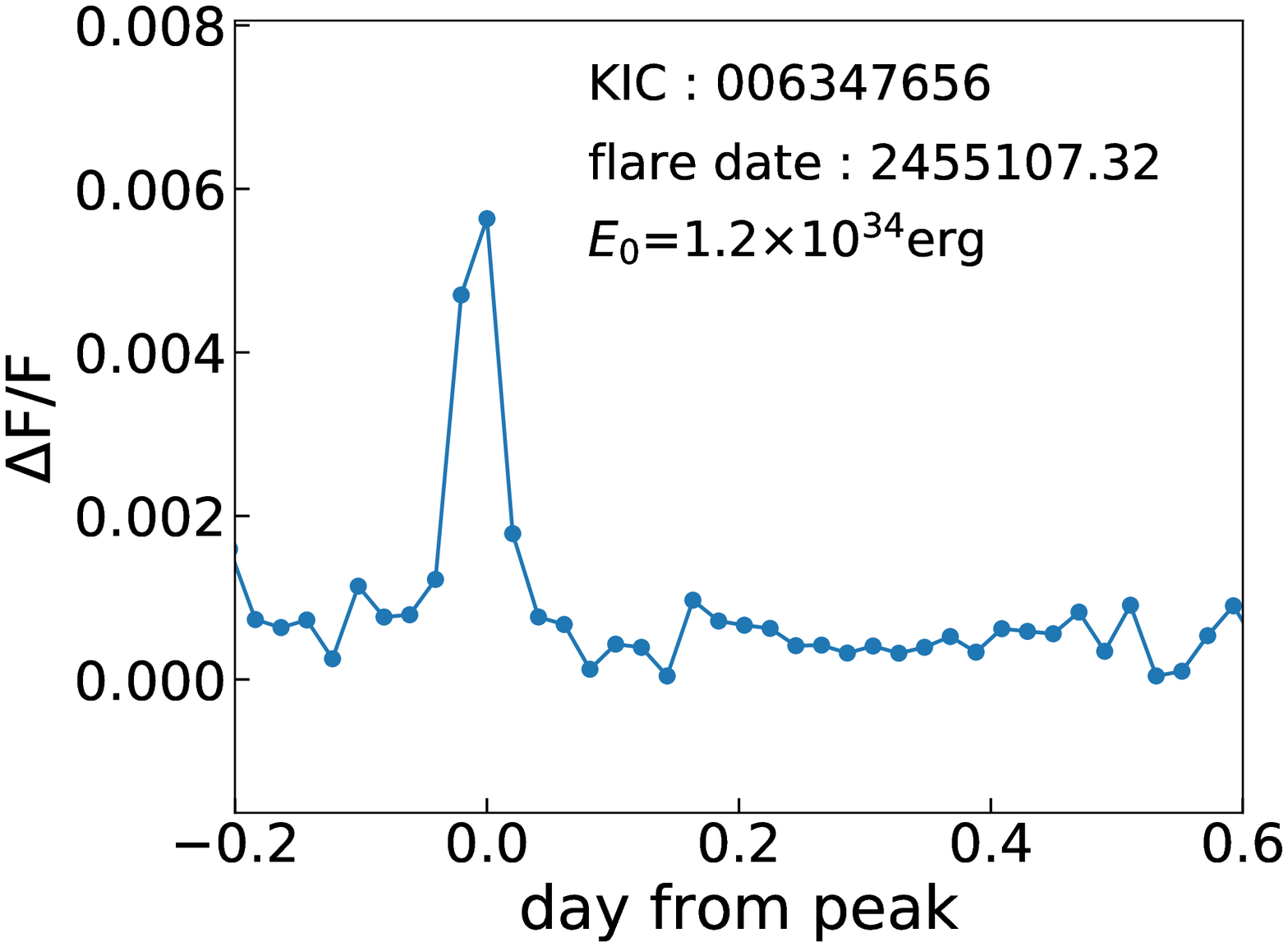}
  \caption{
    Same as Figure \ref{fig:eg_sunlikeflare_noflag} but for the 5 stars that have Flag ``1" in Table \ref{table:params_sunlike}.
   }
  \label{fig:eg_sunlikeflare_flag1}
   \end{figure}
  \addtocounter{figure}{-1}
   \begin{figure}
   \plottwo{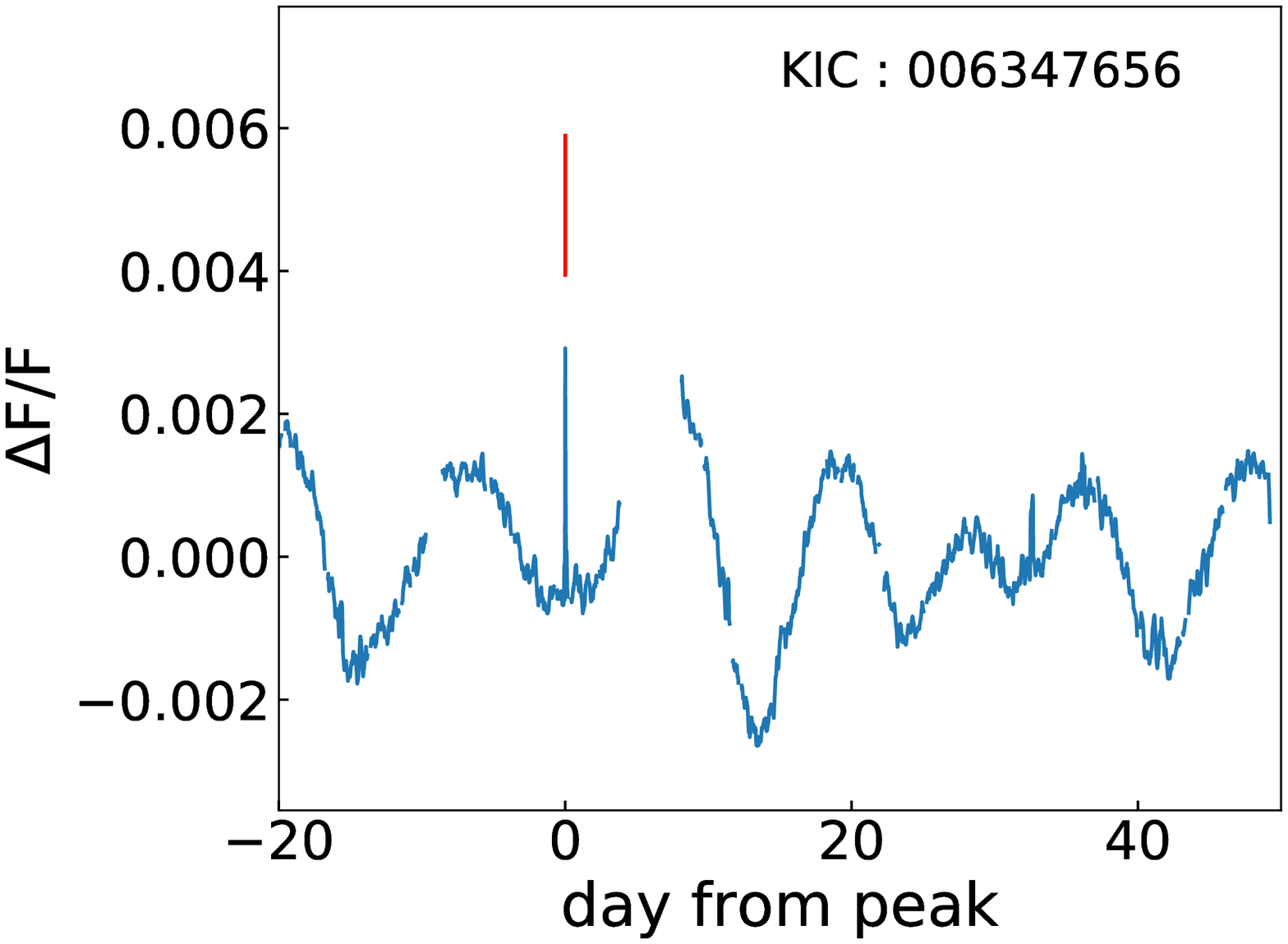}{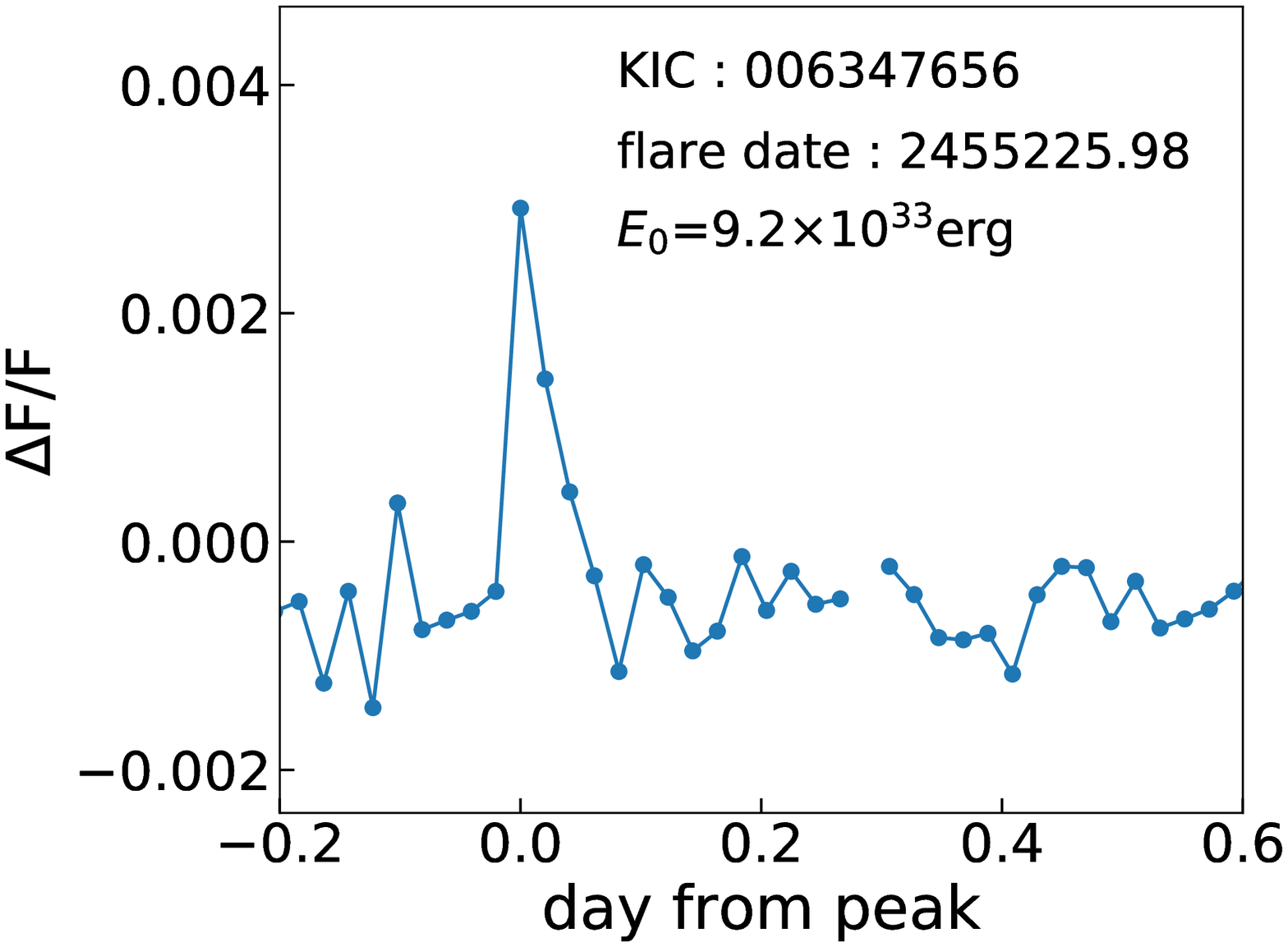}   
   \plottwo{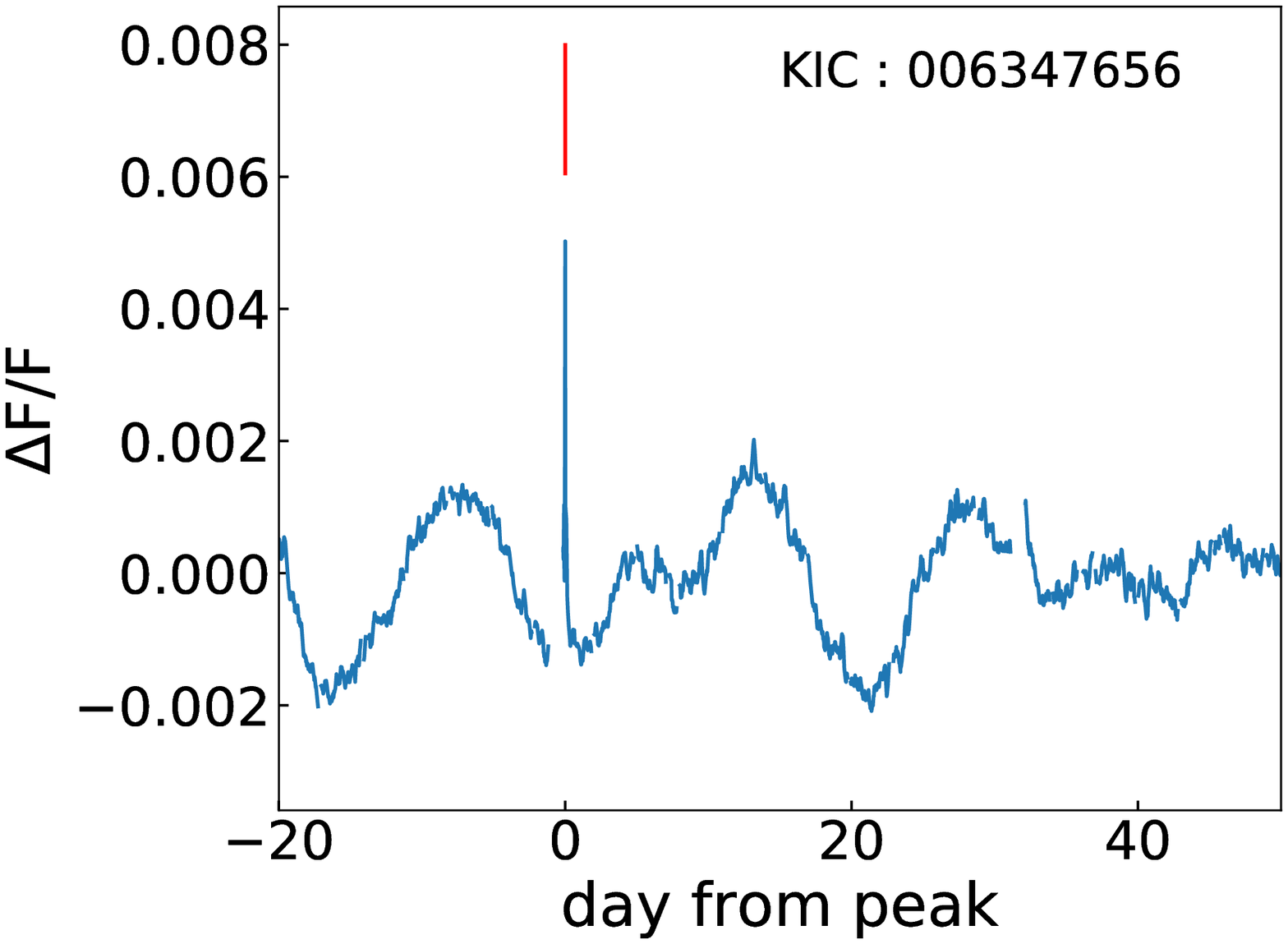}{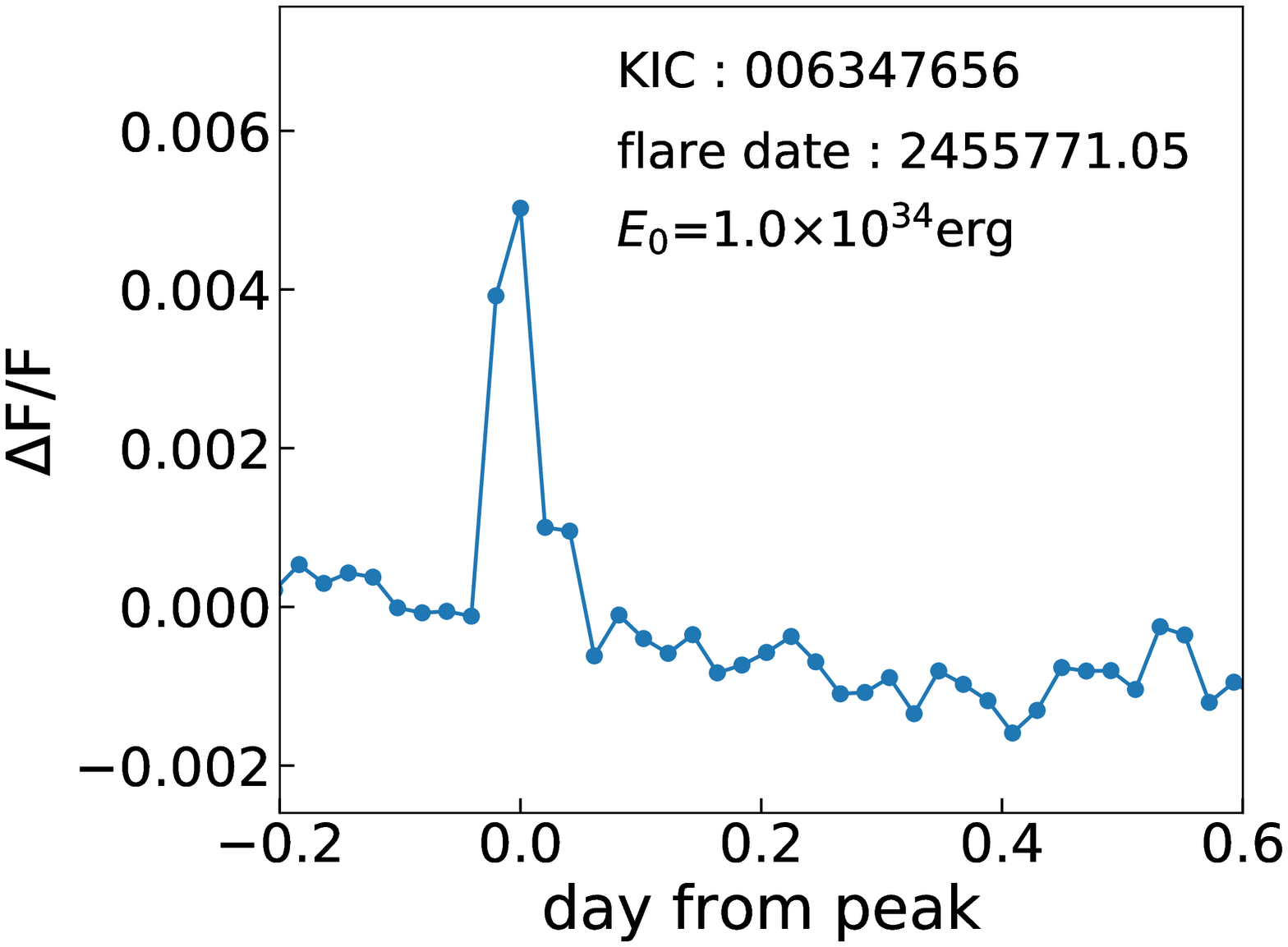}
   \plottwo{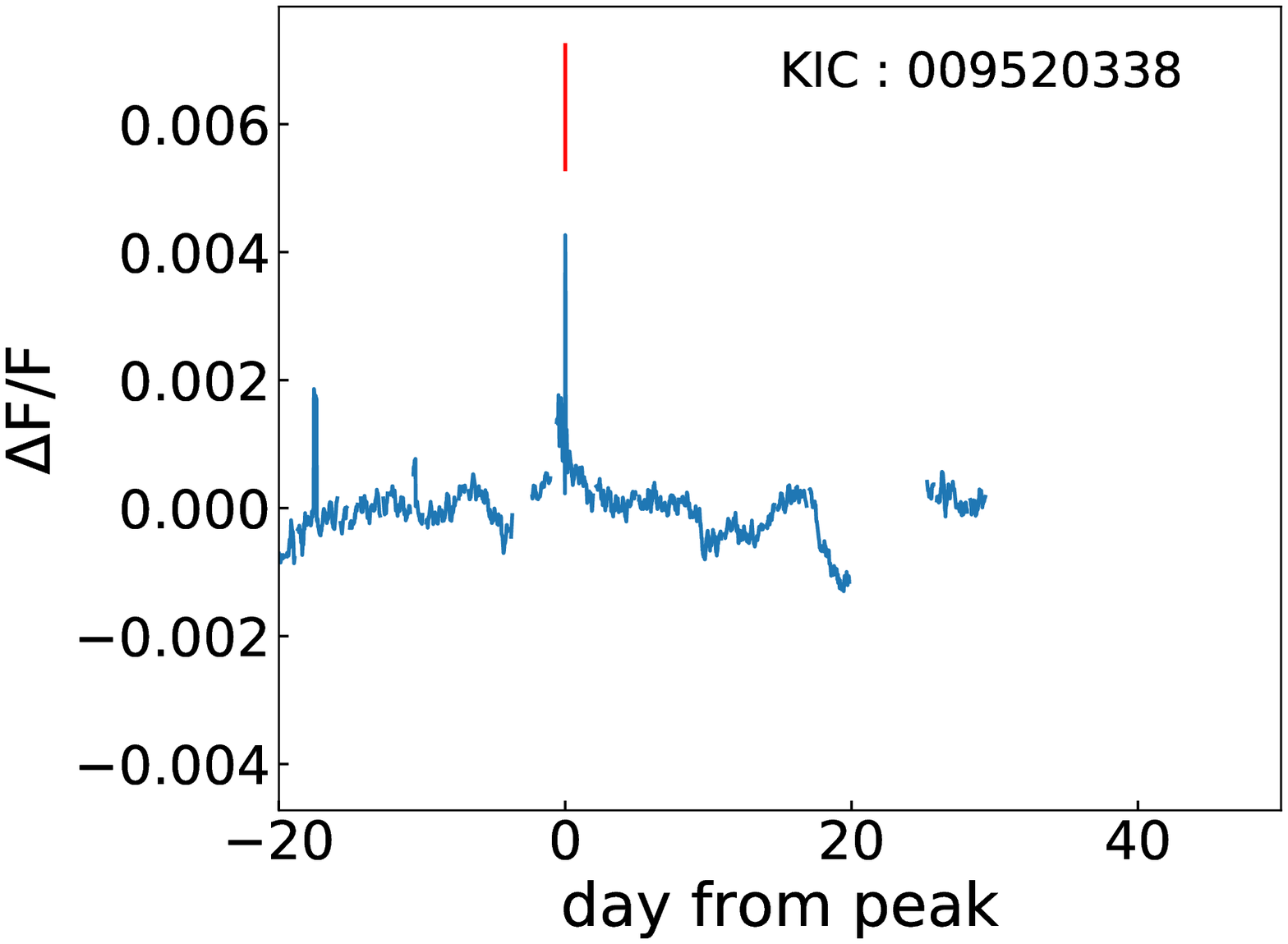}{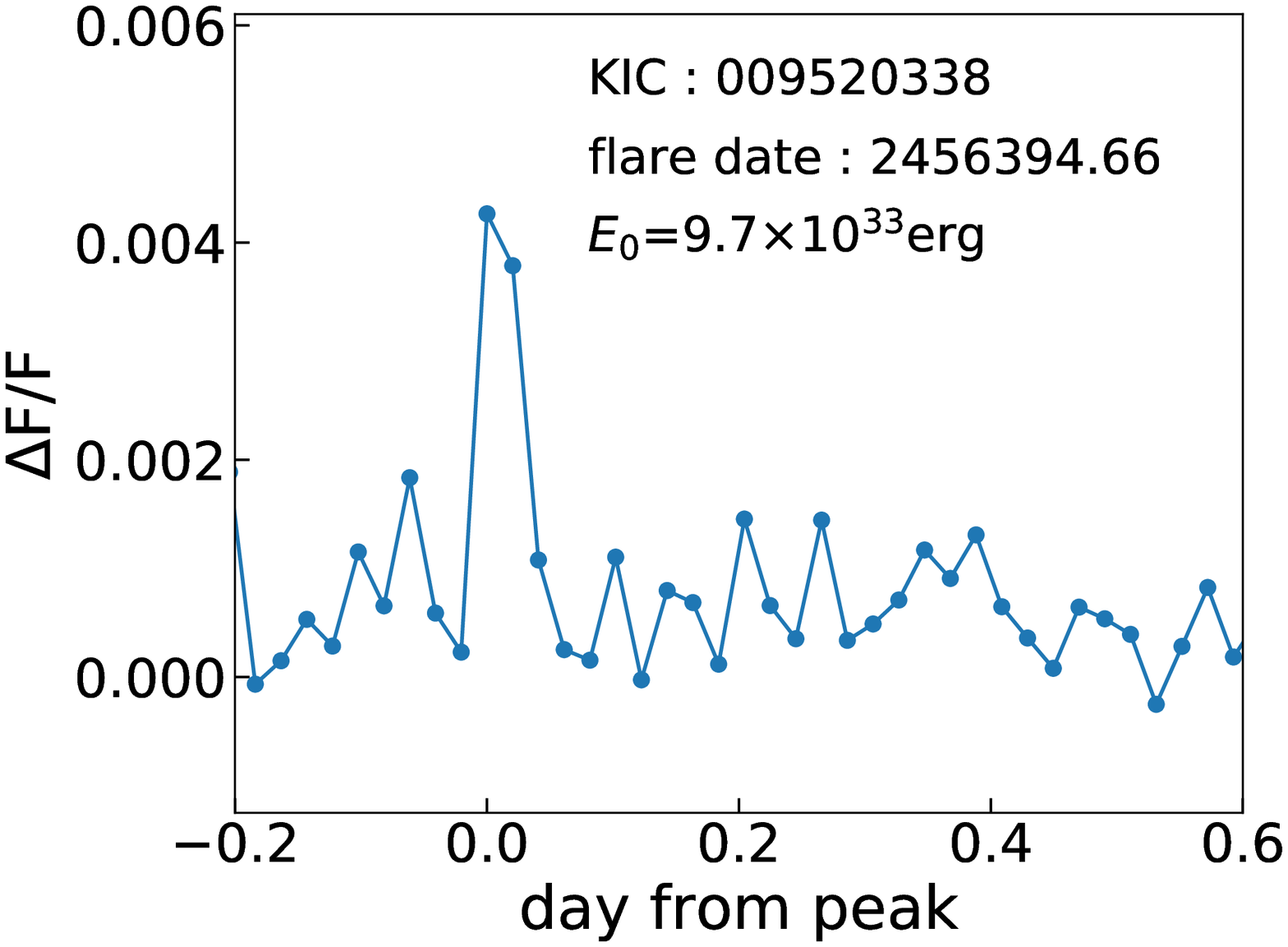}
   \caption{
   (Continued)
   }   
   \end{figure}
  \addtocounter{figure}{-1}
   \begin{figure}   
   \plottwo{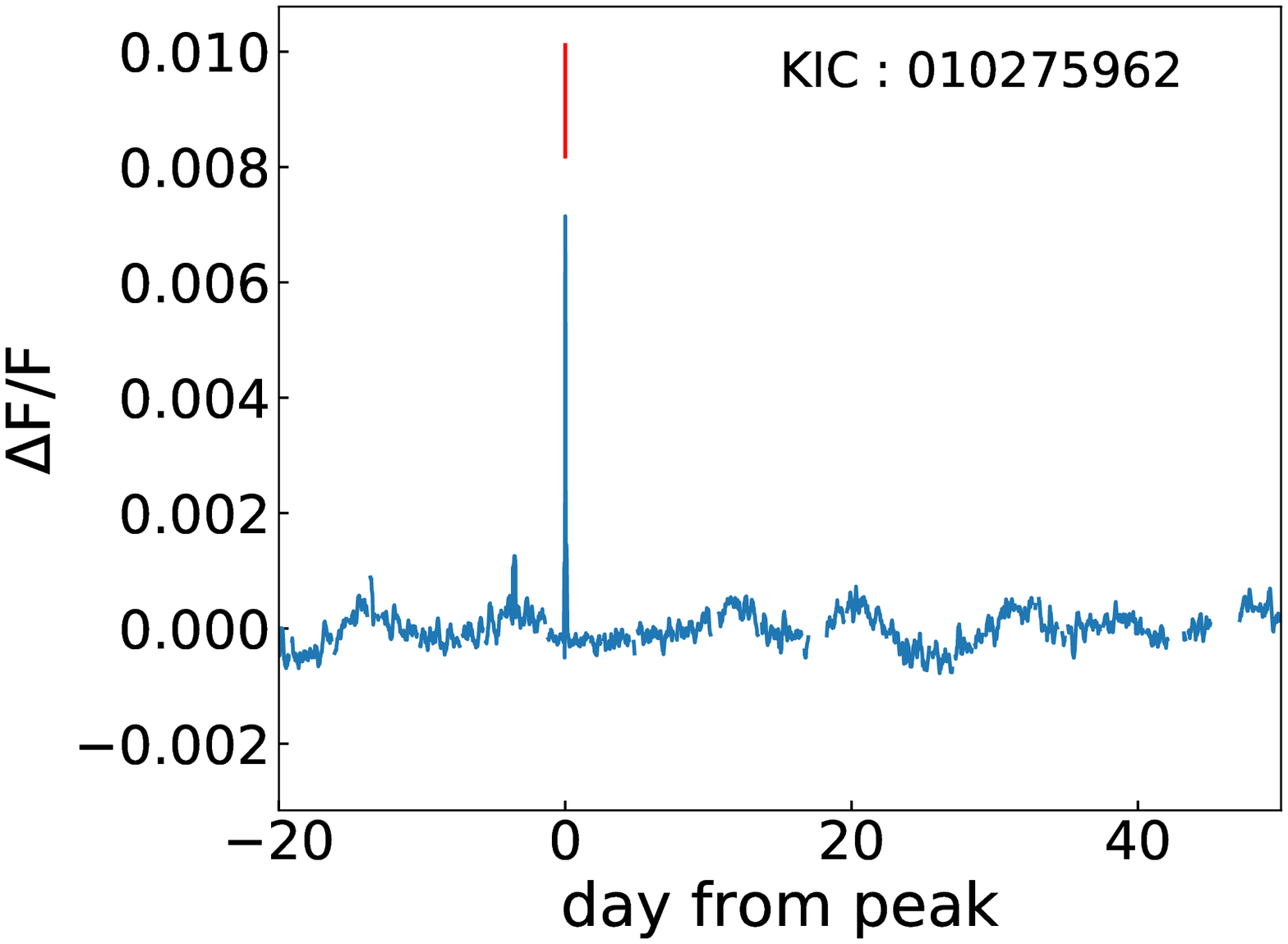}{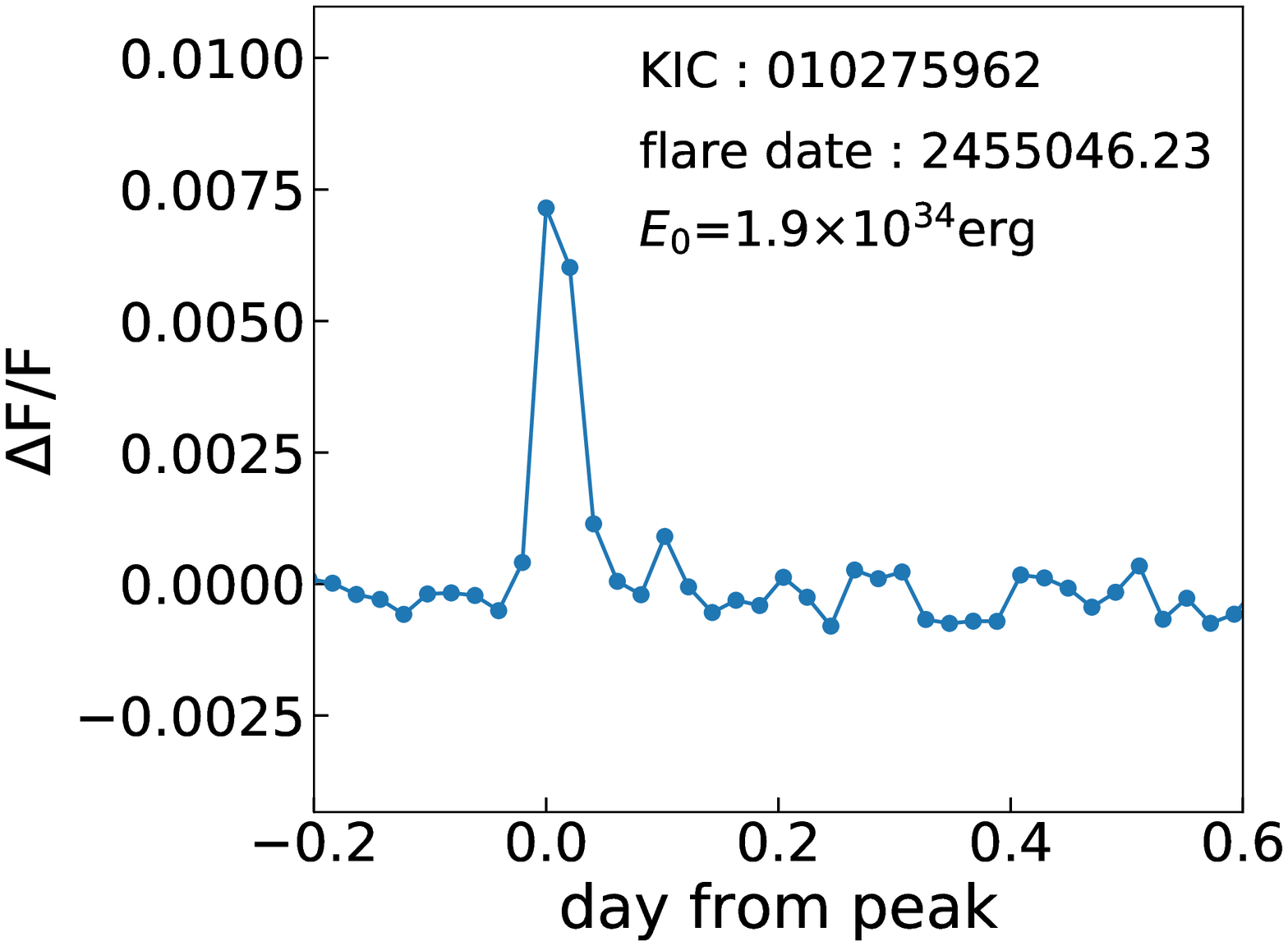}
   \plottwo{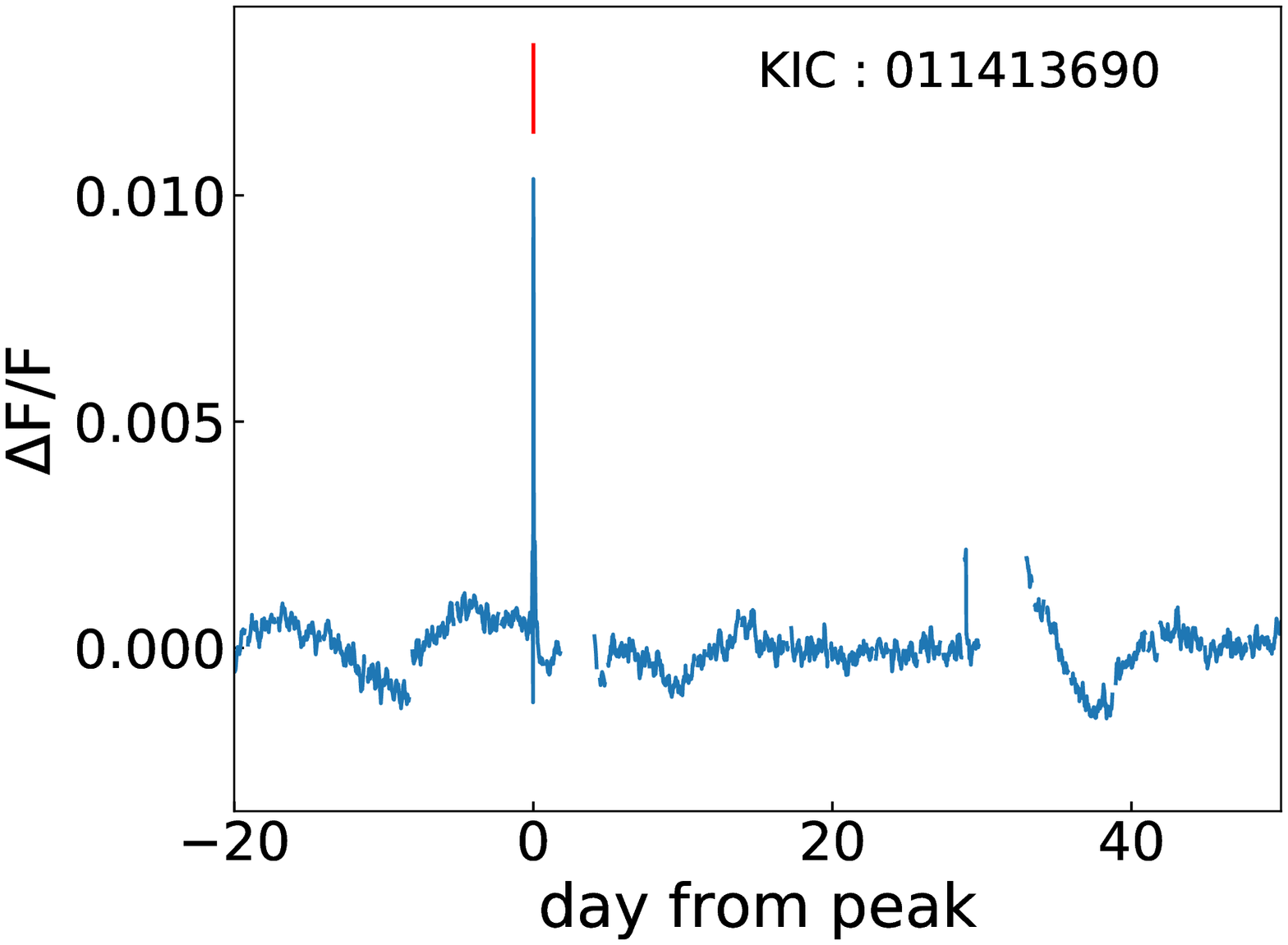}{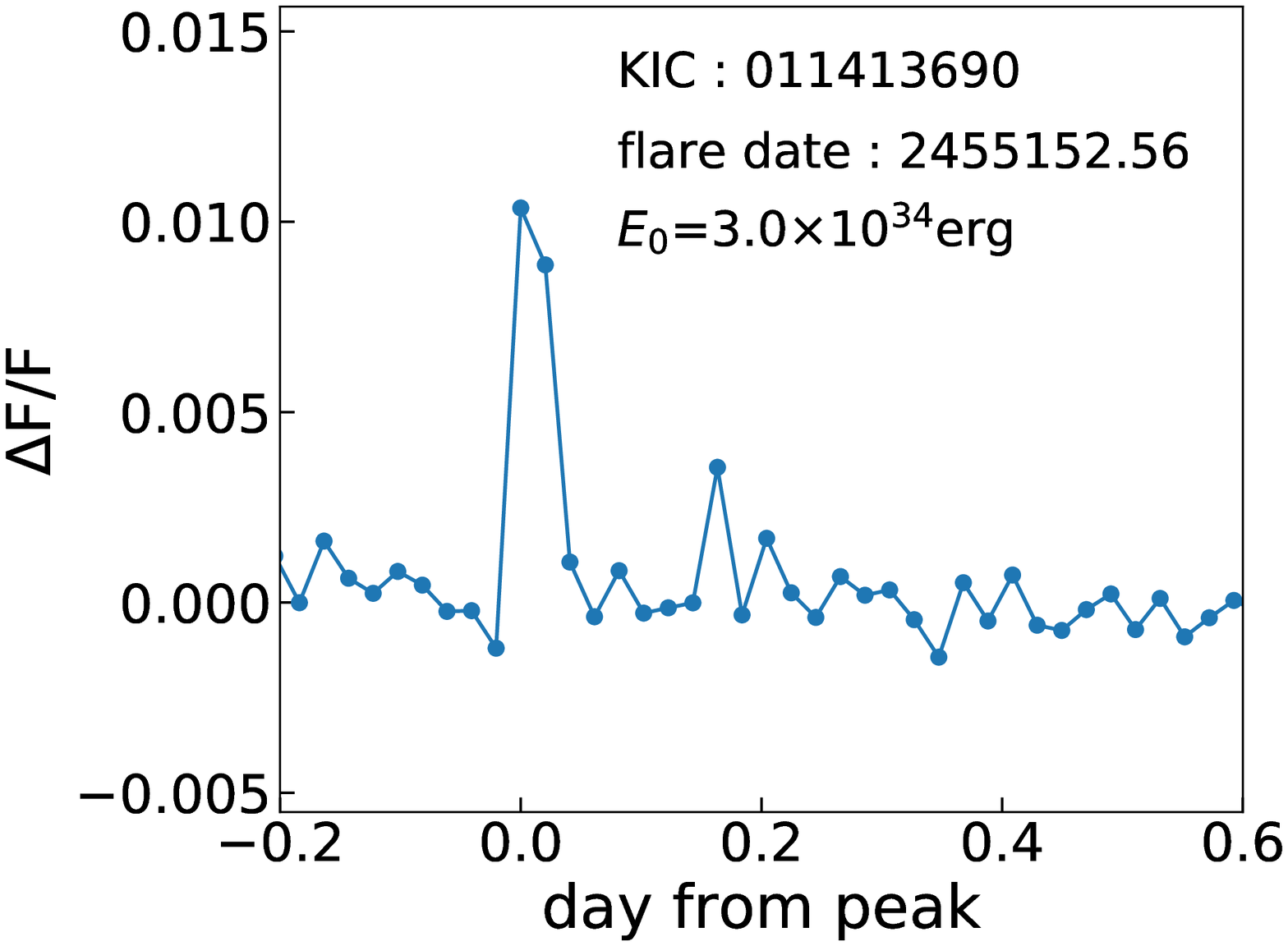}
  \caption{
  (Continued)
   }
  \end{figure}

 \begin{figure}
   \plottwo{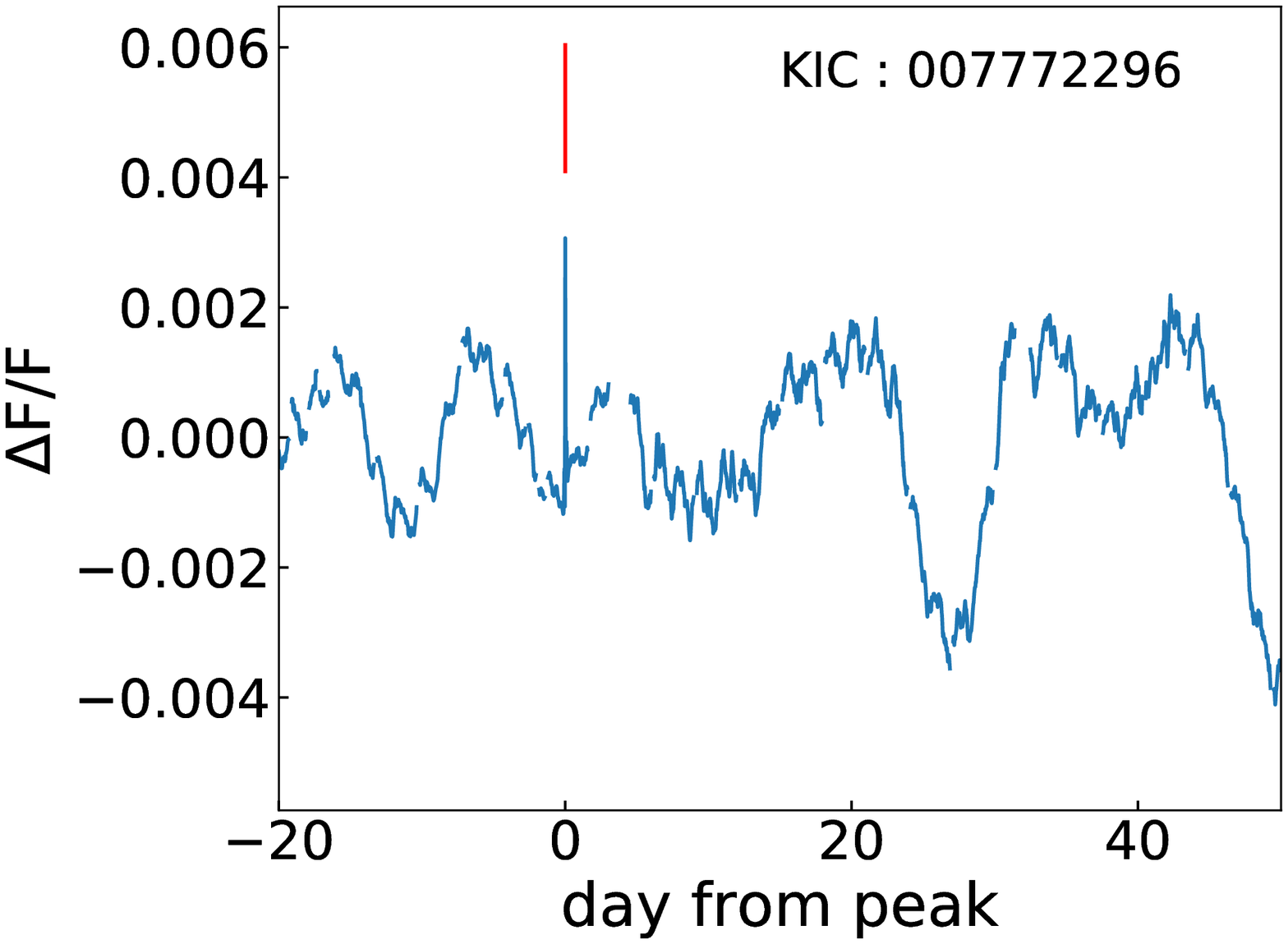}{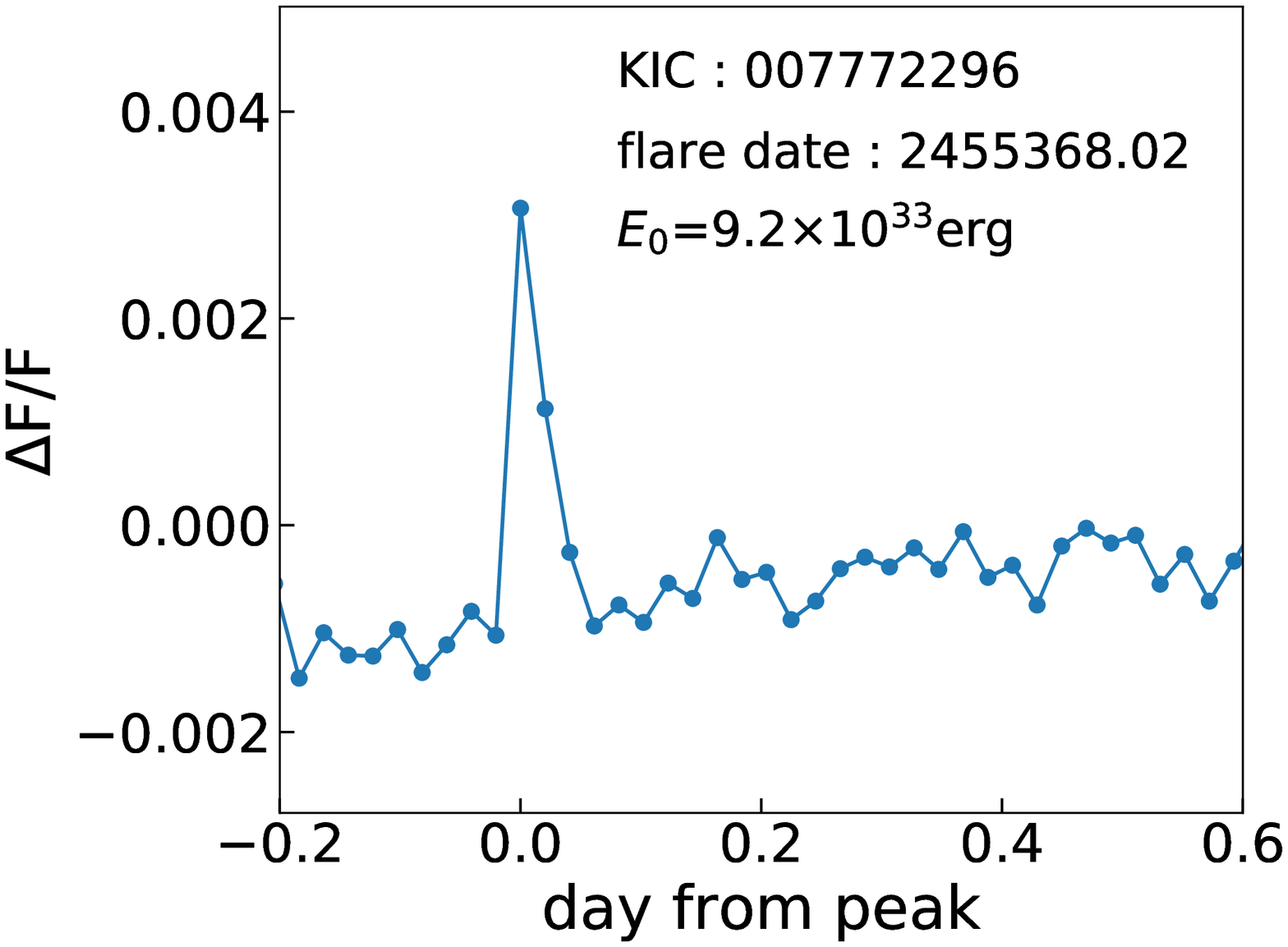}
   \plottwo{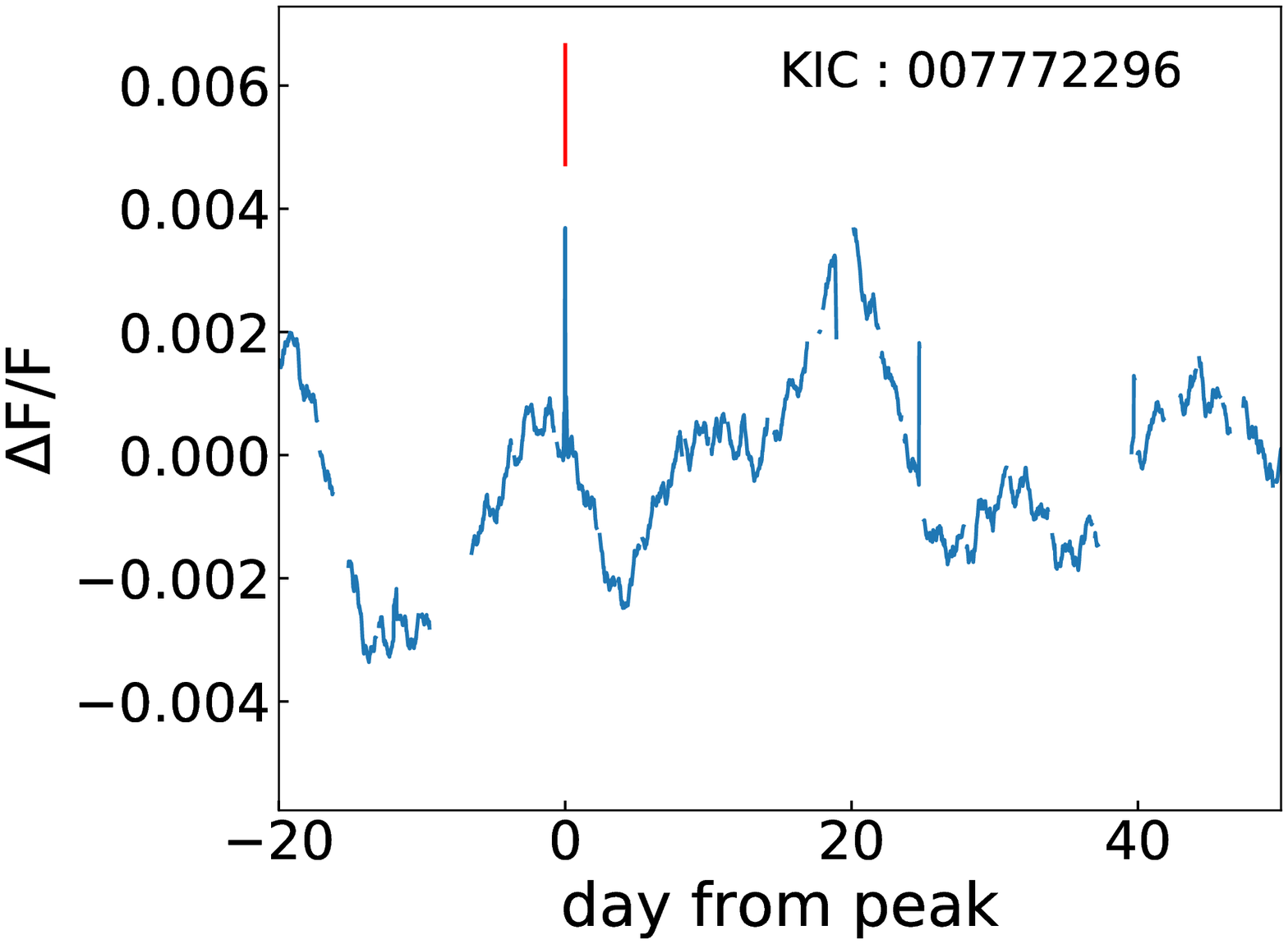}{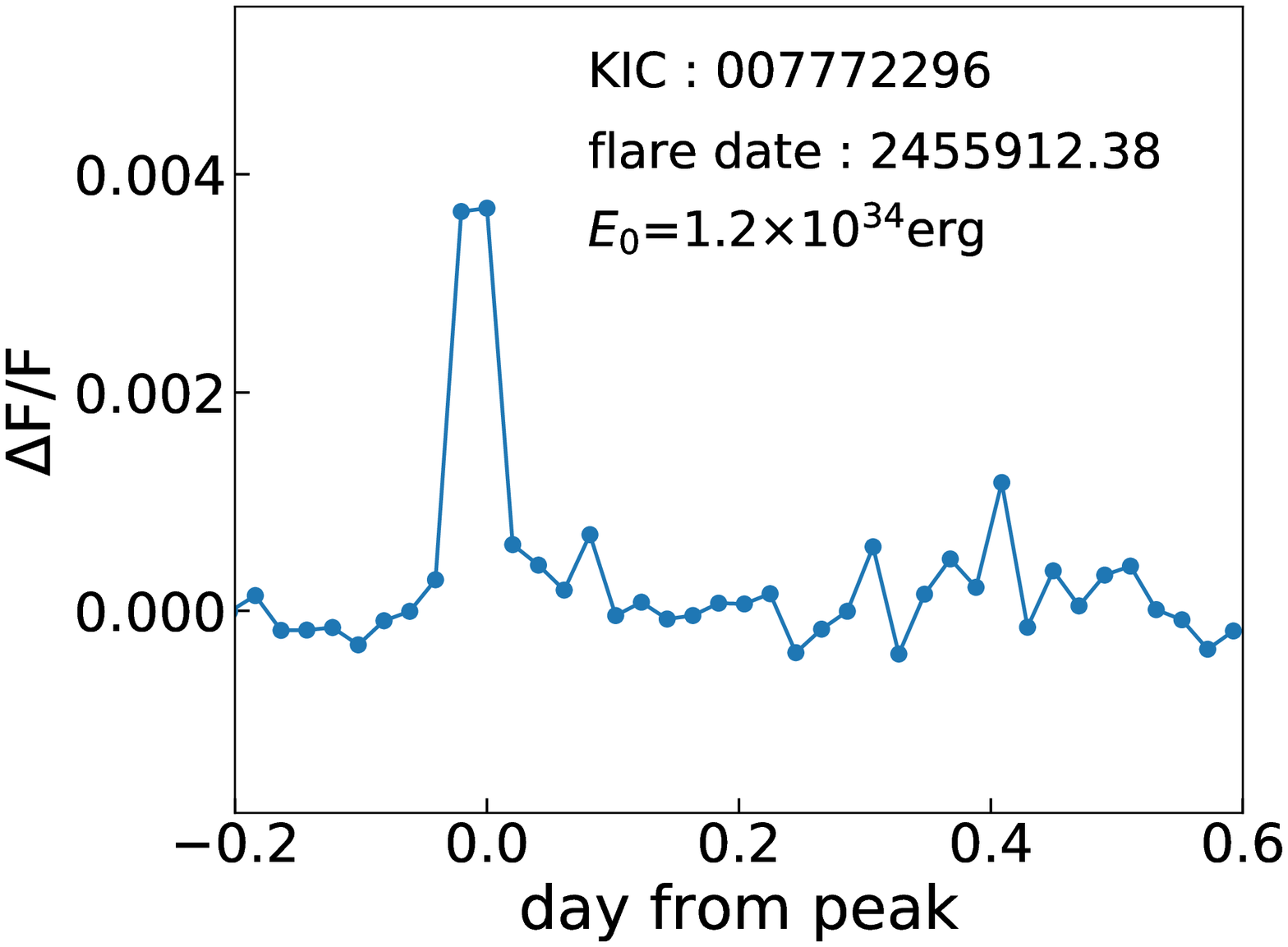}
   \plottwo{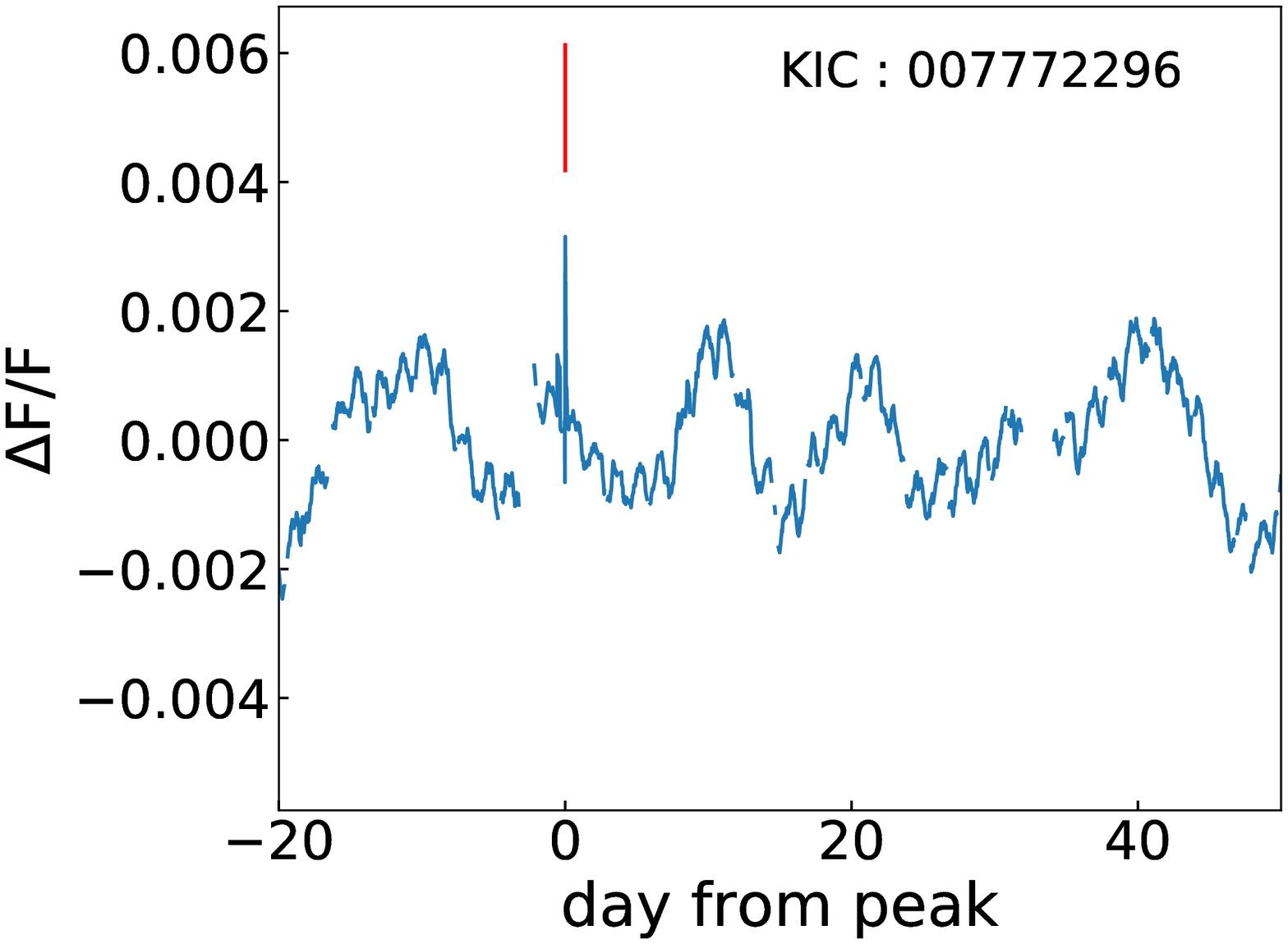}{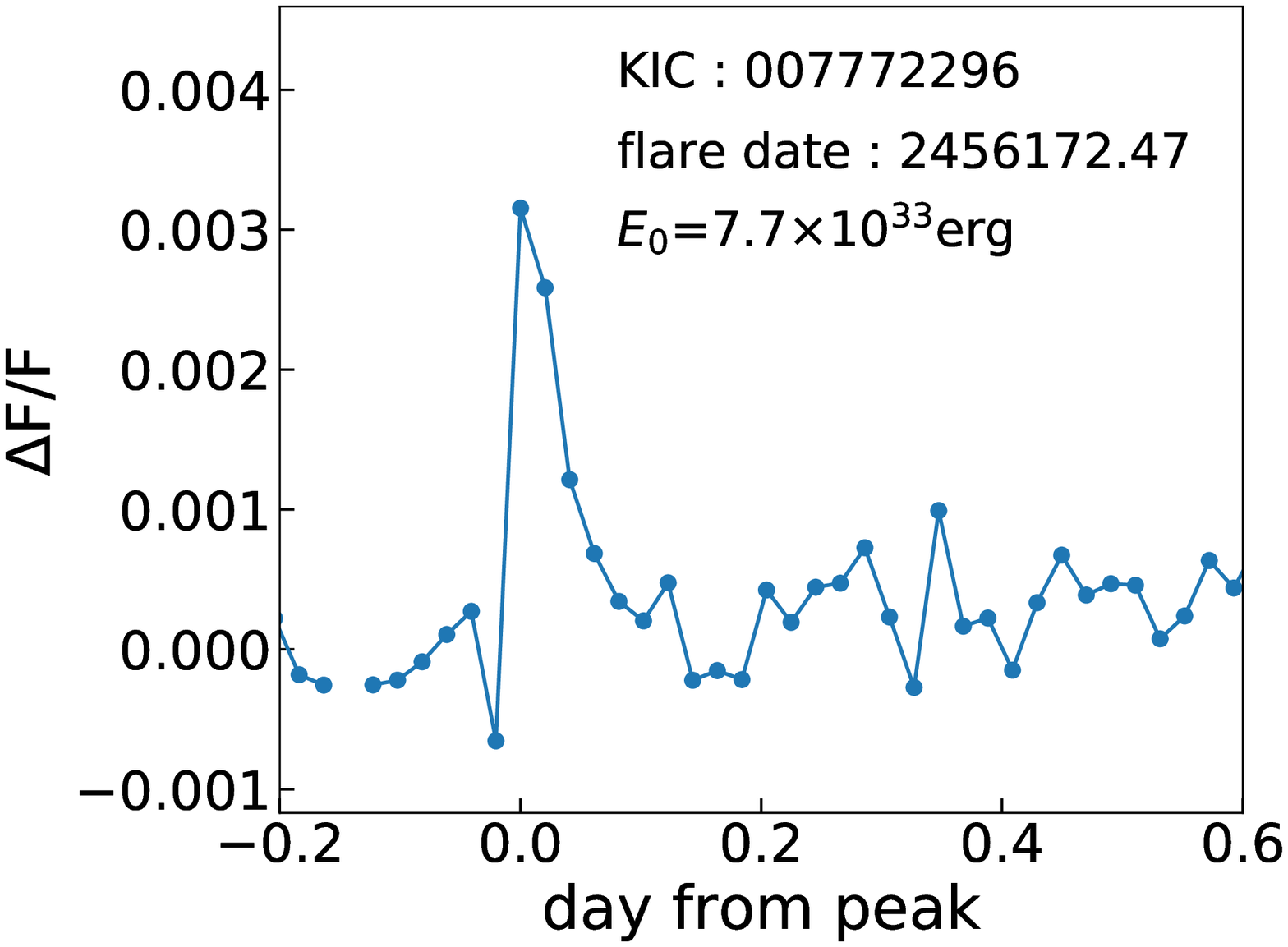}
   \caption{
    Same as Figure \ref{fig:eg_sunlikeflare_noflag} but for the star KIC007772296 that have Flag ``2" in Table \ref{table:params_sunlike}.
   }
  \label{fig:eg_sunlikeflare_flag2}
  \end{figure}

\newpage
  \begin{figure*}[ht!]
   \plottwo{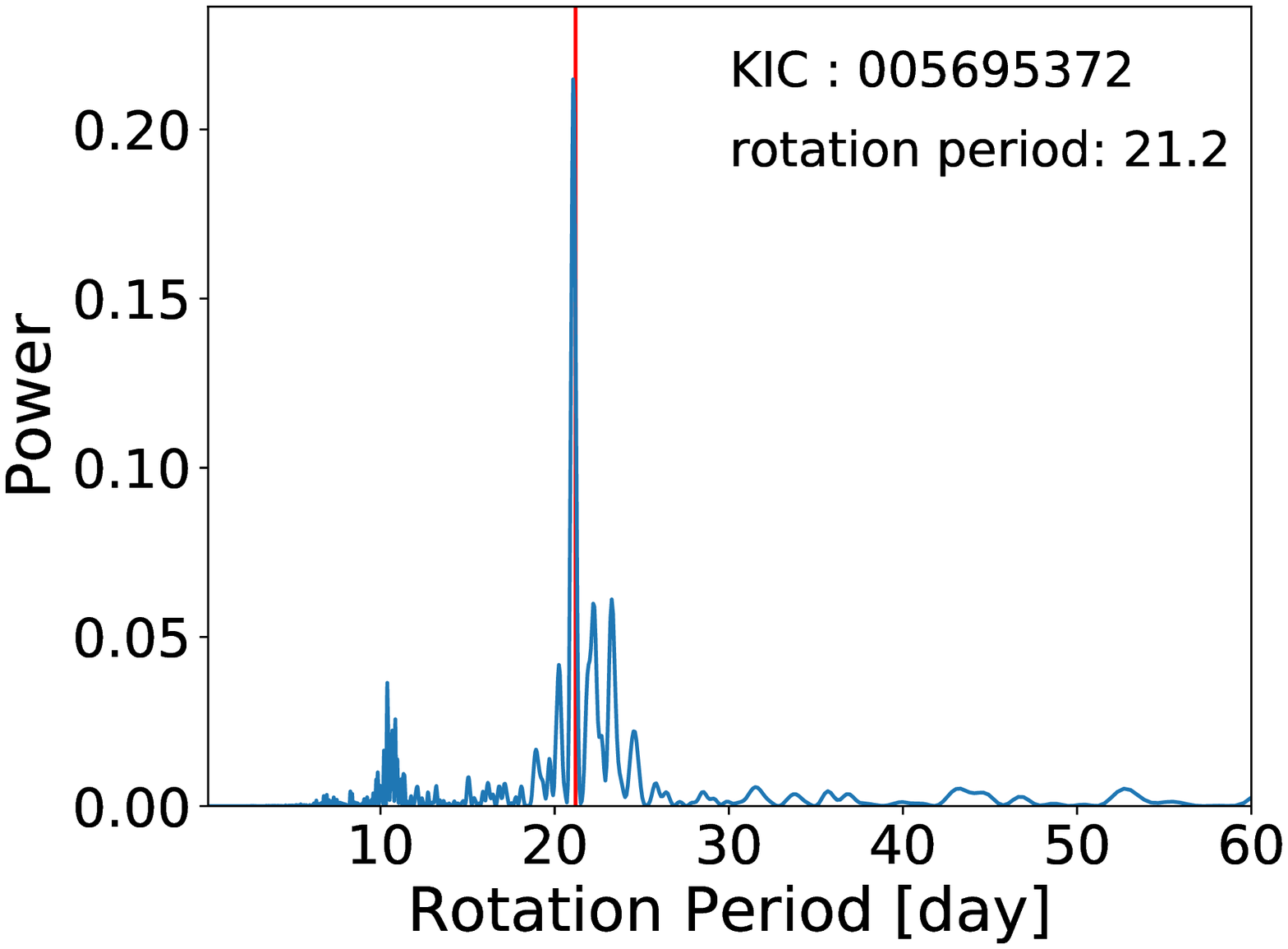}{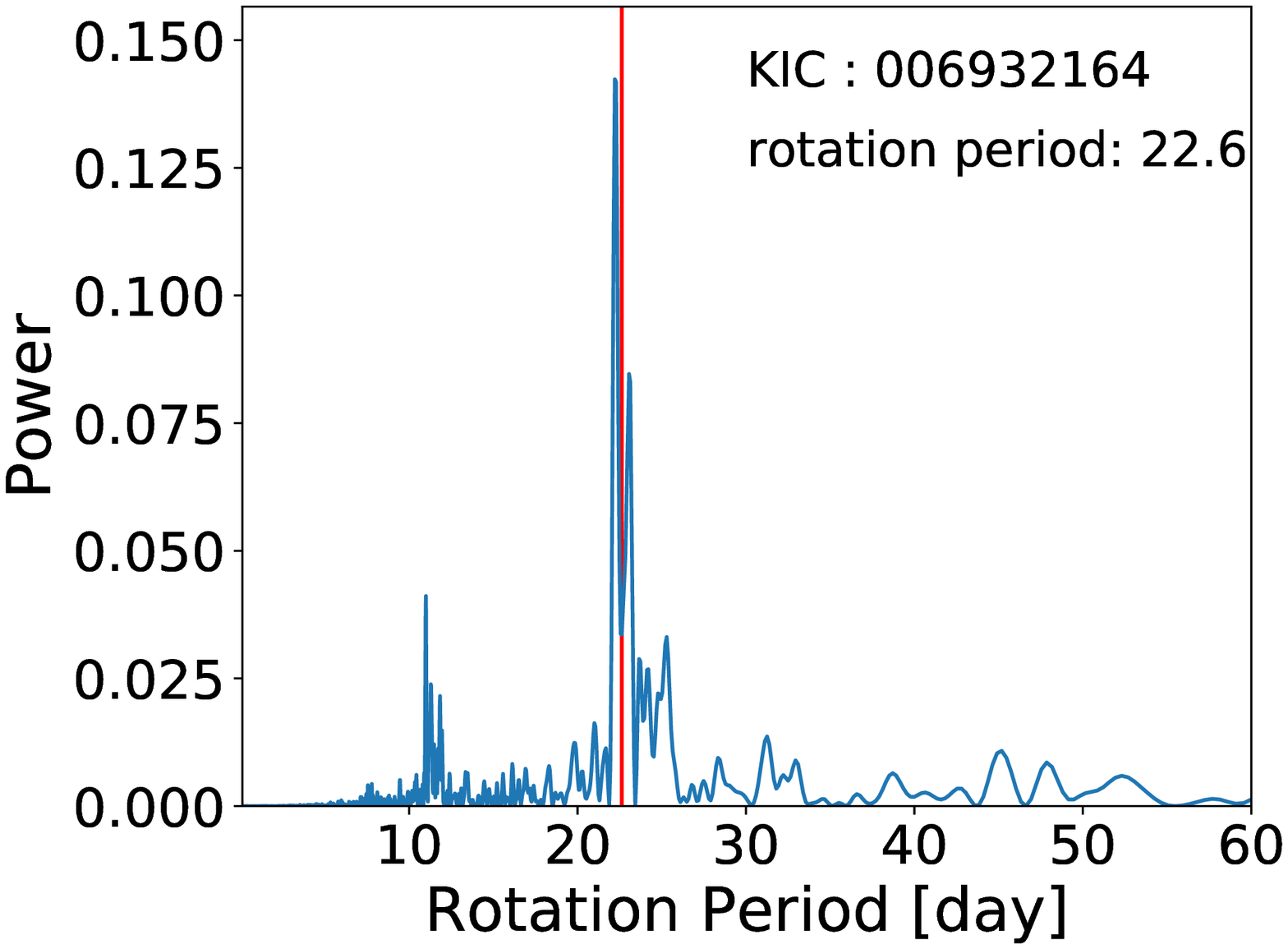}
   \plottwo{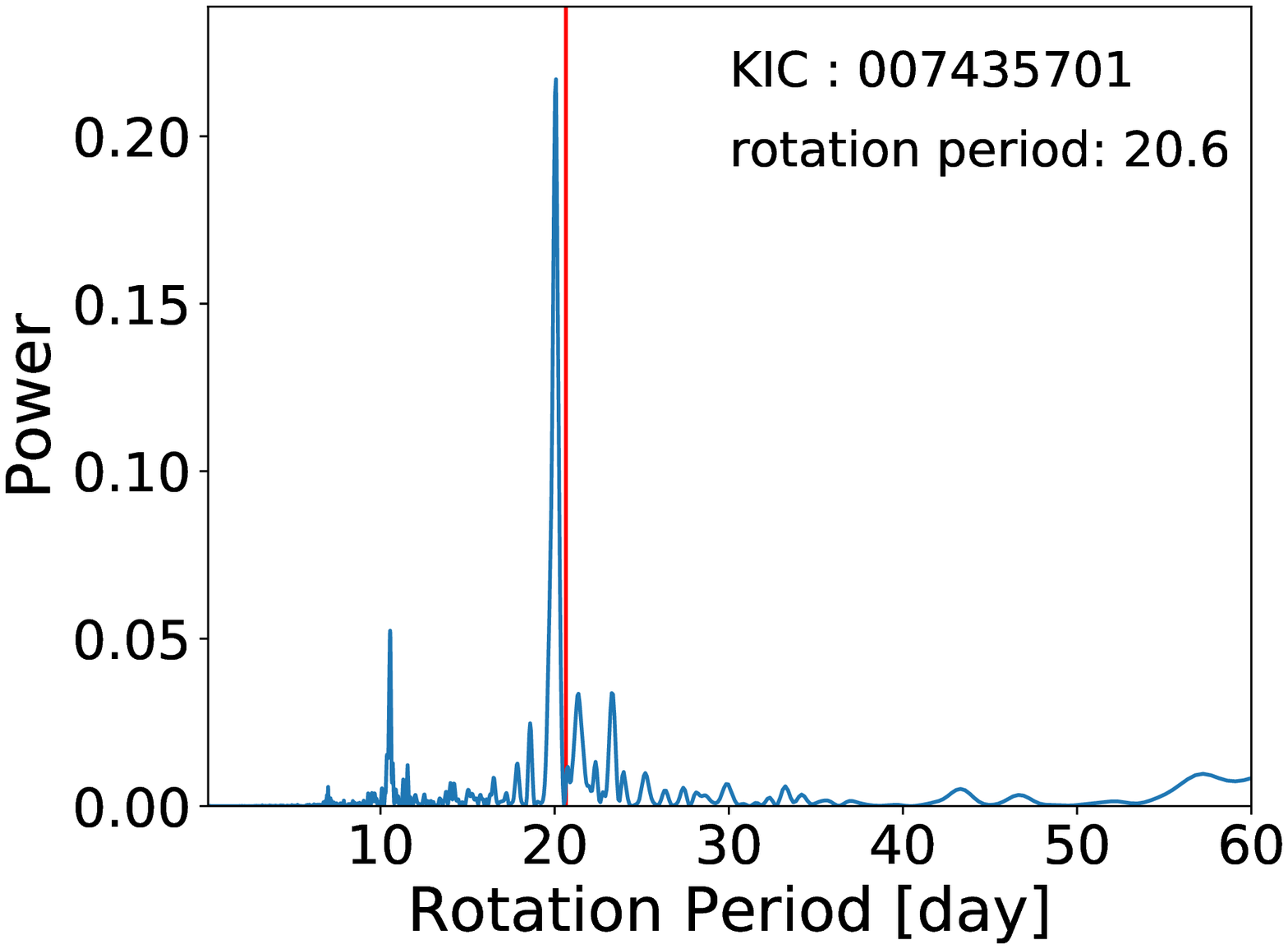}{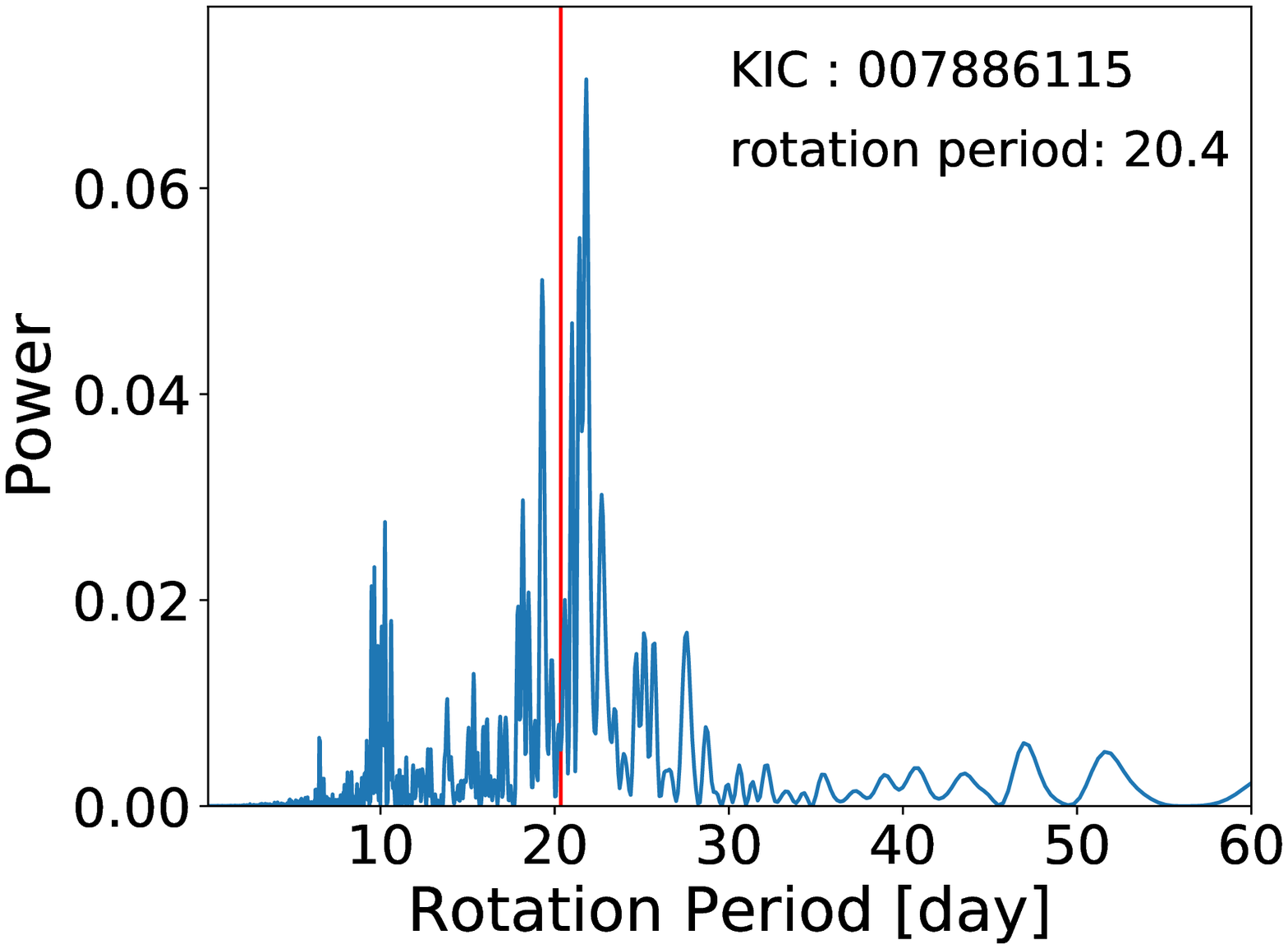}
   \plottwo{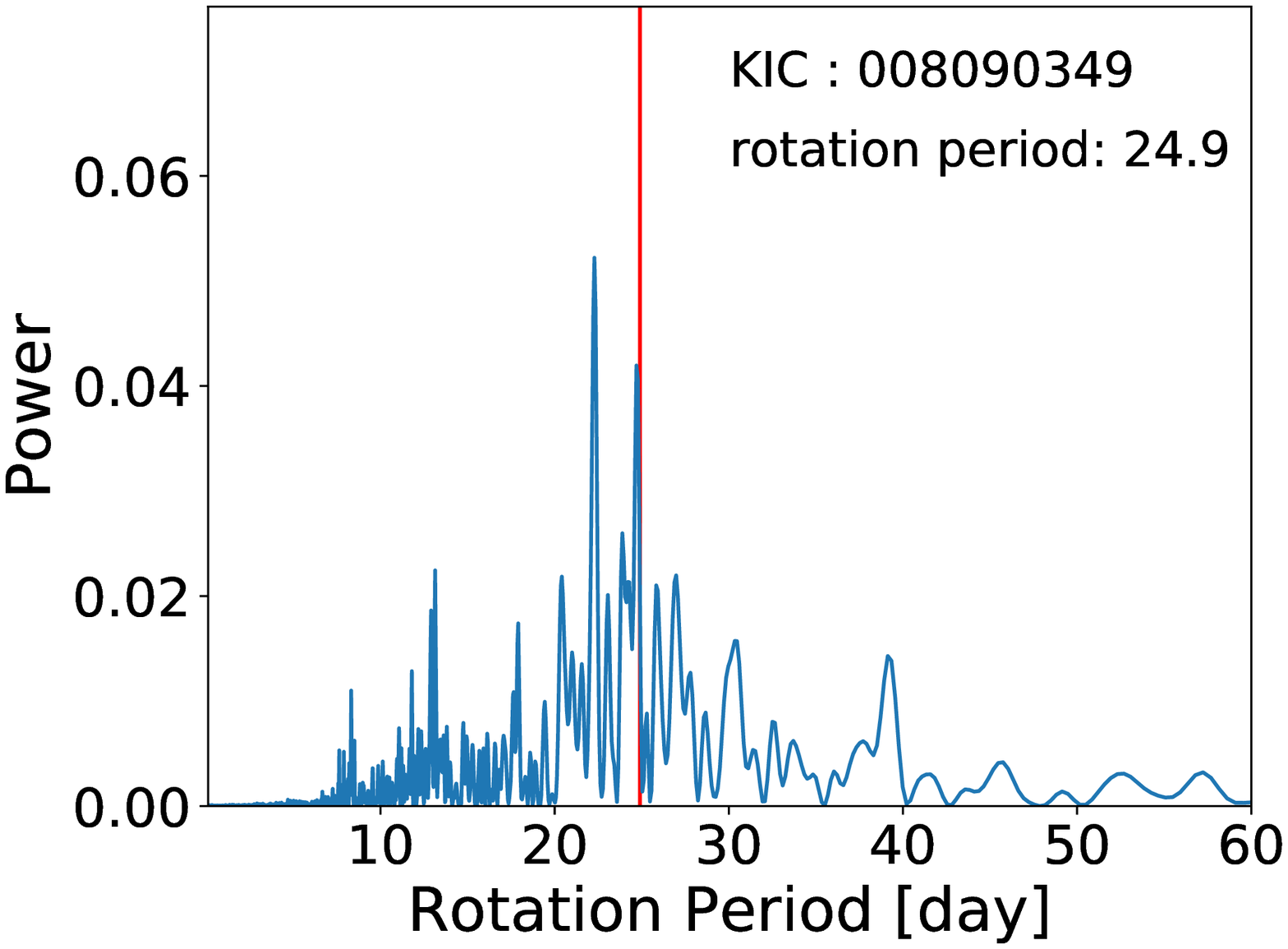}{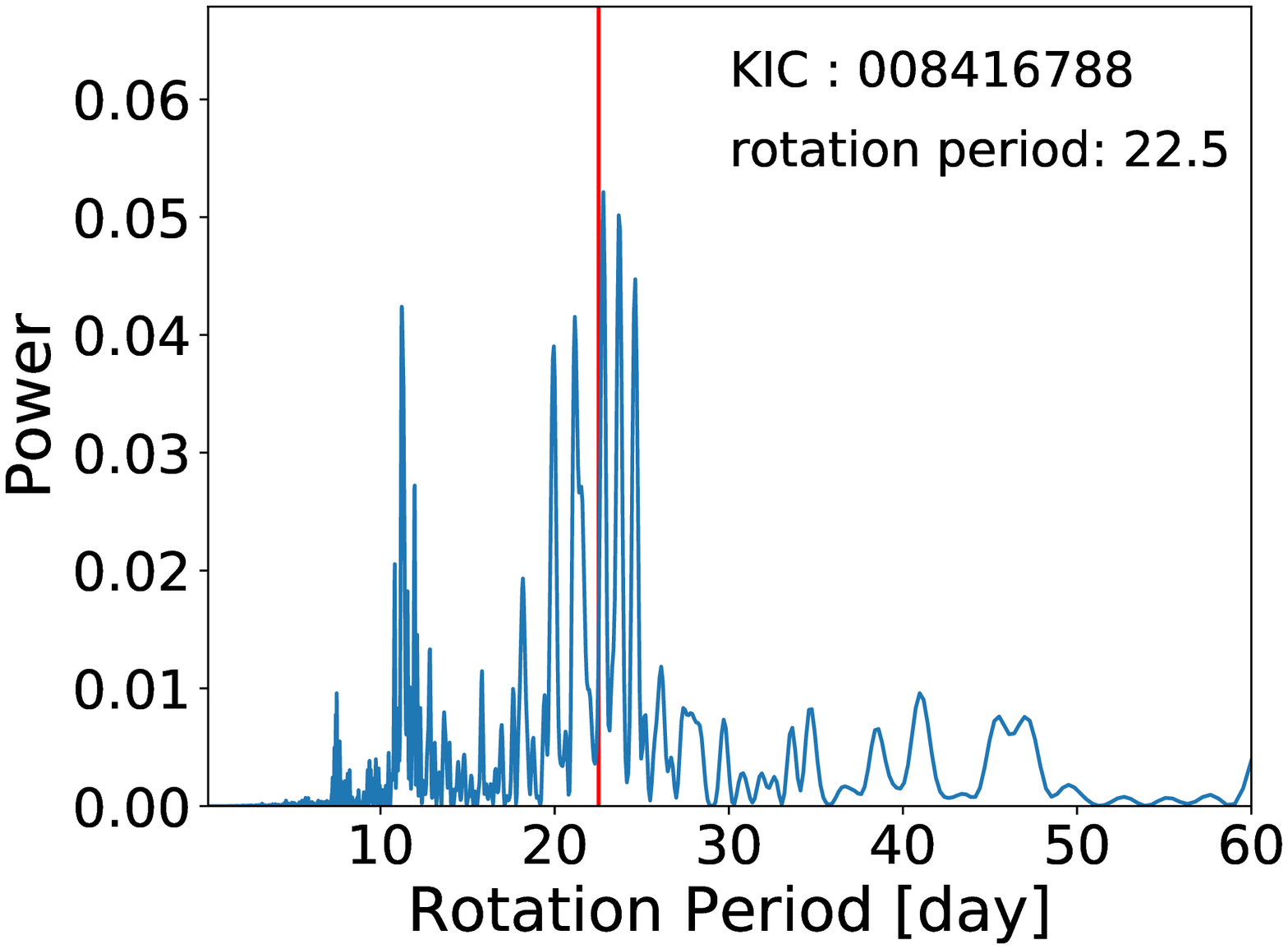}
 \caption{
 The LombScargle Power spectra of the lightcurves of the 10 Sun-like superflare stars that have no flags in Table \ref{table:params_sunlike}.
 The red vertical line in the panels indicate the rotation period $P_{\rm{rot}}$ values reported in \citet{McQuillan+2014}.
 These same $P_{\rm{rot}}$ values are written with the star IDs ($Kepler$ IDs) in the upper right side of the panels.
 }
 \label{fig:lomb_noflag}
 \end{figure*}
\addtocounter{figure}{-1}
  \begin{figure*}
   \plottwo{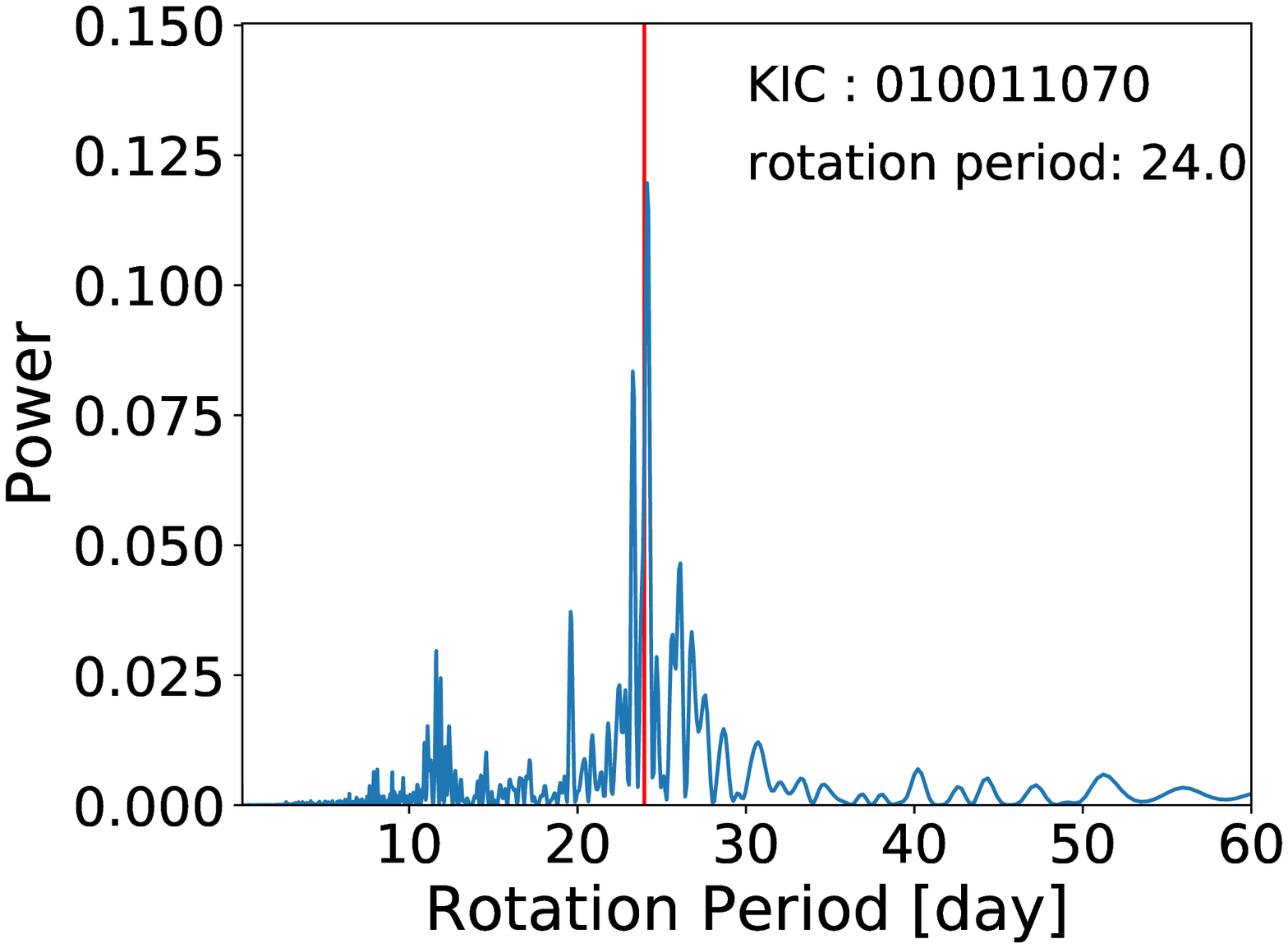}{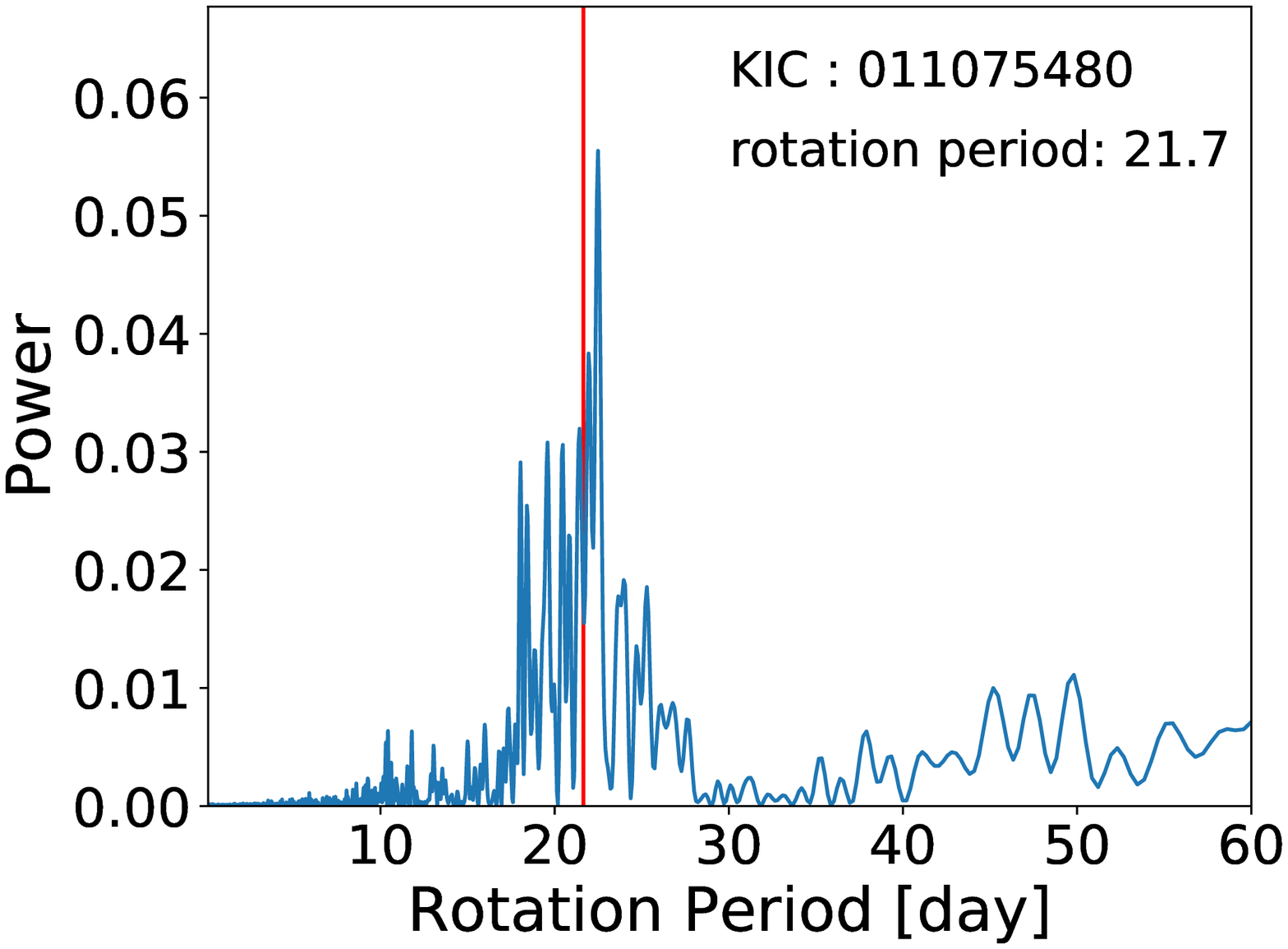}
   \plottwo{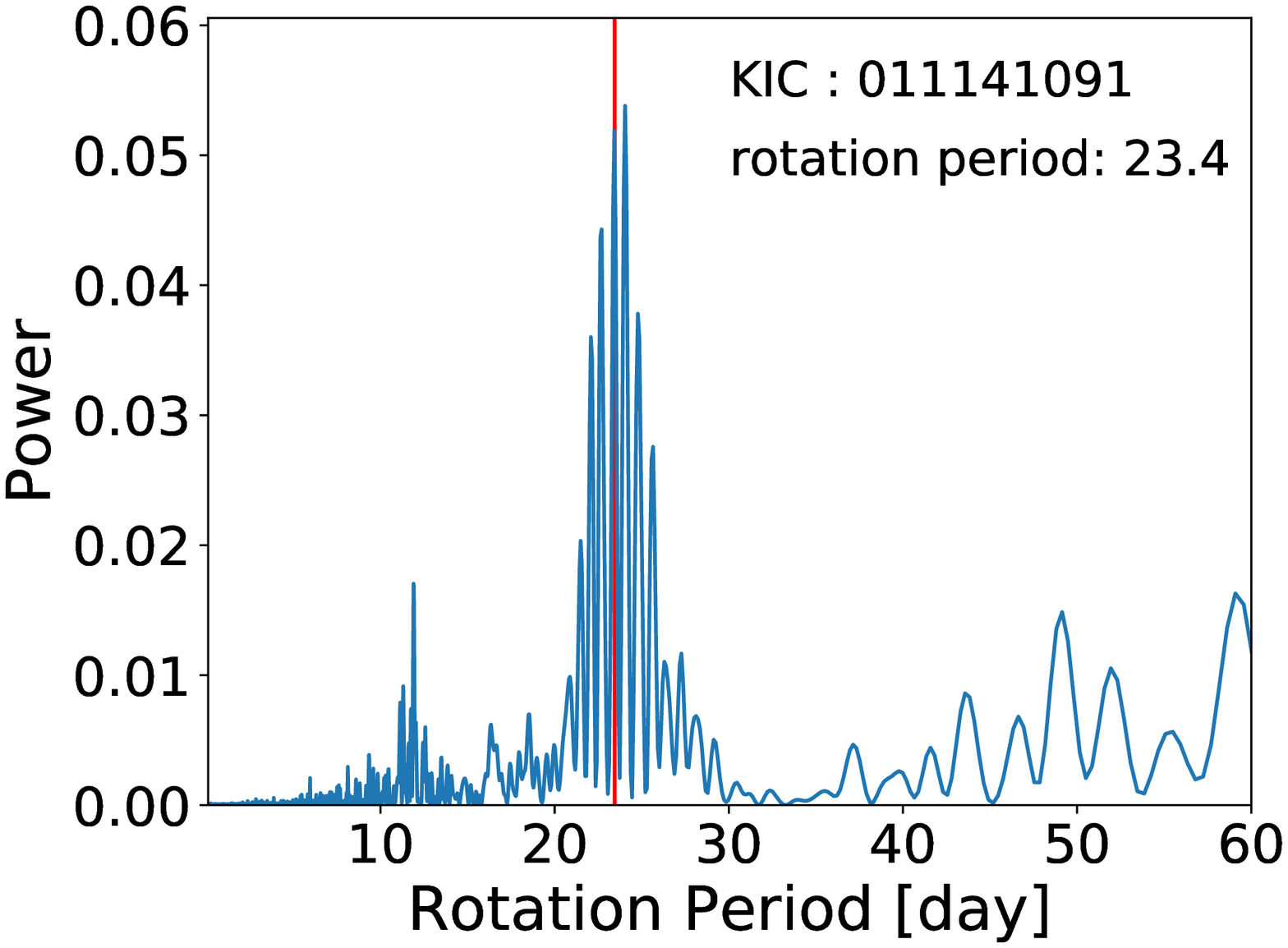}{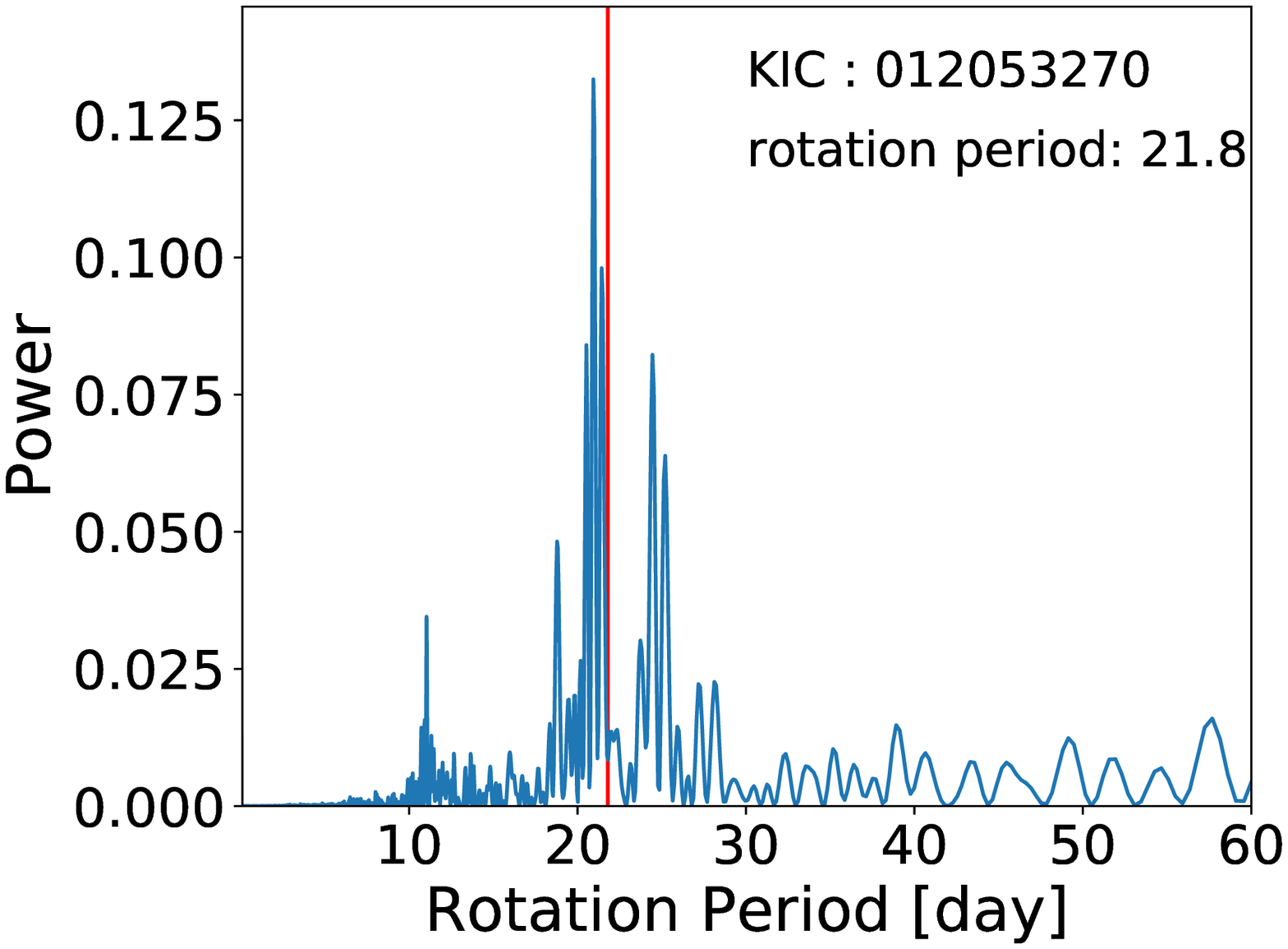}
 \caption{
 (Continued)
 }
 \end{figure*}

  \begin{figure*}
    \plottwo{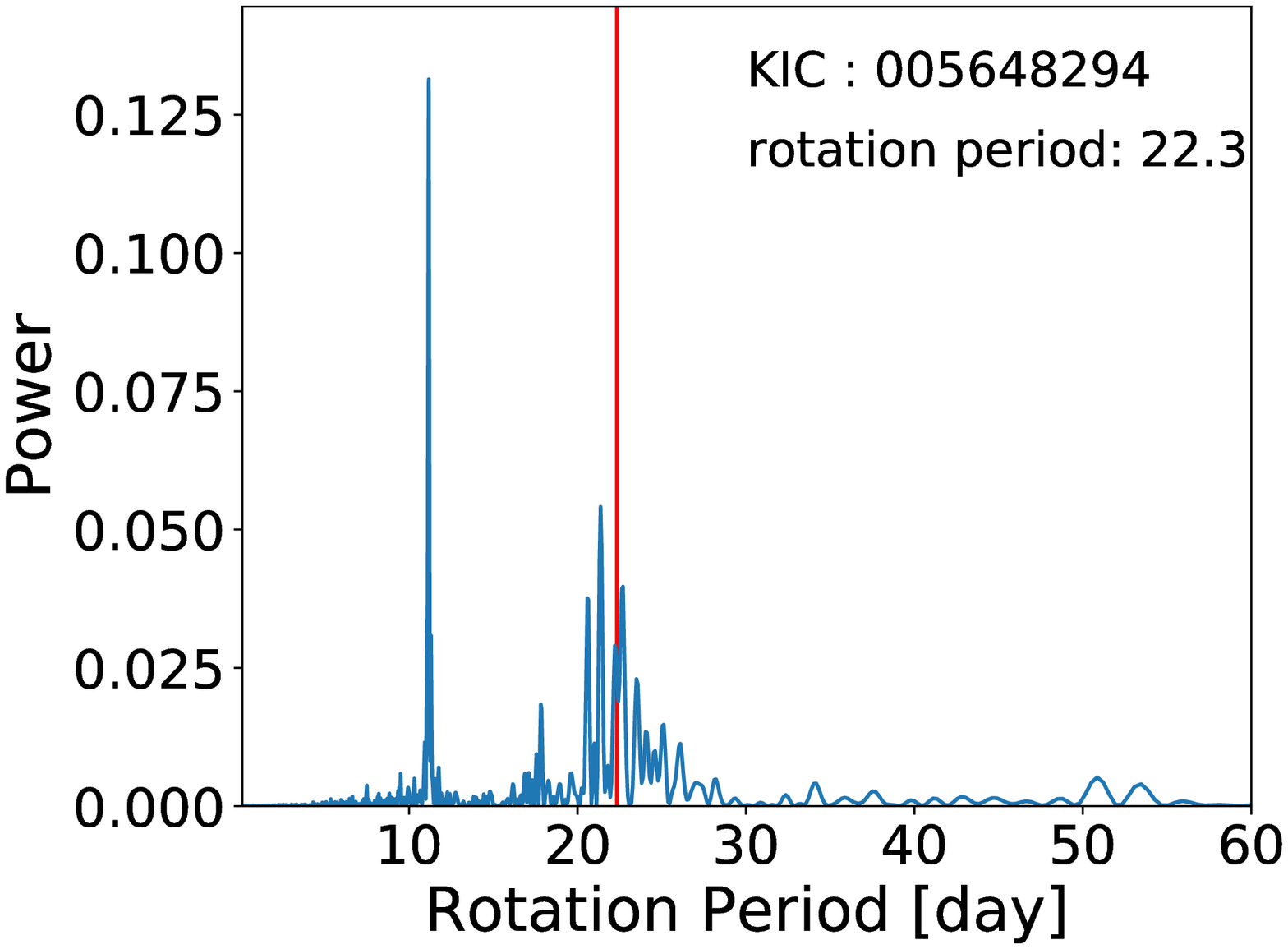}{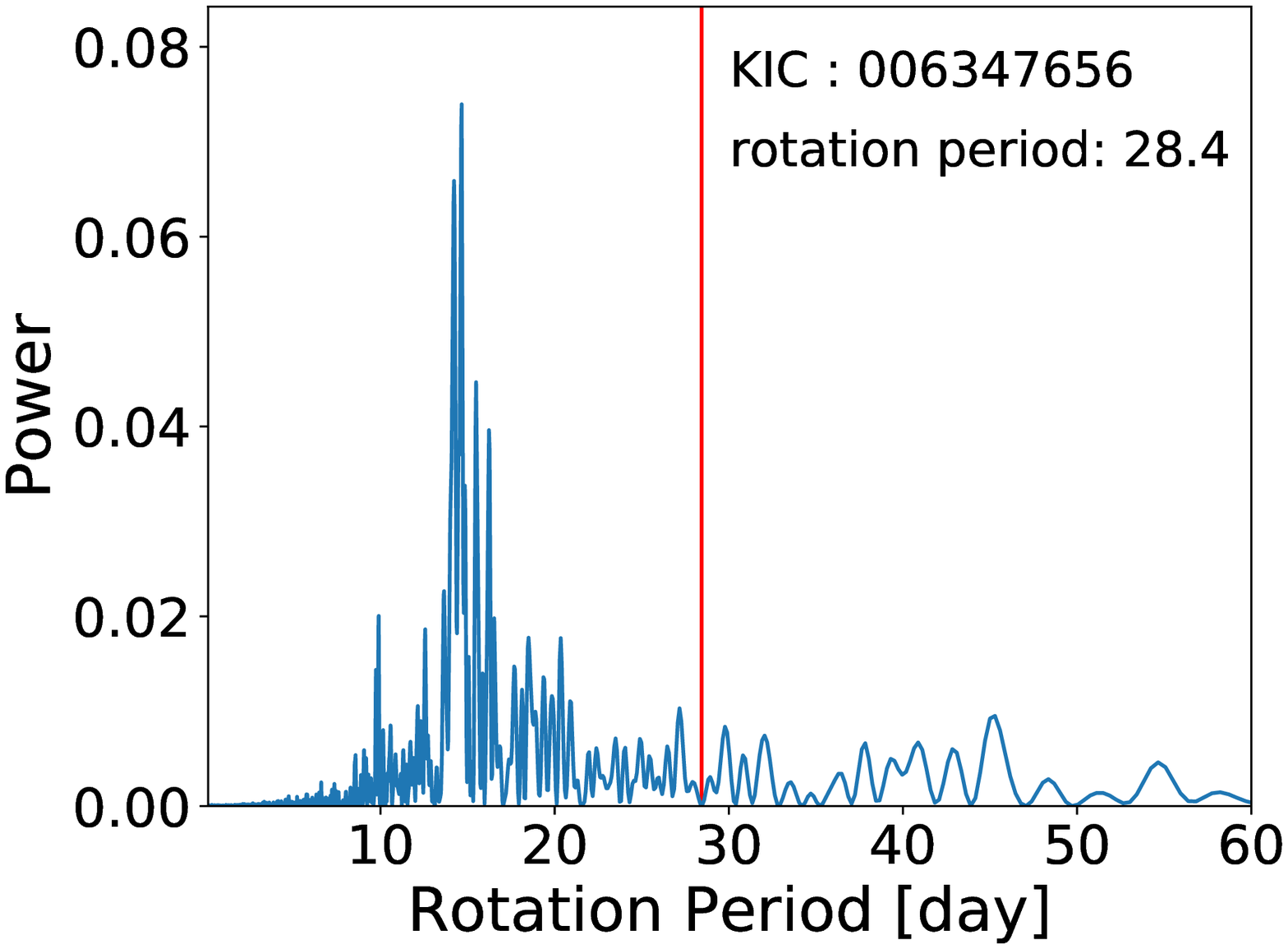}
    \plottwo{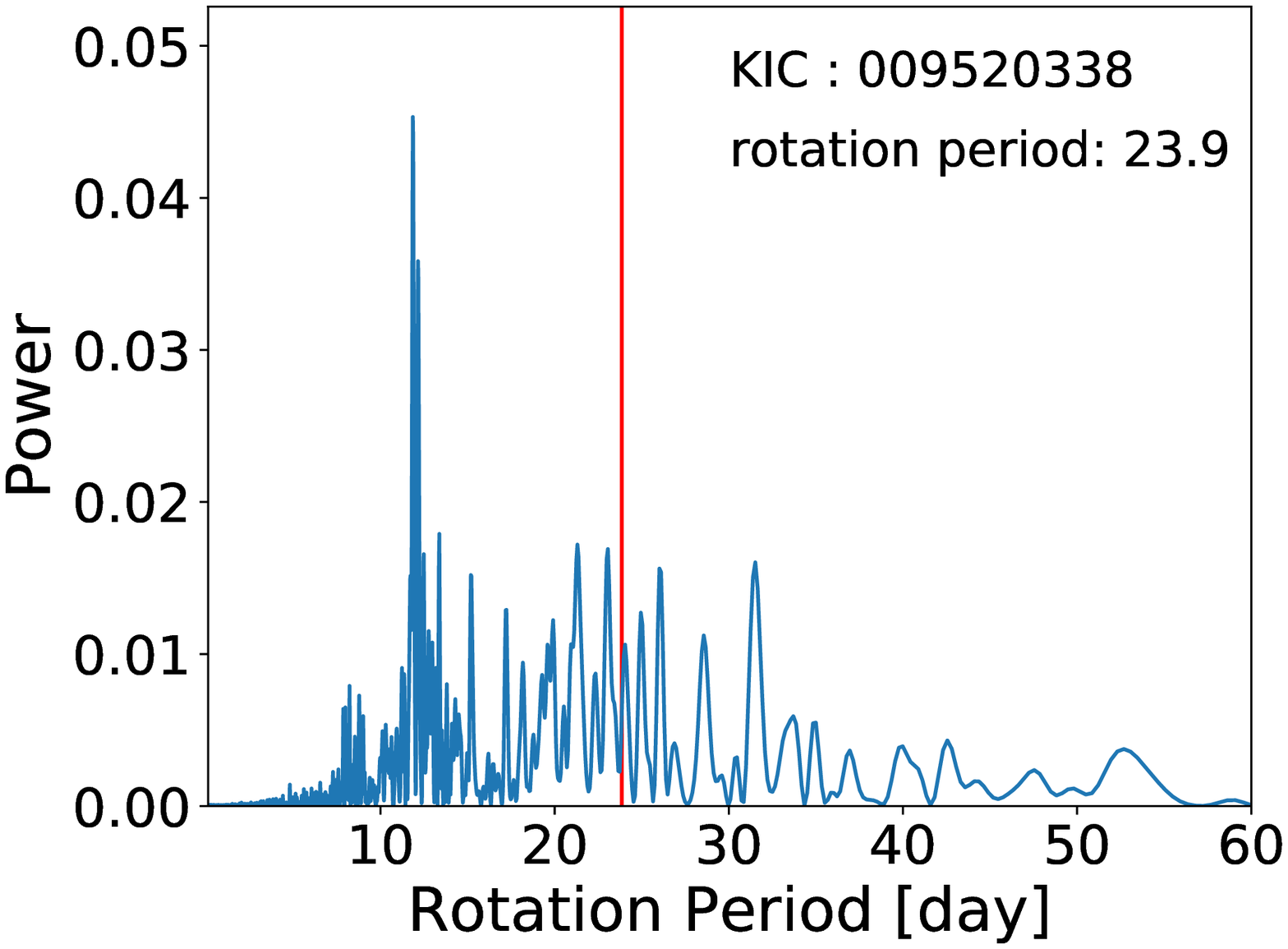}{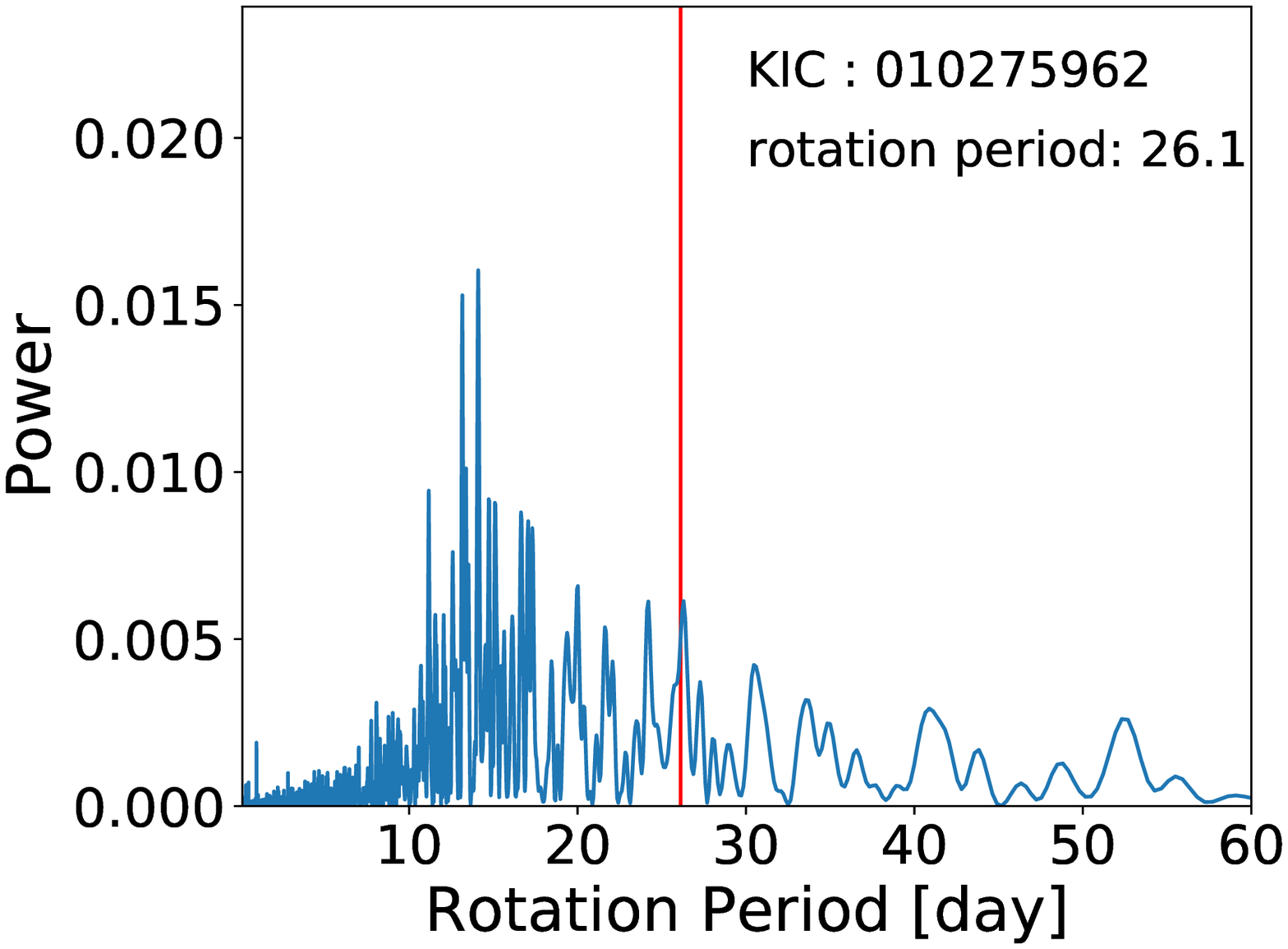}
     \plottwo{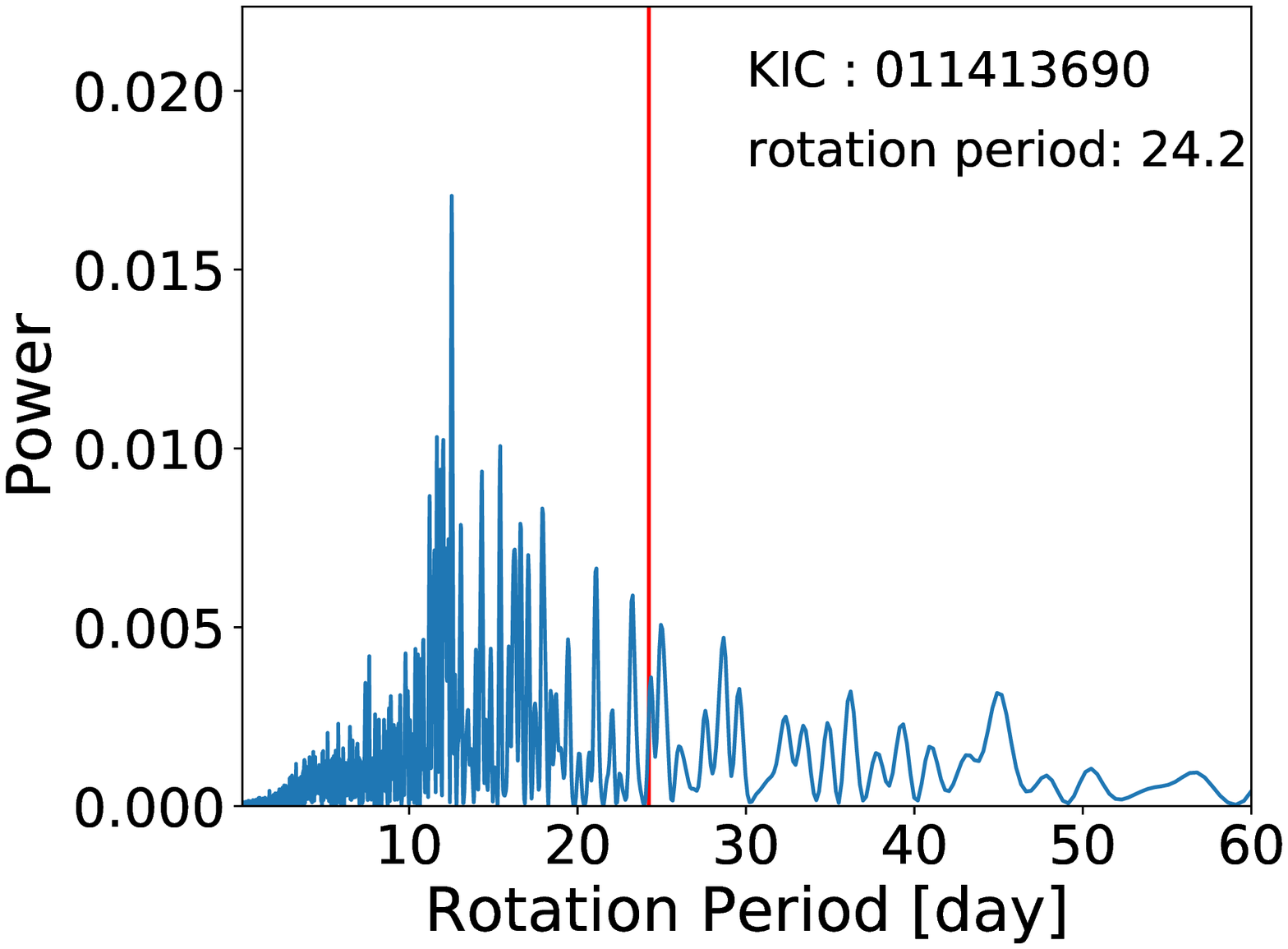}{figure/white.eps}
   \caption{
    Same as Figure \ref{fig:lomb_noflag} but for the 5 stars that have Flag ``1" in Table \ref{table:params_sunlike}.
   }
  \label{fig:lomb_flag1}
  \end{figure*}
  \begin{figure*}
    \fig{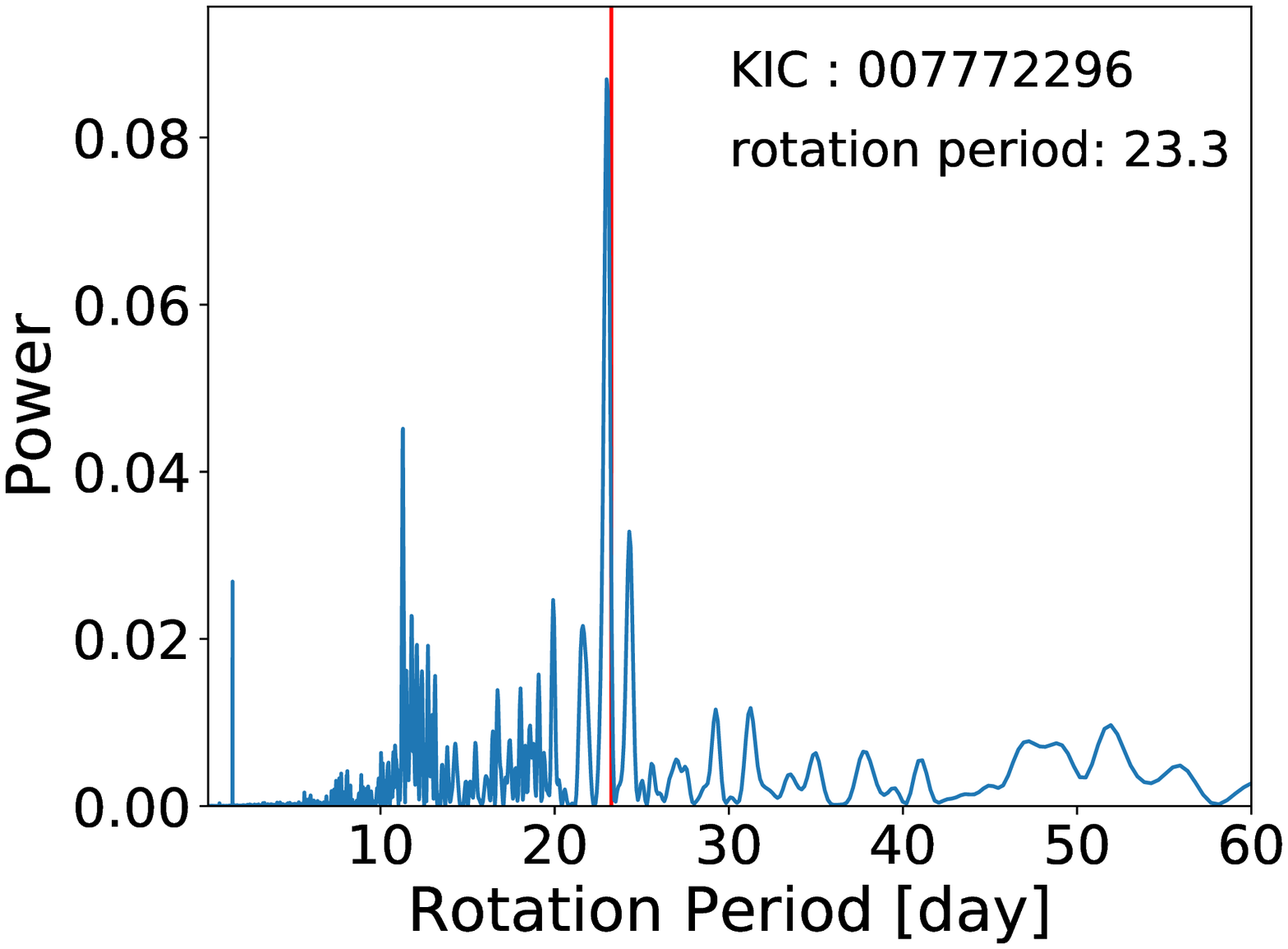}{0.49\textwidth}{}
   \caption{
    Same as Figure \ref{fig:lomb_noflag} but for the star KIC007772296 that have Flag ``2" in Table \ref{table:params_sunlike}.
   }
  \label{fig:lomb_flag2}
  \end{figure*}

\clearpage
 \section{Sun-like superflare stars with spectroscopic observations}\label{app:sunlike-spec-previous}

 \begin{deluxetable*}{llcl}
 \tablecaption{
   Classification changes of 2 superflare stars KIC 009944137 and KIC009766237, which were originally reported as
   Sun-like superflare stars ($T_{\rm eff}$ = 5600 -- 6000 K and $P_{\rm rot}>$ 20 days) in \citet{Nogami+2014}.
  }
   \tablewidth{0pt}
   \tablehead{
        \colhead{Star} & \colhead{Paper} & \colhead{Classification} & \colhead{Reason of Classification}
   }
   \startdata
   KIC 009944137 & \citet{Nogami+2014} & Sun-like star &
    \begin{tabular}{l}
   $T_{\rm{eff}}=5666\pm35$ K\tablenotemark{\dag}, $P_{\rm{rot}}=25.3$ days,\\
   and $v\sin i=1.9\pm0.3$ km s$^{-1}$\tablenotemark{\dag}
    \end{tabular}
   \\
    \cline{2-4}
   & \citeauthor{Notsu+2015a}(\citeyear{Notsu+2015a}\&\citeyear{Notsu+2015b}) &
   Not Sun-like star &
   \begin{tabular}{l}
   $P_{\rm{rot}}$=12.6 days, by estimating the rotation period \\
   using the longer data of Kepler (Quarters 2 -- 16:\\
   $\sim$1500 days), compared with those in \\
   \citet{Nogami+2014} (Quarters 0 -- 6: $\sim$500 days).
   \end{tabular}
   \\
    \cline{2-4}
   & \citet{Notsu+2019}& Sun-like star &
   \begin{tabular}{l}
   $P_{\rm{rot}}$=24.4 days, by using the rotation period value \\
   reported in \citet{McQuillan+2014}, which estimate \\
   rotation periods with the automated \\
   autocorrelation-based method.
   \end{tabular}
   \\
    \cline{2-4}
   & This study & Sun-like star candidate &
      \begin{tabular}{l}
   This star is not included in the classification \\
   of superflare stars in this study (cf. Section \ref{subsec:example}), \\
   since the amplitude of the rotational brightness  \\
   variation is so small that the $Amp$ value is not  \\
   reported in \citet{McQuillan+2014}, and the accuracy \\
   of the $P_{\rm{rot}}$ value is considered to be low.
       \end{tabular}
   \\
   \hline
   KIC 009766237 & \citet{Nogami+2014} & Sun-like star &
\begin{tabular}{l}
   $T_{\rm{eff}}=5606\pm40$ K\tablenotemark{\dag}, $P_{\rm{rot}}=21.8$ days, \\
   and $v\sin i=2.1\pm0.3$ km s$^{-1}$\tablenotemark{\dag}
    \end{tabular}
   \\
    \cline{2-4}
   & \citeauthor{Notsu+2015a}(\citeyear{Notsu+2015a}\&\citeyear{Notsu+2015b}) &
   Not Sun-like star &
   \begin{tabular}{l}
   $P_{\rm{rot}}$=14.2 days, by estimating the rotation period \\
   using the longer data of Kepler (Quarters 2 -- 16: \\
   $\sim$1500 days), compared with those in \\
   \citet{Nogami+2014} (Quarters 0 -- 6: $\sim$500 days).
   \end{tabular}
   \\
    \cline{2-4}
   & \citet{Notsu+2019}& Sun-like star&
     \begin{tabular}{l}
   $P_{\rm{rot}}$=41.5 days, by using the rotation period value \\
   reported in \citet{McQuillan+2014}, which estimate \\
   rotation periods with the automated \\
   autocorrelation-based method.
   \end{tabular}
   \\
    \cline{2-4}
   & This study & Sun-like star candidate &
    \begin{tabular}{l}
   This star is not included in the classification \\
   of superflare stars in this
   study (cf. Section \ref{subsec:example}), \\
   since the amplitude of the rotational brightness \\
   variations is so small that the $Amp$ value is not  \\
   reported in \citet{McQuillan+2014}, and the accuracy \\
   of the $P_{\rm{rot}}$ value is considered to be low.
   \end{tabular}
   \\
   \enddata
   \tablenotetext{\dag}{
    The values of effective temperature $T_{\rm{eff}}$
    and projected rotation velocity $v\sin i$ are
    estimated from our spectroscopic data \citep{Nogami+2014}.
   }
 \label{table:Nogami2014stars}
 \end{deluxetable*}

  As described in Introduction (Section \ref{sec:intro}) of this paper, one of the most important purposes of superflare studies is clarifying whether slowly-rotating Sun-like stars really have superflares or not.
  In this study, we have found 15 Sun-like superflare stars from Kepler data as listed in Table \ref{table:params_sunlike}, but the results only from $Kepler$ photometric data are not enough.
  Our previous spectroscopic observations of 64 solar-type superflare stars with various rotation periods (\citealt{SNotsu+2013}, \citeyear{Notsu+2015a}, \citeyear{Notsu+2015b}, \& \citeyear{Notsu+2019}; \citealt{Nogami+2014}) have suggested
  (i) more than half of solar-type superflare stars are single stars,
  (ii) quasi-periodic brightness variations of the stars can be used for estimations of rotation period and starspot coverage.
  However, most of the observed stars in our previous spectroscopic observations are stars with $P_{\rm{rot}}<$ 20 days and they are not Sun-like stars ($T_{\rm eff}$ = 5600 -- 6000 K and $P_{\rm{rot}}>$ 20 days).
  Because of this, we cannot finally conclude whether the above results (i)\&(ii) can be fully applied to Sun-like superflare stars with $P_{\rm{rot}}>$20 days.
  It is important to investigate each Sun-like superflare star with spectroscopic observations, and confirm that they are really Sun-like superflare stars.

  Among the observed stars in our previous spectroscopic observations, \citet{Nogami+2014} reported that 2 superflare stars (KIC 009944137 and KIC 009766237) were confirmed to be Sun-like stars on the basis of the rotation period values ($P_{\rm{rot}}$) from $Kepler$ photometric data and the projected rotation velocities ($v\sin i$) from the spectroscopic data (see Table \ref{table:Nogami2014stars}).
  However, the remark that these 2 stars are Sun-like superflare stars have changed, as summarized in Table \ref{table:Nogami2014stars}.
  Finally, these 2 stars are not included in 15 Sun-like superflare stars in this study (Table \ref{table:params_sunlike}), since the accuracy of $P_{\rm rot}$ 
  as a result of $Kepler$ photometric observations is considered to be low (see Table \ref{table:Nogami2014stars}).
  Note that the accuracy of $v \sin i$ observations by the high dispersion
  spectrograph aboard Subaru 8.2 m telescope (used in \citealt{Nogami+2014}, \citealt{Notsu+2015a} \& \citeyear{Notsu+2015b}) was very good enough to
  resolve rotation velocity as slow as the Sun ($\sim 2$km s$^{-1}$). 

  KIC 006347656 and KIC 010011070 in Table \ref{table:params_sunlike} are also included in Sun-like superflare stars reported in our previous study using $Kepler$ data \citep{Notsu+2019} 
  \footnote{
  It is noted that the names of these stars were not explicitly written in \citet{Notsu+2019} since we did not show any tables showing the names of stars, but these two stars were included in the data of superflare stars plotted in the figures of \citet{Notsu+2019}.
  }, 
  but they have not been observed in our previous spectroscospic observations (\citealt{SNotsu+2013}, \citeyear{Notsu+2015a}, \citeyear{Notsu+2015b}, \& \citeyear{Notsu+2019}; \citealt{Nogami+2014}).
  Since these previous observations were planned in the initial phase of the researches of superflares on solar-type stars, we first aimed to investigate overall properties of all solar-type superflare stars (e.g., whether the brightness variations can be explained by the rotation of stars with large starspots).
  Therefore, not only slowly-rotating Sun-like superflares but also rapidly-rotating solar-type superflare stars were important targets of these previous observations.
  Considering these things, in order to observe as many as possible superflare stars with various rotation periods in the limited allocated time of Subaru 8.2m and Apache Point 3.5m telescopes,
  we only observed relatively bright superflare stars ($Kepler$-band magnitude brighter than 14.2 mag) in these previous observations.
  As a result, since the two superflare stars KIC 006347656 and KIC 010011070 are faint ($Kepler$-band magnitude fainter than 14.8 mag), they were not included in the target stars of these previous observations.

  As a result, the number of Sun-like superflare stars ($T_{\rm{eff}}$ = 5600 -- 6000 K, $P_{\rm{rot}}\sim$ 25 days, and $t\sim$ 4.6 Gyr) that have been investigated spectroscopically and confirmed to be ``single" Sun-like stars, are now 0.
  Future spectroscopic observations of Sun-like superflare stars are necessary to investigate whether the Sun-like stars really have superflares.
  Although all of the 15 $Kepler$ Sun-like superflare stars listed in Table \ref{table:params_sunlike} are relatively faint (all are fainter than 13.5 mag in $Kepler$-band magnitude and most of them are fainter than 14.5 mag), it is very important to conduct spectroscopic observations of these 15 Sun-like $Kepler$ superflare stars in order to confirm the validity of statistical results of superflares on Sun-like stars discussed in this study using $Kepler$ data (e.g., Figure \ref{fig:flarefreq_sunstar}).
  In addition, nearby bright superflare stars that are found from $TESS$ (\citealt{Ricker+2015}; \citealt{Tu+2020}; \citealt{Doyle+2020})
  and that will be found from $PLATO$ \citep{Rauer+2014} can be also good targets for future spectroscopic studies.

\clearpage
\section{Superflares on Subgiants}\label{app:subgiant}
 In Section \ref{subsec:energy_rotation}, we investigated the relationship between the superflare energy ($E_{\rm{flare}}$) and the rotation period ($P_{\rm{rot}}$) of solar-type superflare stars.
 From Figure \ref{fig:erg_prot_solartype}, the upper limit of $E_{\rm{flare}}$ in a given period bin has a continuous decreasing trend with the rotation period.
 As we also described in Section \ref{subsec:energy_rotation}, this decreasing trend was not reported in our initial study \citep{Notsu+2013}.
 \citet{Notsu+2013} suggested the maximum superflare energy in a given rotation period bin does not have a clear correlation with the rotation period.
 This is because some fraction ($\sim$40\%) of superflare stars used in our initial studies (\citealt{Maehara+2012}; \citealt{Shibayama+2013}; \citealt{Notsu+2013}) are now found to be subgiants \citep{Notsu+2019}, and this contamination of subgiants could affect the statistics.
 In order to check this point, we briefly discuss the superflares on subgiants found from $Kepler$ data.

 We searched for superflares on subgiants from the Kepler 30 minutes (long) time cadence data \citep{Koch+2010} that were taken from 2009 May to 2013 May (quarters 0 -- 17).
 As is listed in Table \ref{table:num_flares_subgiant}, we selected the 36784 G-type subgiant stars on the basis of the evolutionary state classifications (Main Sequence (MS)/ Subgiants / Red giants/ Cool main-sequence binaries) and effective temperature ($T_{\rm{eff}}$) values listed in \citet{Berger+2018}, as done for solar-type stars in Section \ref{subsec:analysis}.
 Among these subgiant stars, 3533 stars having the values of brightness variations amplitude ($Amp$) and rotation period ($P_{\rm{rot}}$) in \citealt{McQuillan+2014} are finally used for the flare search process.
 The method of the flare detection and the flare energy estimation are the same as those done for solar-type stars in Sections \ref{subsec:analysis} and \ref{subsec:flare-energy-estimation}.
 As a result, we detected 3780 flares on 168 G-type subgiant stars (see Table \ref{table:num_flares_subgiant}).
 We should note here that it is possible that among these ``subgiant" stars, not only single evolved ``subgiant" stars, but also pre-main stars or binary stars (e.g. RSCVn-type binary stars), can be included.
 This is because of the simple classifications in \citet{Berger+2018}, which only used the absolute brightness of the star from Gaia-DR2 parallaxes.

\begin{figure}[ht!]
   \plotone{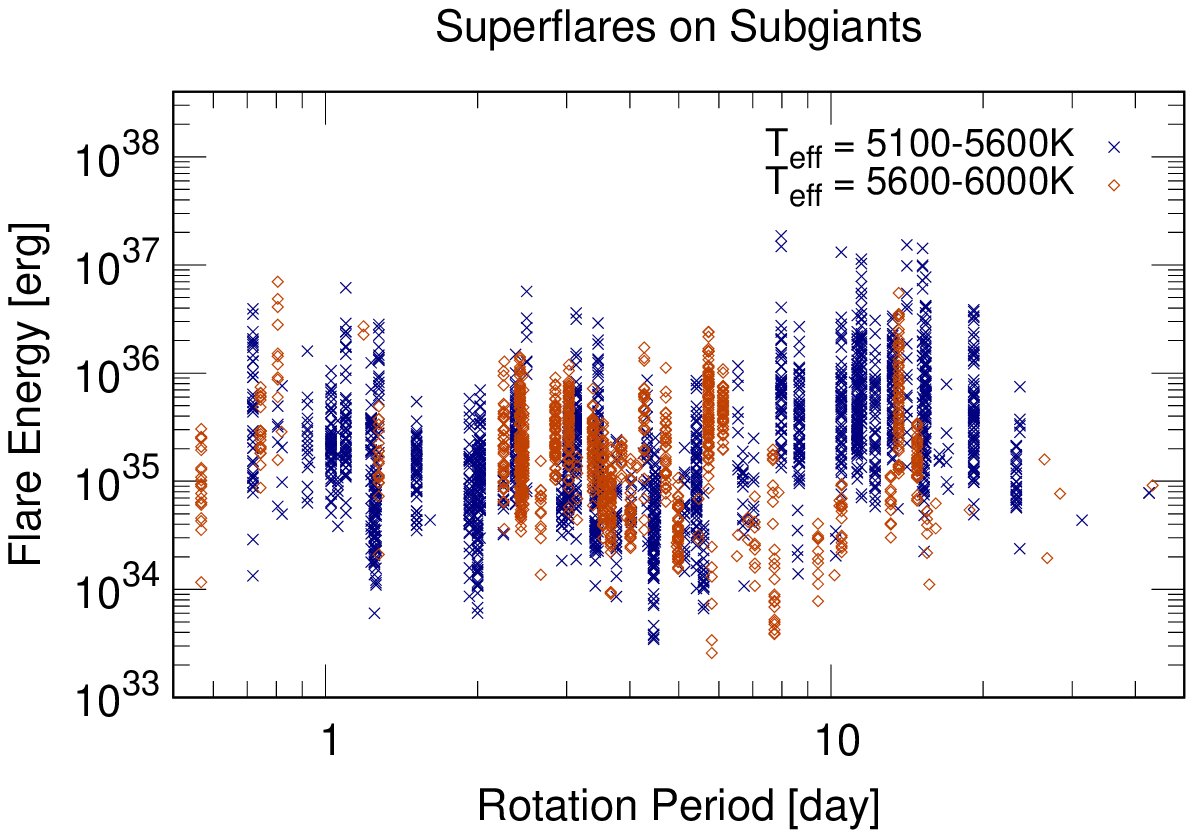}
   \vspace{0mm}
   \caption{
    Scatter plot of the superflare energy ($E_{\rm flare}$) vs. the rotation period ($P_{\rm rot}$).
    Basically the same as Figure \ref{fig:erg_prot_solartype} but for subgiant superflare stars with $T_{\rm eff}=5100$ -- $ 5600$ K (navy crosses) and $T_{\rm eff}=5600-6000$ K (dark orange diamonds).
   }
 \label{fig:erg_prot_subgiant}
 \end{figure}

 Figure \ref{fig:erg_prot_subgiant} shows the relationship between the superflare energy ($E_{\rm{flare}}$) and the rotation period ($P_{\rm{rot}}$) of subgiant superflare stars.
 Different from the results of solar-type stars in Figure \ref{fig:erg_prot_solartype}, the upper limit of $E_{\rm{flare}}$ values in a given period bin does not show a decreasing trend with the rotation period.
 Both rapidly-rotating ($P_{\rm{rot}}\sim 1$ days) and slowly-rotating ($P_{\rm{rot}}>10$ days) have superflares up to $\sim 10^{37}$erg.
 If some of these subgiants were contaminated with solar-type stars, the decreasing trend of solar-type stars in Figure \ref{fig:erg_prot_solartype} might be affected.
 This might be one reason why the decreasing trend was not reported in our initial study \citep{Notsu+2013}.

\begin{figure*}[ht!]
    \gridline{
    \fig{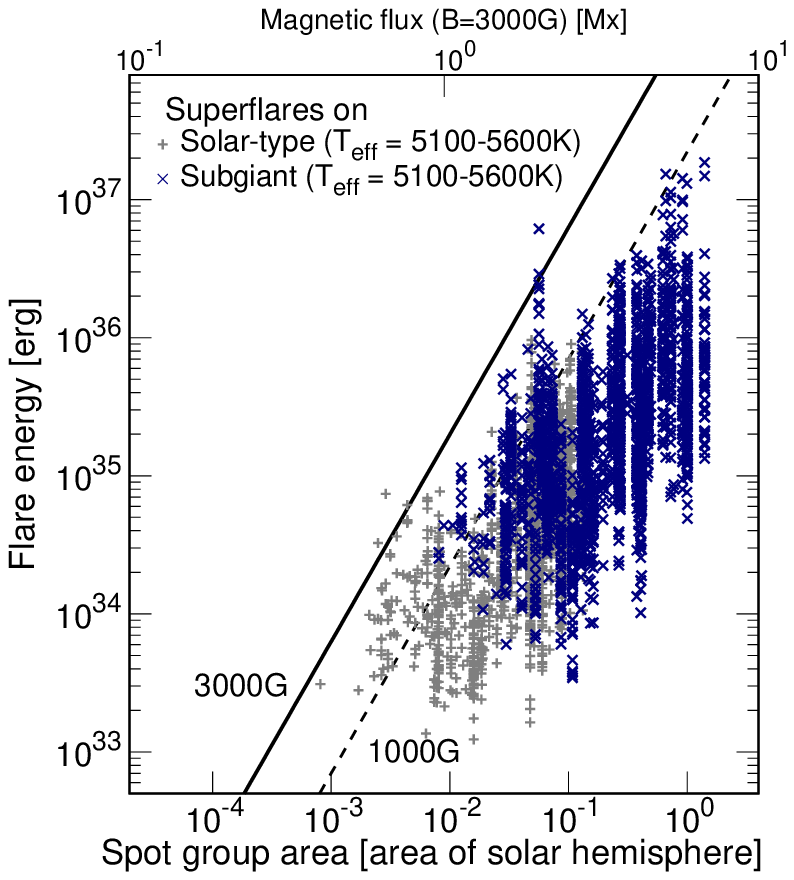}{0.48\textwidth}{\vspace{0mm} (a)}
    \fig{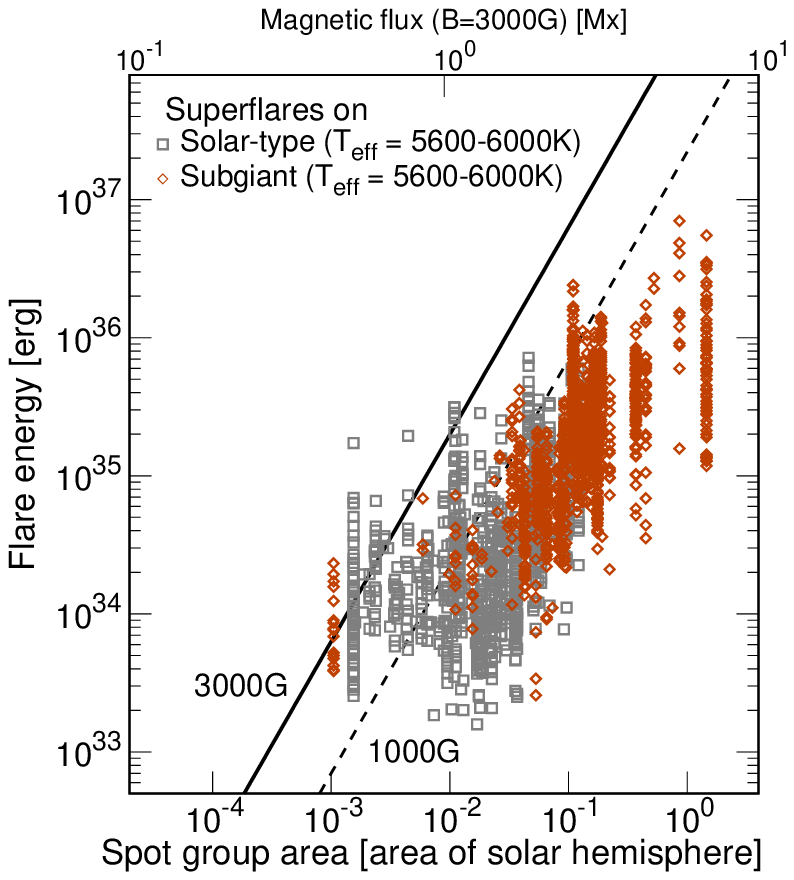}{0.48\textwidth}{\vspace{0mm} (b)}
    }
    \caption{
    Scatter plot of the superflare energy ($E_{\rm flare}$) and the spot group area ($A_{\rm spot}$) of superflares on solar-type stars and G-type subgiants.
    The vertical and horizontal axes, the black filled line, and the black dashed lines are the same as Figure \ref{fig:erg_aspt}.\\
    (a) Stars with $T_{\rm eff} = 5100$ -- $5600$ K.
    Gray plus marks are solar-type stars and navy cross marks are subgiants.
    (b) Stars with $T_{\rm eff} = 5600$ -- $6000$ K.
    Gray squares are solar-type stars and dark orange diamonds are subgiants.
    }
    \label{fig:aspt_erg_subgiant}
 \end{figure*}

 The upper limit of $E_{\rm{flare}}$ values of subgiant superflares in Figure \ref{fig:erg_prot_subgiant} are $\sim 10^{37}$erg, and this is about 10 times larger than the upper limit value of superflares on rapidly-rotating solar-type stars in Figure \ref{fig:erg_prot_solartype}.
 We then discuss this possible difference of the upper limit of $E_{\rm{flare}}$ values between solar-type superflare stars and subgiant superflare stars.
 Figure \ref{fig:aspt_erg_subgiant} shows the scatter plot of data of flare energy ($E_{\rm{flare}}$) as a function of spot group area ($A_{\rm{spot}}$) of solar flares and superflare stars.
 $A_{\rm{spot}}$ values of subgiants are estimated from the brightness variations amplitudes by using the same method as used for solar-type stars in Section \ref{subsec:energy_spot}.
 In Figure \ref{fig:aspt_erg_subgiant}, the data points of subgiant superflare stars tend to be in the right upper side of the panel, compared with those of solar-type superflare stars.
 This means that superflares on subgiants have larger flare energies and occur on stars having larger starspots.
 This is consistent with the idea that flare energy is explained by the magnetic energy stored around large starspots on the stellar surface estimated from Equation (\ref{eq:erg_aspt}).
 As a result, subgiant superflare stars tend to have large starspots and larger energies of superflares, compared with solar-type (main-sequence) stars.
 More detailed discussions on superflares on subgiant stars can be interesting related to dynamo mechanisms on subgiant stars (e.g., \citealt{Katsova+2018}; \citealt{Kovari+2020}), but this is beyond the scope of this paper and we expect future studies.

 \begin{deluxetable*}{lcccc}
   \tablecaption{
   The number of superflares ($N_{\rm{flare, subgiant}}$), subgiant superflare stars ($N_{\rm{flare, subgiant}}$), and all subgiant stars we analyzed ($N_{\rm{star, subgiant}}$).}
   \tablewidth{0pt}
   \tablehead{
     \colhead{} & \colhead{$N_{\rm flare,subgiant}$} & \colhead{$N_{\rm flarestar,subgiant}$} & \colhead{$N_{\rm star, subgiant}$}
   }
   \startdata
   (1) All $Kepler$ stars having $T_{\rm eff}$ and $R_{\rm Gaia}$ values in \citet{Baker+2004} & & & 177911 \\
   (2) All subgiant with $T_{\rm eff} = 5100$ -- $6000$ K & & & 36784\\
   (3) Subgiants having $P_{\rm rot}$ and $Amp$ values in \citet{McQuillan+2014} & 3780 & 168 & 3533 \\ 
   (4) Subgiants with $T_{\rm eff} = 5100$ -- $5600$ K among (3) & 1970 & 66 & 820 \\ 
   (5) Subgiants with $T_{\rm eff} = 5600$ -- $6000$ K among (3) & 1135& 54 & 1212 \\ 
   \enddata
   \label{table:num_flares_subgiant}
   \tablecomments{
     $T_{\rm eff}$ values and evolutionary state classifications (Main Sequence (MS)/Subgiants/Red giants/Cool MS binaries) in \citet{Berger+2018} (cf. Figure 5 of \citealt{Berger+2018}) are used for the classification in this table.
   }
 \end{deluxetable*}

 \clearpage
\bibliography{okamoto_2020}{}

\begin{thebibliography}{}
\expandafter\ifx\csname natexlab\endcsname\relax\def\natexlab#1{#1}\fi
\providecommand{\url}[1]{\href{#1}{#1}}
\providecommand{\dodoi}[1]{doi:~\href{http://doi.org/#1}{\nolinkurl{#1}}}
\providecommand{\doeprint}[1]{\href{http://ascl.net/#1}{\nolinkurl{http://ascl.net/#1}}}
\providecommand{\doarXiv}[1]{\href{https://arxiv.org/abs/#1}{\nolinkurl{https://arxiv.org/abs/#1}}}

\bibitem[{{Airapetian} {et~al.}(2016){Airapetian}, {Glocer}, {Gronoff},
  {H{\'e}brard}, \& {Danchi}}]{Airapetian+2016}
{Airapetian}, V.~S., {Glocer}, A., {Gronoff}, G., {H{\'e}brard}, E., \&
  {Danchi}, W. 2016, Nature Geoscience, 9, 452, \dodoi{10.1038/ngeo2719}

\bibitem[{{Airapetian} {et~al.}(2020){Airapetian}, {Barnes}, {Cohen},
  {Collinson}, {Danchi}, {Dong}, {Del Genio}, {France}, {Garcia-Sage},
  {Glocer}, {Gopalswamy}, {Grenfell}, {Gronoff}, {G{\"u}del}, {Herbst},
  {Henning}, {Jackman}, {Jin}, {Johnstone}, {Kaltenegger}, {Kay}, {Kobayashi},
  {Kuang}, {Li}, {Lynch}, {L{\"u}ftinger}, {Luhmann}, {Maehara}, {Mlynczak},
  {Notsu}, {Osten}, {Ramirez}, {Rugheimer}, {Scheucher}, {Schlieder},
  {Shibata}, {Sousa-Silva}, {Stamenkovi{\'c}}, {Strangeway}, {Usmanov},
  {Vergados}, {Verkhoglyadova}, {Vidotto}, {Voytek}, {Way}, {Zank}, \&
  {Yamashiki}}]{Airapetian+2020}
{Airapetian}, V.~S., {Barnes}, R., {Cohen}, O., {et~al.} 2020, International
  Journal of Astrobiology, 19, 136, \dodoi{10.1017/S1473550419000132}

\bibitem[{{Allen} {et~al.}(1989){Allen}, {Frank}, {Sauer}, \&
  {Reiff}}]{Allen+1989}
{Allen}, J., {Frank}, L., {Sauer}, H., \& {Reiff}, P. 1989, EOS Transactions,
  70, 1479, \dodoi{10.1029/89EO00409}

\bibitem[{{Aschwanden} {et~al.}(2000){Aschwanden}, {Tarbell}, {Nightingale},
  {Schrijver}, {Title}, {Kankelborg}, {Martens}, \& {Warren}}]{Aschwanden+2000}
{Aschwanden}, M.~J., {Tarbell}, T.~D., {Nightingale}, R.~W., {et~al.} 2000,
  \apj, 535, 1047, \dodoi{10.1086/308867}

\bibitem[{{Atri}(2017)}]{Atri+2017}
{Atri}, D. 2017, \mnras, 465, L34, \dodoi{10.1093/mnrasl/slw199}

\bibitem[{{Aulanier} {et~al.}(2013){Aulanier}, {D{\'e}moulin}, {Schrijver},
  {Janvier}, {Pariat}, \& {Schmieder}}]{Aulanier+2013}
{Aulanier}, G., {D{\'e}moulin}, P., {Schrijver}, C.~J., {et~al.} 2013, \aap,
  549, A66, \dodoi{10.1051/0004-6361/201220406}

\bibitem[{{Ayres}(1997)}]{Ayres+1997}
{Ayres}, T.~R. 1997, \jgr, 102, 1641, \dodoi{10.1029/96JE03306}

\bibitem[{{Baker}(2004)}]{Baker+2004}
{Baker}, D.~N. 2004, {Introduction to Space Weather}, Vol. 656 (Springer Berlin
  Heidelberg), 3, \dodoi{10.1007/978-3-540-31534-6_1}

\bibitem[{{Balona}(2015)}]{Balona+2015}
{Balona}, L.~A. 2015, \mnras, 447, 2714, \dodoi{10.1093/mnras/stu2651}

\bibitem[{{Battersby}(2019)}]{Battersby+2019}
{Battersby}, S. 2019, PNAS, 116, 23368, \dodoi{10.1073/pnas.1917356116}

\bibitem[{{Benz} \& {G{\"u}del}(2010)}]{Benz+2010}
{Benz}, A.~O., \& {G{\"u}del}, M. 2010, \araa, 48, 241,
  \dodoi{10.1146/annurev-astro-082708-101757}

\bibitem[{{Berdyugina}(2005)}]{Berdyugina+2005}
{Berdyugina}, S.~V. 2005, Living Reviews in Solar Physics, 2, 8,
  \dodoi{10.12942/lrsp-2005-8}

\bibitem[{{Berger} {et~al.}(2018){Berger}, {Huber}, {Gaidos}, \& {van
  Saders}}]{Berger+2018}
{Berger}, T.~A., {Huber}, D., {Gaidos}, E., \& {van Saders}, J.~L. 2018, \apj,
  866, 99, \dodoi{10.3847/1538-4357/aada83}

\bibitem[{{Brasseur} {et~al.}(2019){Brasseur}, {Osten}, \&
  {Fleming}}]{Brasseur+2019}
{Brasseur}, C.~E., {Osten}, R.~A., \& {Fleming}, S.~W. 2019, \apj, 883, 88,
  \dodoi{10.3847/1538-4357/ab3df8}

\bibitem[{{Brown} {et~al.}(2011){Brown}, {Latham}, {Everett}, \&
  {Esquerdo}}]{Brown+2011}
{Brown}, T.~M., {Latham}, D.~W., {Everett}, M.~E., \& {Esquerdo}, G.~A. 2011,
  \aj, 142, 112, \dodoi{10.1088/0004-6256/142/4/112}

\bibitem[{{Candelaresi} {et~al.}(2014){Candelaresi}, {Hillier}, {Maehara},
  {Brandenburg}, \& {Shibata}}]{Candelaresi+2014}
{Candelaresi}, S., {Hillier}, A., {Maehara}, H., {Brandenburg}, A., \&
  {Shibata}, K. 2014, \apj, 792, 67, \dodoi{10.1088/0004-637X/792/1/67}

\bibitem[{{Carrington}(1859)}]{Carrington+1859}
{Carrington}, R.~C. 1859, \mnras, 20, 13, \dodoi{10.1093/mnras/20.1.13}

\bibitem[{{Cliver} {et~al.}(2020){Cliver}, {Hayakawa}, {Love}, \&
  {Neidig}}]{Cliver+2020}
{Cliver}, E.~W., {Hayakawa}, H., {Love}, J.~J., \& {Neidig}, D.~F. 2020, \apj,
  in press

\bibitem[{{Crosby} {et~al.}(1993){Crosby}, {Aschwanden}, \&
  {Dennis}}]{Crosby+1993}
{Crosby}, N.~B., {Aschwanden}, M.~J., \& {Dennis}, B.~R. 1993, \solphys, 143,
  275, \dodoi{10.1007/BF00646488}

\bibitem[{{Crosley} \& {Osten}(2018)}]{Crosley+2018}
{Crosley}, M.~K., \& {Osten}, R.~A. 2018, \apj, 856, 39,
  \dodoi{10.3847/1538-4357/aaaec2}

\bibitem[{{Davenport}(2016)}]{Davenport+2016}
{Davenport}, J. R.~A. 2016, \apj, 829, 23, \dodoi{10.3847/0004-637X/829/1/23}

\bibitem[{{Davenport} {et~al.}(2020){Davenport}, {Mendoza}, \&
  {Hawley}}]{Davenport+2020}
{Davenport}, J. R.~A., {Mendoza}, G.~T., \& {Hawley}, S.~L. 2020, \aj, 160, 36,
  \dodoi{10.3847/1538-3881/ab9536}

\bibitem[{{Davenport} {et~al.}(2014){Davenport}, {Hawley}, {Hebb},
  {Wisniewski}, {Kowalski}, {Johnson}, {Malatesta}, {Peraza}, {Keil},
  {Silverberg}, {Jansen}, {Scheffler}, {Berdis}, {Larsen}, \&
  {Hilton}}]{Davenport+2014}
{Davenport}, J. R.~A., {Hawley}, S.~L., {Hebb}, L., {et~al.} 2014, \apj, 797,
  122, \dodoi{10.1088/0004-637X/797/2/122}

\bibitem[{{Doyle} {et~al.}(2020){Doyle}, {Ramsay}, \& {Doyle}}]{Doyle+2020}
{Doyle}, L., {Ramsay}, G., \& {Doyle}, J.~G. 2020, \mnras, 494, 3596,
  \dodoi{10.1093/mnras/staa923}

\bibitem[{{Doyle} {et~al.}(2018){Doyle}, {Ramsay}, {Doyle}, {Wu}, \&
  {Scullion}}]{Doyle+2018}
{Doyle}, L., {Ramsay}, G., {Doyle}, J.~G., {Wu}, K., \& {Scullion}, E. 2018,
  \mnras, 480, 2153, \dodoi{10.1093/mnras/sty1963}

\bibitem[{{Emslie} {et~al.}(2012){Emslie}, {Dennis}, {Shih}, {Chamberlin},
  {Mewaldt}, {Moore}, {Share}, {Vourlidas}, \& {Welsch}}]{Emslie+2012}
{Emslie}, A.~G., {Dennis}, B.~R., {Shih}, A.~Y., {et~al.} 2012, \apj, 759, 71,
  \dodoi{10.1088/0004-637X/759/1/71}

\bibitem[{{Feinstein} {et~al.}(2020){Feinstein}, {Montet}, {Ansdell}, {Nord},
  {Bean}, {G{\"u}nther}, {Gully-Santiago}, \& {Schlieder}}]{Feinstein+2020}
{Feinstein}, A.~D., {Montet}, B.~T., {Ansdell}, M., {et~al.} 2020, arXiv
  e-prints, arXiv:2005.07710.
\newblock \doarXiv{2005.07710}

\bibitem[{{Gao} {et~al.}(2016){Gao}, {Xin}, {Liu}, {Zhang}, \&
  {Gao}}]{Gao+2016}
{Gao}, Q., {Xin}, Y., {Liu}, J.-F., {Zhang}, X.-B., \& {Gao}, S. 2016, \apjs,
  224, 37, \dodoi{10.3847/0067-0049/224/2/37}

\bibitem[{{Gershberg}(2005)}]{Gershberg+2005}
{Gershberg}, R.~E. 2005, {Solar-Type Activity in Main-Sequence Stars}
  (Springer-Verlag Berlin Heidelberg), \dodoi{10.1007/3-540-28243-2}

\bibitem[{{Giles} {et~al.}(2017){Giles}, {Collier Cameron}, \&
  {Haywood}}]{Giles+2017}
{Giles}, H. A.~C., {Collier Cameron}, A., \& {Haywood}, R.~D. 2017, \mnras,
  472, 1618

\bibitem[{{G{\"u}del}(2007)}]{Gudel+2007}
{G{\"u}del}, M. 2007, Living Reviews in Solar Physics, 4, 3,
  \dodoi{10.12942/lrsp-2007-3}

\bibitem[{{G{\"u}nther} {et~al.}(2020){G{\"u}nther}, {Zhan}, {Seager},
  {Rimmer}, {Ranjan}, {Stassun}, {Oelkers}, {Daylan}, {Newton}, {Kristiansen},
  {Olah}, {Gillen}, {Rappaport}, {Ricker}, {Vanderspek}, {Latham}, {Winn},
  {Jenkins}, {Glidden}, {Fausnaugh}, {Levine}, {Dittmann}, {Quinn},
  {Krishnamurthy}, \& {Ting}}]{Gunther+2020}
{G{\"u}nther}, M.~N., {Zhan}, Z., {Seager}, S., {et~al.} 2020, \aj, 159, 60,
  \dodoi{10.3847/1538-3881/ab5d3a}

\bibitem[{{Hawley} \& {Fisher}(1992)}]{Hawley+1992}
{Hawley}, S.~L., \& {Fisher}, G.~H. 1992, \apjs, 78, 565,
  \dodoi{10.1086/191640}

\bibitem[{{Hayakawa} {et~al.}(2017{\natexlab{a}}){Hayakawa}, {Tamazawa},
  {Uchiyama}, {Ebihara}, {Miyahara}, {Kosaka}, {Iwahashi}, \&
  {Isobe}}]{Hayakawa+2017b}
{Hayakawa}, H., {Tamazawa}, H., {Uchiyama}, Y., {et~al.} 2017{\natexlab{a}},
  \solphys, 292, 12, \dodoi{10.1007/s11207-016-1039-2}

\bibitem[{{Hayakawa} {et~al.}(2017{\natexlab{b}}){Hayakawa}, {Iwahashi},
  {Ebihara}, {Tamazawa}, {Shibata}, {Knipp}, {Kawamura}, {Hattori}, {Mase},
  {Nakanishi}, \& {Isobe}}]{Hayakawa+2017a}
{Hayakawa}, H., {Iwahashi}, K., {Ebihara}, Y., {et~al.} 2017{\natexlab{b}},
  \apjl, 850, L31, \dodoi{10.3847/2041-8213/aa9661}

\bibitem[{{Hayakawa} {et~al.}(2019){Hayakawa}, {Ebihara}, {Willis}, {Toriumi},
  {Iju}, {Hattori}, {Wild}, {Oliveira}, {Ermolli}, {Ribeiro}, {Correia},
  {Ribeiro}, \& {Knipp}}]{Hayakawa+2019}
{Hayakawa}, H., {Ebihara}, Y., {Willis}, D.~M., {et~al.} 2019, Space Weather,
  17, 1553, \dodoi{10.1029/2019SW002269}

\bibitem[{{Heinzel} \& {Shibata}(2018)}]{Heinzel+2018}
{Heinzel}, P., \& {Shibata}, K. 2018, \apj, 859, 143,
  \dodoi{10.3847/1538-4357/aabe78}

\bibitem[{{Herbst} {et~al.}(2019){Herbst}, {Grenfell}, {Sinnhuber}, {Rauer},
  {Heber}, {Banjac}, {Scheucher}, {Schmidt}, {Gebauer}, {Lehmann}, \&
  {Schreier}}]{Herbst+2019}
{Herbst}, K., {Grenfell}, J.~L., {Sinnhuber}, M., {et~al.} 2019, \aap, 631,
  A101, \dodoi{10.1051/0004-6361/201935888}

\bibitem[{{Honda} {et~al.}(2018){Honda}, {Notsu}, {Namekata}, {Notsu},
  {Maehara}, {Ikuta}, {Nogami}, \& {Shibata}}]{Honda+2018}
{Honda}, S., {Notsu}, Y., {Namekata}, K., {et~al.} 2018, \pasj, 70, 62,
  \dodoi{10.1093/pasj/psy055}

\bibitem[{{Houdebine} {et~al.}(1993){Houdebine}, {Foing}, {Doyle}, \&
  {Rodono}}]{Houdebine+1993}
{Houdebine}, E.~R., {Foing}, B.~H., {Doyle}, J.~G., \& {Rodono}, M. 1993, \aap,
  274, 245

\bibitem[{{Ikuta} {et~al.}(2020){Ikuta}, {Maehara}, {Notsu}, {Namekata},
  {Kato}, {Notsu}, {Okamoto}, {Honda}, {Nogami}, \& {Shibata}}]{Ikuta+2020}
{Ikuta}, K., {Maehara}, H., {Notsu}, Y., {et~al.} 2020, \apj, 902, 73,
  \dodoi{10.3847/1538-4357/abae5f}

\bibitem[{{Jackman} {et~al.}(2018){Jackman}, {Wheatley}, {Pugh},
  {G{\"a}nsicke}, {Gillen}, {Broomhall}, {Armstrong}, {Burleigh}, {Chaushev},
  {Eigm{\"u}ller}, {Erikson}, {Goad}, {Grange}, {G{\"u}nther}, {Jenkins},
  {McCormac}, {Raynard}, {Thompson}, {Udry}, {Walker}, {Watson}, \&
  {West}}]{Jackman+2018}
{Jackman}, J. A.~G., {Wheatley}, P.~J., {Pugh}, C.~E., {et~al.} 2018, \mnras,
  477, 4655, \dodoi{10.1093/mnras/sty897}

\bibitem[{{Karoff} {et~al.}(2016){Karoff}, {Knudsen}, {De Cat}, {Bonanno},
  {Fogtmann-Schulz}, {Fu}, {Frasca}, {Inceoglu}, {Olsen}, {Zhang}, {Hou},
  {Wang}, {Shi}, \& {Zhang}}]{Karoff+2016}
{Karoff}, C., {Knudsen}, M.~F., {De Cat}, P., {et~al.} 2016, Nature
  Communications, 7, 11058, \dodoi{10.1038/ncomms11058}

\bibitem[{{Katsova} {et~al.}(2018){Katsova}, {Kitchatinov}, {Moss}, {Ol{\'a}h},
  \& {Sokoloff}}]{Katsova+2018}
{Katsova}, M.~M., {Kitchatinov}, L.~L., {Moss}, D., {Ol{\'a}h}, K., \&
  {Sokoloff}, D.~D. 2018, Astronomy Reports, 62, 513,
  \dodoi{10.1134/S1063772918080036}

\bibitem[{{Katsova} \& {Livshits}(2015)}]{Katsova+2015}
{Katsova}, M.~M., \& {Livshits}, M.~A. 2015, \solphys, 290, 3663,
  \dodoi{10.1007/s11207-015-0752-6}

\bibitem[{{Kay} {et~al.}(2019){Kay}, {Airapetian}, {L{\"u}ftinger}, \&
  {Kochukhov}}]{Kay+2019}
{Kay}, C., {Airapetian}, V.~S., {L{\"u}ftinger}, T., \& {Kochukhov}, O. 2019,
  \apjl, 886, L37, \dodoi{10.3847/2041-8213/ab551f}

\bibitem[{{K{\H{o}}v{\'a}ri} {et~al.}(2020){K{\H{o}}v{\'a}ri}, {Ol{\'a}h},
  {G{\"u}nther}, {Vida}, {Kriskovics}, {Seli}, {Bakos}, {Hartman}, {Csubry}, \&
  {Bhatti}}]{Kovari+2020}
{K{\H{o}}v{\'a}ri}, Z., {Ol{\'a}h}, K., {G{\"u}nther}, M.~N., {et~al.} 2020,
  \aap, 641, A83, \dodoi{10.1051/0004-6361/202038397}

\bibitem[{{Koch} {et~al.}(2010){Koch}, {Borucki}, {Basri}, {Batalha}, {Brown},
  {Caldwell}, {Christensen-Dalsgaard}, {Cochran}, {DeVore}, {Dunham},
  {Gautier}, {Geary}, {Gilliland}, {Gould}, {Jenkins}, {Kondo}, {Latham},
  {Lissauer}, {Marcy}, {Monet}, {Sasselov}, {Boss}, {Brownlee}, {Caldwell},
  {Dupree}, {Howell}, {Kjeldsen}, {Meibom}, {Morrison}, {Owen}, {Reitsema},
  {Tarter}, {Bryson}, {Dotson}, {Gazis}, {Haas}, {Kolodziejczak}, {Rowe}, {Van
  Cleve}, {Allen}, {Chand rasekaran}, {Clarke}, {Li}, {Quintana}, {Tenenbaum},
  {Twicken}, \& {Wu}}]{Koch+2010}
{Koch}, D.~G., {Borucki}, W.~J., {Basri}, G., {et~al.} 2010, \apjl, 713, L79,
  \dodoi{10.1088/2041-8205/713/2/L79}

\bibitem[{{Kowalski} \& {Allred}(2018)}]{Kowalski+2018}
{Kowalski}, A.~F., \& {Allred}, J.~C. 2018, \apj, 852, 61,
  \dodoi{10.3847/1538-4357/aa9d91}

\bibitem[{{Kowalski} {et~al.}(2010){Kowalski}, {Hawley}, {Holtzman},
  {Wisniewski}, \& {Hilton}}]{Kowalski+2010}
{Kowalski}, A.~F., {Hawley}, S.~L., {Holtzman}, J.~A., {Wisniewski}, J.~P., \&
  {Hilton}, E.~J. 2010, \apjl, 714, L98, \dodoi{10.1088/2041-8205/714/1/L98}

\bibitem[{{Kurita} {et~al.}(2020){Kurita}, {Kino}, {Iwamuro}, {Ohta}, {Nogami},
  {Izumiura}, {Yoshida}, {Matsubayashi}, {Kuroda}, {Nakatani}, {Yamamoto},
  {Tsutsui}, {Iribe}, {Jikuya}, {Ohtani}, {Shibata}, {Takahashi}, {Tokoro},
  {Maihara}, \& {Nagata}}]{Kurita+2020}
{Kurita}, M., {Kino}, M., {Iwamuro}, F., {et~al.} 2020, \pasj, 72, 48,
  \dodoi{10.1093/pasj/psaa036}

\bibitem[{{Leitzinger} {et~al.}(2020){Leitzinger}, {Odert}, {Greimel}, {Vida},
  {Kriskovics}, {Guenther}, {Korhonen}, {Koller}, {Hanslmeier},
  {K{\H{o}}v{\'a}ri}, \& {Lammer}}]{Leitzinger+2020}
{Leitzinger}, M., {Odert}, P., {Greimel}, R., {et~al.} 2020, \mnras, 493, 4570,
  \dodoi{10.1093/mnras/staa504}

\bibitem[{{Lingam} \& {Loeb}(2017)}]{Lingam+2017}
{Lingam}, M., \& {Loeb}, A. 2017, \apj, 848, 41,
  \dodoi{10.3847/1538-4357/aa8e96}

\bibitem[{{Linsky}(2019)}]{Linsky2019}
{Linsky}, J. 2019, {Host Stars and their Effects on Exoplanet Atmospheres},
  Vol. 955 (Springer International Publishing),
  \dodoi{10.1007/978-3-030-11452-7}

\bibitem[{{Loomis}(1861)}]{Loomis+1861}
{Loomis}, E. 1861, American Journal of Science, 32, 318,
  \dodoi{10.2475/ajs.s2-32.96.318}

\bibitem[{{Lynch} {et~al.}(2019){Lynch}, {Airapetian}, {DeVore}, {Kazachenko},
  {L{\"u}ftinger}, {Kochukhov}, {Ros{\'e}n}, \& {Abbett}}]{Lynch+2019}
{Lynch}, B.~J., {Airapetian}, V.~S., {DeVore}, C.~R., {et~al.} 2019, \apj, 880,
  97, \dodoi{10.3847/1538-4357/ab287e}

\bibitem[{{Maehara} {et~al.}(2017){Maehara}, {Notsu}, {Notsu}, {Namekata},
  {Honda}, {Ishii}, {Nogami}, \& {Shibata}}]{Maehara+2017}
{Maehara}, H., {Notsu}, Y., {Notsu}, S., {et~al.} 2017, \pasj, 69, 41,
  \dodoi{10.1093/pasj/psx013}

\bibitem[{{Maehara} {et~al.}(2015){Maehara}, {Shibayama}, {Notsu}, {Notsu},
  {Honda}, {Nogami}, \& {Shibata}}]{Maehara+2015}
{Maehara}, H., {Shibayama}, T., {Notsu}, Y., {et~al.} 2015, Earth, Planets, and
  Space, 67, 59, \dodoi{10.1186/s40623-015-0217-z}

\bibitem[{{Maehara} {et~al.}(2012){Maehara}, {Shibayama}, {Notsu}, {Notsu},
  {Nagao}, {Kusaba}, {Honda}, {Nogami}, \& {Shibata}}]{Maehara+2012}
{Maehara}, H., {Shibayama}, T., {Notsu}, S., {et~al.} 2012, \nat, 485, 478,
  \dodoi{10.1038/nature11063}

\bibitem[{{Mamajek} \& {Hillenbrand}(2008)}]{Mamajek+2008}
{Mamajek}, E.~E., \& {Hillenbrand}, L.~A. 2008, \apj, 687, 1264,
  \dodoi{10.1086/591785}

\bibitem[{{Mathur} {et~al.}(2017){Mathur}, {Huber}, {Batalha}, {Ciardi},
  {Bastien}, {Bieryla}, {Buchhave}, {Cochran}, {Endl}, {Esquerdo}, {Furlan},
  {Howard}, {Howell}, {Isaacson}, {Latham}, {MacQueen}, \&
  {Silva}}]{Mathur+2017}
{Mathur}, S., {Huber}, D., {Batalha}, N.~M., {et~al.} 2017, \apjs, 229, 30,
  \dodoi{10.3847/1538-4365/229/2/30}

\bibitem[{{McQuillan} {et~al.}(2014){McQuillan}, {Mazeh}, \&
  {Aigrain}}]{McQuillan+2014}
{McQuillan}, A., {Mazeh}, T., \& {Aigrain}, S. 2014, \apjs, 211, 24,
  \dodoi{10.1088/0067-0049/211/2/24}

\bibitem[{{Mekhaldi} {et~al.}(2015){Mekhaldi}, {Muscheler}, {Adolphi},
  {Aldahan}, {Beer}, {McConnell}, {Possnert}, {Sigl}, {Svensson}, {Synal},
  {Welten}, \& {Woodruff}}]{Mekhaldi+2015}
{Mekhaldi}, F., {Muscheler}, R., {Adolphi}, F., {et~al.} 2015, Nature
  Communications, 6, 8611, \dodoi{10.1038/ncomms9611}

\bibitem[{{Metcalfe} \& {Egeland}(2019)}]{Metcalfe+2019}
{Metcalfe}, T.~S., \& {Egeland}, R. 2019, \apj, 871, 39,
  \dodoi{10.3847/1538-4357/aaf575}

\bibitem[{{Miyake} {et~al.}(2013){Miyake}, {Masuda}, \&
  {Nakamura}}]{Miyake+2013}
{Miyake}, F., {Masuda}, K., \& {Nakamura}, T. 2013, Nature Communications, 4,
  1748, \dodoi{10.1038/ncomms2783}

\bibitem[{{Miyake} {et~al.}(2012){Miyake}, {Nagaya}, {Masuda}, \&
  {Nakamura}}]{Miyake+2012}
{Miyake}, F., {Nagaya}, K., {Masuda}, K., \& {Nakamura}, T. 2012, \nat, 486,
  240, \dodoi{10.1038/nature11123}

\bibitem[{{Miyake} {et~al.}(2019){Miyake}, {Usoskin}, \&
  {Poluianov}}]{Miyake+2019}
{Miyake}, F., {Usoskin}, I., \& {Poluianov}, S. 2019, {Extreme Solar Particle
  Storms; The hostile Sun} (IOP Publishing), \dodoi{10.1088/2514-3433/ab404a}

\bibitem[{{Mochnacki} \& {Zirin}(1980)}]{Mochnacki+1980}
{Mochnacki}, S.~W., \& {Zirin}, H. 1980, \apjl, 239, L27,
  \dodoi{10.1086/183285}

\bibitem[{{Morris} {et~al.}(2018){Morris}, {Curtis}, {Douglas}, {Hawley},
  {Ag{\"u}eros}, {Bobra}, \& {Agol}}]{Morris+2018}
{Morris}, B.~M., {Curtis}, J.~L., {Douglas}, S.~T., {et~al.} 2018, \aj, 156,
  203, \dodoi{10.3847/1538-3881/aae1ab}

\bibitem[{{Moschou} {et~al.}(2019){Moschou}, {Drake}, {Cohen},
  {Alvarado-G{\'o}mez}, {Garraffo}, \& {Fraschetti}}]{Moschou+2019}
{Moschou}, S.-P., {Drake}, J.~J., {Cohen}, O., {et~al.} 2019, \apj, 877, 105,
  \dodoi{10.3847/1538-4357/ab1b37}

\bibitem[{{Namekata} {et~al.}(2017){Namekata}, {Sakaue}, {Watanabe}, {Asai},
  {Maehara}, {Notsu}, {Notsu}, {Honda}, {Ishii}, {Ikuta}, {Nogami}, \&
  {Shibata}}]{Namekata+2017}
{Namekata}, K., {Sakaue}, T., {Watanabe}, K., {et~al.} 2017, \apj, 851, 91,
  \dodoi{10.3847/1538-4357/aa9b34}

\bibitem[{{Namekata} {et~al.}(2019){Namekata}, {Maehara}, {Notsu}, {Toriumi},
  {Hayakawa}, {Ikuta}, {Notsu}, {Honda}, {Nogami}, \&
  {Shibata}}]{Namekata+2019}
{Namekata}, K., {Maehara}, H., {Notsu}, Y., {et~al.} 2019, \apj, 871, 187,
  \dodoi{10.3847/1538-4357/aaf471}

\bibitem[{{Namekata} {et~al.}(2020a){Namekata}, {Davenport}, {Morris},
  {Hawley}, {Maehara}, {Notsu}, {Toriumi}, {Ikuta}, {Notsu}, {Honda}, {Nogami},
  \& {Shibata}}]{Namekata+2020_ApJ}
{Namekata}, K., {Davenport}, J. R.~A., {Morris}, B.~M., {et~al.} 2020a, \apj,
  891, 103, \dodoi{10.3847/1538-4357/ab7384}

\bibitem[{{Namekata} {et~al.}(2020b){Namekata}, {Maehara}, {Sasaki}, {Kawai},
  {Notsu}, {Kowalski}, {Allred}, {Iwakiri}, {Tsuboi}, {Murata}, {Niwano},
  {Shiraishi}, {Adachi}, {Iida}, {Oeda}, {Honda}, {Tozuka}, {Katoh}, {Onozato},
  {Okamoto}, {Isogai}, {Kimura}, {Kojiguchi}, {Wakamatsu}, {Tampo}, {Nogami},
  \& {Shibata}}]{Namekata+2020_PASJ}
{Namekata}, K., {Maehara}, H., {Sasaki}, R., {et~al.} 2020b, \pasj,
  \dodoi{10.1093/pasj/psaa051}

\bibitem[{{Nizamov}(2019)}]{Nizamov+2019}
{Nizamov}, B.~A. 2019, \mnras, 489, 4338, \dodoi{10.1093/mnras/stz2478}

\bibitem[{{Nogami} {et~al.}(2014){Nogami}, {Notsu}, {Honda}, {Maehara},
  {Notsu}, {Shibayama}, \& {Shibata}}]{Nogami+2014}
{Nogami}, D., {Notsu}, Y., {Honda}, S., {et~al.} 2014, \pasj, 66, L4,
  \dodoi{10.1093/pasj/psu012}

\bibitem[{{Notsu} {et~al.}(2013){Notsu}, {Honda}, {Notsu}, {Nagao},
  {Shibayama}, {Maehara}, {Nogami}, \& {Shibata}}]{SNotsu+2013}
{Notsu}, S., {Honda}, S., {Notsu}, Y., {et~al.} 2013, \pasj, 65, 112,
  \dodoi{10.1093/pasj/65.5.112}

\bibitem[{{Notsu} {et~al.}(2015{\natexlab{a}}){Notsu}, {Honda}, {Maehara},
  {Notsu}, {Shibayama}, {Nogami}, \& {Shibata}}]{Notsu+2015a}
{Notsu}, Y., {Honda}, S., {Maehara}, H., {et~al.} 2015{\natexlab{a}}, \pasj,
  67, 32, \dodoi{10.1093/pasj/psv001}

\bibitem[{{Notsu} {et~al.}(2015{\natexlab{b}}){Notsu}, {Honda}, {Maehara},
  {Notsu}, {Shibayama}, {Nogami}, \& {Shibata}}]{Notsu+2015b}
---. 2015{\natexlab{b}}, \pasj, 67, 33, \dodoi{10.1093/pasj/psv002}

\bibitem[{{Notsu} {et~al.}(2013b){Notsu}, {Shibayama}, {Maehara}, {Notsu},
  {Nagao}, {Honda}, {Ishii}, {Nogami}, \& {Shibata}}]{Notsu+2013}
{Notsu}, Y., {Shibayama}, T., {Maehara}, H., {et~al.} 2013b, \apj, 771, 127,
  \dodoi{10.1088/0004-637X/771/2/127}

\bibitem[{{Notsu} {et~al.}(2019){Notsu}, {Maehara}, {Honda}, {Hawley},
  {Davenport}, {Namekata}, {Notsu}, {Ikuta}, {Nogami}, \&
  {Shibata}}]{Notsu+2019}
{Notsu}, Y., {Maehara}, H., {Honda}, S., {et~al.} 2019, \apj, 876, 58,
  \dodoi{10.3847/1538-4357/ab14e6}

\bibitem[{{Noyes} {et~al.}(1984){Noyes}, {Weiss}, \& {Vaughan}}]{Noyes+1984}
{Noyes}, R.~W., {Weiss}, N.~O., \& {Vaughan}, A.~H. 1984, \apj, 287, 769,
  \dodoi{10.1086/162735}

\bibitem[{{O'Hare} {et~al.}(2019){O'Hare}, {Mekhaldi}, {Adolphi}, {Raisbeck},
  {Aldahan}, {Anderberg}, {Beer}, {Christl}, {Fahrni}, {Synal}, {Park},
  {Possnert}, {Southon}, {Bard}, {Aster Team}, \& {Muscheler}}]{OHare+2019}
{O'Hare}, P., {Mekhaldi}, F., {Adolphi}, F., {et~al.} 2019, Proceedings of the
  National Academy of Science, 116, 5961, \dodoi{10.1073/pnas.1815725116}

\bibitem[{{Osten} {et~al.}(2016){Osten}, {Kowalski}, {Drake}, {Krimm}, {Page},
  {Gazeas}, {Kennea}, {Oates}, {Page}, {de Miguel}, {Nov{\'a}k}, {Apeltauer},
  \& {Gehrels}}]{Osten+2016}
{Osten}, R.~A., {Kowalski}, A., {Drake}, S.~A., {et~al.} 2016, \apj, 832, 174,
  \dodoi{10.3847/0004-637X/832/2/174}

\bibitem[{{Pinsonneault} {et~al.}(2012){Pinsonneault}, {An},
  {Molenda-{\.Z}akowicz}, {Chaplin}, {Metcalfe}, \&
  {Bruntt}}]{Pinsonneault+2012}
{Pinsonneault}, M.~H., {An}, D., {Molenda-{\.Z}akowicz}, J., {et~al.} 2012,
  \apjs, 199, 30, \dodoi{10.1088/0067-0049/199/2/30}

\bibitem[{{Pye} {et~al.}(2015){Pye}, {Rosen}, {Fyfe}, \&
  {Schr{\"o}der}}]{Pye+2015}
{Pye}, J.~P., {Rosen}, S., {Fyfe}, D., \& {Schr{\"o}der}, A.~C. 2015, \aap,
  581, A28, \dodoi{10.1051/0004-6361/201526217}

\bibitem[{{Rauer} {et~al.}(2014){Rauer}, {Catala}, {Aerts}, {Appourchaux},
  {Benz}, {Brandeker}, {Christensen-Dalsgaard}, {Deleuil}, {Gizon}, {Goupil},
  {G{\"u}del}, {Janot-Pacheco}, {Mas-Hesse}, {Pagano}, {Piotto}, {Pollacco},
  {Santos}, {Smith}, {Su{\'a}rez}, {Szab{\'o}}, {Udry}, {Adibekyan}, {Alibert},
  {Almenara}, {Amaro-Seoane}, {Eiff}, {Asplund}, {Antonello}, {Barnes},
  {Baudin}, {Belkacem}, {Bergemann}, {Bihain}, {Birch}, {Bonfils}, {Boisse},
  {Bonomo}, {Borsa}, {Brand {\~a}o}, {Brocato}, {Brun}, {Burleigh}, {Burston},
  {Cabrera}, {Cassisi}, {Chaplin}, {Charpinet}, {Chiappini}, {Church},
  {Csizmadia}, {Cunha}, {Damasso}, {Davies}, {Deeg}, {D{\'\i}az}, {Dreizler},
  {Dreyer}, {Eggenberger}, {Ehrenreich}, {Eigm{\"u}ller}, {Erikson}, {Farmer},
  {Feltzing}, {de Oliveira Fialho}, {Figueira}, {Forveille}, {Fridlund},
  {Garc{\'\i}a}, {Giommi}, {Giuffrida}, {Godolt}, {Gomes da Silva}, {Granzer},
  {Grenfell}, {Grotsch-Noels}, {G{\"u}nther}, {Haswell}, {Hatzes},
  {H{\'e}brard}, {Hekker}, {Helled}, {Heng}, {Jenkins}, {Johansen},
  {Khodachenko}, {Kislyakova}, {Kley}, {Kolb}, {Krivova}, {Kupka}, {Lammer},
  {Lanza}, {Lebreton}, {Magrin}, {Marcos-Arenal}, {Marrese}, {Marques},
  {Martins}, {Mathis}, {Mathur}, {Messina}, {Miglio}, {Montalban}, {Montalto},
  {Monteiro}, {Moradi}, {Moravveji}, {Mordasini}, {Morel}, {Mortier},
  {Nascimbeni}, {Nelson}, {Nielsen}, {Noack}, {Norton}, {Ofir}, {Oshagh},
  {Ouazzani}, {P{\'a}pics}, {Parro}, {Petit}, {Plez}, {Poretti}, {Quirrenbach},
  {Ragazzoni}, {Raimondo}, {Rainer}, {Reese}, {Redmer}, {Reffert},
  {Rojas-Ayala}, {Roxburgh}, {Salmon}, {Santerne}, {Schneider}, {Schou},
  {Schuh}, {Schunker}, {Silva-Valio}, {Silvotti}, {Skillen}, {Snellen}, {Sohl},
  {Sousa}, {Sozzetti}, {Stello}, {Strassmeier}, {{\v{S}}vanda}, {Szab{\'o}},
  {Tkachenko}, {Valencia}, {Van Grootel}, {Vauclair}, {Ventura}, {Wagner},
  {Walton}, {Weingrill}, {Werner}, {Wheatley}, \& {Zwintz}}]{Rauer+2014}
{Rauer}, H., {Catala}, C., {Aerts}, C., {et~al.} 2014, Experimental Astronomy,
  38, 249, \dodoi{10.1007/s10686-014-9383-4}

\bibitem[{{Reid} \& {Hawley}(2005)}]{Reid+2005}
{Reid}, I.~N., \& {Hawley}, S.~L. 2005, {New light on dark stars : red dwarfs,
  low-mass stars, brown dwarfs} (Springer-Verlag Berlin Heidelberg),
  \dodoi{10.1007/3-540-27610-6}

\bibitem[{{Reinhold} {et~al.}(2020){Reinhold}, {Shapiro}, {Solanki}, {Montet},
  {Krivova}, {Cameron}, \& {Amazo-G{\'o}mez}}]{Reinhold+2020}
{Reinhold}, T., {Shapiro}, A.~I., {Solanki}, S.~K., {et~al.} 2020, Science,
  368, 518, \dodoi{10.1126/science.aay3821}

\bibitem[{{Ricker} {et~al.}(2015){Ricker}, {Winn}, {Vanderspek}, {Latham},
  {Bakos}, {Bean}, {Berta-Thompson}, {Brown}, {Buchhave}, {Butler}, {Butler},
  {Chaplin}, {Charbonneau}, {Christensen-Dalsgaard}, {Clampin}, {Deming},
  {Doty}, {De Lee}, {Dressing}, {Dunham}, {Endl}, {Fressin}, {Ge}, {Henning},
  {Holman}, {Howard}, {Ida}, {Jenkins}, {Jernigan}, {Johnson}, {Kaltenegger},
  {Kawai}, {Kjeldsen}, {Laughlin}, {Levine}, {Lin}, {Lissauer}, {MacQueen},
  {Marcy}, {McCullough}, {Morton}, {Narita}, {Paegert}, {Palle}, {Pepe},
  {Pepper}, {Quirrenbach}, {Rinehart}, {Sasselov}, {Sato}, {Seager},
  {Sozzetti}, {Stassun}, {Sullivan}, {Szentgyorgyi}, {Torres}, {Udry}, \&
  {Villasenor}}]{Ricker+2015}
{Ricker}, G.~R., {Winn}, J.~N., {Vanderspek}, R., {et~al.} 2015, Journal of
  Astronomical Telescopes, Instruments, and Systems, 1, 014003,
  \dodoi{10.1117/1.JATIS.1.1.014003}

\bibitem[{{Riley} {et~al.}(2018){Riley}, {Baker}, {Liu}, {Verronen}, {Singer},
  \& {G{\"u}del}}]{Riley+2018}
{Riley}, P., {Baker}, D., {Liu}, Y.~D., {et~al.} 2018, \ssr, 214, 21,
  \dodoi{10.1007/s11214-017-0456-3}

\bibitem[{{Roettenbacher} \& {Vida}(2018)}]{Roettenbacher+2018}
{Roettenbacher}, R.~M., \& {Vida}, K. 2018, \apj, 868, 3,
  \dodoi{10.3847/1538-4357/aae77e}

\bibitem[{{Rubenstein} \& {Schaefer}(2000)}]{Rubenstein+2000}
{Rubenstein}, E.~P., \& {Schaefer}, B.~E. 2000, \apj, 529, 1031,
  \dodoi{10.1086/308326}

\bibitem[{{Sammis} {et~al.}(2000){Sammis}, {Tang}, \& {Zirin}}]{Sammis+2000}
{Sammis}, I., {Tang}, F., \& {Zirin}, H. 2000, \apj, 540, 583,
  \dodoi{10.1086/309303}

\bibitem[{{Schaefer} {et~al.}(2000){Schaefer}, {King}, \&
  {Deliyannis}}]{Schaefer+2000}
{Schaefer}, B.~E., {King}, J.~R., \& {Deliyannis}, C.~P. 2000, \apj, 529, 1026,
  \dodoi{10.1086/308325}

\bibitem[{{Schmieder}(2018)}]{Schmieder+2018}
{Schmieder}, B. 2018, Journal of Atmospheric and Solar-Terrestrial Physics,
  180, 46, \dodoi{10.1016/j.jastp.2017.07.018}

\bibitem[{{Schrijver} {et~al.}(2012){Schrijver}, {Beer}, {Baltensperger},
  {Cliver}, {G{\"u}del}, {Hudson}, {McCracken}, {Osten}, {Peter}, {Soderblom},
  {Usoskin}, \& {Wolff}}]{Schrijver+2012}
{Schrijver}, C.~J., {Beer}, J., {Baltensperger}, U., {et~al.} 2012, Journal of
  Geophysical Research (Space Physics), 117, A08103,
  \dodoi{10.1029/2012JA017706}

\bibitem[{{Segura} {et~al.}(2010){Segura}, {Walkowicz}, {Meadows}, {Kasting},
  \& {Hawley}}]{Segura+2010}
{Segura}, A., {Walkowicz}, L.~M., {Meadows}, V., {Kasting}, J., \& {Hawley}, S.
  2010, Astrobiology, 10, 751, \dodoi{10.1089/ast.2009.0376}

\bibitem[{{Shibata} \& {Magara}(2011)}]{Shibata+2011}
{Shibata}, K., \& {Magara}, T. 2011, Living Reviews in Solar Physics, 8, 6,
  \dodoi{10.12942/lrsp-2011-6}

\bibitem[{{Shibata} \& {Yokoyama}(2002)}]{Shibata+Yokoyama+2002}
{Shibata}, K., \& {Yokoyama}, T. 2002, \apj, 577, 422, \dodoi{10.1086/342141}

\bibitem[{{Shibata} {et~al.}(2013){Shibata}, {Isobe}, {Hillier}, {Choudhuri},
  {Maehara}, {Ishii}, {Shibayama}, {Notsu}, {Notsu}, {Nagao}, {Honda}, \&
  {Nogami}}]{Shibata+2013}
{Shibata}, K., {Isobe}, H., {Hillier}, A., {et~al.} 2013, \pasj, 65, 49,
  \dodoi{10.1093/pasj/65.3.49}

\bibitem[{{Shibayama} {et~al.}(2013){Shibayama}, {Maehara}, {Notsu}, {Notsu},
  {Nagao}, {Honda}, {Ishii}, {Nogami}, \& {Shibata}}]{Shibayama+2013}
{Shibayama}, T., {Maehara}, H., {Notsu}, S., {et~al.} 2013, \apjs, 209, 5,
  \dodoi{10.1088/0067-0049/209/1/5}

\bibitem[{{Shimizu}(1995)}]{Shimizu+1995}
{Shimizu}, T. 1995, \pasj, 47, 251

\bibitem[{{Soderblom} {et~al.}(1993){Soderblom}, {Stauffer}, {MacGregor}, \&
  {Jones}}]{Soderblom+1993}
{Soderblom}, D.~R., {Stauffer}, J.~R., {MacGregor}, K.~B., \& {Jones}, B.~F.
  1993, \apj, 409, 624, \dodoi{10.1086/172694}

\bibitem[{{Takahashi} {et~al.}(2016){Takahashi}, {Mizuno}, \&
  {Shibata}}]{Takahashi+2016}
{Takahashi}, T., {Mizuno}, Y., \& {Shibata}, K. 2016, \apjl, 833, L8,
  \dodoi{10.3847/2041-8205/833/1/L8}

\bibitem[{{Takasao} {et~al.}(2020){Takasao}, {Mitsuishi}, {Shimura}, {Yoshida},
  {Kunitomo}, {Tanaka}, \& {Ishihara}}]{Takasao+2020}
{Takasao}, S., {Mitsuishi}, I., {Shimura}, T., {et~al.} 2020, \apj, 901, 70,
  \dodoi{10.3847/1538-4357/abad34}

\bibitem[{{Thompson} {et~al.}(2016){Thompson}, {Caldwell}, {Jenkins},
  {Barclay}, {Barentsen}, {Bryson}, {Burke}, {Campbell}, {Catanzarite},
  {Christiansen}, {et~al.}}]{Thompson+2016}
{Thompson}, S.~E., {Caldwell}, D.~A., {Jenkins}, J.~M., {et~al.} 2016, Kepler
  Data Release 25 Notes (KSCI-19065-002), NASA Ames Research Center, Moffett
  Field, CA

\bibitem[{{Toriumi} \& {Wang}(2019)}]{Toriumi+2019}
{Toriumi}, S., \& {Wang}, H. 2019, Living Reviews in Solar Physics, 16, 3,
  \dodoi{10.1007/s41116-019-0019-7}

\bibitem[{{Tsurutani} {et~al.}(2003){Tsurutani}, {Gonzalez}, {Lakhina}, \&
  {Alex}}]{Tsurutani+2003}
{Tsurutani}, B.~T., {Gonzalez}, W.~D., {Lakhina}, G.~S., \& {Alex}, S. 2003,
  Journal of Geophysical Research (Space Physics), 108, 1268,
  \dodoi{10.1029/2002JA009504}

\bibitem[{{Tu} {et~al.}(2015){Tu}, {Johnstone}, {G{\"u}del}, \&
  {Lammer}}]{Tu+2015}
{Tu}, L., {Johnstone}, C.~P., {G{\"u}del}, M., \& {Lammer}, H. 2015, \aap, 577,
  L3, \dodoi{10.1051/0004-6361/201526146}

\bibitem[{{Tu} {et~al.}(2020){Tu}, {Yang}, {Zhang}, \& {Wang}}]{Tu+2020}
{Tu}, Z.-L., {Yang}, M., {Zhang}, Z.~J., \& {Wang}, F.~Y. 2020, \apj, 890, 46,
  \dodoi{10.3847/1538-4357/ab6606}

\bibitem[{{Usoskin}(2017)}]{Usoskin+2017}
{Usoskin}, I.~G. 2017, Living Reviews in Solar Physics, 14, 3,
  \dodoi{10.1007/s41116-017-0006-9}

\bibitem[{{Valenti} \& {Fischer}(2005)}]{Valenti+2005}
{Valenti}, J.~A., \& {Fischer}, D.~A. 2005, \apjs, 159, 141,
  \dodoi{10.1086/430500}

\bibitem[{{Van Cleve} \& {Caldwell}(2016)}]{vanCleve+2016}
{Van Cleve}, J.~E., \& {Caldwell}, D.~A. 2016, {Kepler Instrument Handbook},
  Kepler Science Document KSCI-19033-002

\bibitem[{{Van Doorsselaere} {et~al.}(2017){Van Doorsselaere}, {Shariati}, \&
  {Debosscher}}]{Vandoorsselaere+2017}
{Van Doorsselaere}, T., {Shariati}, H., \& {Debosscher}, J. 2017, \apjs, 232,
  26, \dodoi{10.3847/1538-4365/aa8f9a}

\bibitem[{{van Saders} {et~al.}(2016){van Saders}, {Ceillier}, {Metcalfe},
  {Silva Aguirre}, {Pinsonneault}, {Garc{\'\i}a}, {Mathur}, \&
  {Davies}}]{vanSaders+2016}
{van Saders}, J.~L., {Ceillier}, T., {Metcalfe}, T.~S., {et~al.} 2016, \nat,
  529, 181, \dodoi{10.1038/nature16168}

\bibitem[{{Veronig} {et~al.}(2002){Veronig}, {Temmer}, {Hanslmeier}, {Otruba},
  \& {Messerotti}}]{Veronig+2002}
{Veronig}, A., {Temmer}, M., {Hanslmeier}, A., {Otruba}, W., \& {Messerotti},
  M. 2002, \aap, 382, 1070, \dodoi{10.1051/0004-6361:20011694}

\bibitem[{{Vida} {et~al.}(2019){Vida}, {Leitzinger}, {Kriskovics}, {Seli},
  {Odert}, {Kov{\'a}cs}, {Korhonen}, \& {van Driel-Gesztelyi}}]{Vida+2019}
{Vida}, K., {Leitzinger}, M., {Kriskovics}, L., {et~al.} 2019, \aap, 623, A49,
  \dodoi{10.1051/0004-6361/201834264}

\bibitem[{{Wright} {et~al.}(2011){Wright}, {Drake}, {Mamajek}, \&
  {Henry}}]{Wright+2011}
{Wright}, N.~J., {Drake}, J.~J., {Mamajek}, E.~E., \& {Henry}, G.~W. 2011,
  \apj, 743, 48, \dodoi{10.1088/0004-637X/743/1/48}

\bibitem[{{Wu} {et~al.}(2015){Wu}, {Ip}, \& {Huang}}]{Wu+2015}
{Wu}, C.-J., {Ip}, W.-H., \& {Huang}, L.-C. 2015, \apj, 798, 92,
  \dodoi{10.1088/0004-637X/798/2/92}

\bibitem[{{Yamashiki} {et~al.}(2019){Yamashiki}, {Maehara}, {Airapetian},
  {Notsu}, {Sato}, {Notsu}, {Kuroki}, {Murashima}, {Sato}, {Namekata},
  {Sasaki}, {Scott}, {Bando}, {Nashimoto}, {Takagi}, {Ling}, {Nogami}, \&
  {Shibata}}]{Yamashiki+2019}
{Yamashiki}, Y.~A., {Maehara}, H., {Airapetian}, V., {et~al.} 2019, \apj, 881,
  114, \dodoi{10.3847/1538-4357/ab2a71}

\bibitem[{{Yang} {et~al.}(2017){Yang}, {Liu}, {Gao}, {Fang}, {Guo}, {Zhang},
  {Hou}, {Wang}, \& {Cao}}]{Yang+2017}
{Yang}, H., {Liu}, J., {Gao}, Q., {et~al.} 2017, \apj, 849, 36,
  \dodoi{10.3847/1538-4357/aa8ea2}

\end{thebibliography}
\bibliographystyle{aasjournal}

\end{document}